# A dual approach to proving electoral fraud using statistics and forensic evidence

(«Двойное доказательство фальсификаций на выборах статистикой и криминалистикой», полный перевод статьи на русский приложен ниже — full translation into Russian is appended)

Andrey Podlazov[1] and Vadim Makarov[2,3,*]

[1] *RAS Keldysh Institute of Applied Mathematics, Moscow 125047, Russia*
[2] *voting member of Territorial election commission of Vlasikha town, Vlasikha, Moscow region 143010, Russia*
[3] *Vigo Quantum Communication Center, University of Vigo, Vigo E-36310, Spain*

Electoral fraud often manifests itself as statistical anomalies in election results, yet its extent can rarely be reliably confirmed by other evidence. Here we report the complete results of municipal elections in the town of Vlasikha near Moscow, where we observe both statistical irregularities in the vote-counting transcripts and forensic evidence of tampering with ballots during their overnight storage. We evaluate two types of statistical signatures in the vote sequence that can prove batches of fraudulent ballots have been injected. We find that pairs of factory-made security bags with identical serial numbers are used in this fraud scheme. At 8 out of our 9 polling stations, the statistical and forensic evidence agrees (identifying 7 as fraudulent and 1 as honest), while at the remaining station the statistical evidence detects the fraud while the forensic one is insufficient. We also illustrate that the use of tamper-indicating seals at elections is inherently unreliable.

## I. INTRODUCTION

Inclusive social institutions lead to the prosperity of a nation in the long term [1]. In their absence, it is crucial for the society to be informed of its current shortcomings, which is achieved via independent analysis and publicity. A fair election system is one of these institutes. Electoral fraud produces many types of statistical irregularities in election results that reveal it. This has become obvious with the advent of information technologies that allow easy access to and analysis of returns from individual polling stations [2–6]. Fraudulent actions themselves can occasionally be caught but are rarely documented at a scale and coverage comparable with that the statistics predicts, especially that this generally requires cooperation from the authorities [7, 8]. Here we report municipal elections where both statistical analysis and recorded evidence of fraud (or lack thereof) is available for every polling station. They largely corroborate one another, although multiple statistical tests may be required to fully reveal the fraud when counterfeiters try to disguise it.

Statistical methods are based on clearly defined assumptions and thus allow independent verification. Analysing polling station returns allows to detect anomalies in result–turnout distribution [2, 4, 9–17], unwarranted exhaustion of blank ballots [18], dependence of the values in election protocols on the presence of an observer [19–21] or ballot scanner [22–24]. This yields a qualitative estimate of the scale of fraud. More rigorous statistical methods allow to calculate the probability of the anomaly occurring by chance, such as prevalence of integer percentage values in the returns [3, 25–28]. This quantifies its cogency as evidence of fraud.

Our study introduces rigorous statistical analysis of the results of a physical manipulation of ballots, such as their stuffing and swapping. We calculate the probability they form long series of identical votes when they are being tallied and probability of an insufficient number of toggles observed. We analyse vote-counting sequences recorded by observers.

## II. EXPERIMENT

The elections of the town council in Vlasikha took place on 6–8 September 2024. They yielded unexpected data that prompted an otherwise unplanned study.

Vlasikha is a compact (less than 3 km across), restricted-access community that serves the headquarters and main control center of the Russian intercontinental ballistic missile (ICBM) force [29]. The voters are a mix of civilians and military personnel.

The town, populated by 17189 registered voters (of which 35% voted), is divided into 3 electoral districts of about equal size, each represented by 5 seats in the council of 15 total seats. In each district, $c = 11$ individual candidates ran and each voter could select up to five of them in the ballot. The same number of candidates in each district is a coincidence. Although independent candidates were allowed, each candidate here was nominated by a party. The 5 candidates with the largest number of votes in the district got the seats.

Each electoral district is divided into 3 physical polling stations (precincts) and also one virtual e-voting precinct where any voters preferring to cast their vote online could register in advance. The voting in all the precincts pro-ceeded over 3 consecutive days. At the polling stations, each voter used a voting booth to mark his or her paper ballot in secrecy, had an opportunity to fold it to conceal the choices, and cast the ballot into a large transparent box standing in public view. At the end of the first and second day, all ballots cast in that day were extracted from the box, sealed inside a security tamper-evident bag (whose unique number was logged publicly) and stored in a safe in the room of the polling station until the vote counting. All the ballots were counted at the end of the third day.

* makarov@vad1.com

TABLE I. Summary by precinct. $n_1$ ($n_2$), number of voters who obtained ballots in the room during the first (second) day; $n$, total number of valid ballots announced during the vote counting; CP, Communist party; JR, A Just Russia party; LD, Liberal-democratic party; UR, United Russia party. Suspiciously high powers of significances are shown in bold (Sec. B 1).

| Electoral district | $c$ | Precinct number | First day | | Second day | | $n$ | Intentional mixing of ballots? | $\mathrm{p}\alpha'$ | $\mathrm{p}\tilde{\alpha}'$ | Majority in this precinct | Got seats in this district |
|---|---|---|---|---|---|---|---|---|---|---|---|---|
| | | | $n_1$ | Evidence of tampering? | $n_2$ | Evidence of tampering? | | | | | | |
| 1 | 11 | 212 | 157 | positive | 117 | positive | 437 | yes | **4.0** | **4.3** | 5 UR | 5 UR |
| | | 213 | 104 | no record | 68 | positive | 256 | no record | **2.3** | **4.5** | 5 UR | |
| | | 214 | 181 | inconclusive[a,b] | 104 | positive | 388 | no | **9.8** | **18.4** | 5 UR | |
| | | e-voting | | | | | 878 | | | | 2 CP, 3 UR | |
| 2 | 11 | 215 | 183 | negative[c] | 135 | negative[c] | 473 | no | 0.8 | **2.2** | 4 CP, 1 UR | 1 LD, 4 UR |
| | | 216 | 95 | positive | 86 | positive | 318 | yes | **3.6** | **21.0** | 1 LD, 4 UR | |
| | | 217 | 108 | positive | 69 | positive | 290 | yes | 0.4 | **2.6** | 1 LD, 4 UR | |
| | | e-voting | | | | | 726 | | | | 2 CP, 3 UR | |
| 3 | 11 | 218 | 435 | positive | 109 | positive | 577 | no | **10.9** | **9.3** | 1 JR, 4 UR | 1 JR, 4 UR |
| | | 219 | 131 | positive | 93 | inconclusive[a] | 355 | no | **18.7** | **20.2** | 1 JR, 4 UR | |
| | | 220 | 129 | inconclusive[a,b] | 84 | inconclusive[a] | 346 | no | **20.3** | **30.1** | 1 JR, 4 UR | |
| | | e-voting | | | | | 844 | | | | 2 CP, 3 UR | |

[a] No visible changes in geometry of the security tape, with too poor image quality to locate its number.
[b] Vote counting transcript supports a bag swap.
[c] Recently procured bags with short 8-digit numbers are used, for which we have not observed duplicates.

Election officials manned the polling stations during the day. Candidates and observers were also present almost continuously, watched the procedures, made video recordings and took photos. By law, no one was allowed to remain at the polling station at night except police guards stationed outside the sealed room.

Unlike earlier Russian elections [21, 30], the presence of observers no longer impeded the fraud. Analysis of photographs and video recordings shows evidence of tampering with the security bags in 7 out of 9 precincts (Table I). Thus, the ballots inside the bags were likely replaced at night.

During the vote-counting (tallying) procedure, the security bags were opened and their content stacked up with the ballots cast into the box during the third day. A varying amount of mixing between ballots from different days took place during the stacking-up, but was never a complete mixing. The ballots were then displayed and read aloud sequentially, and the votes counted manually, of which we have produced transcripts. These transcripts are amenable to statistical analysis and show a strong evidence that the alleged replacement of ballots during the night has drastically changed the outcome of elections. While local opposition candidates nominated by the Communist party (CP) likely won the majority of the seats, they were all replaced with pro-administration candidates from other parties via the fraud [31].

## III. FORENSIC EVIDENCE

Tamper-indicating seals are known since the antiquity and are ubiquitous [32]. They are unsuitable for securing modern elections [33, 34]. In order for such security devices to be efficient, all the interested participants must be formally trained and adhere to a detailed use protocol for the seal [35, 36]. This is unrealistic at elections where the sides involved—candidates, observers, press, and precinct election officials—essentially consist of members of the general public. They attend the elections infrequently and have no time for such specific training. However, motivated fraudsters have months to prepare and can be well-funded.

There are at least 5 methods for compromising the security bag (Sec. C). Some of them can be done locally and some require the procurement of a factory-made duplicate.

According to our reconstruction of the crime, the precinct commissions in Vlasikha had access to pairs of security bags with identical individual numbers (which were supposed to be unique but were actually made in two copies at the factory). At precinct 219, we have a conclusive photographic evidence of such pair of bags (Sec. D 1). At most other precincts, only tamper-evident sealing tapes were replaced, presumably to avoid forg-ing pen signatures left on the bags by observers, candidates, and election officials. At night, the perpetrators entered the room without the observers or candidates being present (of which evidence is available at 3 precincts) and opened the safe (leaving visible traces on the safe at 2 precincts). They peeled off the original 30-mm-wide security tape from the bag, thus unsealing it, and allegedly swapped the ballots. Tamper-evident adhesive remaining from the tape's removal was cleaned from the bag using solvent, as detailed in Sec. C. Another security tape with identical number, cut off from the unused duplicate bag (losing 5 to 10 mm of its width in the process), was then applied to reseal the bag.

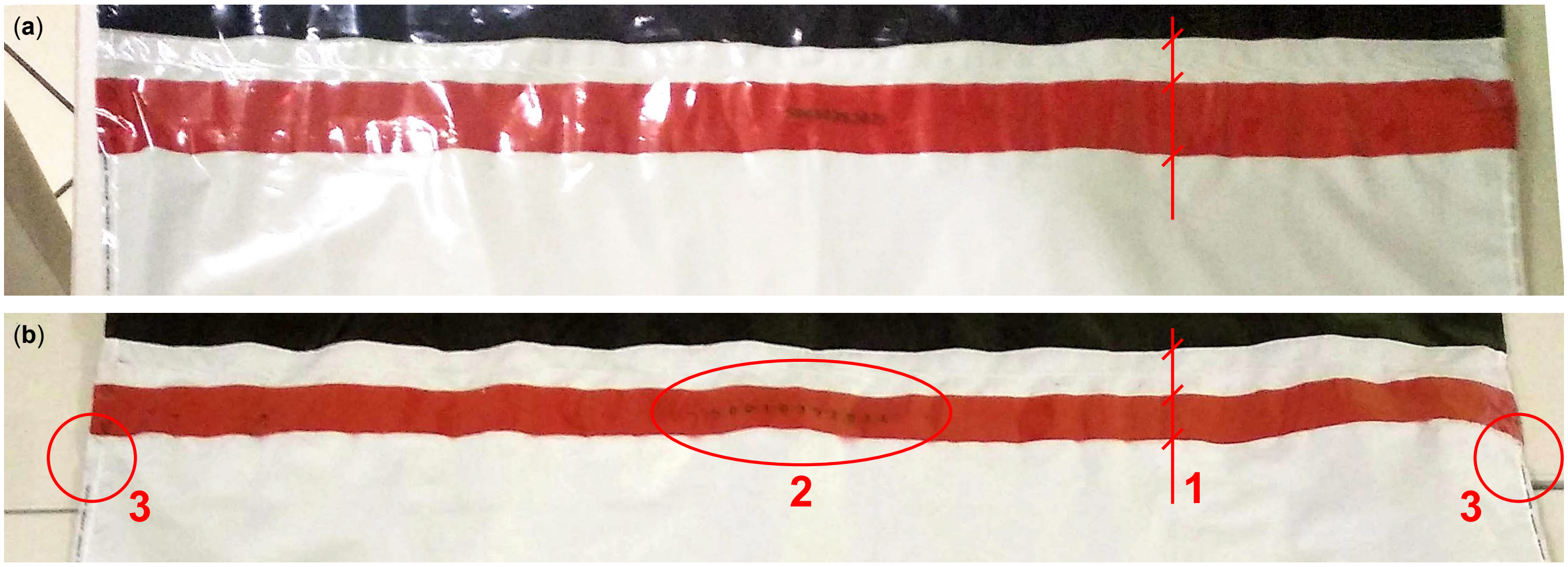

FIG. 1. Red sealing tape on the security bag at precinct 212 (a) after sealing the bag on the second day and (b) before opening it during the vote tallying on the third day. Note (1) changed tape width and placement relative to the upper edge, (2) increased intersymbol space and overall width of the unique number, changing from normally-typeset to sparsely-typeset, and (3) disappeared factory-printed line of text at the bag's edges, presumably removed with solvent.

The fraud succeeded in the sense that the bag numbers matched those logged, while the visual appearance of the tapes and bags raised suspicion from no one during the hectic voting days. The election outcome was certified immediately [31]. However, despite these tamper-evident bags being only a modest security measure [37] not designed to resist the availability of a factory-made duplicate with identical number [38], they eventually worked as intended. A meticulous examination and cross-check of video recordings and photographs taken during the voting days shows the sealing tape and the area around it looking differently at 7 precincts (see example in Fig. 1), is inconclusive at precinct 220, and indicates the lack of tampering at the remaining precinct 215. Remarkably, at precinct 219 we see that not only the tape but the entire bag was swapped for its full factory-made duplicate—complete with its identical individual number—and pen signatures were forged. This was discovered and understood slowly, weeks after the elections. In total, at least 11 bags containing 1479 ballots were tampered with (see Table I and Sec. D).

## IV. STATISTICAL ANALYSIS

The transcript of vote counting for precinct 215 (Fig. 2 left pane) has a random distribution of votes throughout, without any visible features. Transcripts for all the other precincts display irregularities, of which a typical example is precinct 220 shown in Fig. 2 right pane. A random vote sequence (with a clear prevalence of votes for 5 opposition candidates nominated by CP) is interrupted by three packs of identical ballots for 5 pro-administration candidates (1 JR's nominee and 4 UR's nominees). Incidentally, there are 129 ballots in total for these 5 candidates, while the bag from the first day also contains 129 ballots. This raises a suspicion that nobody in reality voted for this particular set of candidates, and the long sequences of identical ballots came from that bag.

Since almost all the other transcripts (shown in Sec. A) display similar visual patterns, we have designed two simple statistical tests.

In the first test, we calculate significances $\alpha_1$ and $\alpha_0$. These are the probabilities of the longest observed continuous series of votes and absences of votes to occur naturally for a candidate at the precinct. Such extremely long series are assumed to result from stuffing batches of identical ballots.

In the second test, we calculate a significance $\tilde{\alpha}$. It is the probability that the number of continuous series arising naturally does not exceed their observed number. A long series can be broken by partial mixing of ballots prior to their counting. However even in this case, the transcript contains fewer toggles between series of votes and absences of votes than statistically expected, which this test should detect.

For these calculations, we make two assumptions. Firstly, we conservatively assume that all the ballots are genuine. This allows us to calculate the probability for a voter to cast or not cast a vote for the candidate. Secondly, we assume that successive ballots are mutually independent. Even if small groups of like-minded voters (such as families and friends) come to vote at the same time, their ballots scatter randomly when dropped into the box and then again when extracted from the box. This should help maintain the independence of ballots in the vote-counting sequence.

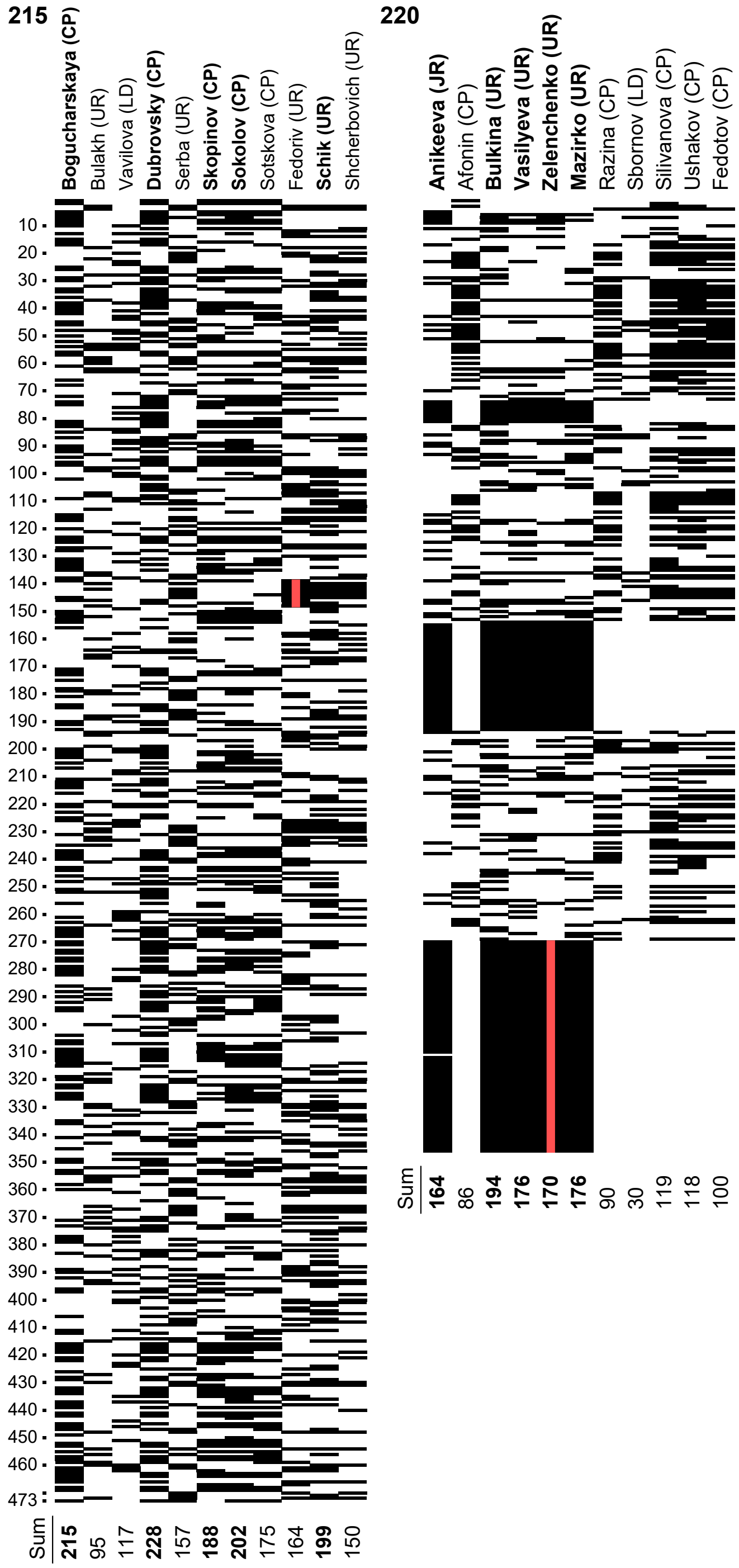

FIG. 2. Transcripts of vote counting for 2 precincts: 215, where we have not observed any evidence of fraud, and 220, where swapping the bag containing 129 ballots is alleged. Top 5 candidates at each precinct are shown in bold. The least probable series is denoted by a red (grey) line.

These significances are computed by an iterative calculation, as detailed in Sec. B, which also lists their values for every precinct and candidate. With these values, we proceed to a summary analysis for the precincts. If any of the probabilities $\alpha_0$, $\alpha_1$, or $\tilde{\alpha}$ turns out to be small, this indicates ballot stuffing. The evidence of such electoral fraud can be provided by the smallest significances obtained for the candidates running in the precinct $\alpha_{\min} = \min_c \{\alpha_0, \alpha_1\}$ and $\tilde{\alpha}_{\min} = \min_c \tilde{\alpha}$. However, in order to correctly characterise the general state of affairs at the precinct, one should take into account multiplicity testing and introduce an appropriate correction. The point is that the more attempts are made to calculate some random variable, the higher the natural chances that it will at least once take on a very small value.

When $c$ candidates run in the precinct, the significance of the hypothesis that the least probable number of series arose naturally doesn't exceed $\tilde{\alpha}' = 1 - (1 - \tilde{\alpha}_{\min})^c \approx c\tilde{\alpha}_{\min}$. Using the approximation here is appropriate if $\tilde{\alpha}_{\min} < 10^{-8}$, to avoid loss of accuracy in computations. Similarly, the significance of the hypothesis that the least probable continuous series arose naturally doesn't exceed $\alpha' = 1 - (1 - \alpha_{\min})^{2c} \approx 2c\alpha_{\min}$. The factor of 2 here accounts for each candidate being counted twice for two types of series (votes and absences of votes).

We ignore the possibility here of obtaining low significances multiple times at the precinct. This eliminates the need to analyze the similarities and differences between candidates. Such simplification may overestimate the overall significances $\alpha'$ and $\tilde{\alpha}'$. This weakens our test, which is acceptable. So we may risk missing violations, but we do not risk making unfounded accusations.

When the fraud is present, the significances tend to have very small values. It is more convenient to list not the values $\alpha$, but their powers $\mathrm{p}\alpha = -\lg \alpha$ (similar to pH in chemistry). Remember that the probability of no fraud drops by an order of magnitude for each increase of the power value by 1. The powers of the overall significances are listed in Table I.

For 5 precincts, the probabilities are extremely low, less than $10^{-9}$ (i.e., $\max\{\mathrm{p}\alpha', \mathrm{p}\tilde{\alpha}'\} > 9$). This certainly indicates the presence of fraud, as the chance the precinct is honest is less than one-in-a-milliard. Higher probabilities in the remaining 4 precincts can either be explained by mixing the ballots before they are counted (such that fraudulent long sequences are broken into sufficiently many fragments), the presence of intentionally introduced random ballots in the original fraudulent sequence, or the precinct being free from this type of fraud.

After extracting the ballots from the bags and boxes on the third day, the commissions checked them to eliminate invalid ballots (those with no marks or more than 5 marks) before the vote counting. To speed up this step, the check was done by several commission members simultaneously, each working a fraction of the ballots. When they added the checked ballots into the final stack, it was done in chunks. This step alone usually introduced some shuffling. In addition, some commissions intentionally mixed the ballots by spreading them into a single heap on a table, stirring the heap and randomly turning over handfuls of ballots (which was recorded in our videos and noted in Table I). We think that moderately low probabilities at precincts 212 and 213, and high at 217 are due to such intentional mixing. The high but still suspicious probability $\tilde{\alpha}' = 6.2 \times 10^{-3}$ at precinct 215 might be owing to one of our assumptions being not entirely correct, e.g., maybe the ballots are not entirely mutually independent. More data from honest precincts should clarify this in future studies.

Evidence of statistical nature with quite high probability of a mistake is admissible in courts [39, 40]. Notably, Russian courts accept a genetic paternity test having the error probability of no more than $10^{-3}$, provided that the mother is known for sure (otherwise, the threshold for the error probability is relaxed to $2.5 \times 10^{-3}$) [41]. Using this approach, electoral fraud should be considered established at all precincts except 215 and 217. However, we advocate a stricter threshold. As discussed above, the fraud is definitely detected at precincts 214, 216, 218, 219, and 220, where $\min\{\alpha', \tilde{\alpha}'\} < 10^{-9}$.

While it may appear that precinct 217 evades fraud detection, our test can be improved to consider ballots with marks for all 5 top candidates. This test reveals the fraud there with $\tilde{\alpha} = 5.9 \times 10^{-15}$ and gives non-suspicious values for the honest precinct 215 (see Sec. B 3).

Overall, after our careful analysis of statistics and transcripts (Sec. A), we know the nocturnal fate of all but one security bag of the 18 used in these elections. In addition to the 11 bags visibly tampered with (Table I and Sec. D), 2 more were replaced, as continuous sequences of identical ballots show: the first bag at precincts 214 and 220. 4 bags remained uncompromised: both bags at precinct 215 and the second bag at precincts 219 and 220. We are only not sure about the first bag at precinct 213.

Unlike the physical polling stations, e-voting precincts allow little insight into their operation. They lack data on the vote counting process and have no evidence of tampering with ballots (Sec. E).

## V. DISCUSSION AND CONCLUSION

We have introduced rigorous statistical tests of the vote-counting sequence that do not make any non-trivial assumptions about the voters' characteristics. These tests detect the ballot-replacement fraud in most cases, which is corroborated by separate forensic evidence.

Results from the 3 e-voting precincts, one honest precinct, and our analysis of the eight fraudulent precincts reveal tightly contested elections, with pro-administration and opposition candidates running closely. The latter de facto won about half or possibly more than half the council seats. The genuine voter turnout is 35%, which is unusually high for Russian municipal elections and shows the people's interest in its outcome. Via the ballot swap, an unlikely combination of pro-administration candidates have been installed in the council and the local opposition lost all the seats it previously held.

We remark that a less rigorous analysis [42] that assumes typical voter's characteristics allows to reconstruct the real results. It predicts a council with 9 CPRF-nominated and 6 UR-nominated members, in agreement with the present study.

Such a flagrant fraud is only possible if its perpetrators enjoy a total legal impunity. Unfortunately, our experience so far shows this is the case. Multiple complaints to the election commissions of all levels were dismissed. Crime reports to the local police and the Investigative committee of the Russian Federation citing the evidence presented in Secs. C and D have been dismissed as not worthy investigation. The police guarding the polling stations at night has reported nothing. Courts dismissed four initial lawsuits on a technicality. The head of the local police, head of the state prosecutor's local office, deputy political officer of the ICBM force, and town major all attended the first meeting of the newly elected council and congratulated its members with a "convincing victory" [43], despite the evidence of fraud covered in the local press and multiple crime reports under consideration by the police. Physical evidence—used security bags—seized by the police after V.M.'s crime report at precinct 217 were later destroyed "as garbage" without examination. By law, every polling station has round-the-clock indoor video surveillance filming the room and the safe. These recordings were stored at the Territorial election commission of Vlasikha town. All requests to view them had been declined and the recordings were eventually destroyed.

A lawsuit asking to declare the elections void and detailing all the available evidence (including this paper and A.P.'s expert report) was lost; the court declined to hear any witness or request procurement documents for the security bags. Its decision simply stated that "no evidence of significant irregularities was presented". The appellate court upheld that decision, declining all motions to call witnesses, request documents, and examine parts of the original bags available, and ruling that statistical inferences and all the images used in this study are inadmissible as evidence. The cassation and Supreme courts upheld the decision.

The data collected by us indicates that two large batches of bags with overlapping number ranges were made for the Election commission of Moscow region, which then distributed pairs of identical bags to Vlasikha (Sec. D 10). The courts declined all motions to investigate this further. The Election commission of Moscow region and the bag manufacturer (Aceplomb Technology located near St. Petersburg [37]) both declined to com-ment in writing. The latter has been supplying security bags to election commissions nationwide since the introduction of multi-day voting in 2020 [44]. Our complaint to the Central election commission about the inadequateness of ballot storage procedures, with a proposal to improve them, was dismissed [45].

The authors—and more than twenty people who have contributed to this study—find it difficult to imagine what motivation and logic drive numerous election officials, all recruited from the local population, to be complicit in stealing votes from their own neighbours. This study also illustrates the extent of corruption and decay of political rights in Russia, even at the most local level of society.

*Acknowledgments.* We thank Vladimir Zaitsev for campaign coordination and support. We thank all 14 candidates nominated by CP, 5 additional observers, and 2 commission members for collecting the evidence and transcribing the vote counts. We thank Andrey Brazhnikov for discussions.

*Author contributions.* A.P. performed the statistical analysis. V.M. coordinated the collection of evidence and its analysis. Both authors wrote the paper.

*Competing interests.* The authors declare that they have no conflicts of interest.

*Data availability.* Video recordings, photographs, and copies of documents are available from the corresponding author on a reasonable request.

## METHODS

### Methods A: Transcripts of vote counting

The transcripts for all 9 physical precincts are shown in Figs. 3 to 5. Ballot ranges in the transcripts that lie outside continuous sequences of identical or almost-identical ballots for the 5 pro-administration candidates clearly show the voters' preference for the opposition candidates. In precinct 214 (Fig. 3), except the 3 packs for the candidates nominated by UR, most of the remaining random-looking votes are for the 4 CP's nominees and one UR's nominee. As discussed in main text, in the honest precinct 215 (Fig. 4), the voter's preferences are for 4 CP's nominees and one UR's nominee. In precincts 219 and 220 (Fig. 5), after we exclude the continuous sequences for the pro-administration candidates, the remaining random-looking votes are clearly in favor of the 5 CP's nominees. In precinct 219, the commission did not mix or rearrange the ballots at all, allowing us to trace the origin of every ballot in the transcript. Ballots 96 through 221 come from the first-day bag that was swapped. These ballots contain a few random votes (probably introduced intentionally in an attempt to mask the fraud) but nevertheless no votes for the popular candidate Razina (CP), forming an extremely improbable sequence (resulting in $\alpha' = 2.2 \times 10^{-19}$).

In electoral district 2 (Fig. 4), the transcripts display another remarkable discrepancy. Precinct 215 has only 6 ballots with marks for all 5 candidates who got the council seats (Vavilova nominated by LD and 4 nominated by UR: Serba, Fedoriv, Schik, Scherbovich). This set of candidates is unlikely to be chosen by a real voter with genuine pro-administration political preferences, because the voter would have to ignore the fifth UR's candidate Bulakh in favor of Vavilova nominated by another party. Indeed, 38 voters here marked all 5 UR candidates including Bulakh. However, in precincts 216 and 217 there are 176 and 175 ballots for all 5 candidates who got the council seats, just a few ballots fewer than the tampered-with security bags contain. Likewise, in precinct 219 (Fig. 5), 94 out of 104 ballots with 5 marks for the unlikely set of candidates who got the council seats (Anikeeva nominated by JR and the 4 nominated by UR) are in the range traced to the swapped bag. To be fair to UR, in district 3 they unexpectedly lost their fifth nominee who withdrew from the race days before the voting, after it became public that he had failed to declare his previous criminal conviction.

A few paper ballots were cast by absentee voters who for medical and other reasons could not come to the polling station and asked the commission members to visit them at home with a portable ballot box. The number of such voters is low, between 0 and 22 per precinct, and does not significantly affect the results.

### Methods B: Calculation of significances

The subject of our analysis is the sequence of valid ballots. They are considered in the order they are retrieved from the ballot box and announced when the votes are being tallied. Invalid ballots are excluded from consideration.

Let $n$ be the number of valid ballots at the precinct. For each candidate running in that precinct, the sequence of ballots is converted into a sequence of flags $f \in \{0;1\}$. Zero means that the candidate was not marked by the voter who filled the ballot, and one means that he or she was.

Let some candidate have a total of $r_0$ zeros and $r_1$ ones ($r_0 + r_1 = n$). The fraction of voters supporting this candidate is $p_1 = q_0 = r_1/n$ and the fraction of voters not supporting him or she is $p_0 = q_1 = r_0/n$. These fractions are treated as the probabilities of the occurrence of ones and zeros in the flag sequence, with the normalisation $p_f + q_f = 1$. Successive ballots are considered to be mutually independent.

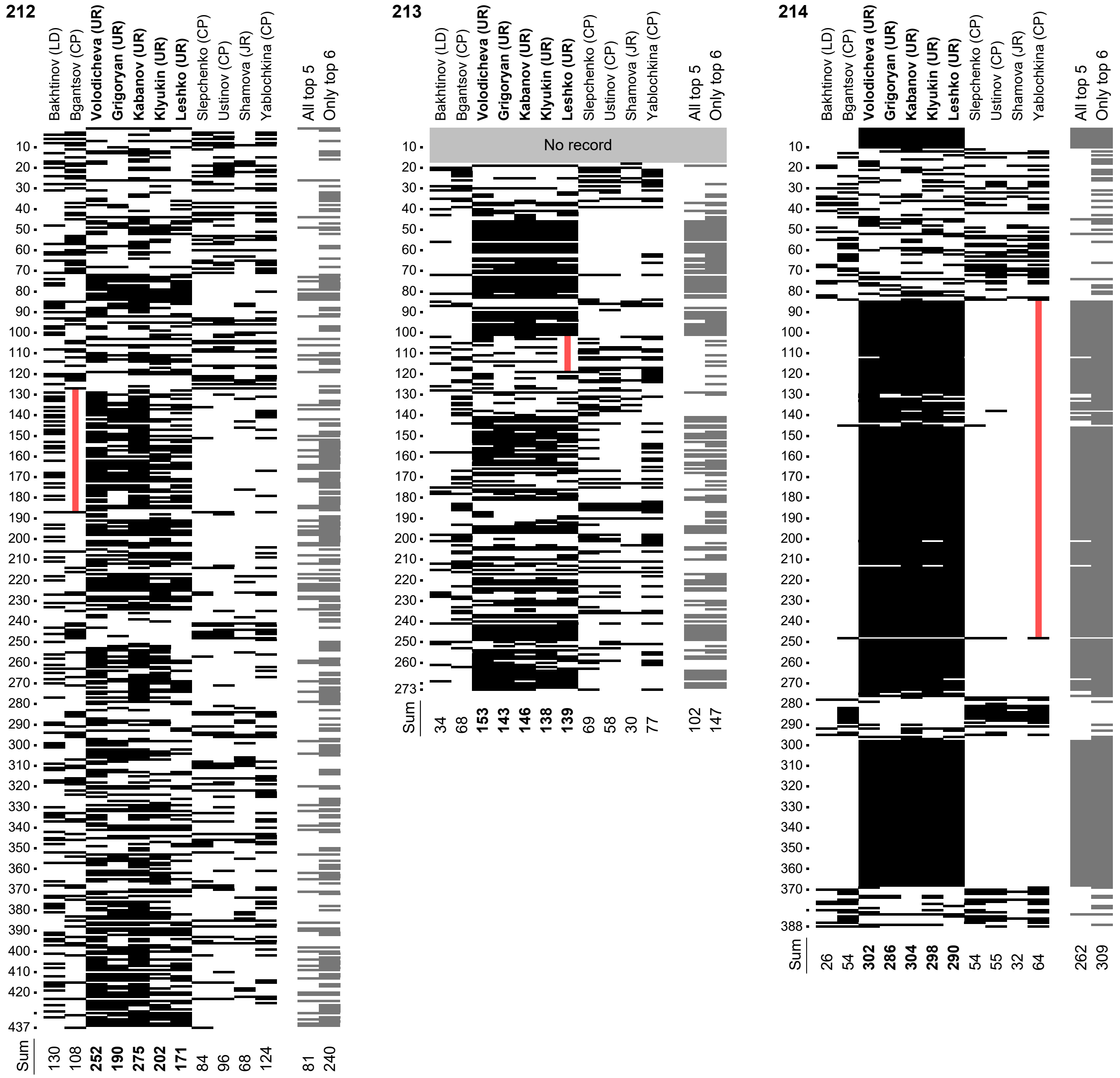

FIG. 3. Transcript of vote counting for precincts 212, 213, and 214 of electoral district 1. Top 5 candidates at each precinct are shown in bold. The least probable series is denoted by a red (grey) line. The last two grey columns are consolidated flags (Sec. B 3). Number of absentee voters per precinct is 4, 17, 0. At precinct 213, our recording starts late and misses about 17 first ballots. Only the recorded 256 ballots are thus used in our calculations.

### 1. Probability of a long series of flags

Let the longest continuous series of zeros and ones for a candidate have the length of $s_0$ and $s_1$ flags. If the probability $p_f$ of occurrence of the flag $f$ happens to be not large enough to explain the length of the series $s_f$, one can suspect falsification of voting results by stuffing a bunch of ballots not filled out by voters. All the ballots in such a batch are most likely filled out identically, while the will of each real voter is individual, so each of them can easily break the series.

Suspicions of fraud can be verified using mathematical statistics. Let us formulate a null-hypothesis that long continuous series of flags arose naturally by chance. An alternative hypothesis is that there was some tampering leading to the emergence of the long series. The choice between these two hypotheses is based on a statistical significance of the null-hypothesis $\alpha$. It quantifies the risk of so-called type I error. That means $\alpha$ is the probabil-ity of mistakenly rejecting the null-hypothesis being true. In other words, if the significance of the null-hypothesis turns out to be small, this means that the fraud took place. The opposite is generally not true: if the significance is large, we cannot reasonably assert anything about the presence or absence of the fraud.

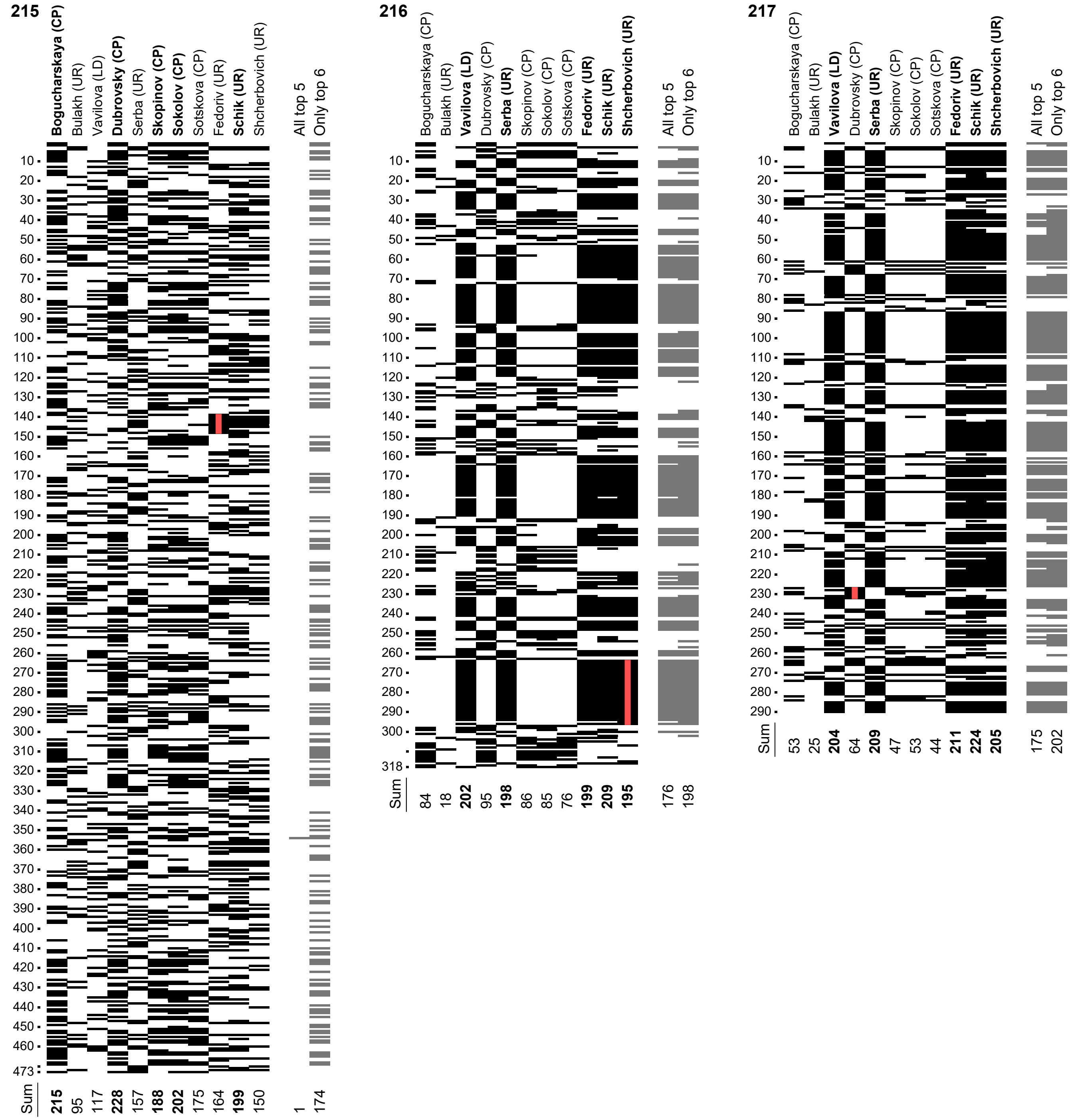

FIG. 4. Transcript of vote counting for precincts 215, 216, and 217 of electoral district 2. Top 5 candidates at each precinct are shown in bold. The least probable series is denoted by a red (grey) line. The last two grey columns are consolidated flags (Sec. B 3). Number of absentee voters per precinct is 4, 6, 9.

The algorithm for precisely calculating the significance for the longest series of flags is quite complex. Therefore, let us first upper-bound $\alpha$ in a simple way. For this, we assume only one series of flags $f$ of length at least $s_f$ to be present. This assumption is precise only for a sufficiently long series, while for short ones it leads to an overestimation of $\alpha_f$, because short series may appear multiple times in the sequence and each series is taken into account as many times as it appears. This approach is acceptable because it only weakens the statistical test.

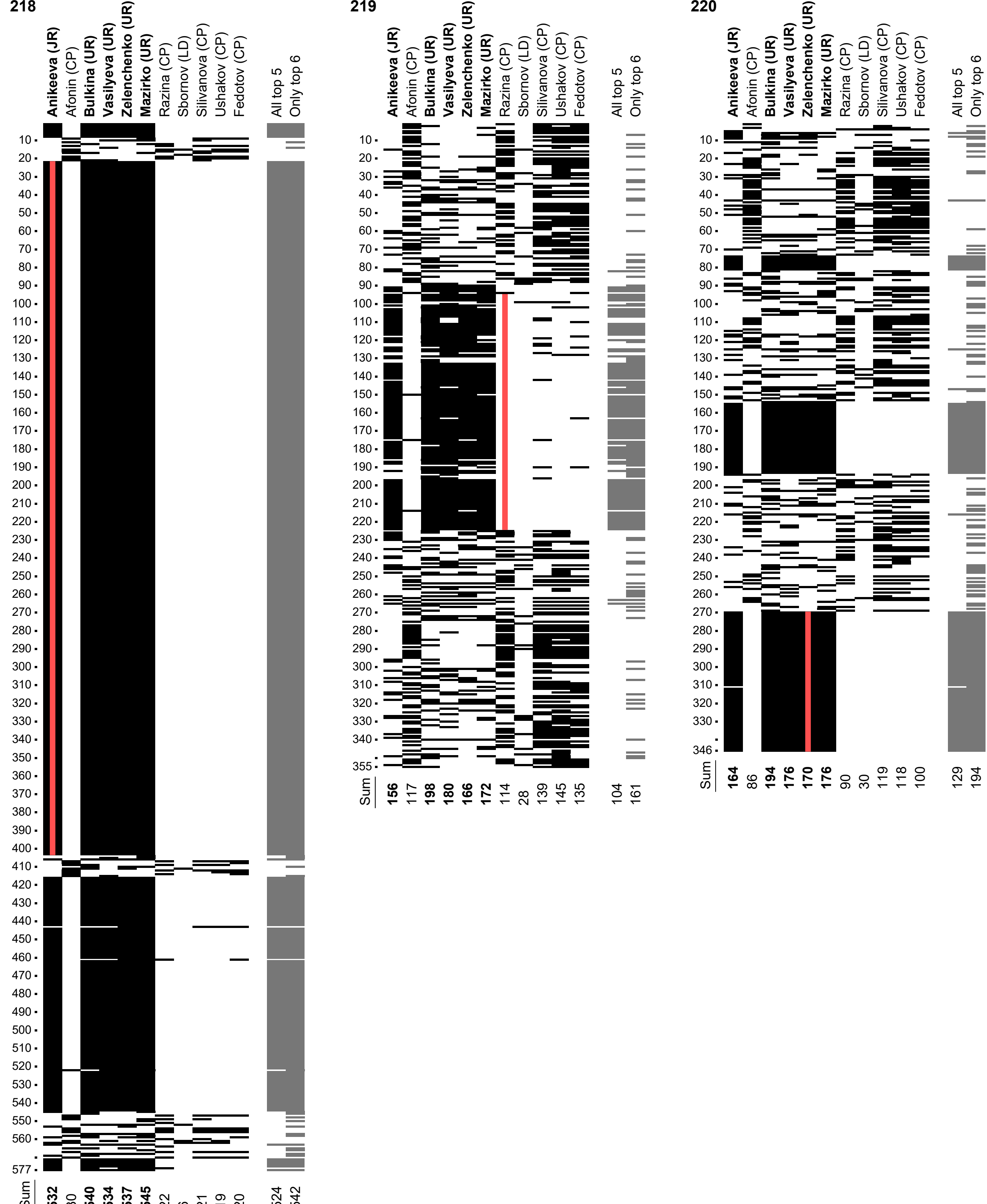

FIG. 5. Transcript of vote counting for precincts 218, 219, and 220 of electoral district 3. Top 5 candidates at each precinct are shown in bold. The least probable series is denoted by a red (grey) line. The last two grey columns are consolidated flags (Sec. B 3). Number of absentee voters per precinct is 0, 22, 1.

Under the assumption made, the probability of observing a series of length at least $s$ and starting with the very first flag is $p^s$ (we omit the index $f$ hereafter). The probability of observing such a series starting with any of the subsequent $n-s$ flags is $qp^s$, because the flag preceding the series must differ from the flags in it. The instances of the series starting with different flag positions are incompatible. That allows us to sum their probabilities. Thus, the significance of the hypothesis about the natural occurrence of the series $\alpha \leq [1+q(n-s)]p^s$. This simple formula can be used for quick estimates. A comparison with results of the exact calculation described below shows that this formula for the data under consideration overestimates $\alpha$ by no more than 5% if $\alpha < 10^{-1}$. But the overestimation rises rapidly for greater $\alpha$.

The exact calculation requires some more effort. Let $u_{n,i,j}$ denote the probability that, in a sequence of $n$ flags, the longest series of flags $f$ has a length of exactly $i$ ballots ($0 \leq i \leq n$), provided that exactly $j$ ($0 \leq j \leq i$) last flags have the same state $f$ as the flags of the series. These probabilities can be found using an iterative process corresponding to adding one more flag to the sequence (i.e., announcing the next ballot):

$$u_{n+1,i,j} = \begin{cases} p(u_{n,i,j-1} + u_{n,i-1,j-1}) & \text{if } j = i \\ pu_{n,i,j-1} & \text{if } 0 < j < i \\ qw_{n,i} & \text{if } j = 0 \end{cases} ,$$

where $w_{n,i} = \sum_{j=0}^{i} u_{n,i,j}$ is the unconditional probability that the longest series of flags $f$ has a length of exactly $s$. The normalisation $u_{0,0,0} = 1$ here is the initial condition and the constraints $u_{n,n+1,j} = 0$ and $w_{n,n+1} = 0$ are the boundary conditions.

When implemented on a computer, the algorithm can be simplified by using only a two-dimensional array for $u$ and only a one-dimensional array for $w$. New probability values obtained by increasing $n$ are written over the old ones. In order to avoid referring to values that are not yet defined or have already been redefined, the arrays should be initialised to zeros, indices $i$ and $j$ should be iterated over in their decreasing order and the $j$-cycle should be nested in the $i$-cycle.

The exact probability that the longest series has a length of at least $s$ is $\alpha = \sum_{i=s}^{n} w_{n,i}$. Tables II to IV show input data and results of this calculation for every candidate in every precinct (see the first 6 columns of every precinct).

In the calculation results, we observe a paradoxical effect of presumably genuine flag series identified as anomalous by the test. The point is that falsifications shift the estimates of the probabilities $p$ and $q$ from their true values. Relatively long series of votes cast for opposition candidates or against pro-administrative candidates may arise given the true probabilities of such an expression of the will of the voters. But these series already become improbable with the shifted probability estimates. At the same time, we note that the significances for such series are still not as small as for series of votes for pro-administration and against opposition candidates.

An event with probability $1/t$ occurs on average once in $t$ independent trials (more precisely, for large $t$ it does not occur at all among the trials with $1/e \approx 36.8\%$ probability, occurs exactly once with $1/e \approx 36.8\%$, twice with $1/2e \approx 18.4\%$, and three or more times with $1 - 2.5/e \approx 8.0\%$ probability). There are $c = 11$ candidates for each of 9 precincts under consideration and 2 possible flag values. So $t = 11 \times 9 \times 2 = 198$. Therefore, only significances $\alpha_f < 1/198$ are suspicious. The powers of such values are shown in bold in the tables. Similarly, the powers of $\alpha' < 1/9$ are shown in bold in Table I, where the multiplicity of candidates and flag states is already taken into account.

It should be especially noted that suspicion is not yet proof. The appearance of one or even more suspicious significances is quite expected in each test. Suspicious values only require attention, while non-suspicious ones can be ignored.

The test does not reveal any candidates with suspicious $\alpha_f$ for precincts 215 and 217 (Table III), it reveals 4 such candidates for precincts 212 and 213 (Table II), 9 for precinct 216 (Table III), and 10 or 11 for precincts 214 (Table II), 218, 219, and 220 (Table IV). Moreover, for the last 4 precincts $\alpha_{\min} < 10^{-11}$, i.e., their probability of the absence of fraud is extremely low.

The low significance of the hypothesis about the natural occurrence of the data anomaly clearly indicates that there was tampering of the type being tested for. However, a high significance may mean either that there was no tampering at all or that there was interference of some other type. In the context of the present analysis, such interference may, in particular, be the additional mixing of ballots before they are tallied (Table I). The mixing is neither required by law nor is a violation, but it makes detecting the ballot stuffing difficult. Therefore, it raises a suspicion.

### 2. Probability of the number of series

When the ballots are partially mixed, they will no longer form a long continuous series even in a hypothetically stuffed batch. However, if the mixing is not thorough, unbroken fragments of this batch will remain. Due to this, the number of series $m$ of consecutive flags with the same state will be lower than it would be in a normal situation. We thus calculate the probability of the occurrence of a certain number of series in the sequence of mutually independent ballots.

Let $a_{n,i}$ and $b_{n,i}$ be the probabilities that a sequence of $n$ flags contains $i$ series, given that the last flag of the sequence has a certain state $f$ or the opposite state $1-f$, respectively (the choice of this states does not affect the final results, so the index $f$ is omitted in calculations below). The probabilities are sequentially calculated us-

TABLE II. Calculation results for precincts of electoral district 1.

| Candidate (nominee of) | Precinct 212 | | | | | | | | | Precinct 213 | | | | | | | | | Precinct 214 | | | | | | | | |
|---|---|---|---|---|---|---|---|---|---|---|---|---|---|---|---|---|---|---|---|---|---|---|---|---|---|---|---|
| | $r_0$ | $s_0$ | p$\alpha_0$ | $r_1$ | $s_1$ | p$\alpha_1$ | $m$ | p$\tilde{\alpha}$ | p$\breve{\alpha}$ | $r_0$ | $s_0$ | p$\alpha_0$ | $r_1$ | $s_1$ | p$\alpha_1$ | $m$ | p$\tilde{\alpha}$ | p$\breve{\alpha}$ | $r_0$ | $s_0$ | p$\alpha_0$ | $r_1$ | $s_1$ | p$\alpha_1$ | $m$ | p$\tilde{\alpha}$ | p$\breve{\alpha}$ |
| Bakhtinov (LD) | 307 | 16 | 0.4 | 130 | 4 | 0.0 | 197 | 0.1 | 0.4 | 222 | 29 | 0.4 | 34 | 3 | 0.4 | 59 | 0.3 | 0.1 | 362 | 194 | **4.7** | 26 | 2 | 0.1 | 38 | 1.0 | **5.5** |
| Bgantsov (CP) | 329 | 59 | **5.3** | 108 | 5 | 0.6 | 138 | 1.7 | **4.7** | 188 | 32 | **2.5** | 68 | 5 | 0.7 | 83 | 1.5 | **4.2** | 334 | 102 | **5.0** | 54 | 8 | **4.3** | 60 | **3.1** | **13.4** |
| Volodicheva (UR) | 185 | 9 | 1.0 | 252 | 12 | 0.7 | 184 | **2.5** | **4.9** | 103 | 10 | 1.8 | 153 | 20 | **2.5** | 92 | **4.1** | **5.6** | 86 | 10 | **4.1** | 302 | 72 | **6.0** | 55 | **12.6** | **21.2** |
| Grigoryan (UR) | 247 | 12 | 0.7 | 190 | 9 | 0.9 | 168 | **5.3** | **2.9** | 113 | 12 | 2.1 | 143 | 16 | 2.0 | 90 | **5.5** | **7.9** | 102 | 19 | **8.6** | 286 | 71 | **7.5** | 52 | **19.4** | **24.7** |
| Kabanov (UR) | 162 | 11 | **2.3** | 275 | 14 | 0.7 | 156 | **5.1** | **3.9** | 110 | 10 | 1.5 | 146 | 14 | 1.4 | 92 | **4.7** | **7.2** | 84 | 13 | **6.2** | 304 | 104 | **9.2** | 54 | **12.5** | **21.8** |
| Klyukin (UR) | 235 | 11 | 0.7 | 202 | 9 | 0.7 | 188 | **2.6** | 1.9 | 118 | 10 | 1.2 | 138 | 11 | 0.9 | 91 | **5.5** | **6.8** | 90 | 15 | **7.1** | 298 | 72 | **6.4** | 62 | **11.9** | **19.5** |
| Leshko (UR) | 266 | 24 | **3.0** | 171 | 12 | **2.5** | 174 | **3.1** | **3.1** | 117 | 17 | **3.7** | 139 | 10 | 0.6 | 95 | **4.5** | **4.4** | 98 | 15 | **6.5** | 290 | 102 | **11.0** | 66 | **13.1** | **23.3** |
| Slepchenko (CP) | 353 | 35 | 1.4 | 84 | 4 | 0.4 | 116 | 1.3 | **2.5** | 187 | 32 | **2.6** | 69 | 4 | 0.2 | 90 | 1.0 | **3.0** | 334 | 102 | **5.0** | 54 | 7 | **3.5** | 65 | **2.4** | **11.5** |
| Ustinov (CP) | 341 | 56 | **4.1** | 96 | 7 | 2.1 | 117 | **2.5** | **4.8** | 198 | 32 | 1.9 | 58 | 4 | 0.4 | 81 | 0.8 | **4.2** | 333 | 109 | **5.6** | 55 | 9 | **5.1** | 54 | **4.4** | **10.6** |
| Shamova (JR) | 369 | 41 | 1.2 | 68 | 5 | 1.5 | 103 | 0.8 | 1.3 | 226 | 45 | 1.0 | 30 | 4 | 1.4 | 48 | 0.6 | 1.6 | 356 | 193 | **6.0** | 32 | 5 | **2.9** | 49 | 0.8 | **9.2** |
| Yablochkina (CP) | 313 | 16 | 0.4 | 124 | 4 | 0.1 | 159 | 1.2 | 1.4 | 179 | 22 | 1.6 | 77 | 8 | 1.9 | 82 | **2.7** | **3.1** | 324 | 163 | **11.2** | 64 | 9 | **4.5** | 60 | **5.3** | **16.3** |

TABLE III. Calculation results for precincts of electoral district 2.

| Candidate (nominee of) | Precinct 215 | | | | | | | | | Precinct 216 | | | | | | | | | Precinct 217 | | | | | | | | |
|---|---|---|---|---|---|---|---|---|---|---|---|---|---|---|---|---|---|---|---|---|---|---|---|---|---|---|---|
| | $r_0$ | $s_0$ | p$\alpha_0$ | $r_1$ | $s_1$ | p$\alpha_1$ | $m$ | p$\tilde{\alpha}$ | p$\breve{\alpha}$ | $r_0$ | $s_0$ | p$\alpha_0$ | $r_1$ | $s_1$ | p$\alpha_1$ | $m$ | p$\tilde{\alpha}$ | p$\breve{\alpha}$ | $r_0$ | $s_0$ | p$\alpha_0$ | $r_1$ | $s_1$ | p$\alpha_1$ | $m$ | p$\tilde{\alpha}$ | p$\breve{\alpha}$ |
| Bogucharskaya (CP) | 258 | 14 | 1.4 | 215 | 7 | 0.2 | 217 | 1.3 | 1.2 | 234 | 34 | **2.6** | 84 | 6 | 1.1 | 93 | **2.9** | **5.3** | 237 | 26 | 0.6 | 53 | 4 | 0.6 | 77 | 0.8 | 0.4 |
| Bulakh (UR) | 378 | 22 | 0.3 | 95 | 3 | 0.0 | 149 | 0.4 | 0.9 | 300 | 62 | 0.4 | 18 | 1 | 0.0 | 37 | 0.2 | 1.0 | 265 | 31 | 0.1 | 25 | 3 | 0.8 | 43 | 0.4 | 0.2 |
| Vavilova (LD) | 356 | 14 | 0.0 | 117 | 4 | 0.1 | 185 | 0.1 | 0.3 | 116 | 13 | **3.4** | 202 | 31 | **4.1** | 76 | **13.8** | **7.7** | 86 | 6 | 0.9 | 204 | 21 | 1.3 | 89 | **3.4** | **2.4** |
| Dubrovskiy (CP) | 245 | 8 | 0.2 | 228 | 8 | 0.3 | 221 | 1.1 | 1.6 | 223 | 33 | **3.2** | 95 | 6 | 0.8 | 94 | **4.5** | **6.4** | 226 | 21 | 0.6 | 64 | 6 | 1.6 | 81 | 1.6 | 0.3 |
| Serba (UR) | 316 | 14 | 0.4 | 157 | 4 | 0.0 | 194 | 1.0 | 0.1 | 120 | 9 | 1.5 | 198 | 31 | **4.3** | 83 | **12.2** | **8.1** | 81 | 6 | 1.0 | 209 | 21 | 1.1 | 83 | **3.7** | 1.4 |
| Skopinov (CP) | 285 | 11 | 0.3 | 188 | 7 | 0.4 | 208 | 1.3 | 1.0 | 232 | 33 | **2.6** | 86 | 5 | 0.5 | 85 | **4.6** | **5.7** | 243 | 28 | 0.6 | 47 | 4 | 0.8 | 69 | 0.8 | 1.1 |
| Sokolov (CP) | 271 | 9 | 0.1 | 202 | 7 | 0.3 | 209 | 1.7 | 0.3 | 233 | 33 | **2.6** | 85 | 5 | 0.6 | 81 | **5.1** | **6.5** | 237 | 28 | 0.8 | 53 | 5 | 1.3 | 71 | 1.3 | 0.2 |
| Sotskova (CP) | 298 | 16 | 1.0 | 175 | 6 | 0.3 | 183 | **3.2** | 1.8 | 242 | 33 | 2.1 | 76 | 6 | 1.4 | 85 | **2.9** | **5.2** | 246 | 28 | 0.5 | 44 | 4 | 0.9 | 73 | 0.4 | 0.5 |
| Fedoriv (UR) | 309 | 10 | 0.0 | 164 | 10 | 2.1 | 193 | 1.5 | 0.1 | 119 | 13 | **3.3** | 199 | 31 | **4.3** | 65 | **18.9** | **8.7** | 79 | 5 | 0.6 | 211 | 21 | 1.0 | 87 | **2.7** | 1.0 |
| Shik (UR) | 274 | 9 | 0.1 | 199 | 6 | 0.1 | 221 | 0.7 | 0.6 | 109 | 11 | **2.8** | 209 | 31 | **3.7** | 91 | **7.6** | **6.2** | 66 | 5 | 0.9 | 224 | 21 | 0.6 | 89 | 1.0 | 0.8 |
| Shcherbovich (UR) | 323 | 13 | 0.2 | 150 | 6 | 0.6 | 203 | 0.4 | 2.0 | 123 | 13 | **3.1** | 195 | 33 | **5.0** | 61 | **22.0** | **10.2** | 85 | 4 | 0.1 | 205 | 21 | 1.3 | 89 | **3.3** | 1.8 |

TABLE IV. Calculation results for precincts of electoral district 3.

| Candidate (nominee of) | Precinct 218 | | | | | | | | | Precinct 219 | | | | | | | | | Precinct 220 | | | | | | | | |
|---|---|---|---|---|---|---|---|---|---|---|---|---|---|---|---|---|---|---|---|---|---|---|---|---|---|---|---|
| | $r_0$ | $s_0$ | p$\alpha_0$ | $r_1$ | $s_1$ | p$\alpha_1$ | $m$ | p$\tilde{\alpha}$ | p$\breve{\alpha}$ | $r_0$ | $s_0$ | p$\alpha_0$ | $r_1$ | $s_1$ | p$\alpha_1$ | $m$ | p$\tilde{\alpha}$ | p$\breve{\alpha}$ | $r_0$ | $s_0$ | p$\alpha_0$ | $r_1$ | $s_1$ | p$\alpha_1$ | $m$ | p$\tilde{\alpha}$ | p$\breve{\alpha}$ |
| Anikeeva (JR) | 45 | 13 | **11.7** | 532 | 382 | **12.3** | 19 | **10.4** | **26.0** | 199 | 30 | **5.4** | 156 | 32 | **9.2** | 85 | **21.2** | **26.5** | 182 | 21 | **3.7** | 164 | 41 | **11.1** | 70 | **29.4** | **25.1** |
| Afonin (CP) | 547 | 385 | **7.9** | 30 | 5 | **3.7** | 31 | **2.6** | **13.9** | 238 | 55 | **7.6** | 117 | 11 | **2.9** | 125 | **3.0** | **17.4** | 260 | 82 | **8.4** | 86 | 6 | 1.2 | 92 | **3.7** | **12.7** |
| Bulkina (UR) | 37 | 5 | **3.2** | 540 | 383 | **9.9** | 31 | **4.4** | **17.3** | 157 | 10 | 1.3 | 198 | 47 | **9.8** | 114 | **10.3** | **20.5** | 152 | 10 | 1.3 | 194 | 78 | **17.5** | 94 | **15.9** | **20.6** |
| Vasilyeva (UR) | 43 | 8 | **6.3** | 534 | 383 | **11.7** | 29 | **6.6** | **26.9** | 175 | 26 | **5.8** | 180 | 29 | **6.3** | 99 | **17.0** | **24.8** | 170 | 13 | 1.8 | 176 | 77 | **20.5** | 86 | **21.3** | **22.0** |
| Zelenchenko (UR) | 40 | 11 | **10.0** | 537 | 382 | **10.8** | 25 | **6.7** | **25.0** | 189 | 26 | **4.9** | 166 | 42 | **11.6** | 89 | **20.9** | **29.6** | 176 | 18 | **3.1** | 170 | 77 | **21.6** | 68 | **31.2** | **25.3** |
| Mazirko (UR) | 32 | 5 | **3.5** | 545 | 385 | **8.5** | 27 | **3.8** | **18.3** | 183 | 13 | 1.5 | 172 | 35 | **8.8** | 101 | **16.0** | **22.5** | 170 | 17 | **3.0** | 176 | 77 | **20.5** | 82 | **23.3** | **26.7** |
| Razina (CP) | 555 | 386 | **5.6** | 22 | 3 | 1.5 | 35 | 0.7 | **11.0** | 241 | 130 | **20.0** | 114 | 8 | 1.6 | 104 | **6.4** | **17.0** | 256 | 77 | **8.2** | 90 | 6 | 1.1 | 101 | **2.9** | **9.0** |
| Sbornov (LD) | 571 | 392 | 1.3 | 6 | 2 | 1.2 | 11 | 0.3 | 1.6 | 327 | 134 | **3.5** | 28 | 4 | 1.9 | 47 | 0.5 | **3.1** | 316 | 84 | 1.9 | 30 | 2 | 0.0 | 53 | 0.4 | 1.3 |
| Silivanova (CP) | 556 | 385 | **5.3** | 21 | 3 | 1.6 | 31 | 0.9 | **9.9** | 216 | 32 | **4.8** | 139 | 14 | **3.4** | 115 | **7.8** | **13.8** | 227 | 77 | **12.1** | 119 | 8 | 1.4 | 107 | **6.3** | **13.5** |
| Ushakov (CP) | 558 | 386 | **4.8** | 19 | 3 | 1.7 | 27 | 0.9 | **7.4** | 210 | 96 | **19.9** | 145 | 12 | **2.4** | 107 | **10.9** | **24.5** | 228 | 77 | **12.0** | 118 | 13 | **3.7** | 105 | **6.6** | **16.1** |
| Fedotov (CP) | 557 | 386 | **5.0** | 20 | 3 | 1.6 | 27 | 1.1 | **6.2** | 220 | 40 | **6.2** | 135 | 7 | 0.7 | 119 | **6.3** | **18.1** | 246 | 77 | **9.5** | 100 | 8 | 1.9 | 109 | **3.1** | **13.8** |

ing iterative equations $a_{n+1,i} = p(a_{n,i} + b_{n,i-1})$ and $b_{n+1,i} = q(b_{n,i} + a_{n,i-1})$ corresponding to the addition of the next flag to the sequence. If the new flag coincides with the last flag of the sequence, the number of series does not change, and if it differs, it increases by 1. The initial conditions are $a_{1,1} = p$ and $b_{1,1} = q$, the boundary conditions are $a_{n,0} = b_{n,0} = a_{n,n+1} = b_{n,n+1} = 0$.

The probability of finding exactly $i$ series in a sequence of $n$ ballots is $v_{n,i} = a_{n,i} + b_{n,i}$. Thus, the significance of the hypothesis that the final number of series $m$ turns out to be small for natural reasons is $\tilde{\alpha} = \sum_{i=1}^{m} v_{n,i}$.

In a computer implementation of this algorithm, just as for the previous one, it is possible to reduce the dimension of the arrays used. To do this, one should initialise the arrays to zeros and iterate over the index $i$ in its decreasing order.

We can also get an analytical estimation $\tilde{\alpha} \approx \Phi(m + 1/2; \mu_n, \sigma_n)$, where $\Phi$ is the Gaussian distribution function. The addition of $1/2$ in the first argument partially compensates for the difference between the sum of discrete probabilities giving the exact significance and the integral of the probability density giving its estimate. The expectation $\mu_n$ and variance $\sigma_n^2$ can be exactly calculated using generating function techniques. Let us denote $g_n(z) = \sum_i a_{n,i} z^i$, $h_n(z) = \sum_i b_{n,i} z^i$, and $y_n(z) = g_n(z) + h_n(z) = \sum_i v_{n,i} z^i$. Then the iterative equations take the form $g_{n+1} = p(g_n + zh_n)$ and $h_{n+1} = q(h_n + zg_n)$. They allow us to express the distribution parameters through the derivatives of the generating functions $\mu_n = y_n'(1) = 1 + (n-1)\theta$ and, for $n > 1$, $\sigma_n^2 = y_n''(1) + \mu_n - \mu_n^2 = (2n-3)\theta - (3n-5)\theta^2$, where $\theta = 2pq$ is the average toggle frequency between series. The intermediate calculations here are quite cumbersome, so we skip them.

This approximation has a considerable drawback when compared with the estimate for the first test. Here the significance can not only be overestimated (which is acceptable), but also underestimated (which is highly undesirable). Therefore, approximate values of $\tilde{\alpha}$ should be treated with caution. For the data under consideration, the greatest overestimation can reach several orders of magnitude (although it reaches at least 1 order only for $\tilde{\alpha} < 10^{-6}$) and the greatest underestimation is less than 6% for $\tilde{\alpha} < 10^{-1}$.

Tables II to IV show the results of exact calculations for every candidate in every precinct (see the third- and second-last columns of every precinct).

Unlike the previous test, only one significance is calculated here, so $t = 11 \times 9 = 99$. Therefore, significances $\tilde{\alpha} < 1/99$ become suspicious. The similar condition $\tilde{\alpha}' < 1/9$ remains unchanged for the overall significance in Table I.

The qualitative results of both tests are generally close for every precinct except 216. The first test leaves some doubt there, while the second one succeeds in proving the fraud, despite the intentional mixing of the ballots by the commission (Table I). This illustrates that the second test is useful. Moreover, the second test is robust to minor human errors in the transcript, whereas the first test is robust only as long as such errors do not affect the least probable series.

However, there is a certain problem with the second test. It gives $\tilde{\alpha}_{\min} = 7.5 \times 10^{-4}$ in precinct 215 that most likely had no fraud. Of course, such significance could also be the result of an unfortunate combination of circumstances. Especially since it is observed only for 1 candidate (for comparison, comparable suspicious values of $\tilde{\alpha}$ for precinct 217 are observed for 4 candidates). But another possible explanation is that the successive ballots are not, in fact, absolutely independent. Perhaps they are not completely mixed for simultaneously voting voters. If this is true, then the second test turns out to be more sensitive to the inaccuracy of our initial assumptions.

### 3. Consolidated tests over groups of candidates

To mitigate the hypothetical effect noted above, the source data for statistical tests are to be consolidated so as to analyse the sequence of flags not for each candidate individually, but for their entire list at once. The state of the flag is now determined by the voter's attitude towards the 5 candidates leading in the precinct. If the voter supported all of them, $f = 1$, and in any other case, $f = 0$ (as visualised in the second-last grey column of each precinct in Figs. 3 to 5). All calculations are performed in the same way as before and the results are shown in Table V. Suspicious values for $t = 9 \times 2 = 18$ independent trials for $\alpha_f$ and $t = 9$ independent trials for $\tilde{\alpha}$ and $\breve{\alpha}$ (see below) are shown in bold.

The consolidated tests not only resolves doubts in favor of precinct 215, but it also conclusively proves fraud at precinct 217. At the same time, precinct 212 somehow manages to avoid suspicion in the consolidated tests, which requires an explanation.

The problem here is the implicit assumption that falsifiers help only the winners and help them all. However, a more complex fraud scheme was apparently implemented at precinct 212. To understand this scheme, a certain political-science analysis is required that goes somewhat beyond the realm of mathematical statistics.

LD nominee Bakhtinov is 6th most popular in precinct 212, while in precinct 213 and e-voting precinct he is 10th (second least popular), and in precinct 214 he is 11th (the least popular) [31]. It is necessary to clarify that LD party is quite unpopular in European Russia, so it nominated only 1 candidate in every electoral district of Vlasikha town. In district 3, LD nominee Sbornov is 11th (the least popular) in every precinct including e-voting, as he had no support from either voters or falsifiers. In district 2, LD nominee Vavilova is 10th (second least popular) in the honest precinct 215 and in e-voting, but 2nd and 5th in precincts 216 and 217, where she was on the list of pro-administration candidates benefitting from the ballot replacement. It is noteworthy that UR nominee Bulakh, eventually not included in this list in district 2, was 11th (the least popular) in all the precincts, which hints at the real popularity of the most powerful party.

TABLE V. Calculation results for the consolidated tests.

| Precinct | $r_0$ | $s_0$ | p$\alpha_0$ | $r_1$ | $s_1$ | p$\alpha_1$ | $m$ | p$\tilde{\alpha}$ | p$\breve{\alpha}$ |
|---|---|---|---|---|---|---|---|---|---|
| 212 | 356 | 24 | 0.4 | 81 | 4 | 0.5 | 126 | 0.5 | **1.6** |
| 213 | 154 | 27 | **4.0** | 102 | 10 | **1.8** | 81 | **6.6** | **7.2** |
| 214 | 126 | 34 | **14.2** | 262 | 71 | **10.1** | 36 | **37.6** | **30.3** |
| 215 | 472 | 353 | 0.2 | 1 | 1 | 0.2 | 3 | 0.1 | 0.4 |
| 216 | 142 | 19 | **4.4** | 176 | 31 | **5.9** | 57 | **29.9** | **12.7** |
| 217 | 115 | 11 | **2.2** | 175 | 21 | **2.6** | 71 | **14.2** | **3.6** |
| 218 | 53 | 18 | **16.0** | 524 | 382 | **14.7** | 19 | **13.4** | **32.8** |
| 219 | 251 | 90 | **11.7** | 104 | 17 | **6.7** | 45 | **23.0** | **35.4** |
| 220 | 217 | 53 | **8.7** | 129 | 41 | **15.3** | 20 | **51.0** | **39.7** |

To sum up, it can be assumed that the LD nominee had support from the falsifiers at precinct 212 as well, despite that he was not among the pro-administration candidates. It should also be noted that voters allegedly supporting the LD nominee gave 64.4% of their remaining votes to the winning pro-administrative candidates at precinct 212. This value is lower, 38.9% (26.9%) at precinct 213 (214).

We can only guess how this happened. On the one hand, it may be a case of silent sabotage. If the falsifiers did not want to carry out their criminal duties, but were afraid to openly protest, they could mark the weakest of the candidates pretending to make a mistake. On the other hand, it could be an attempt to disguise the fraud without risking helping the opposition. Both of these situations seem quite exotic, which is why this phenomenon is rare and affects only one precinct.

None of the proposed explanations exclude the possibility of fewer than 5 candidates being voted on the falsified ballots. Taking this into account, the consolidated test is tuned up as follows. Now $f = 1$ if the ballot contains marks for any of the 6 most popular candidates in that precinct and no other marks, and $f = 0$ otherwise (as visualised in the last grey column of each precinct in Figs. 3 to 5). This allows us to detect fraud at precinct 212 with $\tilde{\alpha} = 6.2 \times 10^{-5}$. At the same time, for other precincts the significances $\alpha_f$ and $\tilde{\alpha}$ rise noticeably, but they still remain low enough to leave no doubt about fraud there (with the exception of precinct 215, of course; Table VI).

The analysis above illustrates some important general considerations. Any rigorous formal test relies on an idea of how the fraud is carried out. Therefore, any such test can be passed with relatively little effort. However, falsifiers as a rule do not make this effort. On the one hand, they are usually confident in their complete impunity, and on the other hand, they are stupid, like most people who commit crimes. This is precisely what allows us, in most cases, to propose effective statistical tests without using any political-science information. However, for a reliable analysis it is necessary to use the whole combination of available statistical tests, because the conditions of applicability of any single test may sometimes be violated.

TABLE VI. Calculation results for the tuned-up consolidated tests.

| Precinct | $r_0$ | $s_0$ | p$\alpha_0$ | $r_1$ | $s_1$ | p$\alpha_1$ | $m$ | p$\tilde{\alpha}$ | p$\breve{\alpha}$ |
|---|---|---|---|---|---|---|---|---|---|
| 212 | 197 | 9 | 0.8 | 240 | 15 | **1.6** | 176 | **4.2** | **3.9** |
| 213 | 109 | 10 | **1.6** | 147 | 15 | **1.6** | 99 | **3.2** | **4.7** |
| 214 | 79 | 10 | **4.4** | 309 | 102 | **8.3** | 62 | **8.7** | **20.3** |
| 215 | 299 | 14 | 0.6 | 174 | 6 | 0.3 | 208 | 0.8 | 0.7 |
| 216 | 120 | 16 | **4.5** | 198 | 33 | **4.8** | 67 | **18.4** | **10.9** |
| 217 | 88 | 7 | **1.3** | 202 | 26 | **2.2** | 71 | **7.7** | 0.7 |
| 218 | 35 | 7 | **5.8** | 542 | 385 | **9.4** | 33 | **3.5** | **17.4** |
| 219 | 194 | 23 | **3.9** | 161 | 17 | **3.6** | 105 | **13.7** | **22.5** |
| 220 | 152 | 15 | **3.1** | 194 | 77 | **17.3** | 98 | **14.4** | **19.3** |

### 4. Test for stationarity of vote flow

The analysis also contains another implicit assumption that probabilities of voting for a candidate are constant over time. But, for example, opposition voters may be more inclined to vote on the third day if they know their votes cast on the first two days might be stolen. In this case, the flow of votes becomes non-stationary. The constancy of probabilities can be verified as follows.

We split the sequence of flags into $l$ segments of equal length (if $n$ is not divisible by $l$, then the boundary flags are distributed between adjacent segments in proportional shares). The number of ones $k_i$ in a segment $i = 1, 2, ..., l$ is binomially distributed with expectation $\mu = r_1/l$ and variance $\sigma^2 = r_0 r_1/nl$. This distribution can be approximated by a Gaussian distribution for a sufficiently large segment length $n/l$. The statistics $\sum_{i=1}^{l} (k_i - \mu)^2/\sigma^2$ has the $\chi^2$-distribution with $l - 1$ degrees of freedom (1 degree is spent on determining the distribution parameters). This allows us to calculate the significance $\breve{\alpha}$ of the hypothesis that all deviations of the vote flow from the stationary one are random. The similar statistics for zeros with $\mu = r_0/l$ is obviously the same.

The specific choice of $l$ is not of great importance, although the presence of a free parameter is in itself a certain shortcoming of this approach. We use the value $l = 12$. It simultaneously gives several segments for each of the 3 voting days and ensures several dozen flags in each segment.

Tables II to IV show the results of these calculations for every candidate in every precinct (see the last columns of every precinct). The probability of voting does not change over time for precinct 215 and perhaps for precinct 217. It changes noticeably for precincts 212 and 213 and very strongly at all the others. However, for the latter, one can no longer rely on the quantitative values of $\breve{\alpha}$, since this asymptotic test tends to underestimate the significance the more, the smaller it is.

We do not doubt the results of voting at precinct 215. So we can consider the assumption discussed to be valid for the voting in question. That also makes it possible to use the check of the stationarity of the vote flow as an auxiliary test, although it does not provide any new results compared to the two main tests. This test is less valuable as evidence, because the nonstationarity of the vote flow can in principle have natural causes. Although, of course, most often it is the result of fraud.

Similar results for the consolidated test are shown in the last column of Table V. The difference between the honest precinct 215 and precinct 217 with a good ballot mixing becomes here more noticeable. For precinct 212, with its unusual approach to falsification, the auxiliary test barely sees anomalies. But if we tune up the consoli-dation (Table VI), we get $\breve{\alpha} = 1.2 \times 10^{-4}$ there, sufficient to have no doubt about the nonstationarity of the flag flow.

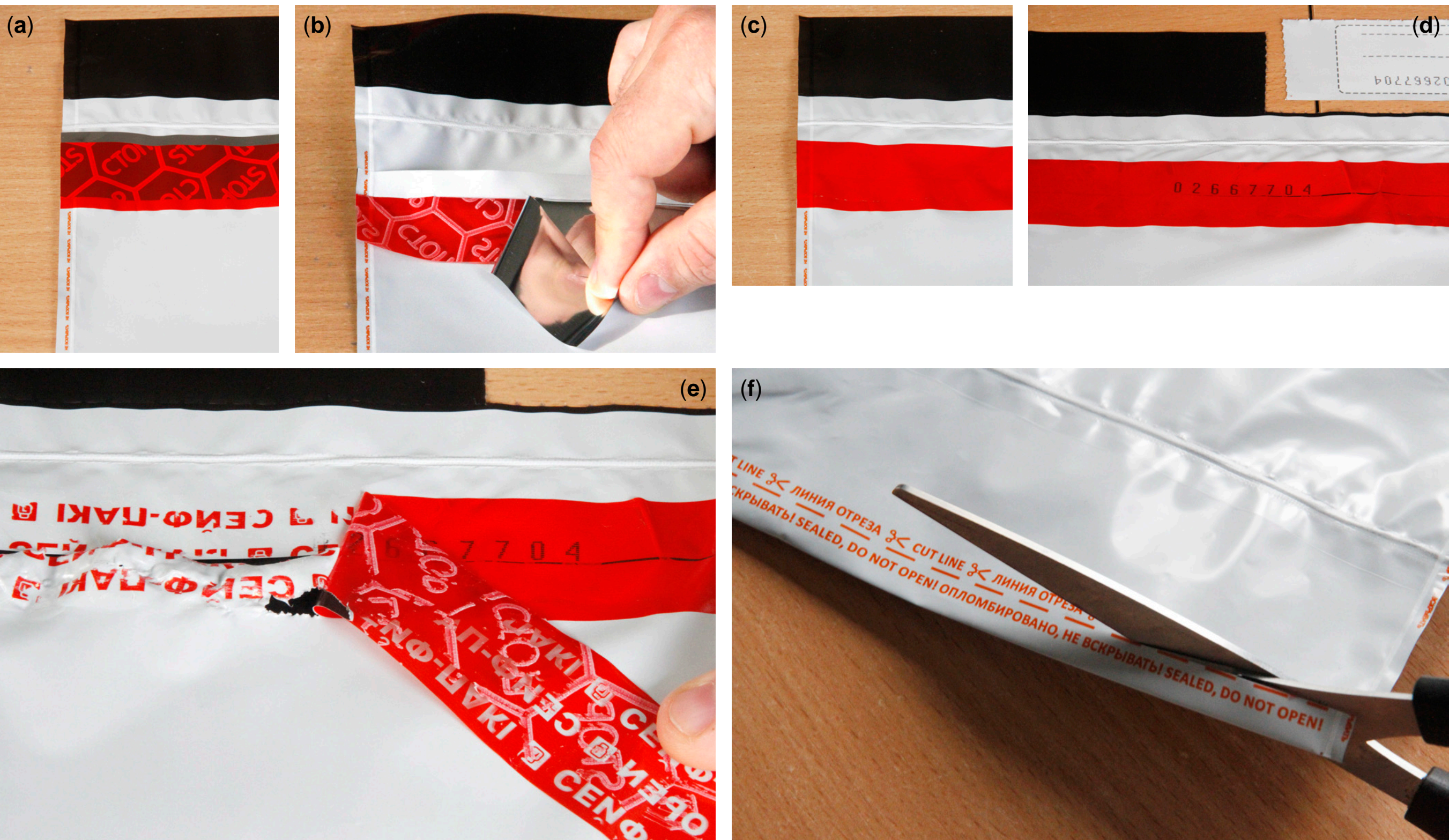

FIG. 6. Normal use of the security bag. (a) In an unused bag, the security tape's bottom edge is already adhered permanently to the bag. A metallised liner protects the glue on its remaining unattached portion. A slice of the liner extends above the tape. The tape is transparent and the liner can be seen through it. (b) To seal the bag, the unattached portion of the tape with the liner is flipped over, such that its unattached adhesive portion faces up, without detaching the tape from the bag. The liner is then removed carefully, without fingers touching the adhesive being exposed. The tape is then flipped back over the bag opening, is pressed against the bag, and sticks to it permanently. (c) and (d) The security tape with the unique bag number seals the opening. The honeycomb pattern with words STOP СТОП is the outermost layer of its multi-layer adhesive. This pattern is made of light-grey adhesive and is clearly visible when the tape is not adhered to the bag or is covered by the reflective liner, but almost disappears once the adhesive attaches to the bag. (d) Control stubs can be detached from the bag. (e) An attempt to remove the security tape separates the multi-layer adhesive and reveals lettering in it. (f) The bag is normally opened by cutting it at the bottom.

### Methods C: Circumventing tamper-indicating measures of security bags

Let us first review how the security bag is normally used. The red security tape initially sticks to the bag by its bottom edge, the rest of its glue-coated width being protected by a silicone-coated liner tape [Fig. 6(a)]. To seal the bag, the liner is removed and the red tape seals the bag opening [Fig. 6(b)–(d)]. An attempt to remove the tape tears its complex multi-layer glue structure apart, leaving visible traces [Fig. 6(e)]. If the tape is removed rapidly by brute force, the bag material stretches and may also tear. Note that the bag's unique number is printed on the adhesive layer of the security tape, thus it cannot be tracelessly removed from the tape or replaced with a different number. The edges of the bag not protected by the tape have a small print running along them (also on the other side not photographed here), to make any attempt to open and reseal the edges visually evident, too [46]. The bag can be opened by cutting or tearing it apart at any place [Fig. 6(f)].

The idea of a tamper-evident bag is that an attempt to access its content does not remain unnoticed. The faster this is discovered upon a visual inspection of the bag, the better. We present 5 possible methods to conceal the unauthorised access. Methods C 1, C 4, and C 5 use readily available materials and do not require a collusion with the bag manufacturer. Methods C 2 and C 3 require access to duplicates of bags with identical numbers made at the factory. A diligent factory should not produce duplicate bags (as the bag numbers must be unique [38]). As discussed in Sec. D, the evidence points that our election commissions have had access to such factory-made duplicates and used methods C 2 and C 3 to rig these particular elections.

### 1. Imitation with red packing tape

Red-colored packing tape mimics the visual appearance of the security tape in its sealed state (Fig. 7), but not in the initial unsealed state. The latter difference can easily be overlooked, as most people observing the elections are not familiar with the security bags.

The preparations would proceed as follows. The original security tape is peeled off in advance, its used portion trimmed, and the remains of its adhesive washed off the bag with a solvent (Fig. 8). While the adhesive is resistant to many common solvents, we have found one that removes it easily while not damaging the print along the bag's edges. The trimmed original tape is stored for future reuse.

A fake sealing tape is then made of suitable red-colored packing tape (costs under $2). The bag's unique number is printed on its adhesive side with a handheld inkjet printer (costs around $100), which contactlessly marks any surface with fast-drying ink (Fig. 9). A liner from another bag is applied to it. The tape is trimmed to correct size and installed on the bag.

The modified bag is then used publicly to seal ballots in a normal way (Fig. 10). At night, the fake tape is easily removed, the ballots are replaced, and the bag is resealed using the original stored tape (Fig. 11). The ballots are then publicly retrieved from the generally normally-looking undamaged bag (Fig. 12). Should anyone try to peel the security tape off at this point, it would display the expected behavior. The only visual clue remaining is the reduced width of the tape, which is almost certain to be overlooked unless one knows specifically what to look for.

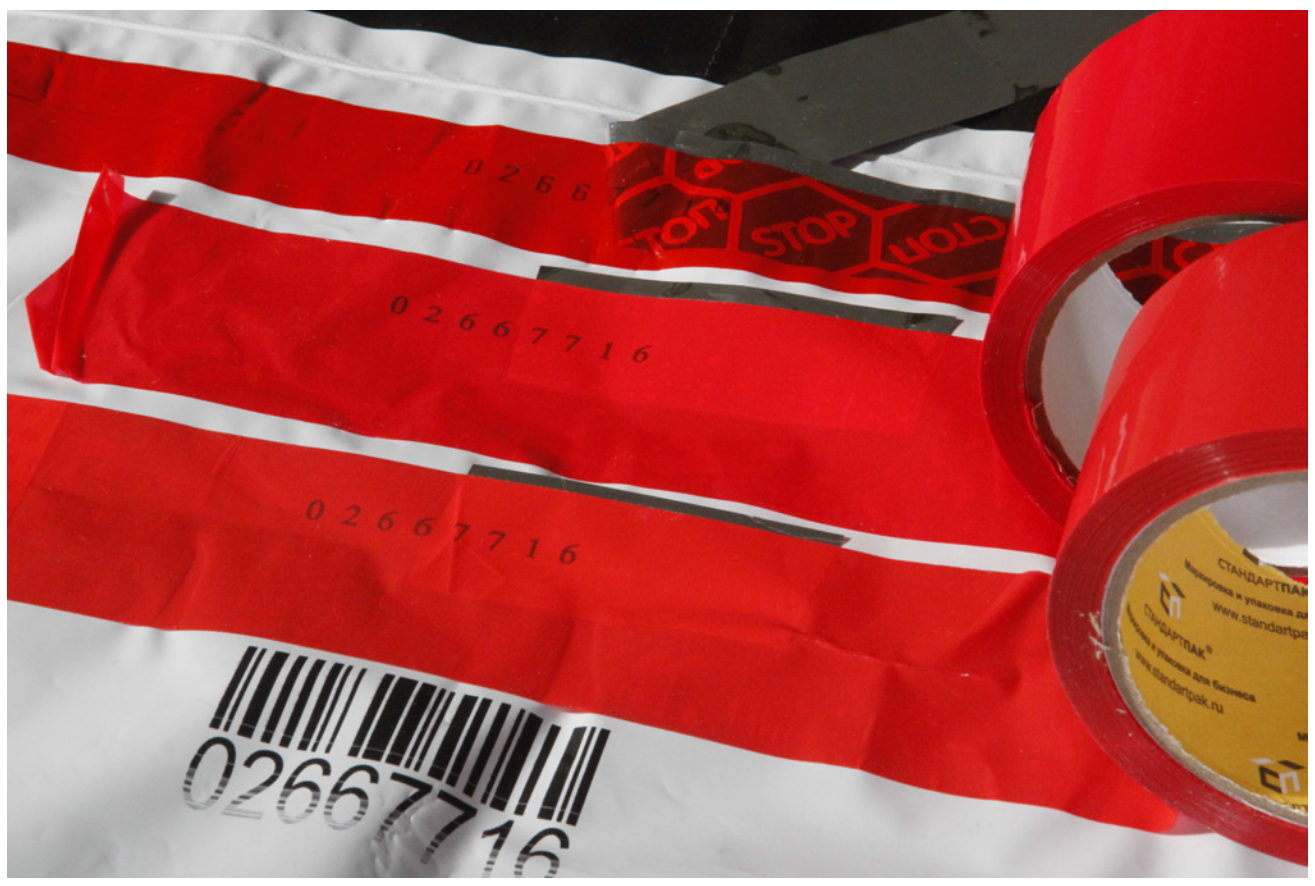

FIG. 7. Red-colored packing tape can mimic the visual appearance of the security tape in its sealed state. The left-hand side of each tape is fully adhered to the bag, while the right-hand side still has some liner under it. The topmost tape is original, while the other two are different brands of packing tape. The numbers on the latter were printed on their adhesive side with a handheld inkjet label printer. All three tapes are transparent and the red color is in their adhesive layer only.

Although this method is feasible, there is no evidence anyone has used it.

### 2. Swapping the bag for its factory-made duplicate

If the perpetrators have a duplicate security bag with the identical number, they can replace the original bag with it at night. The only difficulty would be forging signatures that may be left on the original bag by observers and election officials, and reproducing any imperfections in the tape placement and partial removal of the control stubs. These imperfections at precinct 219 have revealed this whole fraud scheme (Sec. D 1).

### 3. Replacing one security tape with another

To avoid forging signatures, the perpetrators may choose to replace only the tape with its factory-made duplicate, instead of swapping the entire bag. At night, they would peel off the original security tape from the bag with ballots, wash the bag with the solvent, replace the ballots, cut off the tape from the duplicate bag, and install it on the original bag. If any signatures cross the tape, they would still have to be forged or left incomplete, which may be noticed. Also the tape width is reduced. This is what was likely done at most of our precincts, see Sec. D 2 and thereon.

We remark that there seems to be no technical reason why both tapes could not be made the same narrow width. For that, the bag would have to be prepared in advance by cutting off and reinstalling the first original tape, before it appears on public for the first time. However, this would increase the chance that the fraud is spotted early.

At room temperature, the fully adhered tape has to be peeled off carefully and slowly to preclude the risk of stretching and tearing the bag material like that seen in Fig. 6(e). A strong cooling by spraying the content of an inverted duster can on the tape allows faster removal. In either case, it then takes about 10 min to wash the adhesive off the bag with solvent.

### 4. Reducing tape adhesion

Coating the bag with a thin transparent silicone layer lowers the tape's adhesion to the point that the bag becomes tracelessly resealable [45]. While this method has been practiced at many polling stations in 2024 [47–49] and 2025 [50], it has not been employed in Vlasikha. We remark that hiding traces of the backdoor is possible by removing the silicone layer with solvent before resealing the bag [45].

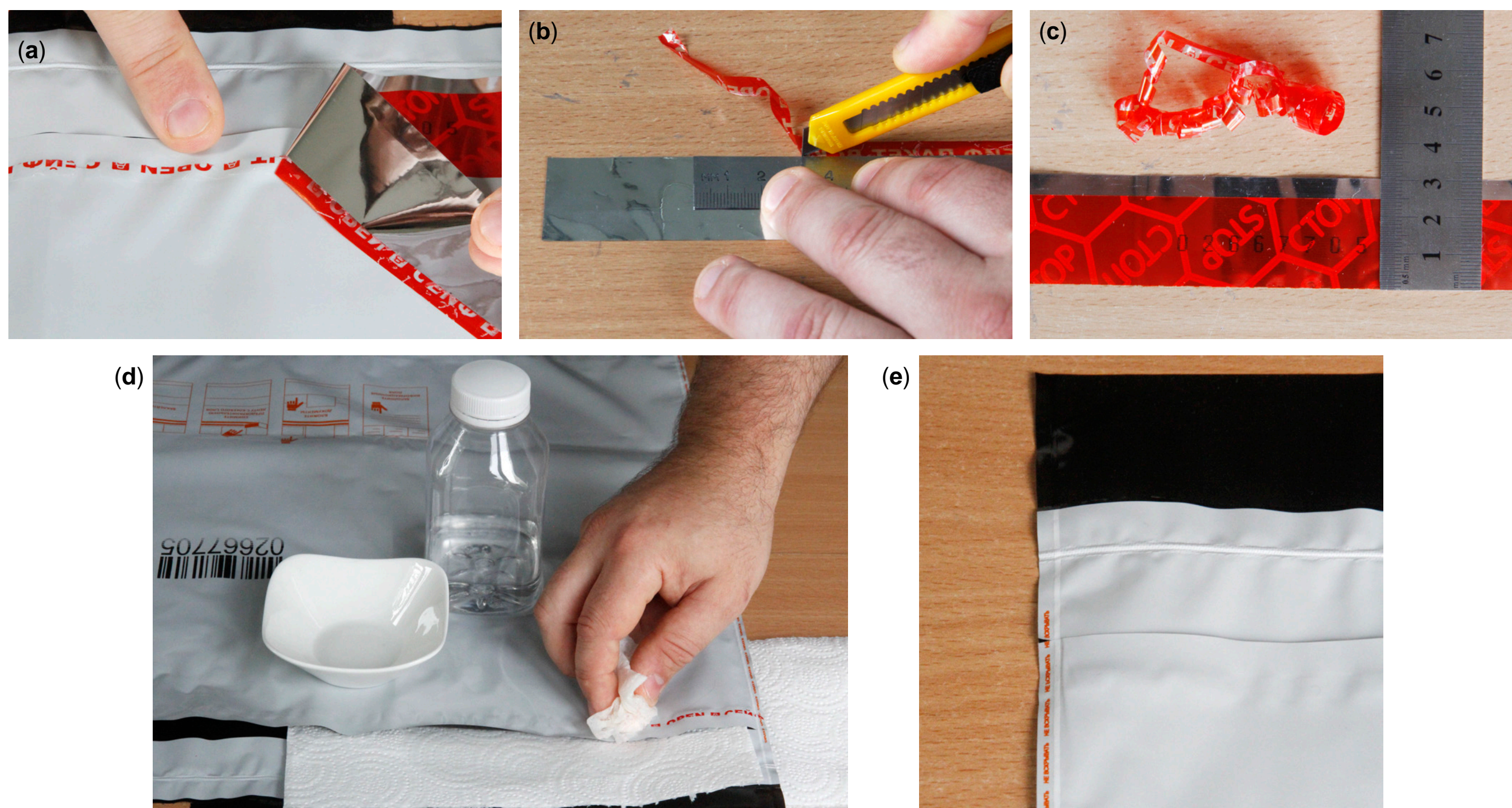

FIG. 8. Removal of the original security tape from an unused bag. (a) The tape is peeled off carefully with the liner still attached, leaving a narrow strip of adhesive behind. (b) Tape is trimmed to remove the used portion of its width. (c) The trimmed original tape is less than 25 mm wide. (d) Adhesive residue is removed from the bag with solvent. (e) The bag is clean and undamaged.

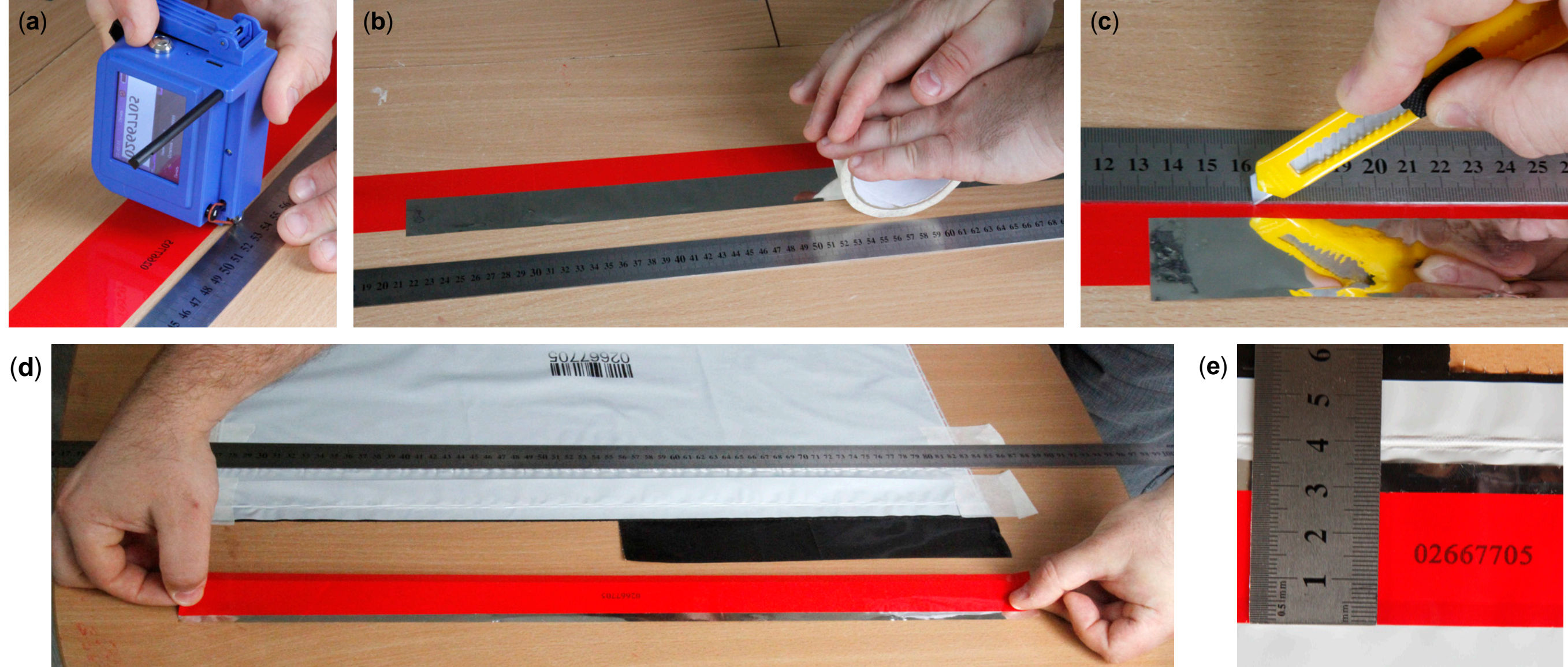

FIG. 9. Installation of a fake tape. (a) A mirrored unique number is printed on the adhesive side of the packing tape. (b) Liner removed from another bag is attached to the tape. A round object is rolled over it to squeeze away bubbles. (c) The tape is trimmed to size. (d) The tape, cut to the size and with the liner attached, is installed on the bag. (e) The installed fake tape has the correct width of about 30 mm.

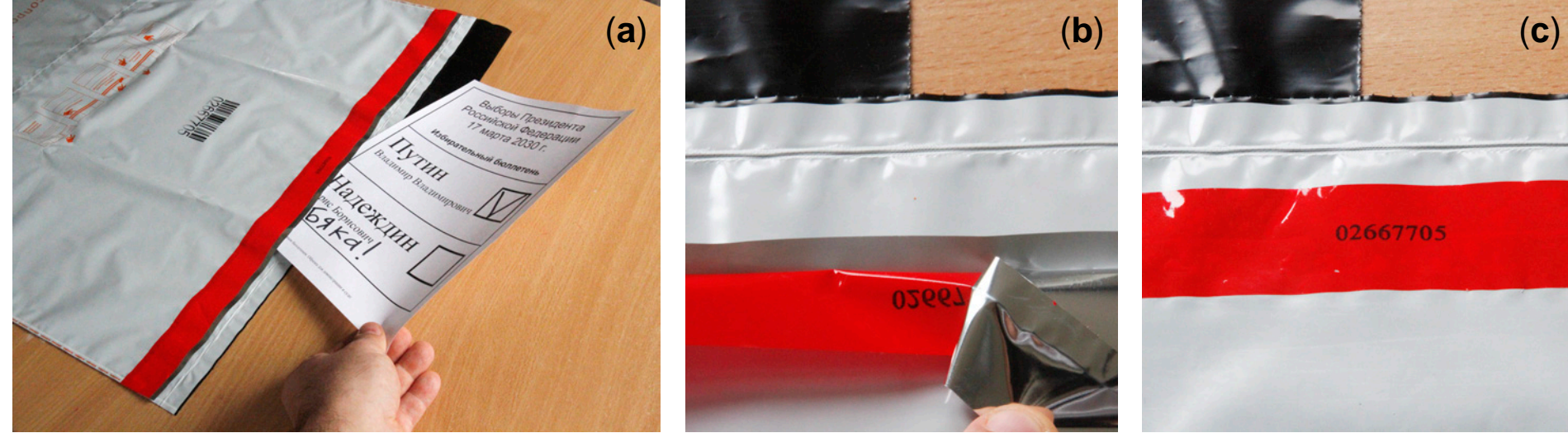

FIG. 10. Sealing ballots in the bag with the fake tape publicly. (a) Genuine ballots are placed inside the bag. (b) The liner is removed normally. (c) The bag is sealed with the tape normally.

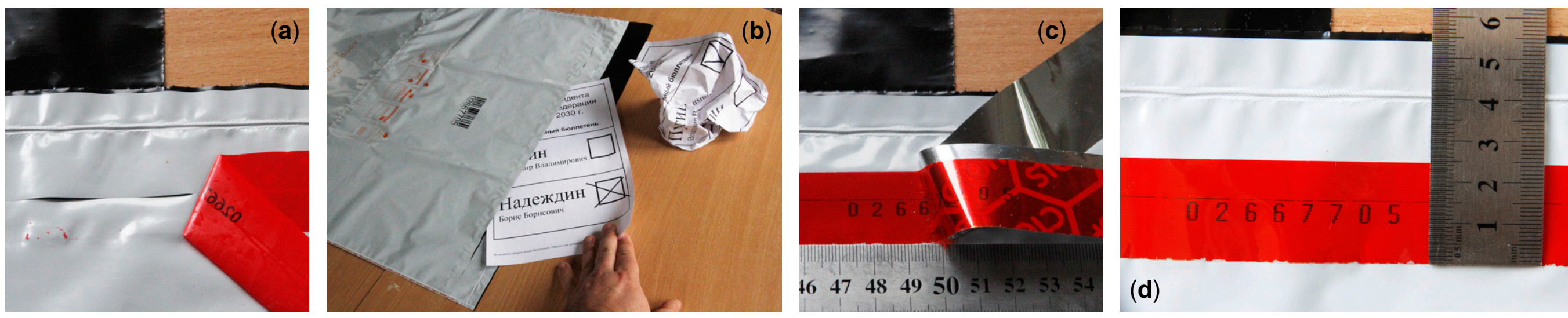

FIG. 11. Opening the bag and replacing the ballots. (a) The fake tape is easily peeled off. If any of its adhesive remains on the bag, it can be removed with solvent. (b) The ballots are replaced with fraudulent ones. (c) The bag is sealed with the stored original tape. (d) The installed tape is however less than 25 mm wide, which is the only visible sign of the tampering. Its uneven bottom edge is due to our slight inaccuracy when making this first sample; trimming it a millimeter narrower produces a perfect edge.

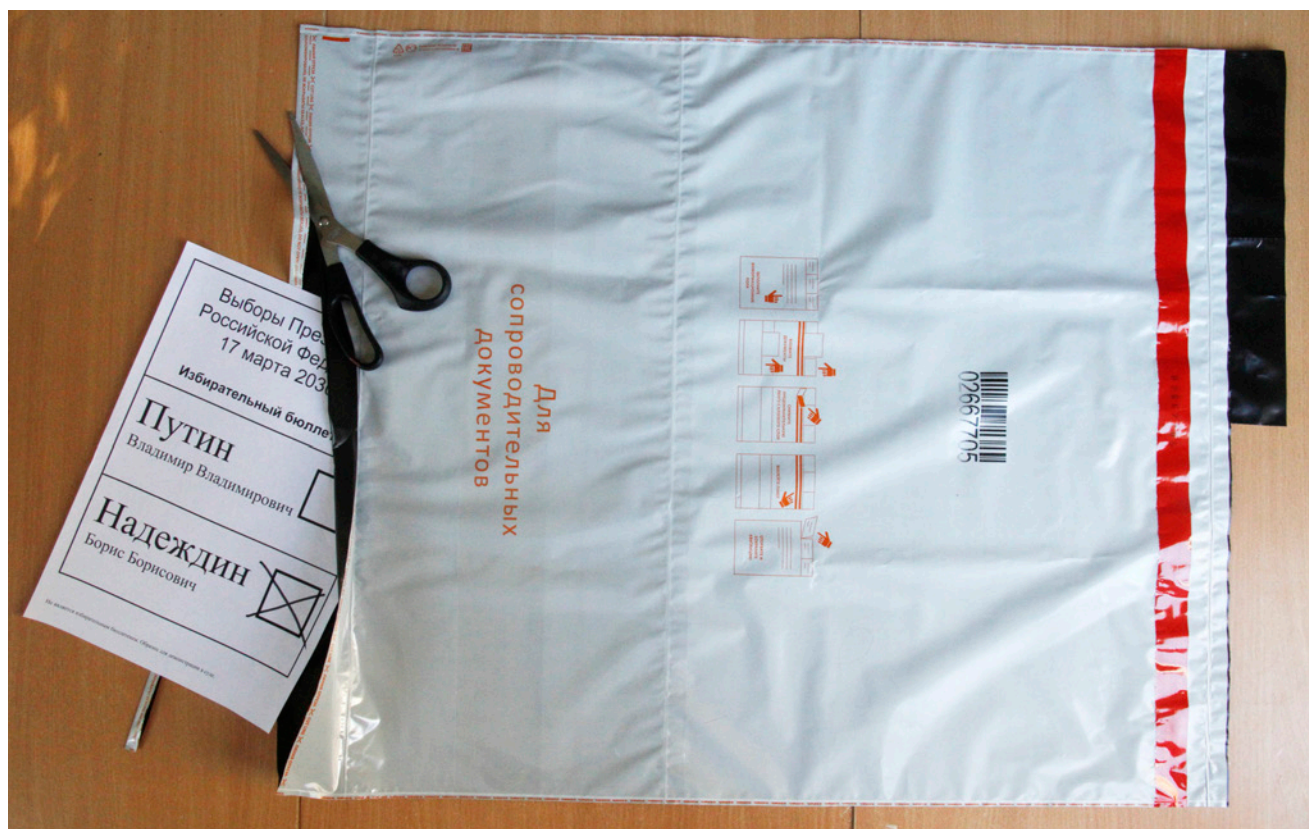

FIG. 12. The bag is opened publicly and the fraudulent ballots contained in it are counted as genuine.

### 5. Inconspicuous slit

The easiest way to access the content of the bag is to cut a carefully-placed slit in it. The slit is then taped over or heat-sealed [46]. Such a slit can always be detected by a meticulous visual inspection of the bag, which includes turning it inside out (after the extraction of its content) and examining its inner surfaces. Unfortunately, Russian election commissions often interfere with attempts to fully inspect the bag and its parts [45]. The slit may be deliberately covered with a paper document (such as a sealing certificate) placed in the outer document pouch of the bag [51]. Alternatively, the slit may be made virtually invisible from the outside by cutting it along a line of printed graphics, seam, or the security tape's edge. It is then closed with a transparent adhesive tape applied at the inner surface and is only readily visible from the inside.

Although this method is widely known, it has not been employed in Vlasikha.

## Methods D: Forensic evidence of bag and tape replacements

The only evidence remaining are photographic images and video recordings made during the voting days. The observers and candidates did not personally seize any physical evidence. Instead, they filed crime reports with the local police and Investigative committee requesting them to seize relevant items for inspection, as well as requested the judges in lawsuits to subpoena for the physical evidence. All these requests have been ignored or declined. All the physical evidence has by now been confirmed destroyed.

The photographs and video recordings discussed below have been taken by more that 15 people and 20 cameras (mostly those in smartphones). To keep our argument clear, we employ strictly uniform image transformations through this paper, such as their rotation and (in collages only) scaling along one of the dimensions. Color reproduction of objects varies with lighting conditions and camera used. The original camera files, exact time and location of each recording have been included in the crime reports and lawsuits filed.

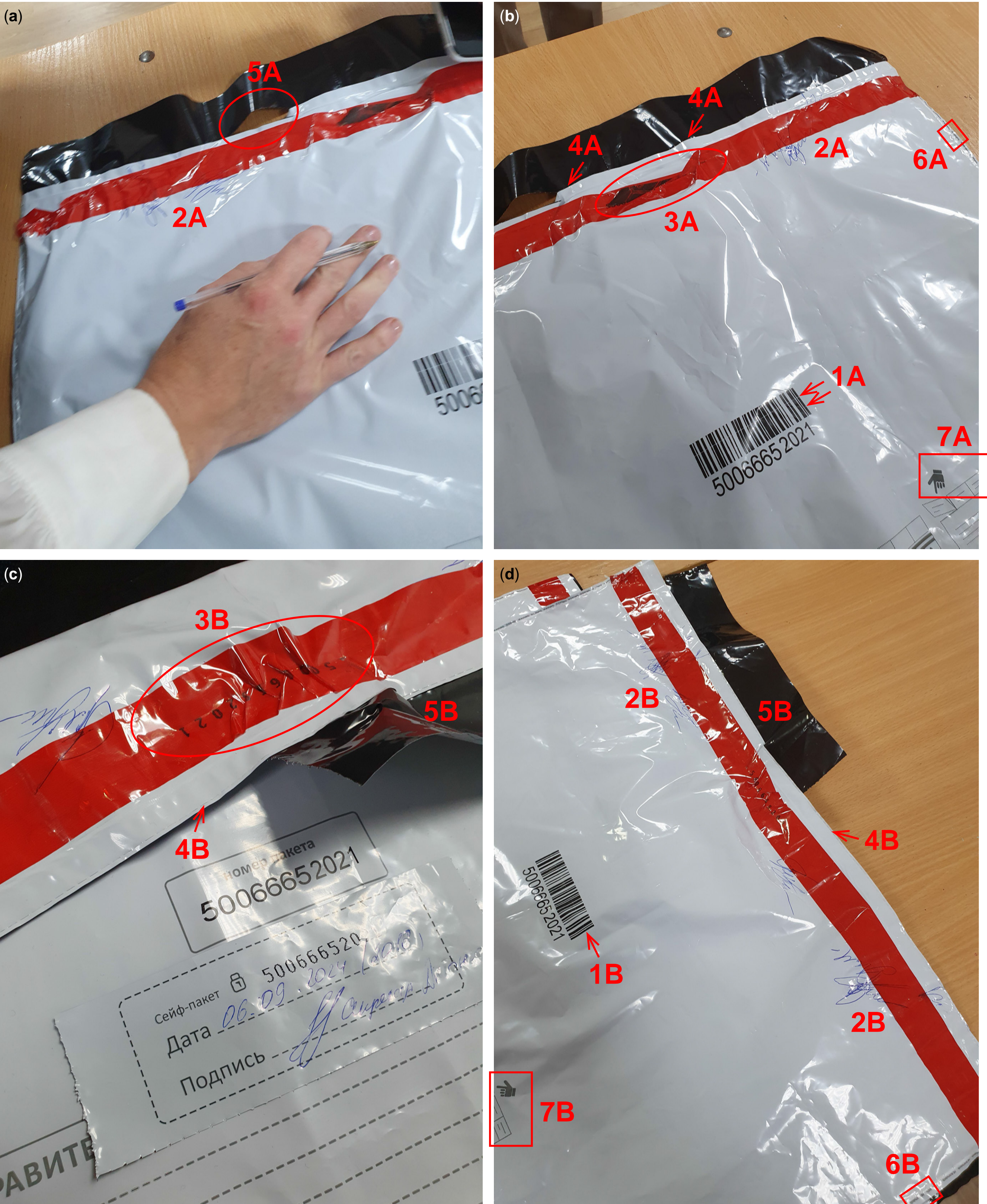

FIG. 13. Photographs of two duplicate bags with identical numbers at precinct 219. (a) and (b) Photographs of bag A taken by an observer on 6 Sep 2024 after ballots were sealed in it. (c) Bag B on 7 Sep 2024 while it was retrieved from the safe for inspection. (d) Bag B on 8 Sep 2024 before it was opened at the vote tallying. Multiple differences prove that not just pen signatures were forged (2B) over a new security tape (3B), but the entire bag A was swapped while in storage on the night of Sep 6 for its nearly-identical factory-made duplicate B bearing the same serial number, complete with the security tape.

### 1. Factory-made duplicate bag at precinct 219

At this precinct, the first bag was swapped for its factory-made duplicate during the first night. High-resolution photographs taken before and after the swap show several features on both bags that lead us to this conclusion (Fig. 13). First, the serial number and barcode printed on the bag is the same (Fig. 14); the barcode decodes correctly. The barcode shows a prominent dead-pixel line (and another fainter line) running across it, which fingerprints the machine and approximate date on which the bags were manufactured. A thermal transfer printer making this number and barcode is inside the machine that assembles the bag. Dead pixels producing white lines in this printer are common, their number growing from none to numerous in random places as the printing head ages. We have found images of barcodes of 5 other bags from the same batch [37] used all over Moscow region in September 2021 elections. They all show either a similar single white line or both lines (depending on the image resolution). Thus, the entire production batch [37] as well as our duplicate bags were made on the same machine.

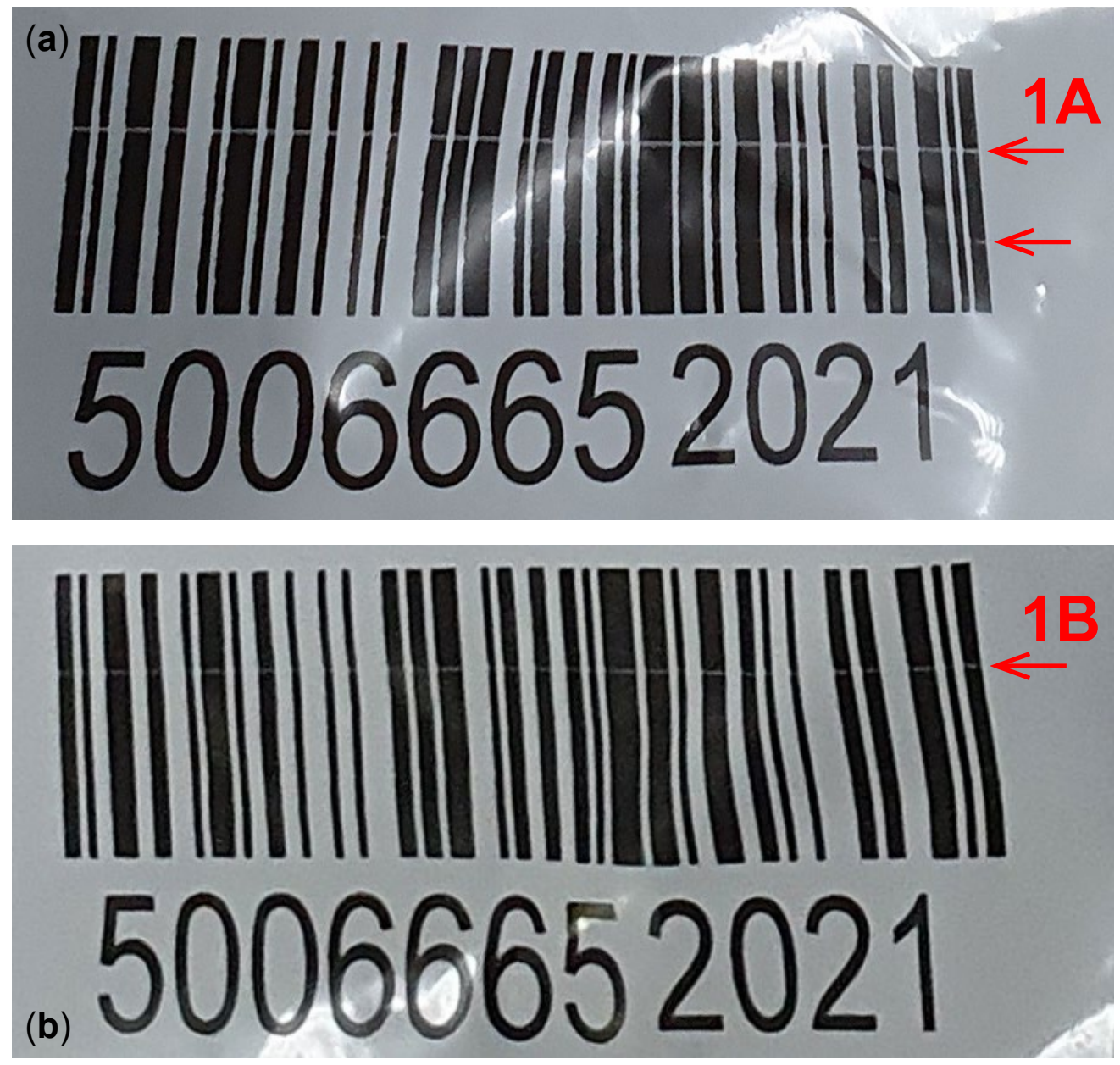

FIG. 14. Barcodes on both bags were printed by the same thermal transfer printer, as evidenced by the same position of a prominent dead-pixel white line. (a) Bag A. (b) Bag B. The second faint line is not visible owing to a lower resolution of the latter image.

Despite the same number and graphic design, these are different bags. A line of perforations that separates the control stubs from the bag punches through a single sheet of material in one bag but both sheets in another (Fig. 16). The second leftmost control stub is partially torn off along this line on 6 Sep [see feature (5A) in Figs. 13 and 16], yet is unseparated in the duplicate bag on the later dates [see feature (5B)]. We also have a video recording of the duplicate bag at the vote tallying, where it is extensively handled for several minutes displaying both sides to the camera, showing no partial separation of this control stub. Line art printed on the bags shows a different amount of ink bleed (Fig. 15). The small print at the bag's edge is further away from the edge on the duplicate bag (Fig. 18). These are all common sample-to-sample variations distinguishing the bags.

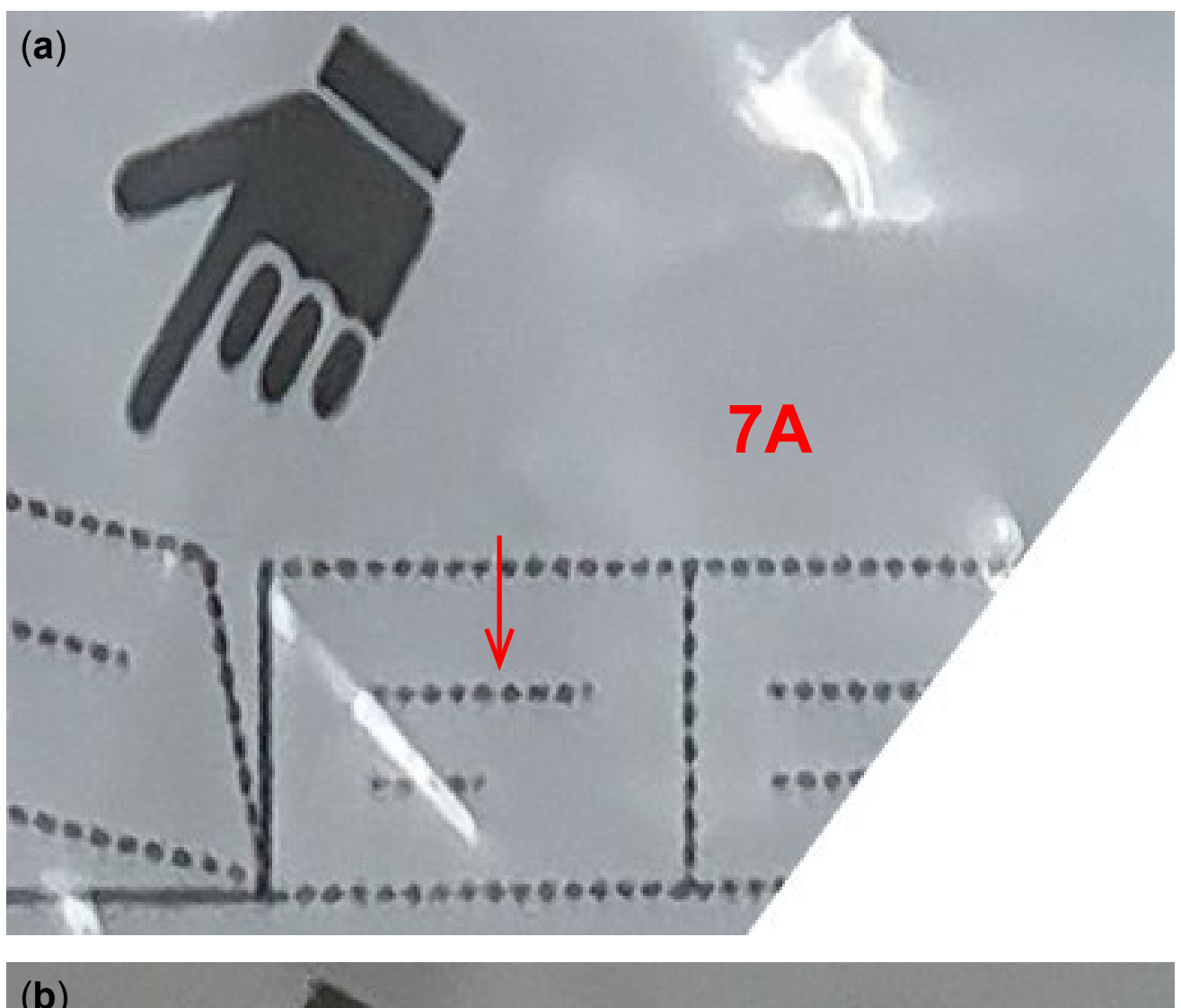

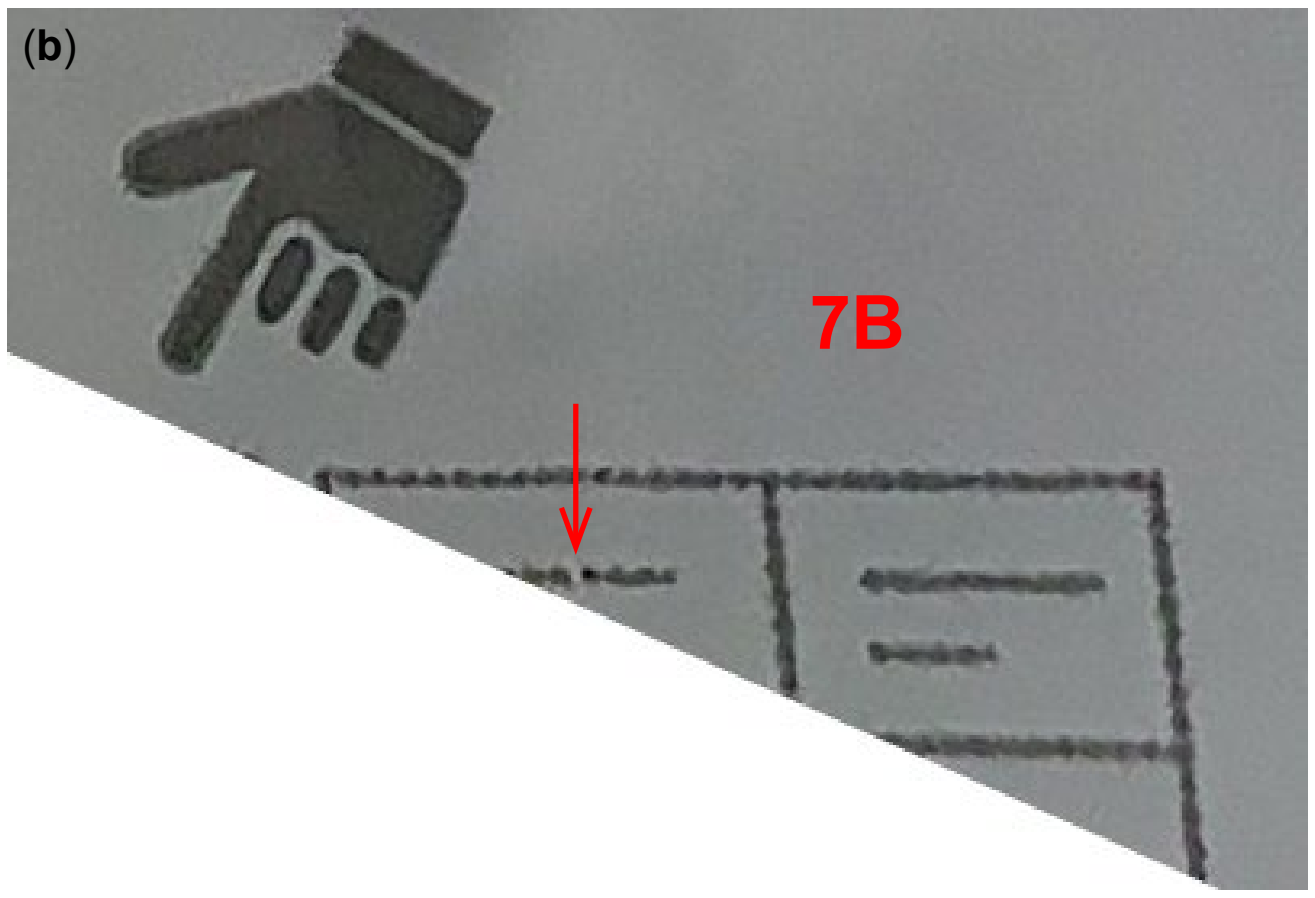

FIG. 15. Ink bleed in line graphics is (a) moderate on bag A keeps lines dotted but (b) stronger on bag B tends to convert them into solid lines. The arrow points at the sole gap between uniformly spaced dots that is not bridged with ink on bag B, which demonstrates that the solid lines in image (b) are in focus and are not a camera artefact.

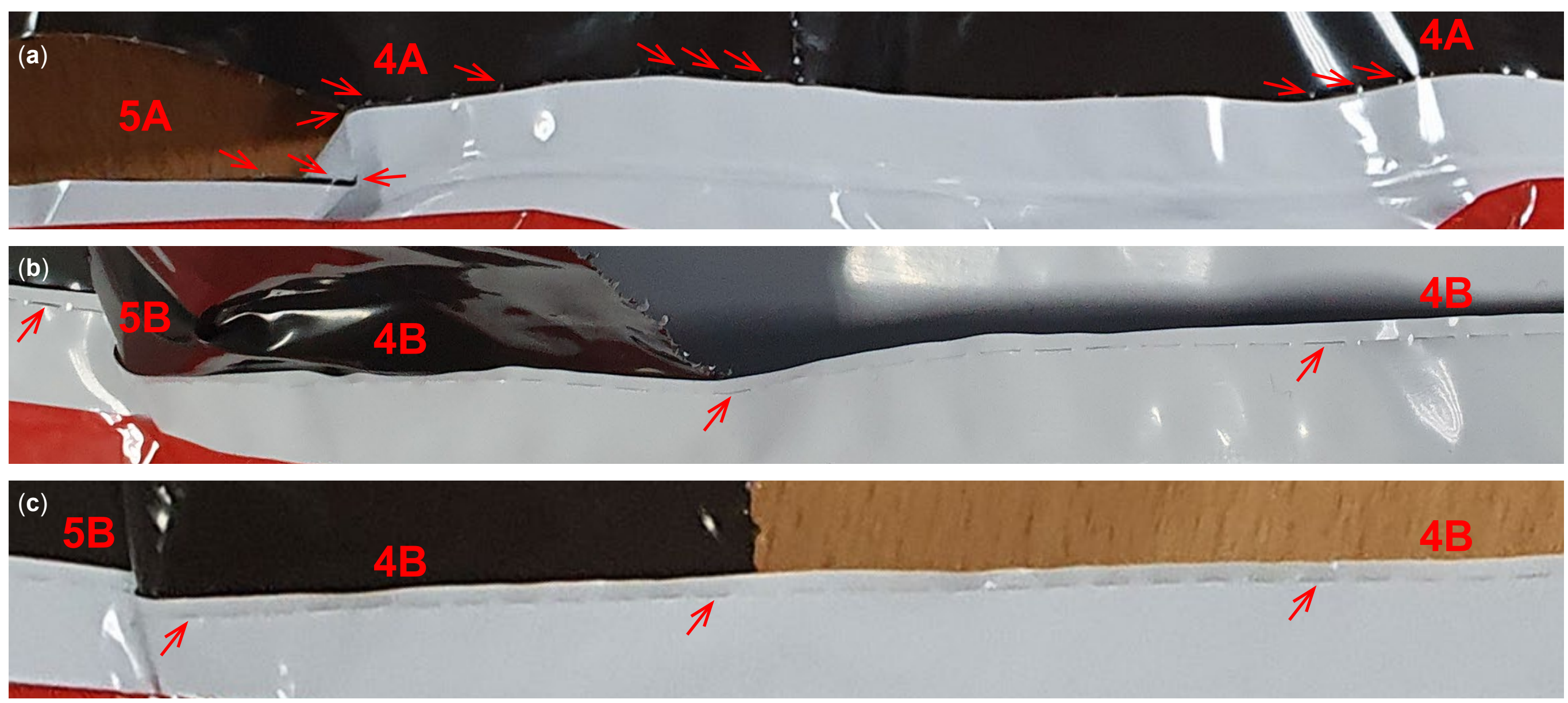

FIG. 16. A line of perforation (marked with arrows) separates the bag from its detachable control stubs. (a) In bag A, the perforation punches through a single sheet of material on which the stubs are printed. Their black side faces the camera. (b) and (c) In bag B, the perforation punches through both sheets of material, which is not an uncommon sample-to-sample variation. Only the top sheet is visible in the images, but the same perforation also pierces the bottom sheet, on which the stubs are printed.

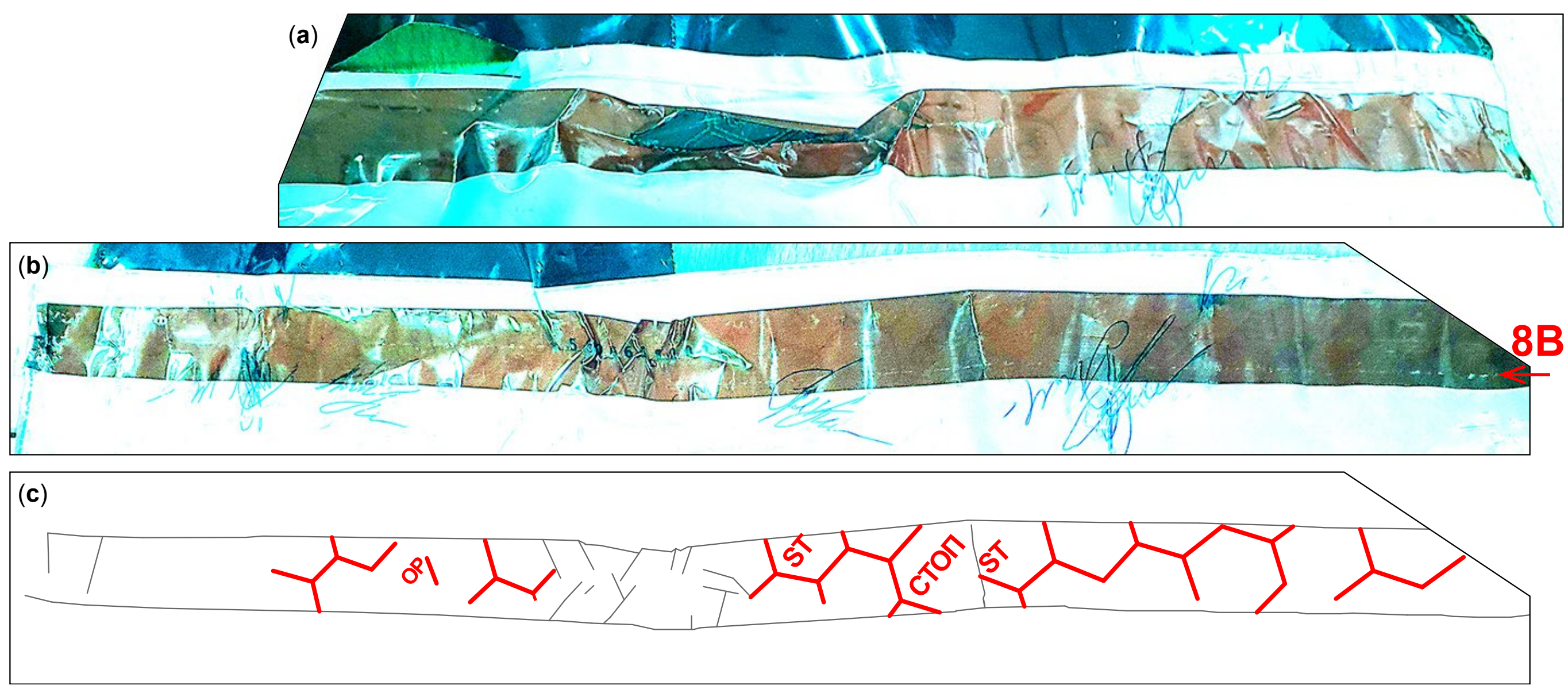

FIG. 17. Photographs of the tapes sealing duplicate bags at precinct 219 processed to maximise contrast. The latter is maximised individually in each primary color channel (red, green, and blue) using a curves tool, applied uniformly over the entire image. The image resolution is then reduced using bicubic interpolation with some sharpening, to further bring out subtle patterns. (a) Bag A. (b) Bag B. Compare the honeycomb pattern visible on the tapes with Figs. 6(a,b) and 7, whose presence here confirms both bags have genuine factory security tapes. For those having difficulty spotting the honeycomb pattern, it is outlined in (c). The size of the visible hexagons also confirms both tapes are about 30 mm wide. The presence of scuff marks (8B) about 5 mm from the tape's bottom edge indicates a normal sealing procedure, with the liner being removed slightly incorrectly [pulled from under the tape as shown in Fig. 7 instead of bending the tape over for easy liner removal as recommended by the manufacturer and shown in Fig. 6(b)]. This tears the adhesive slightly along this line. These marks are typical in normal use, because many people intuitively remove the liner this way instead of reading factory instructions.

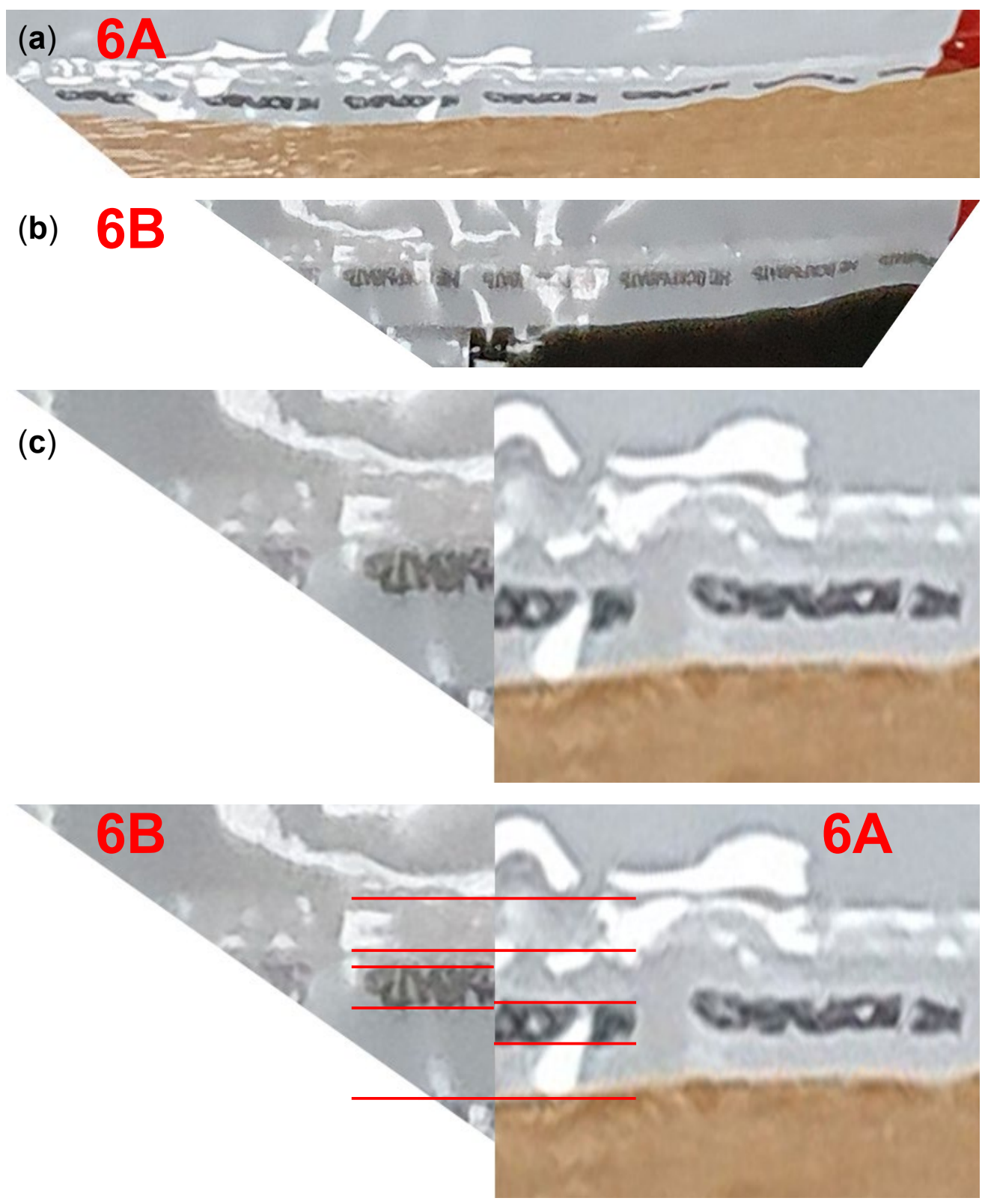

FIG. 18. Position of the small print relative to the bag edge and the seam in (a) bag A and (b) bag B. (c) A collage shows the small print to be at a different distance from the edge on the two bags. In the collage, the image of bag B, taken by the camera at an oblique angle, is uniformly stretched vertically to match that of bag A.

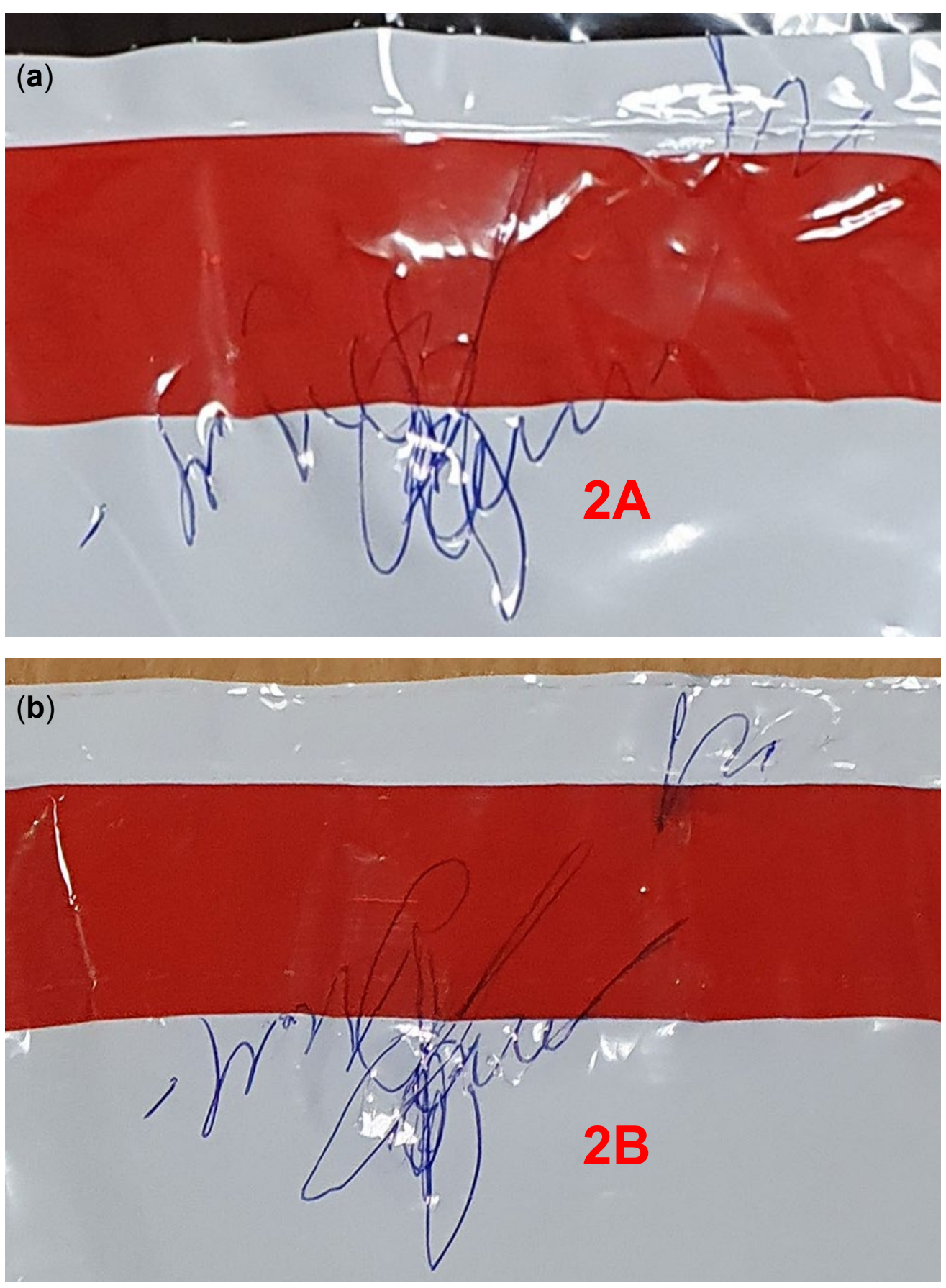

FIG. 19. Signatures of observer Ushakov and candidate Bulkina (who superimposed her signature over his) are (a) original on bag A, according to Ushakov's report, and (b) forged on bag B. Several other signatures are also forged.

As expected, handwritten signatures on the bags are also distinct (Fig. 19). The second set of signatures had been forged, according to reports by observer Ushakov and candidate Razina who both signed the bag on 6 Sep. The tapes themselves are stuck fast in different positions with different creases. Yet both are genuine, untampered security tapes, as evidenced by the presence of faintly visible honeycomb pattern with words STOP СТОП and their correct 30 mm width (Fig. 17). The scuff mark (8B) on the second tape suggests it came attached to the bag and was sealed normally.

The most intriguing feature of this pair of bags is the serial number printed on the tape. While the first number is only partially visible, it is clearly typeset normally with small spacing between the digits, while the second tape bears a sparsely-typeset number (Fig. 20). Why on Earth would the manufacturer, making duplicate bags on the same machine, leave us such an obvious clue? We suppose the most likely answer is that two production batches with the same set of numbers were made on this machine not too far apart in time for the same customer. One of these batches contained 51000 bags ordered by the Election commission of Moscow region [37] and another an unknown number of bags with overlapping but *differently typeset* numbers. Unfortunately, we see several other pairs of red tapes with identical but differently typeset numbers at our other precincts.

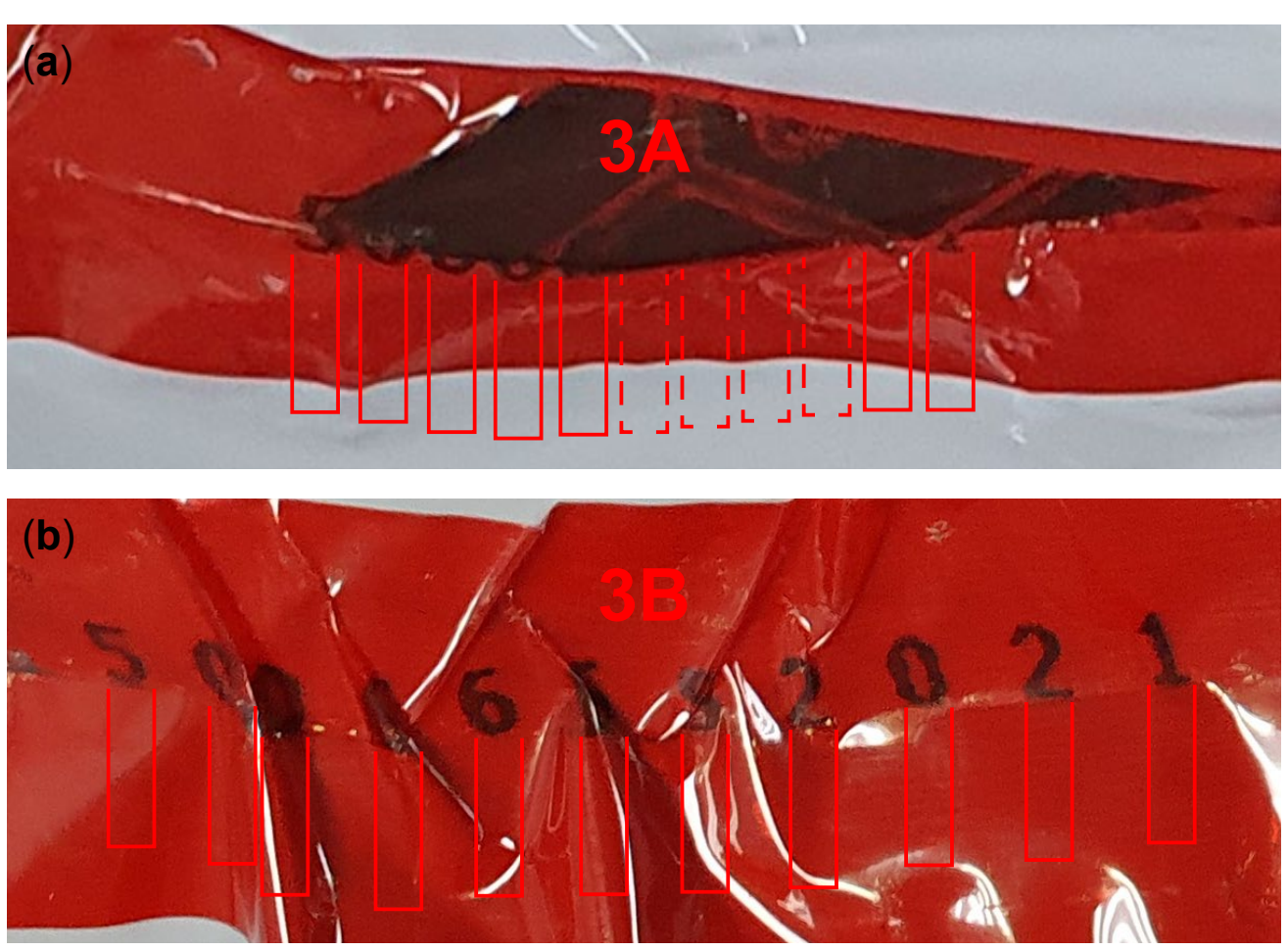

FIG. 20. 11-digit number on the red tape of (a) bag A, where seven digits of it are partially visible, and (b) bag B. The number on bag A is normally typeset, with the inter-digit spacing of less than half the digit width. That on bag B is sparsely typeset, with the inter-digit spacing of about one and a half digit widths.

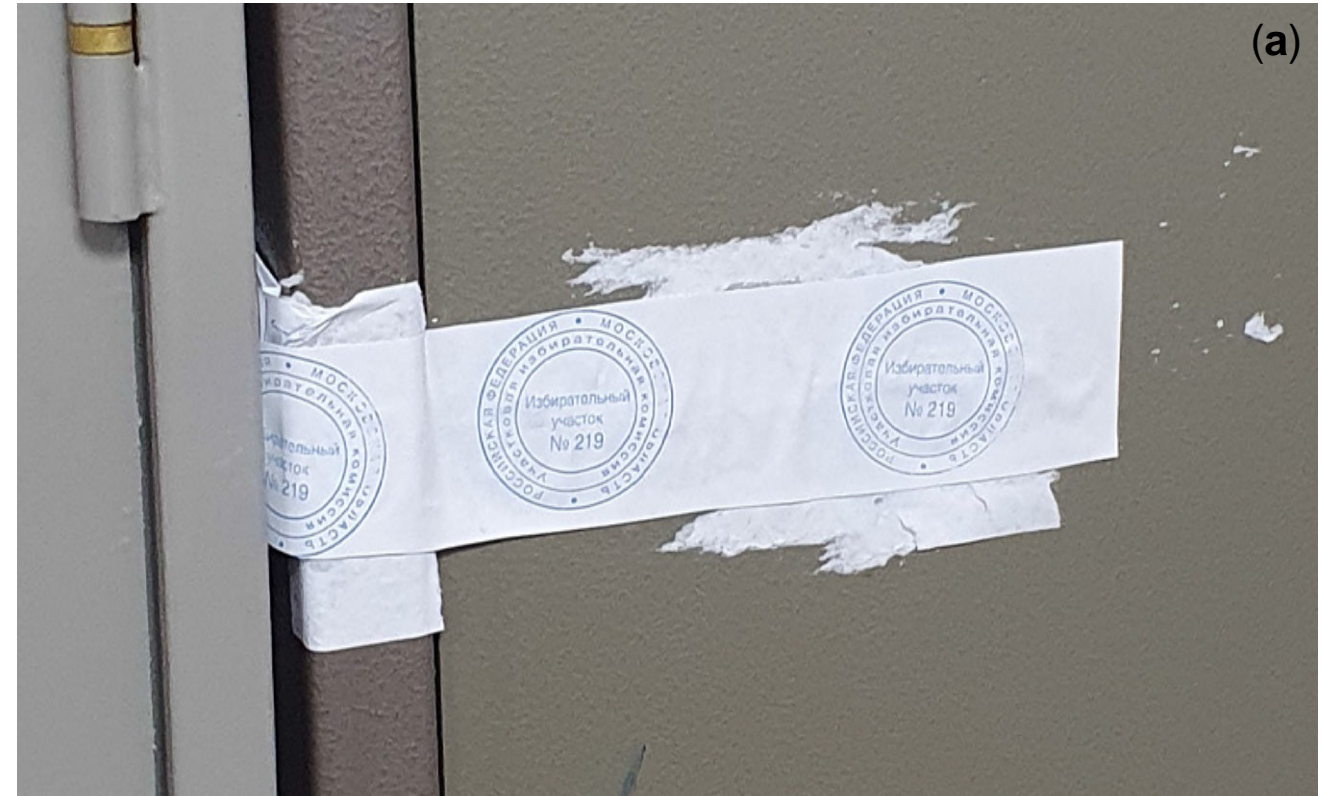

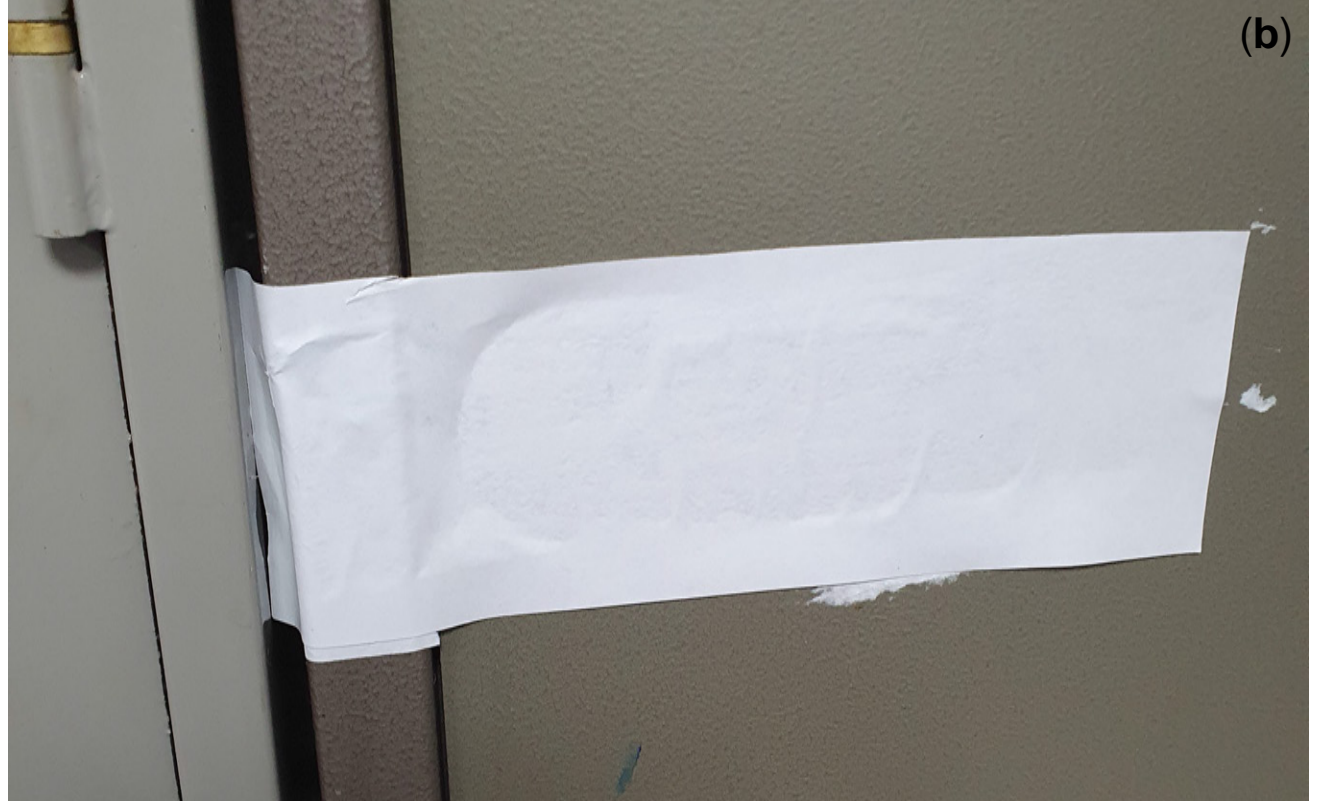

FIG. 21. Safe at precinct 219 (a) before locking the room for the night on 6 Sep and (b) in the morning next day immediately after opening the room. The paper tape on the safe bearing impressions of the precinct's seal had been crudely replaced during the night, while the station was closed.

Other evidence at this precinct includes a replaced paper seal on the safe during the first night (Fig. 21). The room of the polling station was not properly sealed for the night, its front door being sealed diligently but back door merely locked with a key. The safe and the doors were locked by the precinct secretary when closing the station for the night. The recording from an outdoor security camera shows her then waiting on the street for the observers to leave, chatting with a night guard, and returning briefly to the back entrance of the building.

The replacement of the paper seal on the safe (but not the rest of our evidence) was discovered in the morning of the second day and attracted unwanted attention. This might have led to scrapping plans to replace the second bag.

### 2. Replacement of security tapes at precinct 212

At this precinct, security tapes on both bags, originally bearing narrowly-typeset numbers, were replaced at night with tapes bearing sparsely-typeset numbers (Sec. C 3). This was done presumably to avoid forging multiple pen signatures adorning both bags on both sides. The replacement tapes are narrower (Figs. 1 and 22). The small print along the bag's side edges had disappeared near the tapes. Amusingly, the perpetrators at this precinct used a wrong type of solvent that not only removed this factory print, but smeared the adhesive and made the bags very sticky. At the vote tallying, both bags emerged from the safe folded in half, such that the area around the tape was in contact with the bottom of the same bag. These stuck together and had to be separated by an election official with a visible effort, producing a loud tearing sound recorded in our video (Fig. 23). The original tape on the first bag shows the honeycomb pattern indicating it is genuine, as well as its number matching that on the bag [Fig. 22(b)]. The images of the other three tapes are of an insufficient resolution to verify this.

Other evidence at this precinct is a paper seal on the safe shifting its position slightly (by about 1 mm) during the second night.

### 3. Precinct 213

No images of the first bag were recorded.

The second bag displays a tape of the normal width after sealing it on the second day, yet at the vote tallying the tape is narrower and distorted [Fig. 24(a)–(d)]. Its distance from the bag's top edge varies greatly [Fig. 24(d)], suggesting a botched application of the replacement tape. One of its ends folds over to the other side of the bag, forming a rectangular piece not less than 4 mm long there [Fig. 24(e)]. This is impossible in normal use. At the factory, a continuous band of assembled bags with the tape installed is separated into individual bags along this edge by a cutter, which is the final manufacturing step [52]. I.e., the tape's end is always flush with the bag's edge and cannot move in normal use. Indeed, this edge of the bag imaged immediately before its sealing shows no tape extending on the other side [Fig. 24(f)].

### 4. Precinct 214

At this precinct, the cumulative length of three continuous and almost-continuous sequences in the transcript is 273 ballots (Fig. 3). This suggests that both bags were tampered with, as they together contained 285 ballots (Table I). Unfortunately, available images of the bags are of an insufficient resolution to locate the numbers on their tapes. The overall geometry of the first bag's tape does not change overnight, suggesting the tape alone was not replaced (but not ruling out swapping the bag). On the second bag, the tape geometry had visibly changed, becoming narrower and unevenly applied (Fig. 25). The tape's distance from the top edge is uniform on the second day but visibly varies along its length and is larger at the vote tallying. Despite limited resolution of these images, the difference in the tape width is real and measures both in a side-by-side comparison [Fig. 25(f)] and using the height of the barcode as a reference. Thus, the tape was replaced (Sec. C 3).

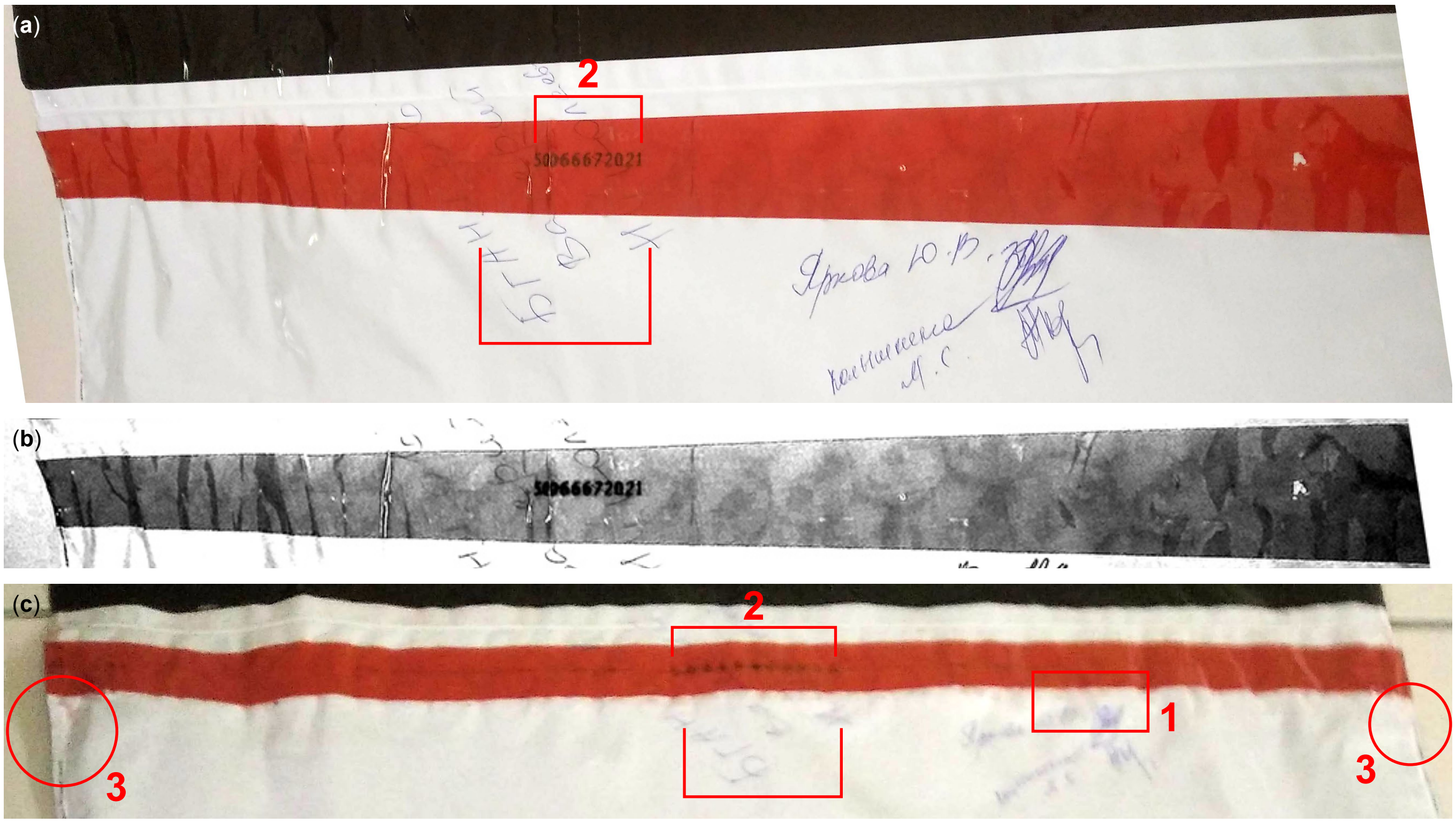

FIG. 22. Red sealing tape on the security bag at precinct 212 (a) after sealing the bag on the first day, (b) red color channel from the same image with its contrast maximised, showing the honeycomb pattern on the tape, and (c) before opening this bag during the vote counting on the third day. Note (1) reduced tape width and shifted placement relative to the pen signatures, (2) increased intersymbol space and overall width of the 11-digit number, changing from normally-typeset to sparsely-typeset, which is also apparent in comparison with the width of the nearby pen signature, and (3) disappeared factory-printed line of text at the bag's edges, presumably removed with solvent.

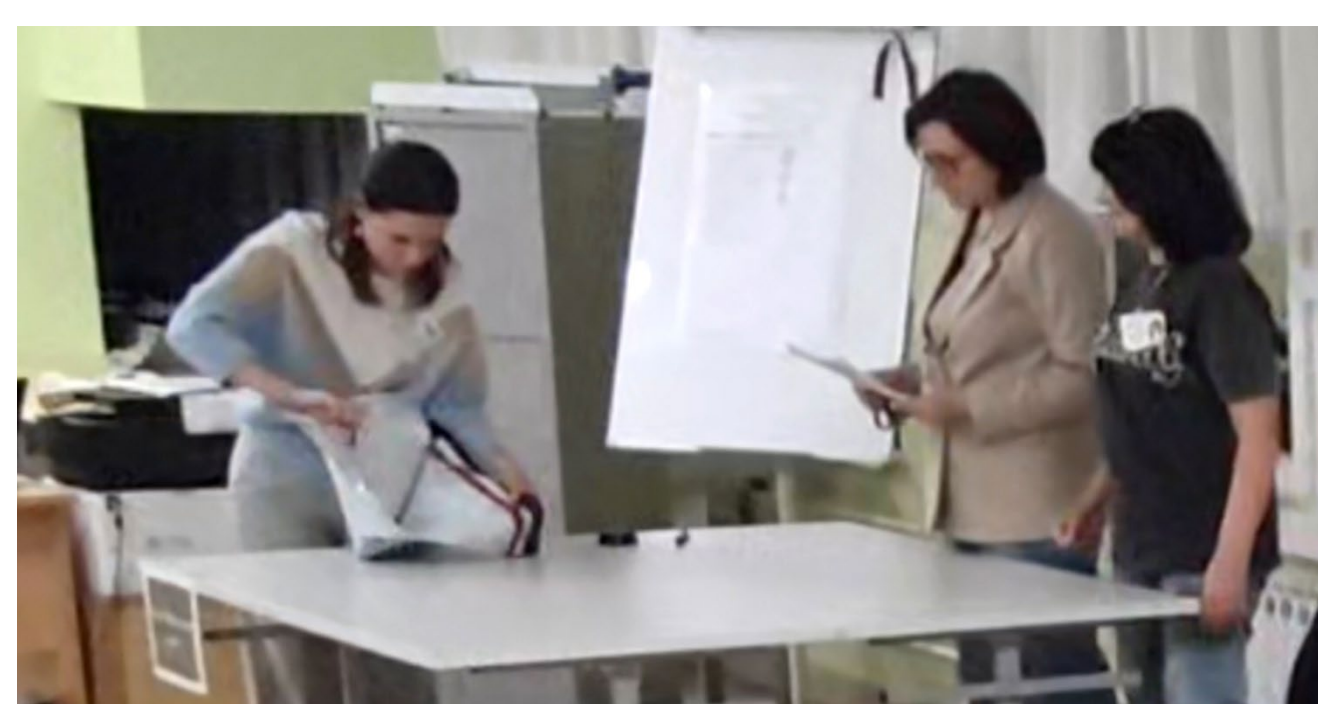

FIG. 23. The secretary of precinct 212 is applying force to separate the sticky area around the tape clinging to the rest of the bag, while the chair and vice-chair of the precinct are calmly looking.

Other evidence at this precinct is that its election officials were entering and leaving the building several times after the precinct closed for the first night and entered the room well before the legal observer admission time (i.e., before 7:00) in the morning. They were there alone.

### 5. Precinct 215

At this precinct, unlike all the other precincts, bags bearing 8-digit numbers were used to store ballots. We do not see any change in their tape geometry when we compare their high-resolution photographs taken on the first two days with a low-resolution video recording made at the vote tallying. Factory-made duplicates of bags and tapes with 8-digit numbers are not yet known to exist. The commission did not intentionally mix the ballots. The vote counting transcript and its statistical analysis (Sec. IV) do not indicate significant irregularities. We conclude this precinct did not tamper with the ballots.

Despite no tampering, we could not help but notice that this precinct cheated openly at the vote counting in favor of the ER nominees. They recorded 20 more votes for Fedoriv and 16 more votes for Schik [31] than are found in our transcript made from a complete video recording. Our commissions truly are obstinate.

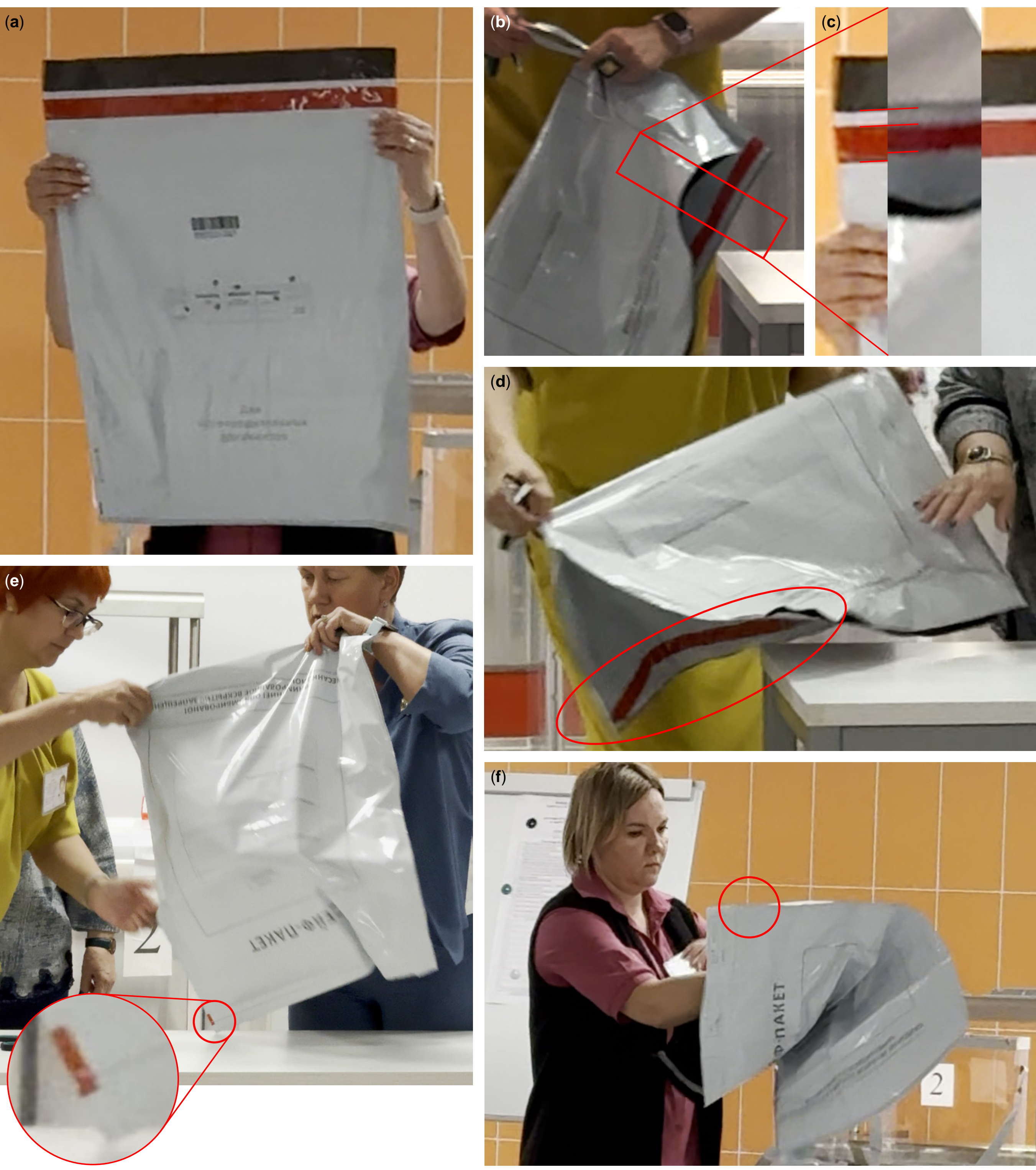

FIG. 24. Frames from video recordings of the second security bag at precinct 213 (a) after sealing the bag on the second day, (b) and (d) after opening this bag during the vote counting on the third day. (c) A collage between fragments of (a) and (b) illustrates that the tape width and/or placement have changed during the night along its entire length. The tape width in image (a) measures about 31 mm, using 20-mm height of the barcode as a reference [52]. The same tape width is obtained using the bag height of 695 mm (not including the detachable control stubs) as a reference. (e) The other side of the bag after opening it shows a rectangular end of the tape folded over to this side (see inset). (f) Before closing the bag on the second day, no tape is visible at this side.

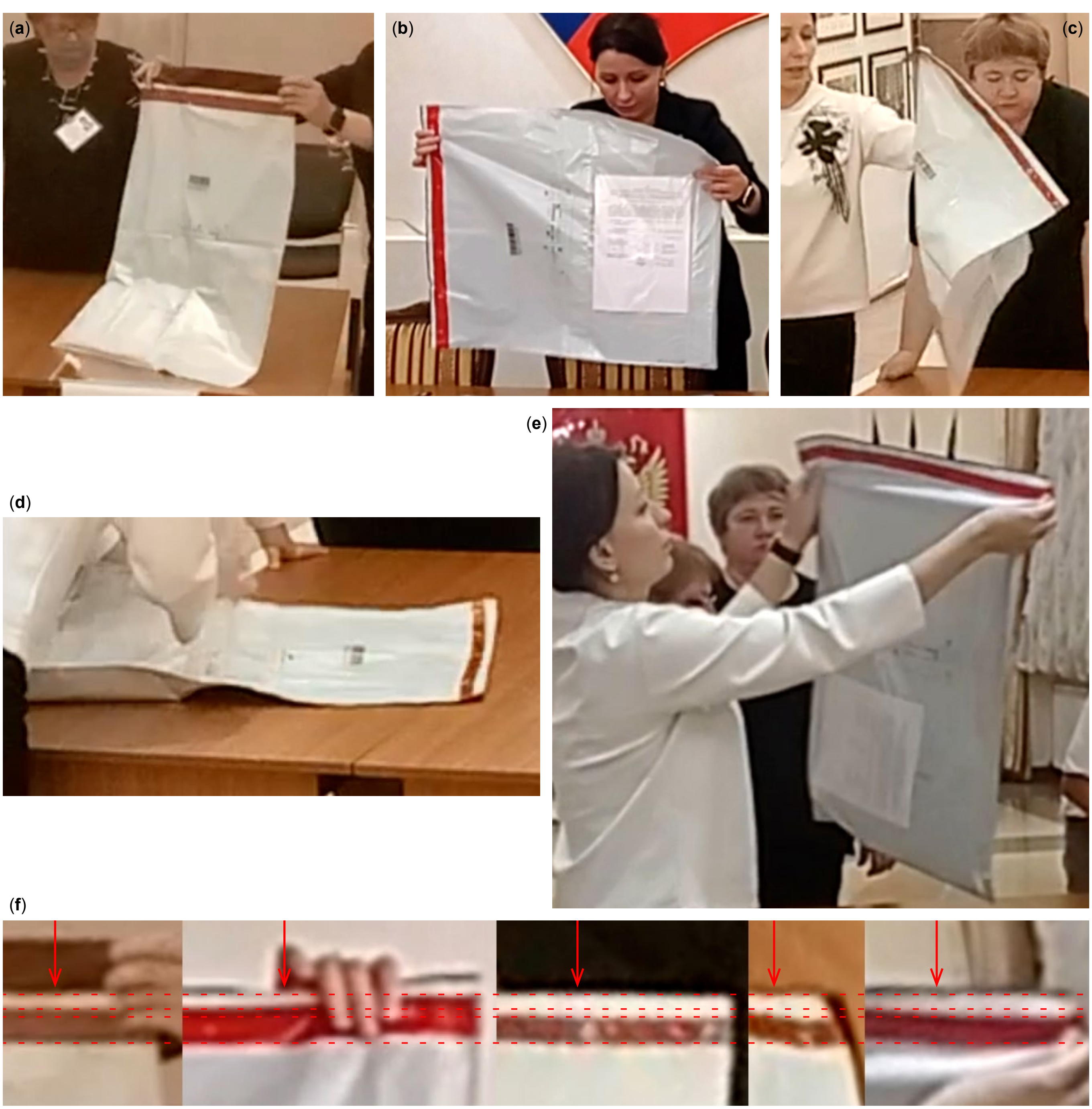

FIG. 25. Frames from video recordings of the second security bag at precinct 214 after sealing the bag on the second day made by (a) camera 1 and (b) camera 2, then before opening this bag during the vote counting on the third day made by (c) camera 1, (d) camera 1, and (e) camera 2. (f) A collage between rotated fragments of (a)–(e) shows the changed tape width and distance from the edge, as well as consistency of our geometry estimation between different cameras and viewing angles. The arrow denotes a point about 1/4 the bag width from its right edge in each collage segment. The tape width can also be compared to the height of the barcode visible in the images, which is 20 mm, according to the factory specification [52]. Using that as a reference, the tape width in images (a) and (b) measures no less than 30 mm, while it is close to 20 mm in images (c) and (d).

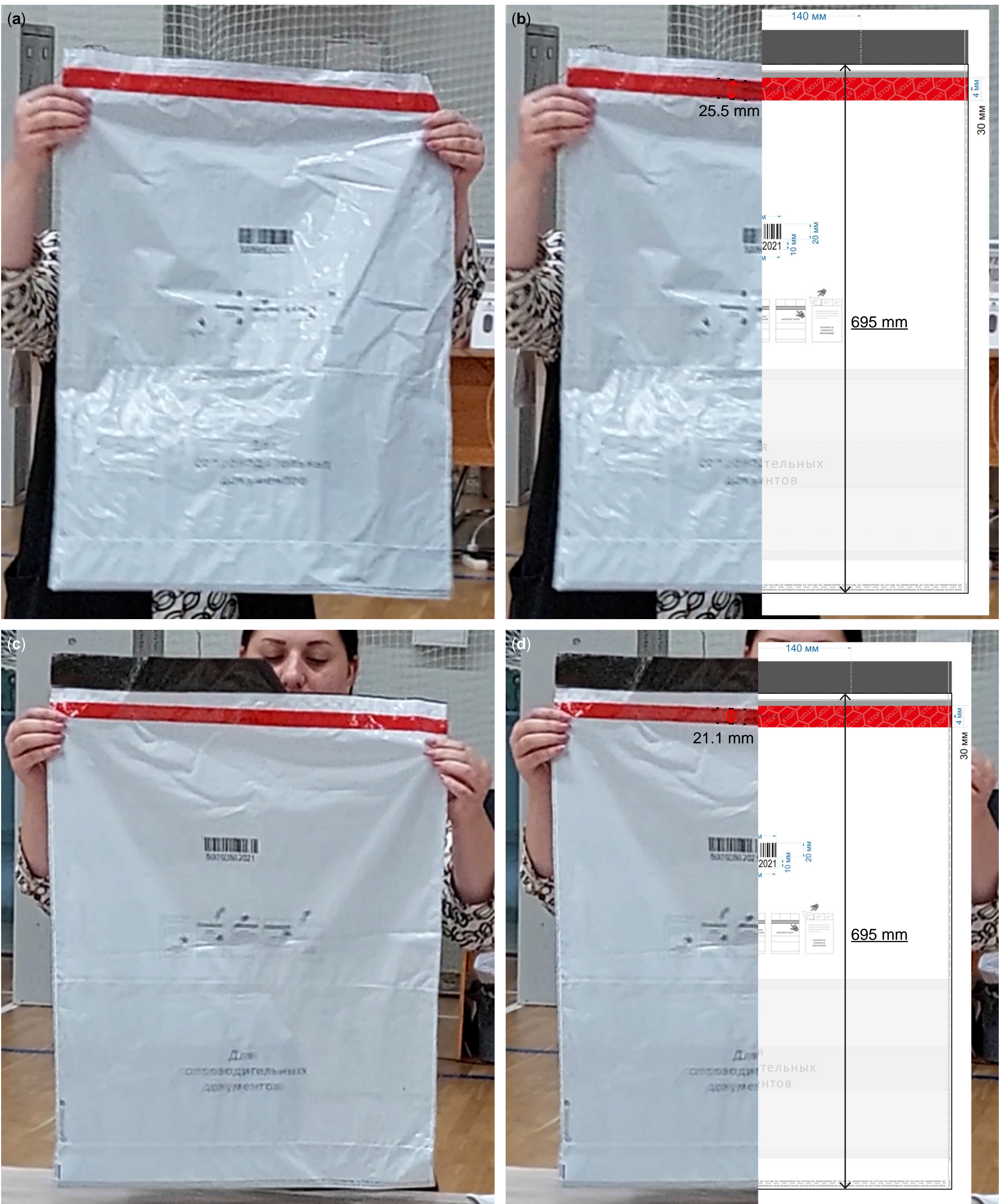

FIG. 26. Frames from a video recording of the security bags at precinct 216 before opening them at the vote tallying. (a) The first security bag, (b) with half the engineering drawing superimposed. The tape width measures 25.5 mm using the specification dimensions of the bag as a reference. (c) The second security bag, (d) with half the engineering drawing superimposed. The tape width measures 21.1 mm using the same reference.

## 6. Precinct 216

At this precinct, available video recordings from the first two days are of too low resolution to measure the tape width (Fig. 27). They only leave a subjective impression that the tapes are of a normal width. However, at the vote tallying both bags are displayed flat on the camera in good resolution. This allows us to compare their undistorted geometry with an engineering drawing of this production batch provided by the factory [52] (Fig. 26). The tape width measures 25.5 mm (21.1 mm) on the first (second) bag, indicating the replacement of both tapes. A glimpse of the number on the first replacement tape, visible in a few frames, shows it sparsely typeset (Fig. 28).

Oddly, this precinct cheated openly against the ER candidates during the vote counting. They accrued 12 fewer votes for Fedoriv and 8 fewer for Schik [31] than we count in our transcript.

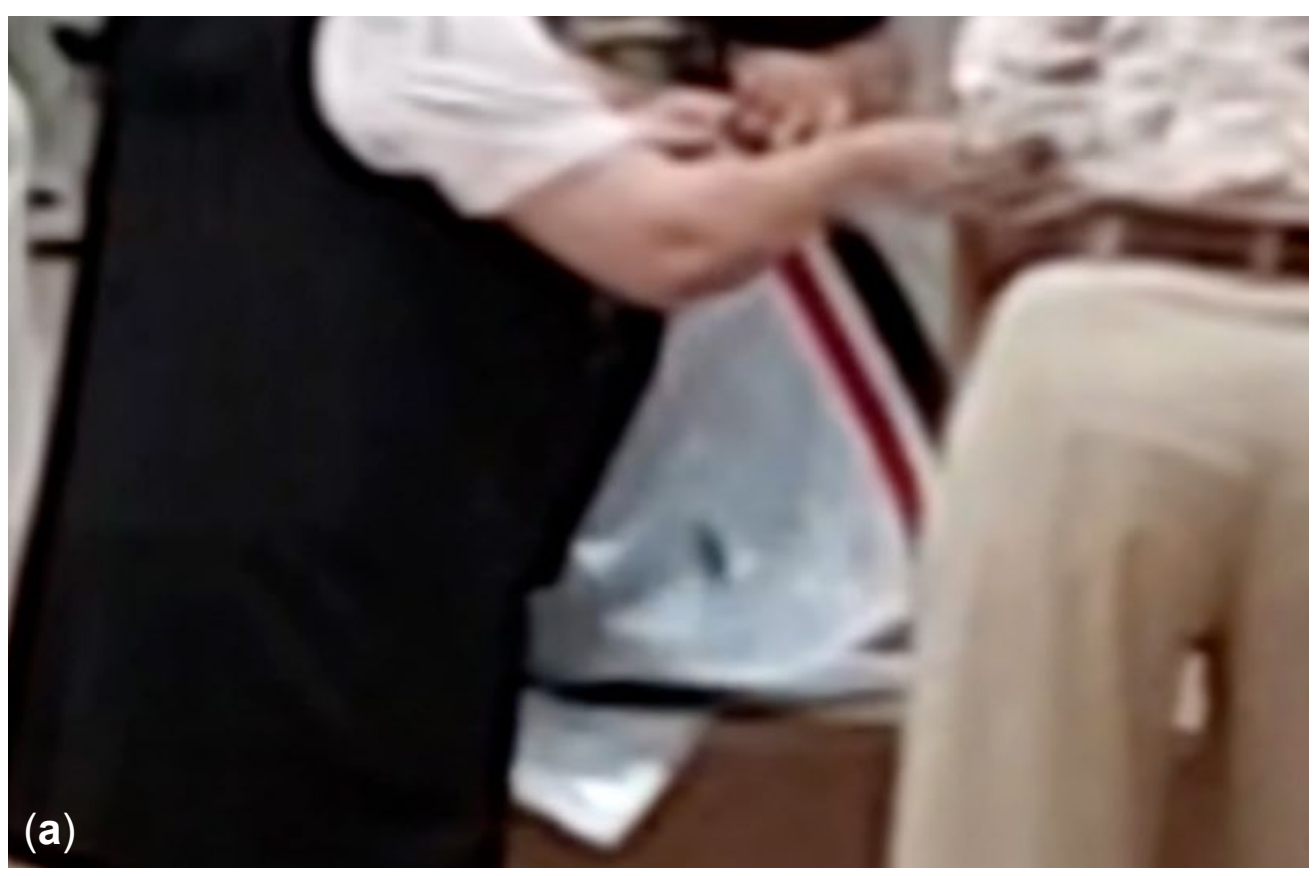

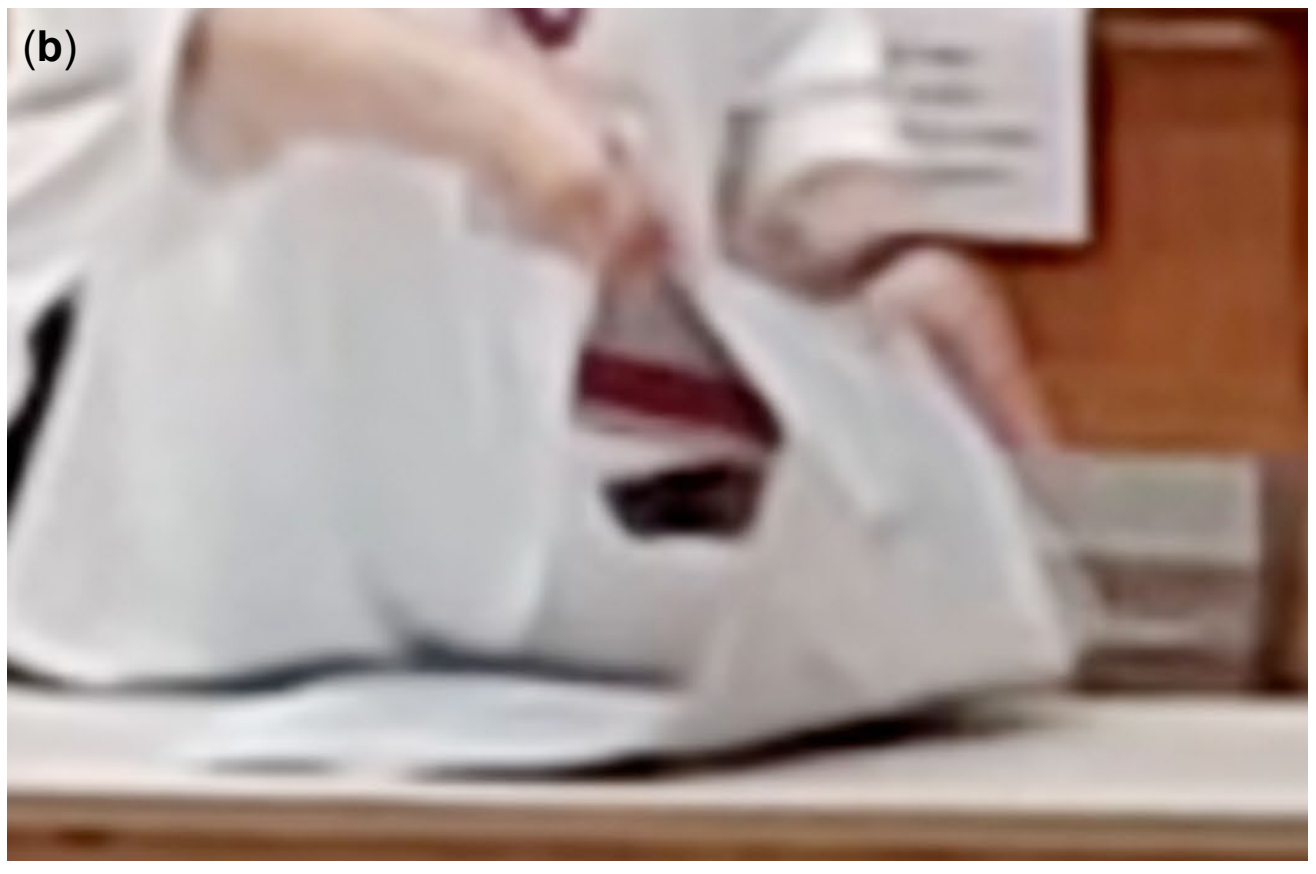

FIG. 27. Frames from video recordings at precinct 216 of (a) the first security bag after sealing it on the first day and (b) the second security bag after sealing it on the second day.

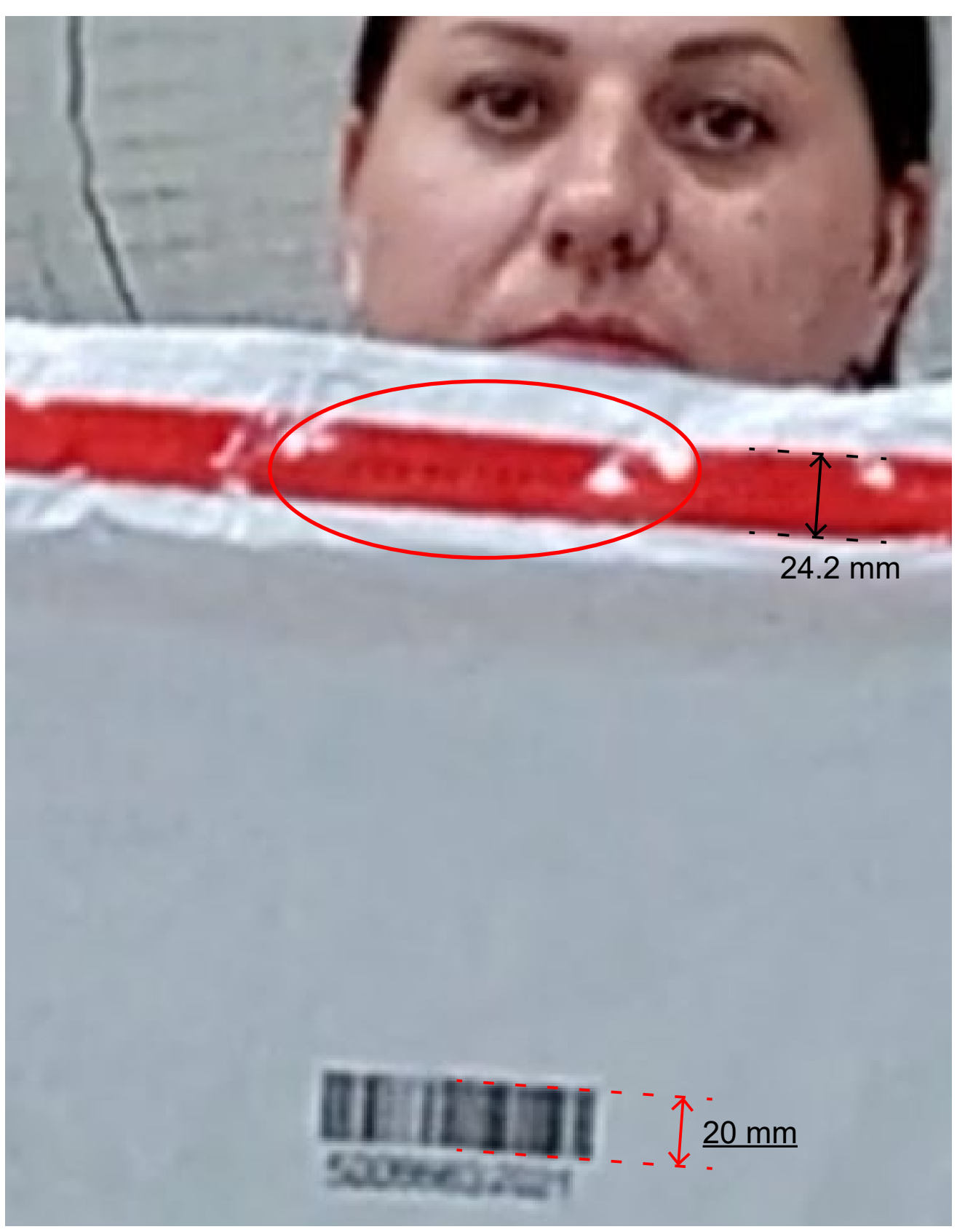

FIG. 28. Frame from a video recording of the first security bag at precinct 216 before opening it at the vote tallying. Ten out of eleven digits of the sparsely-typeset number are faintly visible on the tape. The tape width can also be compared to the height of the barcode, which is 20 mm, according to the factory specification. Using that as a reference, the tape width measures 24.2 mm.

## 7. Precinct 217

Similarly to precinct 216, both tapes were replaced. Using the bag height as a reference, we measure the tape on the first bag to be less than 24.9 mm wide [Fig. 29(a)]. The latter image has a notable amount of perspective distortion, being a significant portion of a wide-angle camera frame. This distortion visually compressed the bottom part of the bag in the measurement direction, thus overestimating the tape width. A more accurate measurement would be possible if we applied a perspective-correction transformation to the image. However there is no need to resort to it, because the tape width already measures well under 30 mm. Using the 20-mm-high barcode as a reference—which suffers from similar but lesser perspective distortion—the tape measures narrower than 24.7 mm. A collage [Fig. 29(b)] visually illustrates this as well.

The second bag was only imaged at the vote tallying while it was being extracted from the safe (Fig. 30). Using its barcode as a reference, the tape width measures

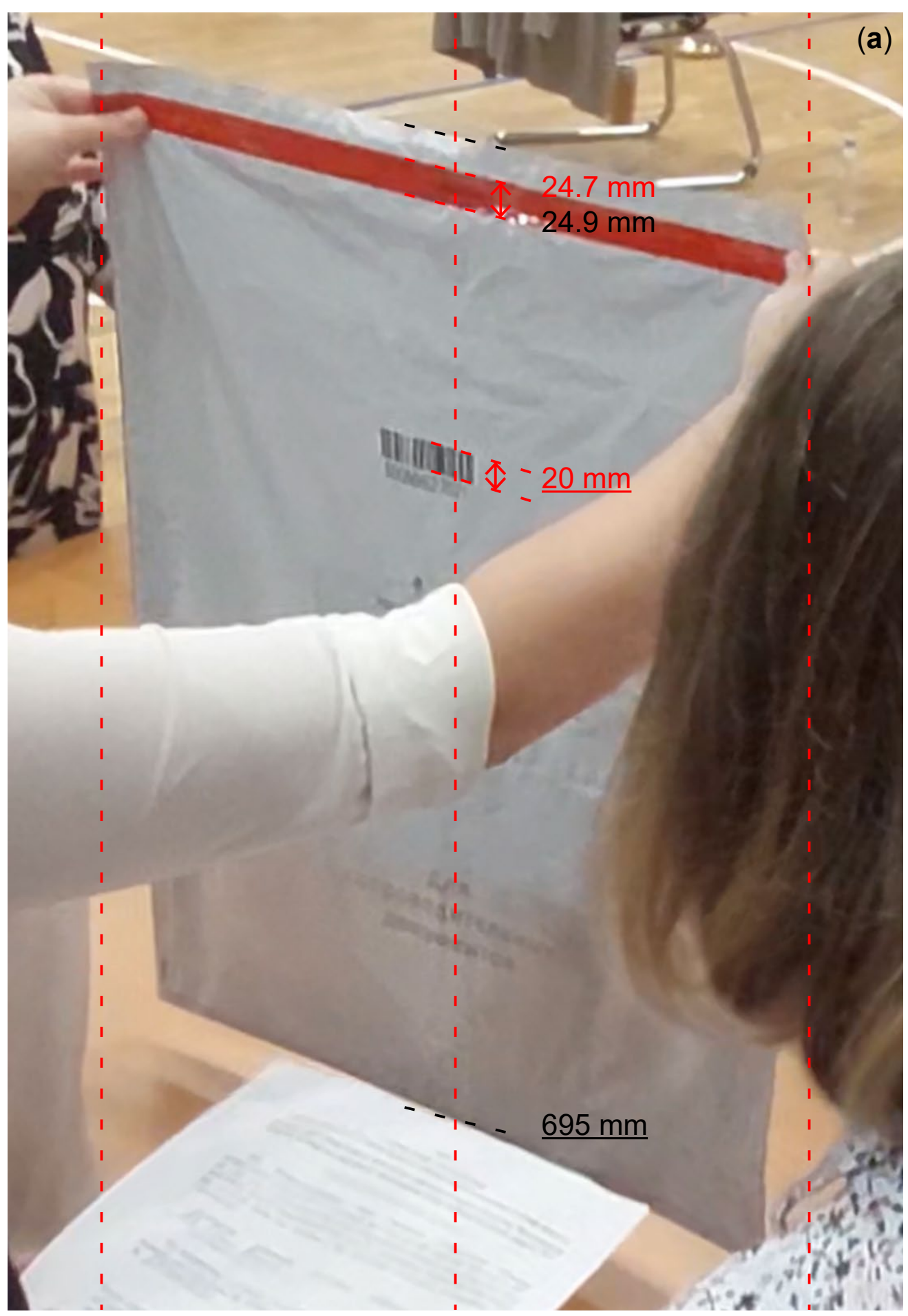

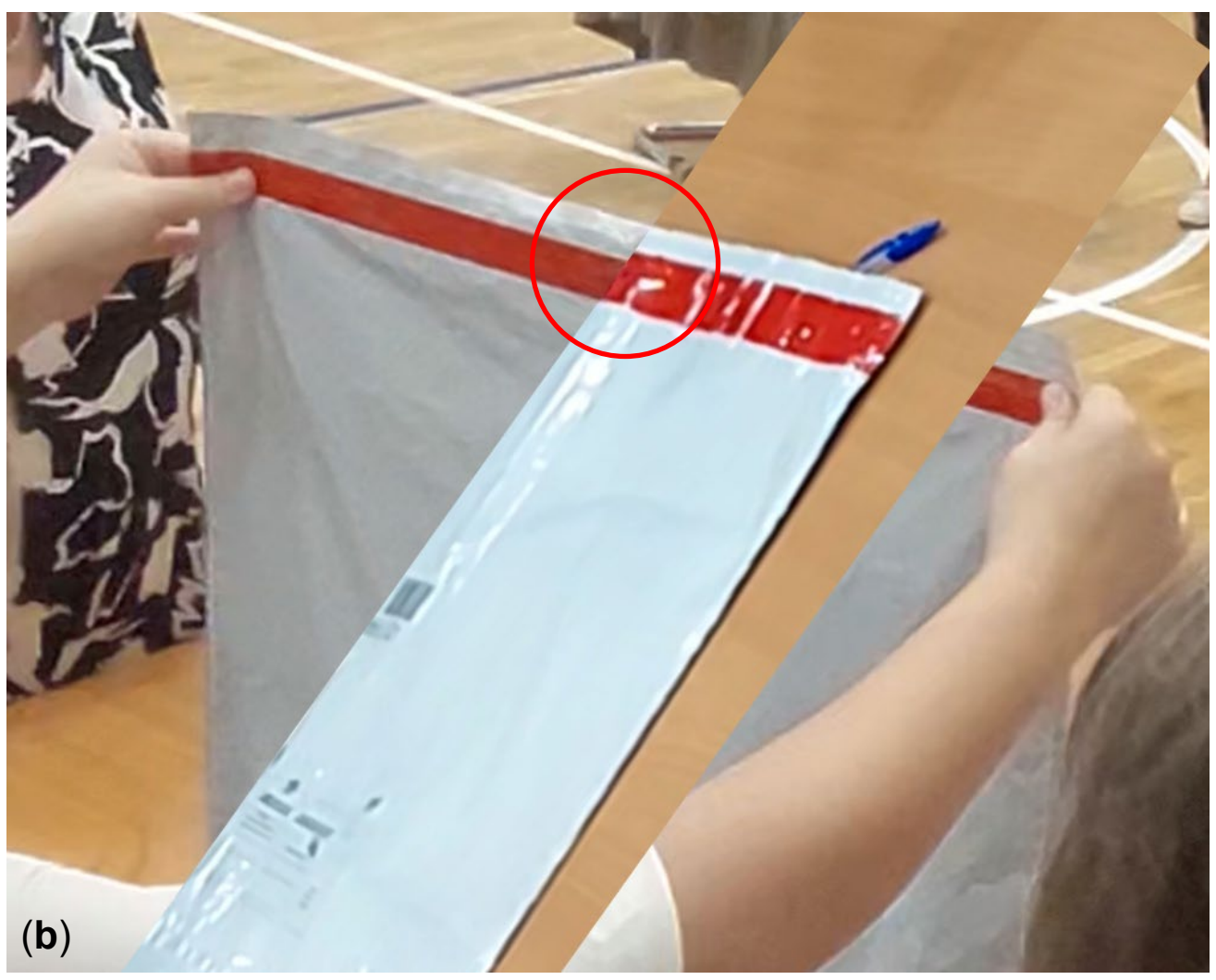

FIG. 29. Frames from video recordings of the first security bag at precinct 217. (a) The bag is displayed before opening it at the vote tallying. Using the barcode (bag) height as a reference, the tape width measures 24.7 mm (24.9 mm). Both measurements are an overestimate of the tape width, owing to perspective distortion. (b) A collage with a rotated image taken on the first day after sealing this bag. This side-by-side comparison shows that the tape has become narrower and its distance from the top edge has increased.

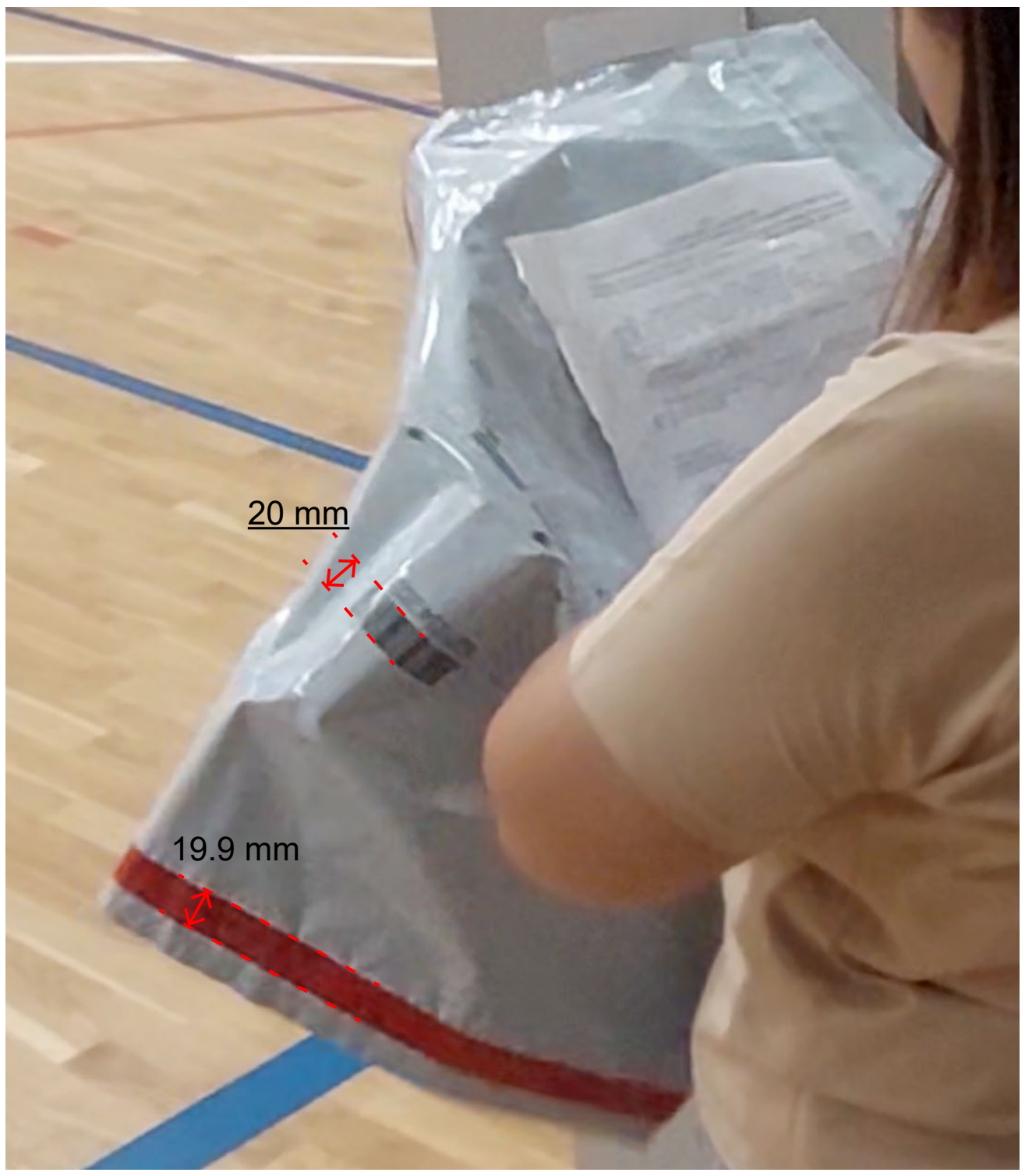

FIG. 30. Frame from a video recording of the second security bag at precinct 217 before opening it at the vote tallying. Using the barcode height as a reference, the tape width measures 19.9 mm.

19.9 mm.

Other evidence at this precinct is that its election officials entered the building well before the legal observer admission time (i.e., before 7:00) in the morning of the second day. They were there alone.

## 8. Precinct 218

At this precinct, both bags were loaded into the safe unsealed, as our video recordings show. On the first day, the precinct chair publicly tore away the red tape off the bag completely instead of sealing it, and put the open bag with the ballots into the safe before closing the precinct for the night. On the second day, she did nothing to seal the bag, leaving the tape and liner in place on the open bag. At the vote tallying, both bags emerged from the safe properly sealed with red tape. We suppose that either the first bag was swapped for its duplicate or another tape was re-installed on it at night. The second bag was simply sealed at night normally.

The transcript for this precinct contains an extremely long sequence of 382 identical ballots for the 5 winning candidates, as well as two shorter sequences (Fig. 5). Very few votes appear random.

### 9. Precinct 220

At this precinct, ballots were sealed into a bag with narrowly-typeset number on the tape on the first day, but sparsely-typeset number on the second day (Fig. 31). At the vote tallying, the geometry of both tapes appears unchanged, complete with visible creases on the second tape. The image resolution is insufficient to locate the numbers. This does not rule out swapping the first bag, however. The total length of the three continuous sequences of identical ballots in the transcript (Fig. 5) equals 124 and the total number of ballots for the 5 winning candidates equals 129. This is consistent with the possible replacement of 129 ballots in the first bag during the first night. Any plans to replace the second bag were probably scrapped, because of the public attention to nearby precinct 219 on the second day (Sec. D 1), which was located 50 m away and shared the gate entrance with this precinct.

Here the commission also cheated openly during the vote counting. They recorded more votes for the 5 winning candidates than was read aloud from the ballots and undercounted votes for most opposition candidates. E.g., Anikeeva (JR) accrued 24 more votes and Fedotov (CP) 17 fewer votes [31] than we count ourselves in our transcript made from a complete video recording.

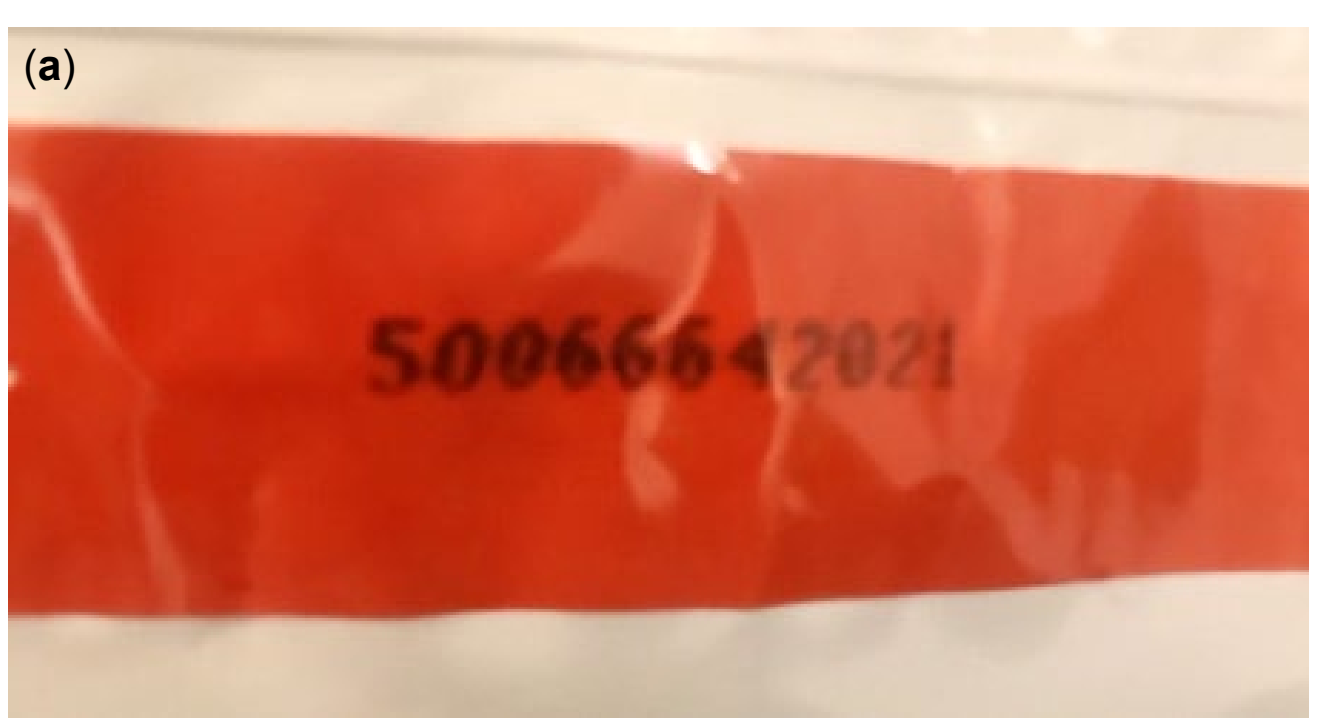

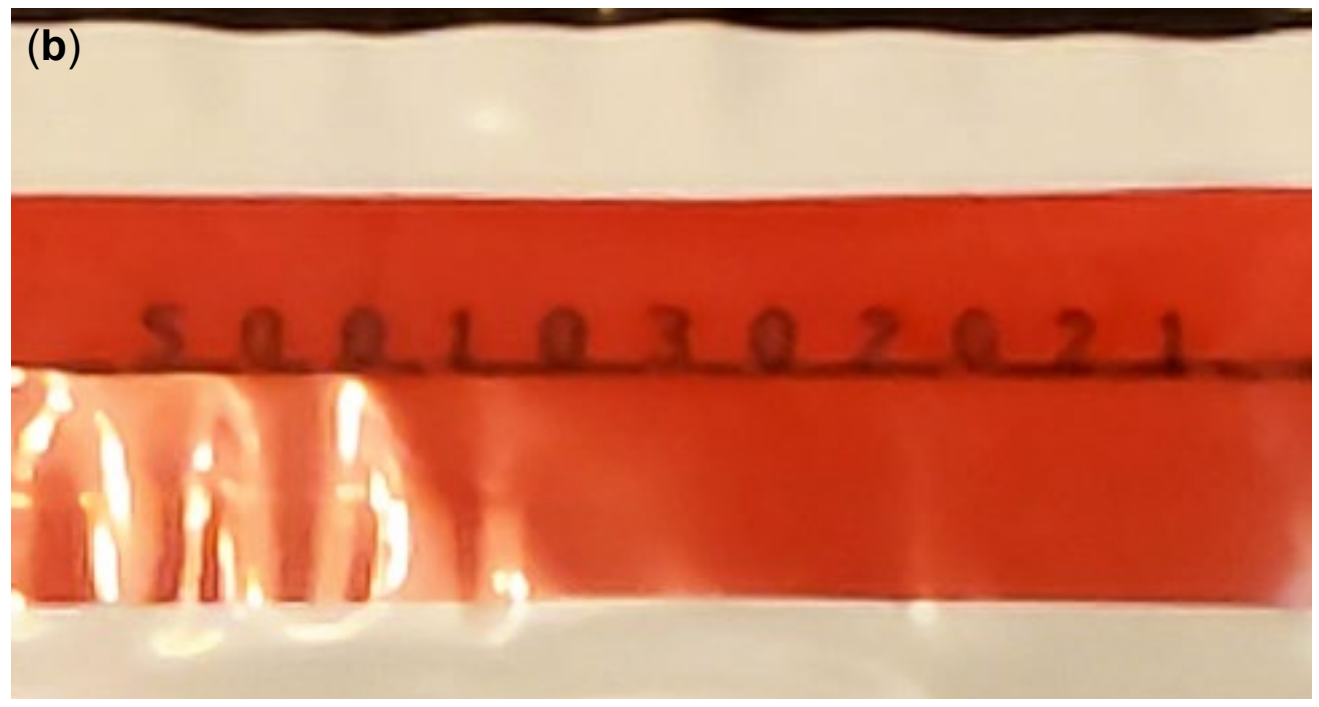

FIG. 31. Frames from video recordings of tapes at precinct 220 on (a) the first security bag after sealing it on the first day and (b) the second security bag after sealing it on the second day.

### 10. How many duplicates of bags were made at the factory?

According to information from the manufacturer, the batch delivered to the Election commission of Moscow region in August 2021 contained 51000 bags with custom 11-digit numbers 50*xxxxx*2021, where *xxxxx* is a variable part ranging 00001 through 53749 (not every number in this range was delivered) [52]. The regional commission subsequently distributed these bags to its territorial election commissions. We remark that bags of the same size but partially different graphics design sold in retail have 8-digit numbers and are also widely used in elections. The custom 11-digit numbers thus stand out. We have searched in observer discussion groups and found images of several such bags in other municipalities of Moscow region. Table VII lists all 11-digit tape numbers found.

If we look at the tapes with normally-typeset numbers imaged in high enough resolution to read them, they have the variable part ranging 06664 to 43625. Those with sparsely-typeset numbers range 01030 to 06665. This suggests both bags were produced in a large quantity. Curiously, the specification shared with us by the manufacturer explicitly states a *sparsely-typeset* number of an overall size $70 \times 4$ mm on the tape. Yet, the 4 bags found outside Vlasikha have normally-typeset numbers sized about $40 \times 4$ mm. While the manufacturer has declined to provide us further documentation, the available data suggests that two large batches were delivered to the Election commission of Moscow region, containing on the order of $10^4$ bag pairs with identical numbers.

The table reveals a curious aspect of organisation in Vlasikha. The fraudsters had duplicate bags with two ranges of consecutive numbers: 8 bag pairs 50**01029**2021–50**01036**2021 and 8 bag pairs 50**06660**2021–50**06667**2021. We guess that each fraudulent precinct got one bag pair from each range. Not every bag was used in the end, but we infer that each precinct (except non-fraudulent 215) had two pairs available.

### Methods E: Electronic online voting

No transcripts are available for our e-voting precincts. The nationwide online e-voting system (developed by Rostelecom) deployed in Moscow region does not allow observing marks in individual ballots [53, 54]. Only the totals are published. The only additional data available on e-voting is the time instants each ballot is issued and received [54], which is useless for our analysis. For completeness, we show a plot of voting frequency versus time in Fig. 32. Although fraud at e-voting is a risk, its results in our elections appear to be realistic and we suspect no tampering with electronic ballots at our e-voting precincts. In each of them, the voters preferred two CP's and 3 UR's nominees.

Opinions on e-voting in Russia are ambiguous. This

TABLE VII. All 11-digit bag numbers that have been observed, sorted in ascending order. The variable part of the number is shown in bold. A number in parentheses means its digits were not directly verified on the red tape, owing to insufficient image resolution. A number placed between columns means at least one of the two tapes was used, but we do not know which one. A number in square brackets means the tape was not recorded, but we infer it was used in a tape replacement or bag swap based on other evidence.

<table>
<tr><th rowspan="2">Election date</th><th rowspan="2">Municipality</th><th rowspan="2">Precinct number</th><th colspan="2">Number on tape</th></tr>
<tr><th>Normally typeset</th><th>Sparsely typeset</th></tr>
<tr><td>Sep 2024</td><td>Vlasikha</td><td>214</td><td>(50<b>01029</b>2021)</td><td>(50<b>01029</b>2021)</td></tr>
<tr><td></td><td></td><td>220</td><td></td><td>50<b>01030</b>2021</td></tr>
<tr><td></td><td></td><td>213</td><td>(50<b>01031</b>2021)</td><td>(50<b>01031</b>2021)</td></tr>
<tr><td></td><td></td><td>219</td><td></td><td>50<b>01032</b>2021</td></tr>
<tr><td></td><td></td><td>218</td><td colspan="2">(50<b>01033</b>2021)</td></tr>
<tr><td></td><td></td><td>217</td><td>[50<b>01034</b>2021]</td><td>(50<b>01034</b>2021)</td></tr>
<tr><td></td><td></td><td>212</td><td>(50<b>01035</b>2021)</td><td>(50<b>01035</b>2021)</td></tr>
<tr><td></td><td></td><td>216</td><td>(50<b>01036</b>2021)</td><td>(50<b>01036</b>2021)</td></tr>
<tr><td></td><td></td><td>218</td><td>(50<b>06660</b>2021)</td><td>(50<b>06660</b>2021)</td></tr>
<tr><td></td><td></td><td>213</td><td colspan="2">(50<b>06661</b>2021)</td></tr>
<tr><td></td><td></td><td>217</td><td>(50<b>06662</b>2021)</td><td>(50<b>06662</b>2021)</td></tr>
<tr><td></td><td></td><td>216</td><td>(50<b>06663</b>2021)</td><td>(50<b>06663</b>2021)</td></tr>
<tr><td></td><td></td><td>220</td><td>50<b>06664</b>2021</td><td>[50<b>06664</b>2021]</td></tr>
<tr><td></td><td></td><td>219</td><td>50<b>06665</b>2021</td><td>50<b>06665</b>2021</td></tr>
<tr><td></td><td></td><td>214</td><td>[50<b>06666</b>2021]</td><td>[50<b>06666</b>2021]</td></tr>
<tr><td></td><td></td><td>212</td><td>50<b>06667</b>2021</td><td>(50<b>06667</b>2021)</td></tr>
<tr><td>Sep 2021</td><td>Lyubertsy</td><td>1558</td><td>50<b>13317</b>2021</td><td></td></tr>
<tr><td></td><td></td><td></td><td>50<b>18815</b>2021</td><td></td></tr>
<tr><td></td><td>Balashikha</td><td>3667</td><td>50<b>23439</b>2021</td><td></td></tr>
<tr><td></td><td>Reutov</td><td>2642</td><td>50<b>43625</b>2021</td><td></td></tr>
</table>

technology is almost completely unregulated in the legal sense [55]. The voting is actually operated by various organisations other than the election commissions. The results of e-voting in Moscow (which uses a system different from the rest of the country) are known to be completely fraudulent [53, 56, 57]. Fraud at regional e-voting is also possible, however from the available data one can only claim that there is an administrative coercion to participate [53, 58]. Its traces are quite clear in Vlasikha (Fig. 32), where the bulk of e-voters participated either at the beginning of or during the lunch break on the first day, which was a work day. The shape of our curves is generally consistent with e-voting patterns observed at other recent elections in Moscow region [54].

Since e-voting participants are easily amenable to administrative coercion, one can conclude that they are more loyal than the voters in general. Thus pro-administration candidates cannot fare better at the physical precincts than at the e-voting ones. The returns for different precincts (Table I) do not agree with this conclusion, which in itself may serve as an indirect evidence of fraud.

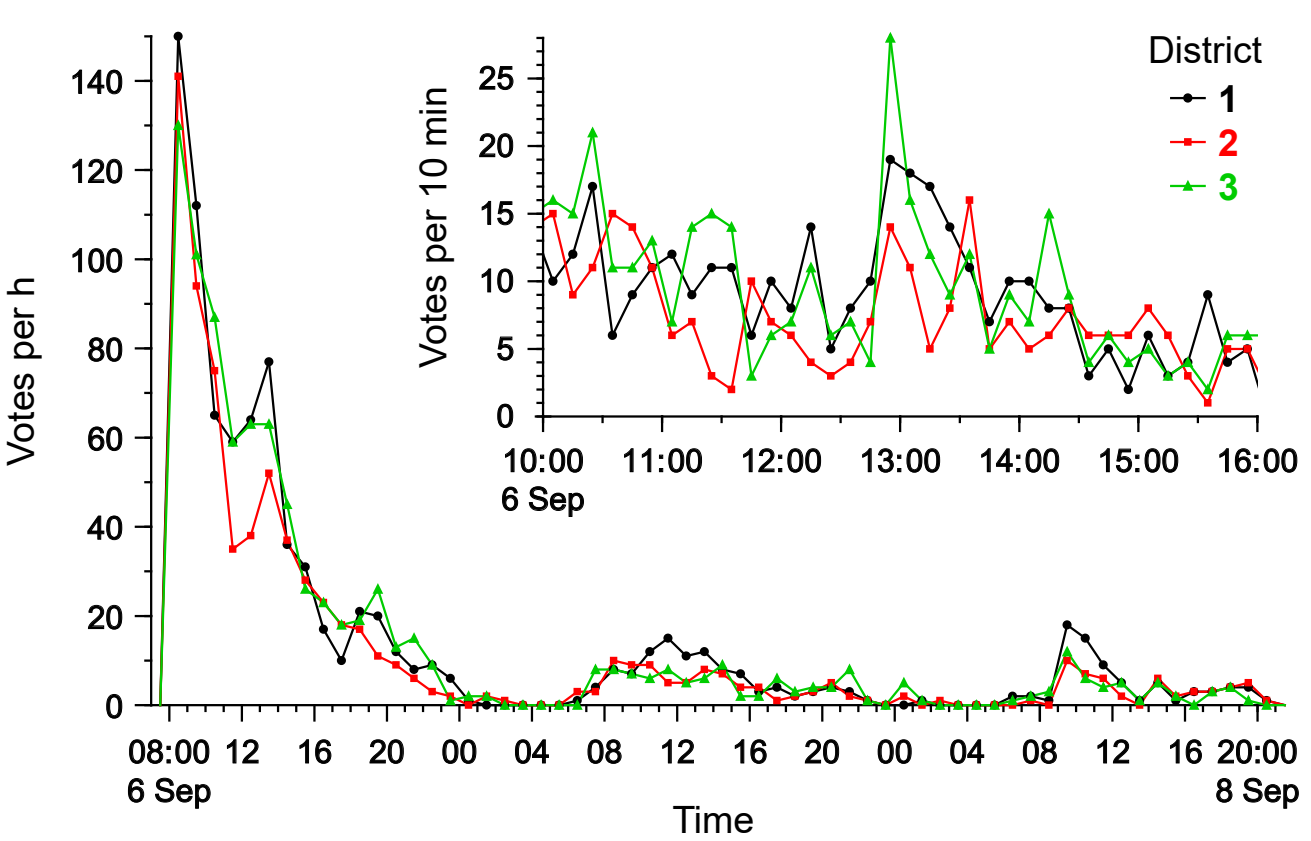

FIG. 32. Voting frequency in e-voting precincts. The online voting started at 08:00 on 6 Sep and continued until 20:00 on 8 Sep. The two ballots cast after that time were issued to the voters before 20:00. The hourly time bins adequately show all the histogram features except a sharp peak just before 13:00 on the first day in district 3, which is shown with 10-minute bins in the inset. Large peaks may be explained by voter coercion.

# Двойное доказательство фальсификаций на выборах статистикой и криминалистикой

(авторский перевод статьи с английского языка — translation of original article “A dual approach to proving electoral fraud using statistics and forensic evidence”)

Андрей Подлазов[1] and Вадим Макаров[2,3,*]

[1] *Институт прикладной математики им. М. В. Келдыша РАН, г. Москва*

[2] *член территориальной избирательной комиссии пос. Власиха Московской области с правом решающего голоса*

[3] *Виговский исследовательский центр квантовой связи, университет г. Виго, Испания*

Фальсификации на выборах обычно ведут к появлению статистических аномалий в итогах голосования, однако масштаб фальсификаций редко может быть надежно подтвержден иными доказательствами. В статье подробно рассмотрены выборы совета депутатов поселка Власиха Московской области, где имеются как статистические аномалии в стенограммах подсчета голосов, так и криминалистические доказательства подмены бюллетеней во время их ночного хранения на избирательных участках. Введены два статистических теста последовательности подсчета голосов, позволяющие строго доказать, что были вброшены партии подложных бюллетеней. Показано, что для подмены бюллетеней использовались изготовленные на фабрике пары сейф-пакетов с идентичными серийными номерами. На 8 из 9 избирательных участков поселка статистические и криминалистические доказательства совпадают (идентифицируя 7 фальсифицированных участков и 1 честный), а на оставшемся участке статистический тест обнаруживает фальсификации, в то время как криминалистических доказательств недостаточно. Показано, что пломбировка с индикацией попытки вскрытия принципиально не может обеспечить надежной защиты при использовании ее на выборах.

## I. ВВЕДЕНИЕ

Инклюзивные социальные институты в долгосрочной перспективе ведут к процветанию нации [1]. При их отсутствии крайне важно, чтобы общество было информировано об имеющихся у него проблемах, что достигается посредством независимого анализа и публичности. Справедливая избирательная система является одним из таких институтов. Электоральные фальсификации порождают множество типов статистических аномалий в итогах голосования, которые их и выдают. Это стало очевидным с появлением информационных технологий, которые позволяют легко получать и анализировать итоги с отдельных избирательных участков [2–6]. Сами фальсификационные действия иногда можно поймать, но они редко документируются в масштабах и охвате, сопоставимых с теми, которые показывает статистика, тем более, что для этого обычно требуется сотрудничество со стороны властей [7, 8]. Здесь мы рассказываем о выборах местного самоуправления, где и статистический анализ, и зафиксированные доказательства фальсификаций (либо их отсутствия) доступны для каждого избирательного участка. Они в значительной степени подтверждают друг друга. Но чтобы полностью раскрыть фальсификации в случаях, когда фальсификаторы пытаются их замаскировать, может потребоваться несколько разных статистических тестов.

Статистические методы основаны на четко определенных предположениях и, следовательно, допускают независимую проверку. Анализ итоговых протоколов избирательных участков позволяет обнаружить аномальные зависимости результата от явки [2, 4, 9–17], неоправданное исчерпание бюллетеней [18], зависимость чисел в протоколах от присутствия наблюдателя [19–21] или использования электронных урн [22–24]. Это дает качественную оценку масштаба фальсификаций. Более строгие статистические методы позволяют рассчитать вероятность случайного возникновения аномалии, например, избытка целочисленных процентных значений в протоколах [3, 25–28]. Это количественно оценивает ее весомость как доказательства фальсификации.

В нашем исследовании предложен строгий статистический анализ результатов физического манипулирования бюллетенями, такого как их вброса и подмены. Мы рассчитываем вероятность того, что при подсчете бюллетени образуют длинные серии идентичных голосов, и вероятность того, что будет зафиксировано недостаточное количество переключений между сериями наличия и отсутствия голосов. Мы анализируем стенограммы подсчета голосов, записанные наблюдателями.

* makarov@vad1.com

Таблица I. Сводка по участкам. $n_1$ ($n_2$) — количество избирателей, получивших бюллетени в помещении для голосования в течение первого (второго) дня голосования; $n$ — общее количество действительных бюллетеней, оглашенных на подсчете голосов; КПРФ — Коммунистическая партия Российской Федерации; СРЗП — Социалистическая политическая партия «Справедливая Россия — Патриоты — За правду»; ЛДПР — Либерально-демократическая партия России; ЕР — Всероссийская политическая партия «Единая Россия». Подозрительно высокие показатели значимости выделены жирным шрифтом (приложение В 1)

| Избирательный округ | $c$ | Номер участка | Первый день | | Второй день | | $n$ | Намеренное перемешивание бюллетеней? | $\mathrm{p}\alpha'$ | $\mathrm{p}\tilde{\alpha}'$ | Лидеры на этом участке | Получили мандаты в этом округе |
|---|---|---|---|---|---|---|---|---|---|---|---|---|
| | | | $n_1$ | Следы вскрытия пакета? | $n_2$ | Следы вскрытия пакета? | | | | | | |
| 1 | 11 | 212 | 157 | да | 117 | да | 437 | да | **4,0** | **4,3** | 5 ЕР | 5 ЕР |
| | | 213 | 104 | нет записи | 68 | да | 256 | нет записи | **2,3** | **4,5** | 5 ЕР | |
| | | 214 | 181 | неизвестно[a,b] | 104 | да | 388 | нет | **9,8** | **18,4** | 5 ЕР | |
| | | ДЭГ | | | | | 878 | | | | 2 КПРФ, 3 ЕР | |
| 2 | 11 | 215 | 183 | нет[c] | 135 | нет[c] | 473 | нет | 0,8 | **2,2** | 4 КПРФ, 1 ЕР | 1 ЛДПР, 4 ЕР |
| | | 216 | 95 | да | 86 | да | 318 | да | **3,6** | **21,0** | 1 ЛДПР, 4 ЕР | |
| | | 217 | 108 | да | 69 | да | 290 | да | 0,4 | **2,6** | 1 ЛДПР, 4 ЕР | |
| | | ДЭГ | | | | | 726 | | | | 2 КПРФ, 3 ЕР | |
| 3 | 11 | 218 | 435 | да | 109 | да | 577 | нет | **10,9** | **9,3** | 1 СРЗП, 4 ЕР | 1 СРЗП, 4 ЕР |
| | | 219 | 131 | да | 93 | неизвестно[a] | 355 | нет | **18,7** | **20,2** | 1 СРЗП, 4 ЕР | |
| | | 220 | 129 | неизвестно[a,b] | 84 | неизвестно[a] | 346 | нет | **20,3** | **30,1** | 1 СРЗП, 4 ЕР | |
| | | ДЭГ | | | | | 844 | | | | 2 КПРФ, 3 ЕР | |

[a] Не обнаружено видимых изменений в геометрии сейф-ленты, но качество съемки слишком низкое, чтобы осмотреть ее номер.
[b] Стенограмма подсчета голосов поддерживает подмену сейф-пакета.
[c] Использовались недавно закупленные сейф-пакеты с короткими 8-значными номерами, дубликатов которых мы не наблюдали.

## II. ЭКСПЕРИМЕНТ

Выборы совета депутатов поселка Власиха состоялись 6–8 сентября 2024 г. Они неожиданно дали экспериментальные данные, позволившие выполнить наше незапланированное исследование.

Власиха — компактное (менее 3 км в поперечнике) закрытое административно-территориальное образование с пропускным режимом, которое обслуживает главный штаб Ракетных войск стратегического назначения (РВСН) [29]. Избиратели представляют собой смесь гражданских лиц и военнослужащих.

В поселке проживают 17 189 избирателей, из которых 35% проголосовали. Он разделен на 3 избирательных округа примерно равного размера, каждый из которых представлен 5 депутатами в совете из 15 депутатов. В каждом округе баллотировалось $c = 11$ кандидатов и каждый избиратель мог отметить до 5 из них в бюллетене. Одинаковое количество кандидатов в каждом округе — совпадение. Хотя самовыдвижение было возможным, каждый кандидат на этих выборах был выдвинут какой-либо партией. В каждом округе мандаты получили 5 кандидатов, набравшие наибольшие числа голосов.

Каждый избирательный округ состоял из 3-х обычных избирательных участков и одного участка дистанционного электронного голосования (ДЭГ), на который любой избиратель, предпочитающий проголосовать удаленно по интернету, мог зарегистрироваться заранее. Голосование на всех участках проходило в течение трех дней подряд. На обычных участках каждый избиратель использовал кабинку для голосования, чтобы тайно заполнить свой бумажный бюллетень, имел возможность сложить его, чтобы скрыть свой выбор, и опускал бюллетень в большую прозрачную урну, стоящую у всех на виду. В конце первого и второго дней голосования все бюллетени извлекались из урны и запечатывались в пластиковый сейф-пакет, обеспечивающий защиту от незаметного вскрытия. Индивидуальный номер каждого сейф-пакета фиксировался в акте и сейф-пакет убирался в металлический сейф, стоящий в помещении для голосования, до подсчета голосов в конце третьего дня. Все бюллетени были подсчитаны после окончания голосования в конце третьего дня.

Днем на избирательных участках работали члены избирательных комиссий. Кандидаты и наблюдатели также присутствовали практически непрерывно, следили за процедурами, делали видеозаписи и фотографии. По закону никому не разрешалось находиться на избирательном участке ночью, за исключением полицейских, охраняющих опечатанные помещения для голосования снаружи.

В отличие от некоторых предыдущих российских выборов [21, 30], присутствие наблюдателей на этот раз не помешало фальсификациям. Анализ фотографий и видеозаписей показывает, что на 7 из 9 участков видны следы вскрытия сейф-пакетов (таблица I). Следовательно и вероятно, бюллетени в пакетах были ночью подменены.

На подсчете голосов сейф-пакеты вскрывались и бюллетени из них складывались в одну стопку с бюл-

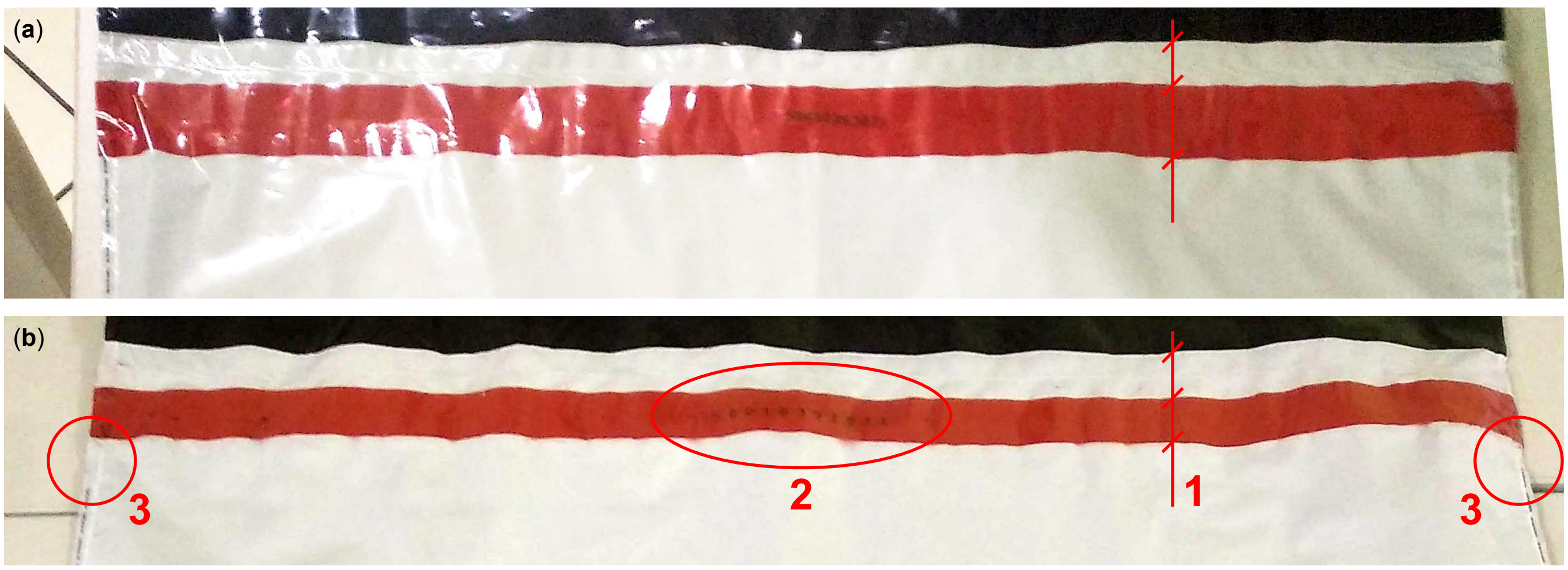

Рис. 1. Красная индикаторная лента на сейф-пакете на избирательном участке 212 (a) после запечатывания пакета во второй день и (b) перед его вскрытием во время подсчета голосов на третий день. Обозначены (1) изменившаяся ширина ленты и ее положение относительно верхнего края, (2) увеличившиеся интервалы между цифр и общая ширина индивидуального номера на ленте, типографский набор которого изменился с нормального на разреженный, (3) исчезновение части строки микрошрифта по краям пакета, предположительно смытой растворителем летенями, опущенными в урну в течение третьего дня. В процессе формирования стопки бюллетени разных дней в той или иной мере перемешивались, но полного смешивания никогда не было. Затем бюллетени последовательно предъявлялись для обозрения присутствующим и отметки в них зачитывались вслух, а голоса за кандидатов суммировались вручную членами комиссии. По видео- и звукозаписям оглашения бюллетеней мы сделали стенограммы. Они поддаются статистическому анализу и дают убедительные доказательства того, что предполагаемая подмена бюллетеней ночью радикально повлияла на итоги выборов. Хотя местные оппозиционные кандидаты, выдвинутые Коммунистической партией Российской Федерации (КПРФ), вероятно, выиграли большинство мест в совете, все они с помощью фальсификаций были заменены административными кандидатами из других партий [31].

## III. КРИМИНАЛИСТИЧЕСКИЕ ДОКАЗАТЕЛЬСТВА

Различные способы пломбировки, выявляющие несанкционированный доступ, применяются с античных времен и распространены повсеместно [32]. Они не подходят для обеспечения безопасности современных выборов [33, 34]. Для того, чтобы меры безопасности такого рода были эффективны, все заинтересованные участники должны быть должным образом обучены и досконально придерживаться протокола использования пломбировки [35, 36]. Это нереально обеспечить на выборах, где все заинтересованные стороны — кандидаты, наблюдатели, пресса и члены участковых избирательных комиссий — по сути состоят из представителей широкой общественности. Они нечасто занимаются выборами и не имеют времени на такое специальное обучение. В то же время, у мотивированных фальсификаторов есть месяцы на подготовку и они могут хорошо финансироваться.

Существует не менее 5 способов малозаметного вскрытия сейф-пакета (приложение C). Некоторые из них могут быть выполнены самостоятельно, а некоторые требуют поставки заводской подделки.

Согласно нашей реконструкции преступления, участковые комиссии Власихи имели доступ к парам сейф-пакетов с идентичными индивидуальными номерами (которые должны были быть уникальными, но на самом деле были изготовлены производителем в двух экземплярах). На участке 219 у нас есть убедительное фотографическое доказательство наличия такой пары пакетов (приложение D 1). На большинстве других участков были подменены только контрольные пломбировочные ленты (сейф-ленты), предположительно для того, чтобы избежать необходимости подделывать подписи, оставленные на пакетах наблюдателями, кандидатами и членами избирательных комиссий. Ночью преступники проникли в помещение для голосования в отсутствие наблюдателей или кандидатов (доказательства чего имеются на 3 участках) и открыли сейфы (оставив на 2 участках видимые следы на сейфах). Они отклеили оригинальную сейф-ленту шириной 30 мм от пакета, таким образом распечатав его, и предположительно подменили бюллетени. Оставшийся после удаления ленты на пакете окрашенный клей, показывающий факт вскрытия, был смыт с пакета с помощью растворителя, как подробно описано в приложении C. Затем пакет был запечатан повторно другой сейф-лентой с идентичным номером, отрезанной от неиспользованного дубликата пакета и потерявшей в процессе отрезания от 5 до 10 мм своей ширины.

Фальсификации удались в том смысле, что номера пакетов совпали с зафиксированными в актах, а внешний вид лент и пакетов ни у кого не вызвал подозрений в суматошные дни голосования. Итоги выборов были немедленно утверждены [31]. Однако, несмотря на то, что эти сейф-пакеты считаются лишь защитой умеренного уровня от подмены и подделки [37] и их использование не предполагает наличия заводской подделки с идентичным номером [38], в конечном итоге они выполнили свою функцию. Тщательное изучение и сверка видеозаписей и фотографий, сделанных в дни голосования, показывает, что пломбировочная лента и окружающая ее область выглядят по-разному на 7 участках (см. пример на рис. 1), не дает нам информации на участке 220 и указывает на отсутствие фальсификации на оставшемся участке 215. Примечательно, что на участке 219 мы видим, что не только лента, но и весь пакет были заменены на его точную заводскую копию — с идентичным индивидуальным номером, — а подписи авторучкой на пакете были подделаны. Всё это было обнаружено и понято не сразу, лишь через несколько недель после выборов. В общей сложности имел место несанкционированный доступ к содержимому не менее 11 пакетов, содержащих 1 479 бюллетеней (см. таблицу I и приложение D).

## IV. СТАТИСТИЧЕСКИЙ АНАЛИЗ

Стенограмма подсчета голосов для избирательного участка 215 (левая половина рис. 2) демонстрирует случайное распределение голосов по всей ее протяженности, без каких-либо видимых особенностей. Стенограммы для всех остальных участков содержат аномалии, типичным примером которых является участок 220, стенограмма которого показана на правой половине рис. 2. Случайная последовательность голосов (с явным преобладанием голосов за 5 оппозиционных кандидатов от КПРФ) прерывается тремя пачками одинаковых бюллетеней за 5 административных кандидатов (1 кандидат от СРЗП и 4 кандидата от ЕР). Кстати, всего за этих 5 кандидатов подано 129 бюллетеней, в то время как в пакете с первого дня голосования также содержатся 129 бюллетеней. Возникает подозрение, что на самом деле никто не голосовал за этот конкретный набор кандидатов, а длинные серии одинаковых бюллетеней взялись из этого пакета.

Поскольку почти все остальные стенограммы, приведенные в приложении A, демонстрируют схожие видимые глазом аномалии, мы разработали два простых статистических теста.

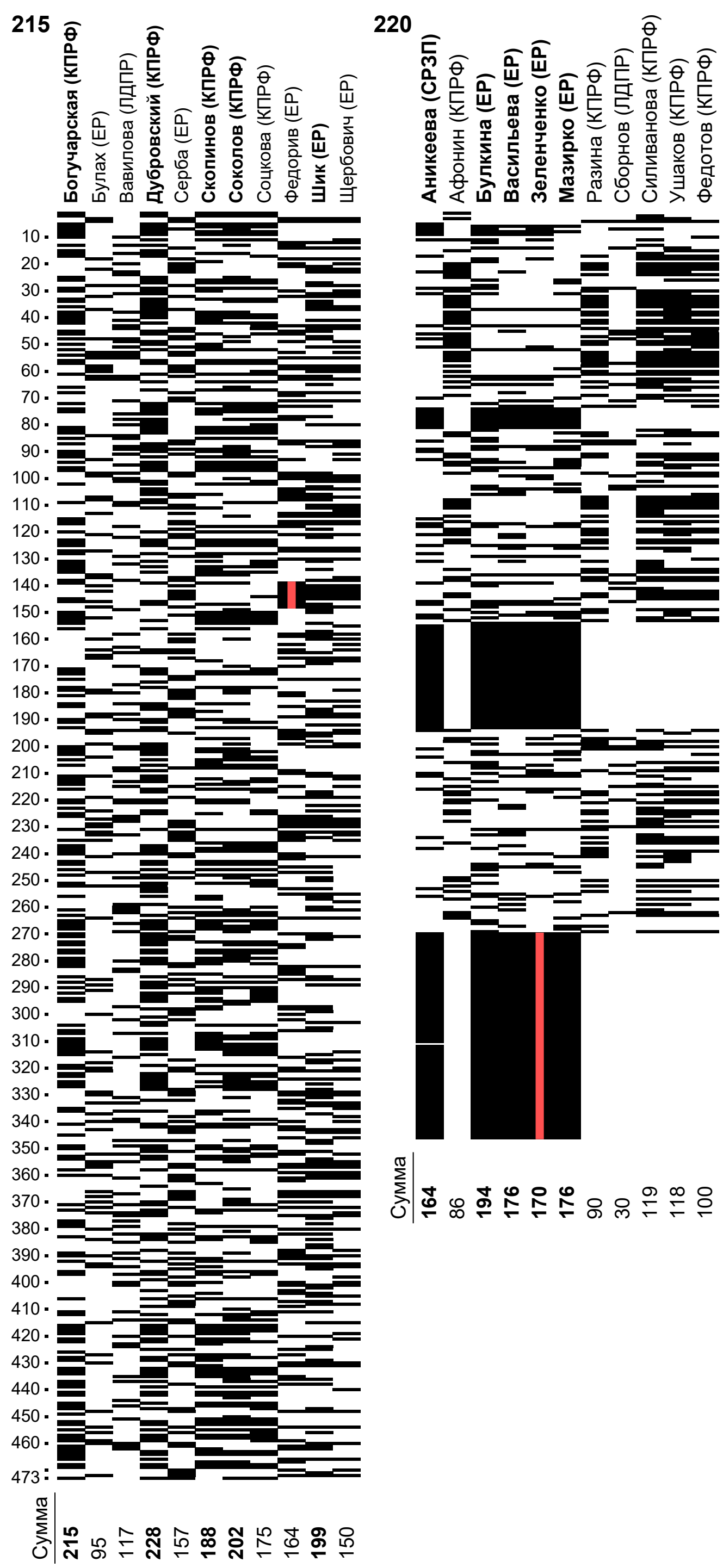

Рис. 2. Стенограммы подсчета голосов на двух избирательных участках: 215, где мы не обнаружили никаких признаков фальсификаций, и 220, где подозревается подмена пакета, содержащего 129 бюллетеней. Жирным шрифтом выделены 5 лидирующих на участке кандидатов. Наименее вероятная серия обозначена красной (серой) линией

В первом тесте вычисляются значимости $\alpha_1$ и $\alpha_0$. Это вероятности того, что самая длинная непрерывная серия голосов за и против кандидата на избирательном участке возникла естественным путем. Предполагается, что такие чрезвычайно длинные серии мо-гут быть результатом вброса пачек идентичных бюллетеней.

Во втором тесте вычисляется значимость $\tilde{\alpha}$. Это вероятность того, что число непрерывных серий, возникающих естественным образом, не превышает их наблюдаемого числа. Длинная серия может быть разорвана частичным перемешиванием бюллетеней перед их подсчетом. Однако даже в этом случае стенограмма содержит меньше переключений между сериями голосов и отсутствия голосов, чем статистически ожидается, что этот тест и призван обнаружить.

Для этих расчетов делаются два предположения. Во-первых, мы консервативно полагаем, что все бюллетени подлинные. Это позволяет рассчитывать вероятность того, что избиратель отдаст или не отдаст голос за кандидата. Во-вторых, мы полагаем, что последовательные бюллетени взаимно независимы. Даже если небольшие группы единомышленников (например, родственники или друзья) приходят голосовать одновременно, их бюллетени разлетаются случайным образом, когда их бросают в урну, а затем снова, когда их вынимают из урны. Это должно способствовать сохранению независимости бюллетеней в последовательности подсчета голосов.

Указанные значимости вычисляются с помощью итерационных алгоритмов, подробно описанных в приложении B, где также приведены их значения для всех избирательных участков и кандидатов. Зная эти значения, мы переходим к сводному анализу для избирательных участков. Если любая из вероятностей $\alpha_0$, $\alpha_1$ или $\tilde{\alpha}$ оказывается малой, это указывает на вброс бюллетеней. Доказательством такой электоральной фальсификации могут служить наименьшие значимости, полученные для кандидатов, баллотирующихся на избирательном участке $\alpha_{\min} = \min_c \{\alpha_0, \alpha_1\}$ и $\tilde{\alpha}_{\min} = \min_c \tilde{\alpha}$. Однако для того, чтобы корректно характеризовать общее положение дел на участке, следует учитывать множественность тестирования, введя соответствующую поправку. Дело в том, что чем больше попыток вычислить некоторую случайную величину, тем выше естественные шансы того, что она хотя бы один раз примет очень малое значение.

Когда в округе баллотируется $c$ кандидатов, значимость гипотезы о том, что наименее вероятное число серий возникло естественным путем, не превышает $\tilde{\alpha}' = 1 - (1 - \tilde{\alpha}_{\min})^c \approx c\tilde{\alpha}_{\min}$. Использование приближения здесь уместно, чтобы избежать потери точности при вычислении, если $\tilde{\alpha}_{\min} < 10^{-8}$. Аналогично значимость гипотезы о том, что наименее вероятная непрерывная серия возникла естественным путем, не превышает $\alpha' = 1 - (1 - \alpha_{\min})^{2c} \approx 2c\alpha_{\min}$. Множитель 2 здесь учитывает, что каждый кандидат подсчитывается дважды, для двух типов серий (наличие и отсутствие за него голосов).

Мы игнорируем здесь возможность многократного появления низких значимостей на участке. Это устраняет необходимость анализировать сходства и различия между кандидатами. Такое упрощение может переоценить общие значимости $\alpha'$ и $\tilde{\alpha}'$. Это ослабляет наш тест, что приемлемо. Так мы рискуем пропустить нарушения, но не рискуем выдвинуть необоснованные обвинения.

При наличии фальсификаций значимости, как правило, очень малы. Удобнее приводить не значения $\alpha$, а их десятичные показатели $\mathrm{p}\alpha = -\lg \alpha$ (аналогично pH в химии). Следует подчеркнуть, что вероятность отсутствия фальсификаций падает на порядок при каждом увеличении показателя на 1. Показатели общих значимостей приведены в таблице I.

Для 5 участков вероятности крайне низки, менее $10^{-9}$ (т. е., $\max\{\mathrm{p}\alpha', \mathrm{p}\tilde{\alpha}'\} > 9$). Это, безусловно, указывает на наличие фальсификаций, поскольку вероятность того, что участок честен, составляет менее одного случая на миллиард. Более высокие вероятности на оставшихся 4 участках можно объяснить либо перемешиванием бюллетеней перед подсчетом (при этом фальсифицированные длинные серии разбиваются на достаточно большое количество фрагментов), либо наличием в исходной фальсифицированной серии намеренно вставленных в нее случайных бюллетеней, либо отсутствием на участке такого рода фальсификаций.

После извлечения бюллетеней из пакетов и урн во время подсчета голосов, комиссии просмотрели их на предмет исключения недействительных бюллетеней (без отметок или с более чем 5 отметками) перед оглашением. Чтобы ускорить этот шаг, просмотр велся несколькими членами комиссии одновременно, каждый из которых обрабатывал часть бюллетеней. Когда они добавляли просмотренные бюллетени в общую стопку, они это делали частями. Этот шаг сам по себе обычно приводил к некоторой перетасовке бюллетеней в общей стопке. Вдобавок, некоторые комиссии намеренно перемешивали бюллетени, высыпав их россыпью на стол, мешая массу бюллетеней руками, выхватывая из нее и переворачивая кипы бюллетеней (что зафиксировано на наших видео и отмечено в таблице I). Мы считаем, что умеренно низкие вероятности на участках 212 и 213 и высокая на участке 217 обусловлены таким намеренным перемешиванием. Высокая, но все еще подозрительная вероятность $\tilde{\alpha}' = 6{,}2 \times 10^{-3}$ на участке 215 может быть связана с тем, что одно из наших предположений не совсем верно, например, возможно, бюллетени не полностью взаимно независимы. Стенограммы с большего количества честных избирательных участков должны прояснить этот вопрос в будущих исследованиях.

Статистические доказательства с довольно высокой вероятностью ошибки допустимы в судах [39, 40]. Примечательно, что российские суды принимают как доказательство генетический тест на отцовство с вероятностью ошибки не более $10^{-3}$ при условии, что мать известна наверняка (в противном случае пороговая вероятность ошибки ослабляется до $2{,}5 \times 10^{-3}$) [41]. При таком подходе, фальсификация выборов должна считаться установленной на всех участках, кроме 215 и 217. Однако мы выступаем за более строгий порог. Как обсуждалось выше, фальсификации определенно обнаружены на участках 214, 216, 218, 219 и 220, где $\min\{\alpha', \tilde{\alpha}'\} < 10^{-9}$.

Хотя может показаться, что участок 217 избегает обнаружения фальсификаций, тест можно улучшить, чтобы он учитывал бюллетени с оценками за всех 5 лидирующих кандидатов. Этот тест выявляет здесь фальсификации с $\tilde{\alpha} = 5{,}9 \times 10^{-15}$ и дает не вызывающие подозрений значения для честного избирательного участка 215 (см. приложение B 3).

В целом, после тщательного анализа статистики и стенограмм (приложение A), мы знаем ночную участь всех, кроме одного, сейф-пакетов из 18 использованных на этих выборах. В дополнение к 11 пакетам с зафиксированными следами подделки (таблица I и приложение D), еще 2 были подменены целиком — как показывают непрерывные серии идентичных бюллетеней, это пакеты с бюллетенями первого дня голосования на участках 214 и 220. 4 пакета во время хранения не вскрывали: оба пакета на участке 215 и пакеты с бюллетенями второго дня голосования на участках 219 и 220. Мы не уверены только относительно пакета первого дня голосования на участке 213.

В отличие от обычных избирательных участков, работу участков ДЭГ проследить практически невозможно. Недоступны наблюдению как процесс подсчета голосов, так и какие-либо сведения, по которым можно было бы судить об отсутствии или наличии незаконных манипуляций с бюллетенями (приложение E).

## V. ОБСУЖДЕНИЕ И ЗАКЛЮЧЕНИЕ

Нами предложены строгие статистические тесты последовательности подсчета голосов, не делающие никаких нетривиальных предположений об особенностях избирателей. Данные тесты в большинстве случаев обнаруживают фальсификации, связанные с подменой бюллетеней. Это подтверждается отдельным криминалистическим анализом.

Результаты с 3 участков ДЭГ, 1 честного участка и наш анализ 8 участков, где были обнаружены следы вскрытия сейф-пакетов, показывают, что выборы были очень конкурентными и кандидаты от администрации и оппозиции шли вровень. Последние фактически выиграли около половины или, возможно, более половины мандатов в совете. Реальная явка избирателей составила 35%, что необычно высоко для российских муниципальных выборов и показывает заинтересованность людей в их результатах. С помощью фальсификаций в совет была посажена неправдоподобная комбинация кандидатов от администрации, а местная оппозиция потеряла все мандаты, которые у нее были в предыдущем составе совета.

Заметим, что менее строгий анализ [42], предполагающий типичные характеристики избирателей, позволяет реконструировать истинные итоги выборов. Он предсказывает, что в совет прошло 9 депутатов от КПРФ и 6 от ЕР. Это согласуется с результатами настоящего исследования.

Такие вопиющие фальсификации возможны только в том случае, если виновные в них пользуются полной юридической безнаказанностью. К сожалению, наш опыт показывает, что это так и есть. Многочисленные жалобы в избирательные комиссии всех уровней были отклонены. Заявления о преступлениях в местное отделение полиции и Следственный комитет Российской Федерации с приложением доказательств, представленных в приложениях C и D, не удостоились проверки и были отклонены как «не содержащие объективных данных, свидетельствующих о совершении преступления». Полиция, охраняющая избирательные участки ночью, ничего не сообщала. Суды отказались рассматривать доказательства истцов в 4 изначально поданных исках, отклонив их по формальной причине. Глава поселкового управления полиции, глава местного отделения прокуратуры, заместитель командующего РВСН по военно-политической работе и глава поселка присутствовали на первом заседании новоизбранного совета и поздравили депутатов с «убедительной победой на выборах» [43], при том, что на тот момент доказательства фальсификаций были освещены в местной прессе, а в местном отделении полиции находились на рассмотрении многочисленные заявления о преступлениях. Вещественные доказательства — использованные сейф-пакеты, — изъятые полицией после заявления В.М. о преступлении на участке 217, были позже уничтожены «как мусор» без их исследования. По закону на каждом избирательном участке ведется круглосуточная видеофиксация помещения для голосования и сейфа. Эти видеозаписи хранились в территориальной избирательной комиссии пос. Власиха. Все запросы на их просмотр были отклонены и записи через некоторое время были уничтожены.

Судебный иск на отмену итогов выборов, включающий подробно все имеющиеся доказательства (в том числе эту статью и заключение специалиста А.П.), был проигран; суд первой инстанции отказался вызывать свидетелей и истребовать документы по поставкам сейф-пакетов. В решении суда лишь написано, что «доказательств... не представлено». Апелляционный суд оставил это решение в силе, отклонив все ходатайства о вызове свидетелей, истребовании документов и экспертизе сохранившихся частей сейф-пакетов, а также постановив, что выводы, основанные на статистике, и все изображения, использованные в данном исследовании, являются недопустимыми доказательствами. Кассационный и Верховный суды оставили решение без изменений.

Собранные нами данные показывают, что для избирательной комиссии Московской области были изготовлены две большие партии пакетов с пересекающимися диапазонами номеров, идентичные пары которых затем передали во Власиху (приложение D 10).

Суды отклонили все ходатайства о дальнейшем исследовании этого вопроса. Избирательная комиссия Московской области и производитель сейф-пакетов (ООО «Аспломб Технолоджи», фабрика находится в Ленинградской области [37]) отказались давать нам письменные пояснения по этому вопросу. Эта фабрика поставляет сейф-пакеты избиркомам по всей России с момента введения многодневного голосования в 2020 г. [44]. Наша жалоба в Центральную избирательную комиссию на недостаточность действующей процедуры хранения бюллетеней для обеспечения их сохранности с предложениями по ее улучшению была отклонена [45].

Авторы — и более двадцати человек, которые внесли свой вклад в это исследование, — не могут понять, какие мотивы и логика движут многочисленными членами избирательных комиссий, набранными из местного населения, крадущими голоса своих соседей и знакомых. Это исследование также иллюстрирует размах коррупции и угасание политических прав в России даже на местном уровне.

*Благодарности.* Авторы признательны Владимиру Зайцеву за координацию общественного контроля за выборами и поддержку избирательной кампании КПРФ. Мы благодарны всем 14 кандидатам, выдвинутым КПРФ, 5 добровольным наблюдателям и 2 членам комиссий за сбор доказательств и расшифровку стенограмм подсчета голосов. Спасибо Андрею Бражникову за обсуждения.

*Вклады авторов.* А.П. провел статистический анализ стенограмм. В.М. координировал сбор и криминалистический анализ доказательств. Статью писали и переводили на русский язык оба автора.

*Конфликты интересов.* Авторы заявляют об отсутствии конфликта интересов.

*Доступность исходных данных.* Оригиналы видеозаписей и фотографий, а также копии документов, использованных в данном исследовании, могут быть предоставлены по обоснованному запросу к В.М.

## ПРИЛОЖЕНИЯ

### Приложение A: Стенограммы подсчета голосов

Стенограммы для всех 9 обычных избирательных участков показаны на рис. 3–5. Диапазоны бюллетеней в стенограммах, которые лежат за пределами непрерывных серий идентичных или почти идентичных бюллетеней для 5 административных кандидатов, ясно показывают предпочтение избирателями кандидатов от оппозиции. На участке 214 (рис. 3), за исключением 3 блоков бюллетеней за кандидатов от ЕР, большинство оставшихся случайных голосов принадлежат 4 кандидатам от КПРФ и одному кандидату от ЕР. Как мы уже отмечали в основном тексте, на честном участке 215 (рис. 4) предпочтения избирателей принадлежат 4 кандидатам от КПРФ и 1 кандидату от ЕР. На участках 219 и 220 (рис. 5) после исключения непрерывных серий за административных кандидатов остаются случайные голоса явно в пользу 5 кандидатов от КПРФ. На участке 219 комиссия вообще не перемешивала бюллетени разных дней, что позволило нам проследить происхождение каждого бюллетеня в стенограмме. Бюллетени с 96 по 221 взяты из пакета первого дня, который был подменен. Эти бюллетени содержат несколько случайных голосов (вероятно, вставленных намеренно в попытке скрыть фальсификацию), но тем не менее, среди них нет ни одного голоса за популярного кандидата Разину (КПРФ), что формирует крайне маловероятную серию, приводя к $\alpha' = 2{,}2 \times 10^{-19}$.

В избирательном округе № 2 (рис. 4) стенограммы демонстрируют еще одно примечательное несоответствие. На участке 215 имеются всего 6 бюллетеней с отметками за всех 5 кандидатов, получивших мандаты: Вавилову от ЛДПР и 4 кандидатов (Сербу, Федорива, Шик и Щербович) от ЕР. Этот набор кандидатов вряд ли будет выбран реальным избирателем с искренними про-административными политическими предпочтениями, потому что избирателю пришлось бы проигнорировать Булаха — пятого кандидата от ЕР — в пользу Вавиловой от другой партии. Действительно, 38 человек проголосовали здесь за всех 5 кандидатов от ЕР, включая Булаха. Однако, на участках 216 и 217 поданы соответственно 176 и 175 бюллетеней за всех 5 кандидатов, получивших мандаты, — всего на несколько бюллетеней меньше, чем содержится во вскрывавшихся ночью сейф-пакетах. Аналогично, на избирательном участке 219 (рис. 5), 94 из 104 бюллетеней с 5 отметками за маловероятный набор кандидатов, получивших мандаты (Аникееву от СРЗП и 4 кандидатов от ЕР), были извлечены из подмененного сейф-пакета. Справедливости ради следует сказать, что в округе № 3 ЕР неожиданно потеряла своего пятого кандидата, снявшего свою кандидатуру за несколько дней до голосования после того, как стало публично известно, что он утаил свою судимость.

Несколько бумажных бюллетеней были поданы избирателями, которые по медицинским и иным причинам не смогли явиться на избирательный участок и попросили членов комиссии посетить их на дому с переносной урной для голосования. Количество таких избирателей невелико (от 0 до 22 на участок) и не оказывает существенного влияния на итоги.

### Приложение B: Вычисление значимостей

Предметом анализа является последовательность действительных бюллетеней, рассматриваемых в том порядке, в котором они были извлечены из урны и оглашены на подсчете голосов. Недействительные бюллетени из рассмотрения исключаются.

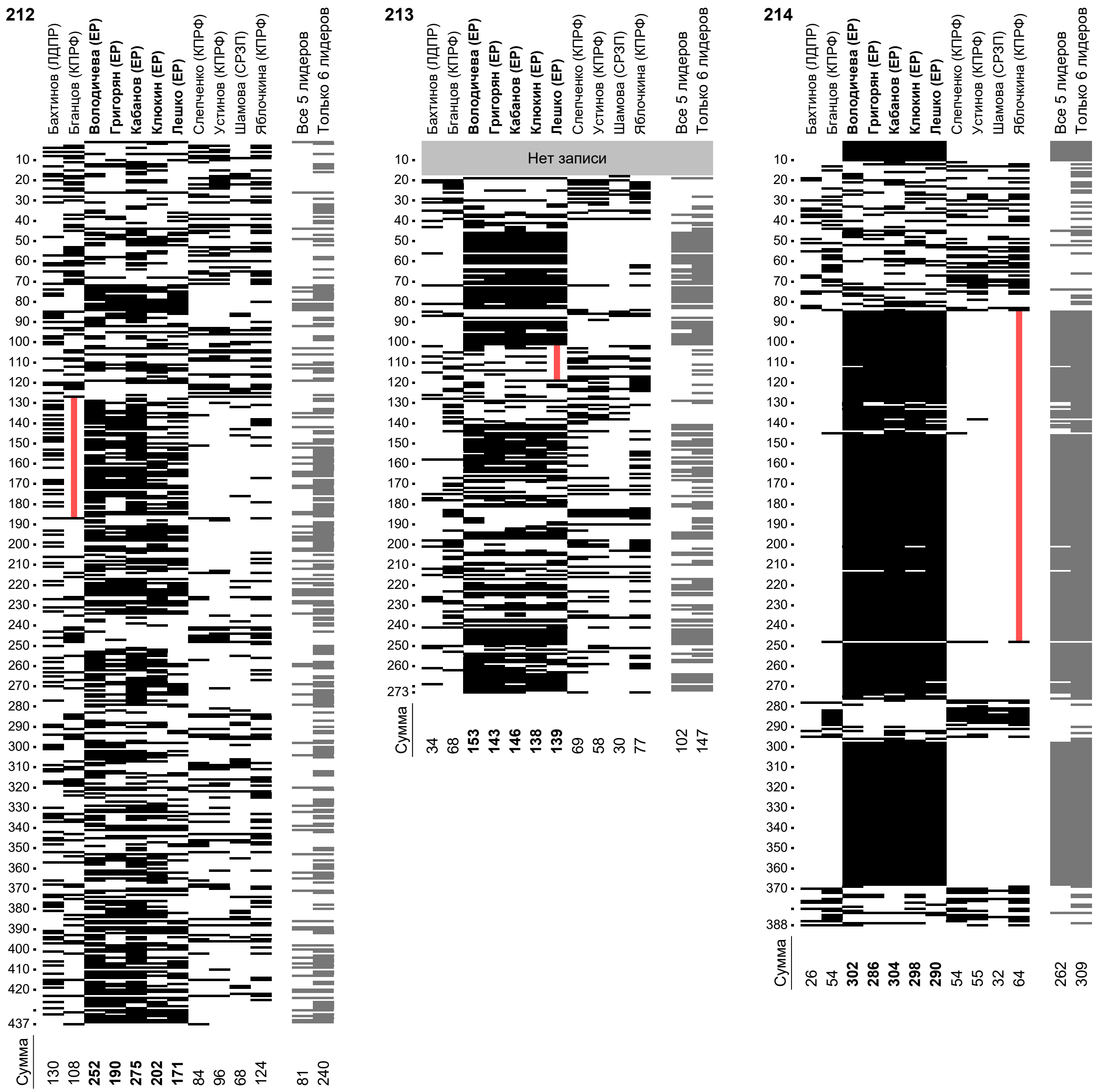

Рис. 3. Стенограммы подсчета голосов на участках 212, 213 и 214 избирательного округа № 1. Жирным шрифтом выделены 5 лидирующих на участке кандидатов. Наименее вероятная серия обозначена красной (серой) линией. Последние два серых столбца — консолидированные флаги (приложение В 3). Вне помещения для голосования проголосовали 4, 17, 0 избирателей соответственно. На участке 213 наша запись начинается с опозданием и пропускает около 17 бюллетеней в начале оглашения, поэтому в наших расчетах используются только записанные 256 бюллетеней

Пусть $n$ — число действительных бюллетеней на избирательном участке. Для каждого кандидата, баллотирующегося на этом участке, последовательность бюллетеней преобразуется в последовательность флагов $f \in \{0; 1\}$. Ноль означает, что кандидат не был отмечен избирателем, заполнившим бюллетень, а единица, — что был.

Пусть у некоторого кандидата имеются в общей сложности $r_0$ нулей и $r_1$ единиц ($r_0 + r_1 = n$). Доля избирателей, поддержавших этого кандидата, равна $p_1 = q_0 = r_1/n$, а доля избирателей, не поддерживающих его, равна $p_0 = q_1 = r_0/n$. Эти доли рассматриваются как вероятности появления единиц и нулей в последовательности флагов с нормировкой

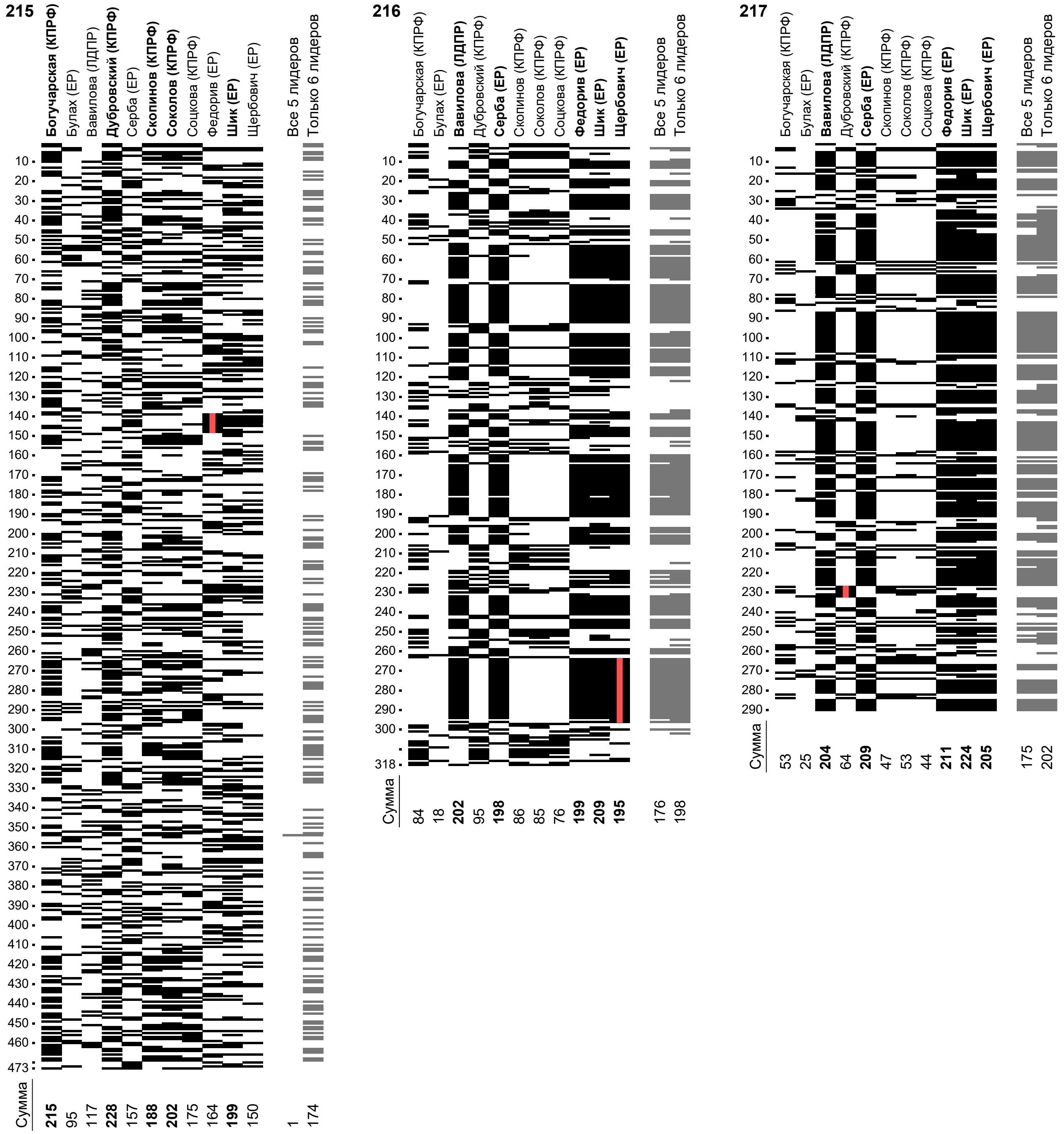

Рис. 4. Стенограммы подсчета голосов на участках 215, 216 и 217 избирательного округа № 2. Жирным шрифтом выделены 5 лидирующих на участке кандидатов. Наименее вероятная серия обозначена красной (серой) линией. Последние два серых столбца — консолидированные флаги (приложение В 3). Вне помещения для голосования проголосовали 4, 6, 9 избирателей соответственно

$p_f + q_f = 1$. Последовательные бюллетени считаются взаимно независимыми.

### 1. Вероятность длинной серии флагов

Пусть самые длинные непрерывные серии нулей и единиц для кандидата имеют длину $s_0$ и $s_1$ флагов соответственно. Если вероятность $p_f$ появления флага

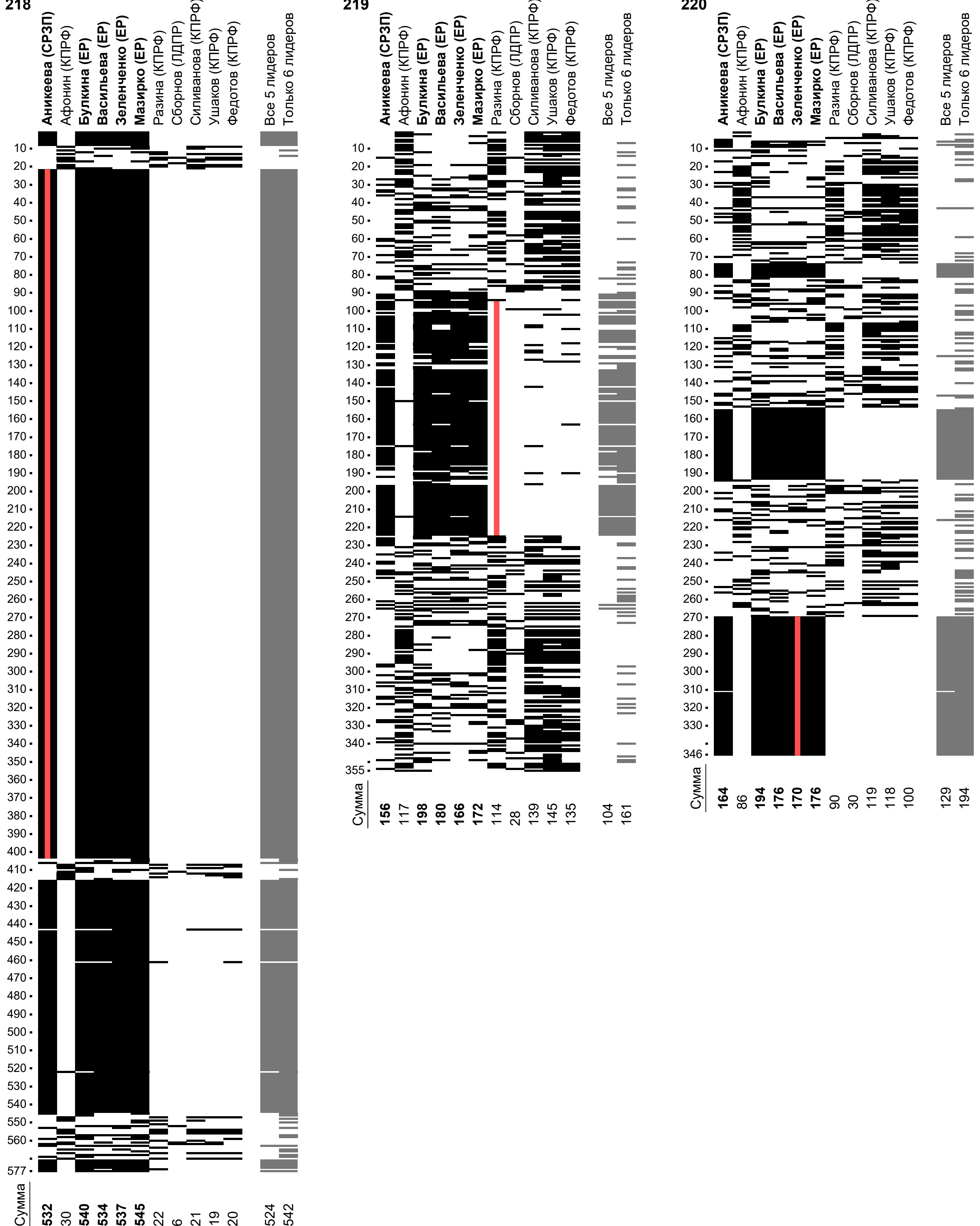

Рис. 5. Стенограммы подсчета голосов на участках 218, 219 и 220 избирательного округа № 3. Жирным шрифтом выделены 5 лидирующих на участке кандидатов. Наименее вероятная серия обозначена красной (серой) линией. Последние два серых столбца — консолидированные флаги (приложение В 3). Вне помещения для голосования проголосовали 0, 22, 1 избирателей соответственно $f$ недостаточно велика, чтобы объяснить длину серии $s_f$, можно заподозрить фальсификацию итогов голосования путем вброса пачки бюллетеней, не заполненных избирателями. Все бюллетени в такой пачке, скорее всего, заполнены одинаково, а волеизъявление каждого реального избирателя индивидуально, поэтому каждый из них может легко разорвать серию.

Подозрения в фальсификации можно проверить с помощью математической статистики. Сформулируем нулевую гипотезу о том, что длинная непрерывная серия флагов случайно возникла естественным путем. Альтернативная гипотеза заключается в том, что имело место некоторое вмешательство, приведшее к появлению длинной серии. Выбор между этими двумя гипотезами основан на статистической значимости нулевой гипотезы $\alpha$. Она измеряет риск так называемой ошибки первого рода. Это означает, что $\alpha$ — это вероятность ошибочно отклонить верную нулевую гипотезу. Другими словами, если значимость нулевой гипотезы оказывается малой, это означает, что имела место фальсификация. Обратное, как правило, неверно: если значимость велика, мы не можем обоснованно утверждать что-либо о наличии или отсутствии фальсификации.

Алгоритм точного вычисления значимости для самой длинной серии флагов довольно сложен. Поэтому сначала оценим $\alpha$ сверху простым способом. Для этого предположим, что присутствует только одна серия флагов $f$ длины не менее $s_f$. Это предположение является точным только для достаточно длинных серий, тогда как для коротких оно приводит к переоценке $\alpha_f$, поскольку короткие серии могут встречаться в последовательности несколько раз, и каждая серия учитывается столько раз, сколько она встречается. Такой подход приемлем, поскольку он только ослабляет статистический тест.

При сделанном предположении вероятность наблюдения серии длиной не менее $s$, начинающейся с самого первого флага, равна $p^s$ (здесь и далее мы опускаем индекс $f$). Вероятность наблюдения такой серии, начинающейся с любого из последующих $n - s$ флагов, равна $qp^s$, поскольку флаг, предшествующий серии, должен отличаться от флагов в ней. События наблюдения серий, начинающихся с разных позиций флагов, несовместимы, что позволяет просуммировать их вероятности. Таким образом, значимость гипотезы о естественном возникновении серии $\alpha \le [1+q(n-s)]p^s$. Эту простую формулу можно использовать для быстрых оценок. Сравнение с результатами точного расчета, описанного ниже, показывает, что для рассматриваемых данных она завышает $\alpha$ не более, чем на 5% при $\alpha < 10^{-1}$. Но завышение быстро растет для бо́льших $\alpha$.

Точный расчет требует некоторых дополнительных усилий. Пусть $u_{n,i,j}$ — вероятность того, что в последовательности из $n$ флагов самая длинная серия флагов $f$ имеет длину ровно $i$ бюллетеней $(0 \le i \le n)$ при условии, что ровно $j$ $(0 \le j \le i)$ последних флагов имеют то же состояние $f$, что и флаги серии. Эти вероятности можно найти с помощью итерационного процесса, соответствующего добавлению еще одного флага к последовательности (т. е., объявлению следующего бюллетеня):

$$u_{n+1,i,j} = \begin{cases} p(u_{n,i,j-1} + u_{n,i-1,j-1}) & \text{если } j = i \\ pu_{n,i,j-1} & \text{если } 0 < j < i \\ qw_{n,i} & \text{если } j = 0 \end{cases},$$

где $w_{n,i} = \sum_{j=0}^{i} u_{n,i,j}$ — безусловная вероятность того, что самая длинная серия флагов $f$ имеет длину ровно $s$. Нормировка $u_{0,0,0} = 1$ здесь служит начальным условием, а ограничения $u_{n,n+1,j} = 0$ и $w_{n,n+1} = 0$ — граничными условиями.

При реализации на компьютере алгоритм можно упростить, используя только двумерный массив для $u$ и только одномерный массив для $w$. Новые значения вероятности, полученные при увеличении $n$, записываются поверх старых. Чтобы избежать ссылок на значения, которые еще не определены или уже переопределены, массивы следует инициализировать нулями, индексы $i$ и $j$ следует перебирать в порядке убывания, а цикл по $j$ вложить в цикл по $i$.

Точная вероятность того, что самая длинная серия имеет длину не менее $s$, равна $\alpha = \sum_{i=s}^{n} w_{n,i}$. Таблицы II–IV содержат исходные данные и результаты этого расчета для каждого кандидата на каждом избирательном участке (см. первые 6 столбцов каждого участка).

Здесь мы сталкиваемся с парадоксальным эффектом, когда предположительно подлинные серии флагов определяются тестом как аномальные. Дело в том, что фальсификации смещают оценки вероятностей $p$ и $q$ от их истинных значений. Относительно длинные серии голосов за оппозиционных или против административных кандидатов могут возникнуть при истинных вероятностях такого волеизъявления избирателей. Но эти же серии становятся маловероятными при смещенных оценках вероятностей. Впрочем, следует отметить, что значимости для таких серий всё еще не столь малы, как для серий голосов за административных и против оппозиционных кандидатов.

Событие с вероятностью $1/t$ происходит в среднем один раз на $t$ независимых испытаний (точнее, для больши́х $t$ оно вообще не происходит с вероятностью $1/e \approx 36{,}8\%$, происходит ровно один раз с вероятностью $1/e \approx 36{,}8\%$, дважды — с вероятностью $1/2e \approx 18{,}4\%$ и трижды и чаще — с вероятностью $1 - 2{,}5/e \approx 8{,}0\%$). Имеется $c = 11$ кандидатов для каждого из 9 рассматриваемых участков и 2 возможных значения флага. Таким образом, $t = 11 \times 9 \times 2 = 198$. Следовательно, подозрительными являются только значимости $\alpha_f < 1/198$. Десятичные показатели таких значимостей выделены в таблицах жирным шрифтом. Аналогичным образом, десятичные показатели $\alpha' < 1/9$ выделены жирным шрифтом в таблице I, где множественность кандидатов и

Таблица II. Результаты расчетов по участкам избирательного округа № 1

| Кандидат (партия) | Избирательный участок 212 | | | | | | | | | Избирательный участок 213 | | | | | | | | | Избирательный участок 214 | | | | | | | | |
|---|---|---|---|---|---|---|---|---|---|---|---|---|---|---|---|---|---|---|---|---|---|---|---|---|---|---|---|
| | $r_0$ | $s_0$ | $\mathrm{p}\alpha_0$ | $r_1$ | $s_1$ | $\mathrm{p}\alpha_1$ | $m$ | $\mathrm{p}\tilde{\alpha}$ | $\mathrm{p}\breve{\alpha}$ | $r_0$ | $s_0$ | $\mathrm{p}\alpha_0$ | $r_1$ | $s_1$ | $\mathrm{p}\alpha_1$ | $m$ | $\mathrm{p}\tilde{\alpha}$ | $\mathrm{p}\breve{\alpha}$ | $r_0$ | $s_0$ | $\mathrm{p}\alpha_0$ | $r_1$ | $s_1$ | $\mathrm{p}\alpha_1$ | $m$ | $\mathrm{p}\tilde{\alpha}$ | $\mathrm{p}\breve{\alpha}$ |
| Бахтинов (ЛДПР) | 307 | 16 | 0,4 | 130 | 4 | 0,0 | 197 | 0,1 | 0,4 | 222 | 29 | 0,4 | 34 | 3 | 0,4 | 59 | 0,3 | 0,1 | 362 | 194 | **4,7** | 26 | 2 | 0,1 | 38 | 1,0 | **5,5** |
| Бганцов (КПРФ) | 329 | 59 | **5,3** | 108 | 5 | 0,6 | 138 | 1,7 | **4,7** | 188 | 32 | **2,5** | 68 | 5 | 0,7 | 83 | 1,5 | **4,2** | 334 | 102 | **5,0** | 54 | 8 | **4,3** | 60 | **3,1** | **13,4** |
| Володичева (ЕР) | 185 | 9 | 1,0 | 252 | 12 | 0,7 | 184 | **2,5** | **4,9** | 103 | 10 | 1,8 | 153 | 20 | **2,5** | 92 | **4,1** | **5,6** | 86 | 10 | **4,1** | 302 | 72 | **6,0** | 55 | **12,6** | **21,2** |
| Григорян (ЕР) | 247 | 12 | 0,7 | 190 | 9 | 0,9 | 168 | **5,3** | **2,9** | 113 | 12 | 2,1 | 143 | 16 | 2,0 | 90 | **5,5** | **7,9** | 102 | 19 | **8,6** | 286 | 71 | **7,5** | 52 | **19,4** | **24,7** |
| Кабанов (ЕР) | 162 | 11 | **2,3** | 275 | 14 | 0,7 | 156 | **5,1** | **3,9** | 110 | 10 | 1,5 | 146 | 14 | 1,4 | 92 | **4,7** | **7,2** | 84 | 13 | **6,2** | 304 | 104 | **9,2** | 54 | **12,5** | **21,8** |
| Клюкин (ЕР) | 235 | 11 | 0,7 | 202 | 9 | 0,7 | 188 | **2,6** | 1,9 | 118 | 10 | 1,2 | 138 | 11 | 0,9 | 91 | **5,5** | **6,8** | 90 | 15 | **7,1** | 298 | 72 | **6,4** | 62 | **11,9** | **19,5** |
| Лешко (ЕР) | 266 | 24 | **3,0** | 171 | 12 | **2,5** | 174 | **3,1** | **3,1** | 117 | 17 | **3,7** | 139 | 10 | 0,6 | 95 | **4,5** | **4,4** | 98 | 15 | **6,5** | 290 | 102 | **11,0** | 66 | **13,1** | **23,3** |
| Слепченко (КПРФ) | 353 | 35 | 1,4 | 84 | 4 | 0,4 | 116 | 1,3 | **2,5** | 187 | 32 | **2,6** | 69 | 4 | 0,2 | 90 | 1,0 | **3,0** | 334 | 102 | **5,0** | 54 | 7 | **3,5** | 65 | **2,4** | **11,5** |
| Устинов (КПРФ) | 341 | 56 | **4,1** | 96 | 7 | 2,1 | 117 | **2,5** | **4,8** | 198 | 32 | 1,9 | 58 | 4 | 0,4 | 81 | 0,8 | **4,2** | 333 | 109 | **5,6** | 55 | 9 | **5,1** | 54 | **4,4** | **10,6** |
| Шамова (СРЗП) | 369 | 41 | 1,2 | 68 | 5 | 1,5 | 103 | 0,8 | 1,3 | 226 | 45 | 1,0 | 30 | 4 | 1,4 | 48 | 0,6 | 1,6 | 356 | 193 | **6,0** | 32 | 5 | **2,9** | 49 | 0,8 | **9,2** |
| Яблочкина (КПРФ) | 313 | 16 | 0,4 | 124 | 4 | 0,1 | 159 | 1,2 | 1,4 | 179 | 22 | 1,6 | 77 | 8 | 1,9 | 82 | **2,7** | **3,1** | 324 | 163 | **11,2** | 64 | 9 | **4,5** | 60 | **5,3** | **16,3** |

Таблица III. Результаты расчетов по участкам избирательного округа № 2

| Кандидат (партия) | Избирательный участок 215 | | | | | | | | | Избирательный участок 216 | | | | | | | | | Избирательный участок 217 | | | | | | | | |
|---|---|---|---|---|---|---|---|---|---|---|---|---|---|---|---|---|---|---|---|---|---|---|---|---|---|---|---|
| | $r_0$ | $s_0$ | $\mathrm{p}\alpha_0$ | $r_1$ | $s_1$ | $\mathrm{p}\alpha_1$ | $m$ | $\mathrm{p}\tilde{\alpha}$ | $\mathrm{p}\breve{\alpha}$ | $r_0$ | $s_0$ | $\mathrm{p}\alpha_0$ | $r_1$ | $s_1$ | $\mathrm{p}\alpha_1$ | $m$ | $\mathrm{p}\tilde{\alpha}$ | $\mathrm{p}\breve{\alpha}$ | $r_0$ | $s_0$ | $\mathrm{p}\alpha_0$ | $r_1$ | $s_1$ | $\mathrm{p}\alpha_1$ | $m$ | $\mathrm{p}\tilde{\alpha}$ | $\mathrm{p}\breve{\alpha}$ |
| Богучарская (КПРФ) | 258 | 14 | 1,4 | 215 | 7 | 0,2 | 217 | 1,3 | 1,2 | 234 | 34 | **2,6** | 84 | 6 | 1,1 | 93 | **2,9** | **5,3** | 237 | 26 | 0,6 | 53 | 4 | 0,6 | 77 | 0,8 | 0,4 |
| Булах (ЕР) | 378 | 22 | 0,3 | 95 | 3 | 0,0 | 149 | 0,4 | 0,9 | 300 | 62 | 0,4 | 18 | 1 | 0,0 | 37 | 0,2 | 1,0 | 265 | 31 | 0,1 | 25 | 3 | 0,8 | 43 | 0,4 | 0,2 |
| Вавилова (ЛДПР) | 356 | 14 | 0,0 | 117 | 4 | 0,1 | 185 | 0,1 | 0,3 | 116 | 13 | **3,4** | 202 | 31 | **4,1** | 76 | **13,8** | **7,7** | 86 | 6 | 0,9 | 204 | 21 | 1,3 | 89 | **3,4** | **2,4** |
| Дубровский (КПРФ) | 245 | 8 | 0,2 | 228 | 8 | 0,3 | 221 | 1,1 | 1,6 | 223 | 33 | **3,2** | 95 | 6 | 0,8 | 94 | **4,5** | **6,4** | 226 | 21 | 0,6 | 64 | 6 | 1,6 | 81 | 1,6 | 0,3 |
| Серба (ЕР) | 316 | 14 | 0,4 | 157 | 4 | 0,0 | 194 | 1,0 | 0,1 | 120 | 9 | 1,5 | 198 | 31 | **4,3** | 83 | **12,2** | **8,1** | 81 | 6 | 1,0 | 209 | 21 | 1,1 | 83 | **3,7** | 1,4 |
| Скопинов (КПРФ) | 285 | 11 | 0,3 | 188 | 7 | 0,4 | 208 | 1,3 | 1,0 | 232 | 33 | **2,6** | 86 | 5 | 0,5 | 85 | **4,6** | **5,7** | 243 | 28 | 0,6 | 47 | 4 | 0,8 | 69 | 0,8 | 1,1 |
| Соколов (КПРФ) | 271 | 9 | 0,1 | 202 | 7 | 0,3 | 209 | 1,7 | 0,3 | 233 | 33 | **2,6** | 85 | 5 | 0,6 | 81 | **5,1** | **6,5** | 237 | 28 | 0,8 | 53 | 5 | 1,3 | 71 | 1,3 | 0,2 |
| Соцкова (КПРФ) | 298 | 16 | 1,0 | 175 | 6 | 0,3 | 183 | **3,2** | 1,8 | 242 | 33 | 2,1 | 76 | 6 | 1,4 | 85 | **2,9** | **5,2** | 246 | 28 | 0,5 | 44 | 4 | 0,9 | 73 | 0,4 | 0,5 |
| Федорив (ЕР) | 309 | 10 | 0,0 | 164 | 10 | 2,1 | 193 | 1,5 | 0,1 | 119 | 13 | **3,3** | 199 | 31 | **4,3** | 65 | **18,9** | **8,7** | 79 | 5 | 0,6 | 211 | 21 | 1,0 | 87 | **2,7** | 1,0 |
| Шик (ЕР) | 274 | 9 | 0,1 | 199 | 6 | 0,1 | 221 | 0,7 | 0,6 | 109 | 11 | **2,8** | 209 | 31 | **3,7** | 91 | **7,6** | **6,2** | 66 | 5 | 0,9 | 224 | 21 | 0,6 | 89 | 1,0 | 0,8 |
| Щербович (ЕР) | 323 | 13 | 0,2 | 150 | 6 | 0,6 | 203 | 0,4 | 2,0 | 123 | 13 | **3,1** | 195 | 33 | **5,0** | 61 | **22,0** | **10,2** | 85 | 4 | 0,1 | 205 | 21 | 1,3 | 89 | **3,3** | 1,8 |

Таблица IV. Результаты расчетов по участкам избирательного округа № 3

| Кандидат (партия) | Избирательный участок 218 | | | | | | | | | Избирательный участок 219 | | | | | | | | | Избирательный участок 220 | | | | | | | | |
|---|---|---|---|---|---|---|---|---|---|---|---|---|---|---|---|---|---|---|---|---|---|---|---|---|---|---|---|
| | $r_0$ | $s_0$ | $\mathrm{p}\alpha_0$ | $r_1$ | $s_1$ | $\mathrm{p}\alpha_1$ | $m$ | $\mathrm{p}\tilde{\alpha}$ | $\mathrm{p}\breve{\alpha}$ | $r_0$ | $s_0$ | $\mathrm{p}\alpha_0$ | $r_1$ | $s_1$ | $\mathrm{p}\alpha_1$ | $m$ | $\mathrm{p}\tilde{\alpha}$ | $\mathrm{p}\breve{\alpha}$ | $r_0$ | $s_0$ | $\mathrm{p}\alpha_0$ | $r_1$ | $s_1$ | $\mathrm{p}\alpha_1$ | $m$ | $\mathrm{p}\tilde{\alpha}$ | $\mathrm{p}\breve{\alpha}$ |
| Аникеева (СРЗП) | 45 | 13 | **11,7** | 532 | 382 | **12,3** | 19 | **10,4** | **26,0** | 199 | 30 | **5,4** | 156 | 32 | **9,2** | 85 | **21,2** | **26,5** | 182 | 21 | **3,7** | 164 | 41 | **11,1** | 70 | **29,4** | **25,1** |
| Афонин (КПРФ) | 547 | 385 | **7,9** | 30 | 5 | **3,7** | 31 | **2,6** | **13,9** | 238 | 55 | **7,6** | 117 | 11 | **2,9** | 125 | **3,0** | **17,4** | 260 | 82 | **8,4** | 86 | 6 | 1,2 | 92 | **3,7** | **12,7** |
| Булкина (ЕР) | 37 | 5 | **3,2** | 540 | 383 | **9,9** | 31 | **4,4** | **17,3** | 157 | 10 | 1,3 | 198 | 47 | **9,8** | 114 | **10,3** | **20,5** | 152 | 10 | 1,3 | 194 | 78 | **17,5** | 94 | **15,9** | **20,6** |
| Васильева (ЕР) | 43 | 8 | **6,3** | 534 | 383 | **11,7** | 29 | **6,6** | **26,9** | 175 | 26 | **5,8** | 180 | 29 | **6,3** | 99 | **17,0** | **24,8** | 170 | 13 | 1,8 | 176 | 77 | **20,5** | 86 | **21,3** | **22,0** |
| Зеленченко (ЕР) | 40 | 11 | **10,0** | 537 | 382 | **10,8** | 25 | **6,7** | **25,0** | 189 | 26 | **4,9** | 166 | 42 | **11,6** | 89 | **20,9** | **29,6** | 176 | 18 | **3,1** | 170 | 77 | **21,6** | 68 | **31,2** | **25,3** |
| Мазирко (ЕР) | 32 | 5 | **3,5** | 545 | 385 | **8,5** | 27 | **3,8** | **18,3** | 183 | 13 | 1,5 | 172 | 35 | **8,8** | 101 | **16,0** | **22,5** | 170 | 17 | **3,0** | 176 | 77 | **20,5** | 82 | **23,3** | **26,7** |
| Разина (КПРФ) | 555 | 386 | **5,6** | 22 | 3 | 1,5 | 35 | 0,7 | **11,0** | 241 | 130 | **20,0** | 114 | 8 | 1,6 | 104 | **6,4** | **17,0** | 256 | 77 | **8,2** | 90 | 6 | 1,1 | 101 | **2,9** | **9,0** |
| Сборнов (ЛДПР) | 571 | 392 | 1,3 | 6 | 2 | 1,2 | 11 | 0,3 | 1,6 | 327 | 134 | **3,5** | 28 | 4 | 1,9 | 47 | 0,5 | **3,1** | 316 | 84 | 1,9 | 30 | 2 | 0,0 | 53 | 0,4 | 1,3 |
| Силиванова (КПРФ) | 556 | 385 | **5,3** | 21 | 3 | 1,6 | 31 | 0,9 | **9,9** | 216 | 32 | **4,8** | 139 | 14 | **3,4** | 115 | **7,8** | **13,8** | 227 | 77 | **12,1** | 119 | 8 | 1,4 | 107 | **6,3** | **13,5** |
| Ушаков (КПРФ) | 558 | 386 | **4,8** | 19 | 3 | 1,7 | 27 | 0,9 | **7,4** | 210 | 96 | **19,9** | 145 | 12 | **2,4** | 107 | **10,9** | **24,5** | 228 | 77 | **12,0** | 118 | 13 | **3,7** | 105 | **6,6** | **16,1** |
| Федотов (КПРФ) | 557 | 386 | **5,0** | 20 | 3 | 1,6 | 27 | 1,1 | **6,2** | 220 | 40 | **6,2** | 135 | 7 | 0,7 | 119 | **6,3** | **18,1** | 246 | 77 | **9,5** | 100 | 8 | 1,9 | 109 | **3,1** | **13,8** |

состояний флага уже учтена.

Следует особо подчеркнуть, что подозрение — это еще не доказательство. Появление одного или даже нескольких подозрительных значений вполне ожидаемо в каждом тесте. Подозрительные значения требуют только внимания, тогда как неподозрительные можно игнорировать.

Тест не обнаруживает ни одного кандидата с подозрительными $\alpha_f$ для участков 215 и 217 (таблица III), обнаруживает 4 таких кандидатов для участков 212 и 213 (таблица II), 9 — для участка 216 (таблица III) и 10 или 11 — для участков 214 (таблица II), 218, 219 и 220 (таблица IV). Более того, для последних 4 участков $\alpha_{\min} < 10^{-11}$, т. е., вероятность отсутствия фальсификаций на них крайне низка.

Низкая значимость гипотезы о естественном возникновении аномалии данных явно указывает на то, что имело место фальсификация того типа, на который ориентирован тест. Однако высокая значимость может означать либо отсутствие фальсификаций вообще, либо вмешательство какого-то иного типа. В контексте настоящего анализа таким вмешательством может быть, в частности, дополнительное перемешивание бюллетеней перед их подсчетом (таблица I). Перемешивание не требуется по закону и не является нарушением, но оно затрудняет обнаружение вбросов бюллетеней и потому вызывает подозрения.

## 2. Вероятность количества серий флагов

При частичном перемешивании бюллетеней они уже не образуют непрерывную длинную серию даже в гипотетически вброшенной пачке. Однако, если перемешивание не слишком тщательное, останутся неразорванные фрагменты этой пачки. Из-за этого количество серий $m$ последовательных флагов одинакового состояния будет ниже, чем было бы в нормальной ситуации. Поэтому давайте вычислим вероятность появления определенного количества серий в последовательности взаимно независимых бюллетеней.

Пусть $a_{n,i}$ и $b_{n,i}$ — вероятности того, что последовательность из $n$ флагов содержит $i$ серий, при условии, что последний флаг последовательности имеет определенное состояние $f$ или противоположное состояние $1-f$ соответственно (выбор этих состояний не влияет на конечные результаты, поэтому индекс $f$ в вычислениях ниже опущен). Вероятности последовательно вычисляются с использованием итерационных уравнений $a_{n+1,i} = p(a_{n,i}+b_{n,i-1})$ и $b_{n+1,i} = q(b_{n,i}+a_{n,i-1})$, соответствующих добавлению очередного флага в последовательность. Если новый флаг совпадает с последним флагом последовательности, то количество серий не меняется, а если отличается, то увеличивается на 1. Начальные условия: $a_{1,1} = p$ и $b_{1,1} = q$, граничные условия: $a_{n,0} = b_{n,0} = a_{n,n+1} = b_{n,n+1} = 0$.

Вероятность обнаружения ровно $i$ серий в последовательности из $n$ бюллетеней равна $v_{n,i} = a_{n,i} + b_{n,i}$. Таким образом, значимость гипотезы о том, что итоговое число серий $m$ окажется малым по естественным причинам, $\tilde{\alpha} = \sum_{i=1}^{m} v_{n,i}$.

При компьютерной реализации этого алгоритма, как и для предыдущего, можно уменьшить размерность используемых массивов. Для этого следует инициализировать массивы нулями и перебрать индекс $i$ в порядке убывания.

Мы также можем получить аналитическую оценку $\tilde{\alpha} \approx \Phi(m+1/2; \mu_n, \sigma_n)$, где $\Phi$ — гауссова функция распределения. Добавка $1/2$ в первом аргументе частично компенсирует разницу между суммой дискретных вероятностей, дающей точную значимость, и интегралом от плотности вероятности, дающим оценку значимости. Математическое ожидание $\mu_n$ и дисперсию $\sigma_n^2$ можно точно вычислить с помощью производящих функций. Пусть $g_n(z) = \sum_i a_{n,i} z^i$, $h_n(z) = \sum_i b_{n,i} z^i$, и $y_n(z) = g_n(z) + h_n(z) = \sum_i v_{n,i} z^i$. Тогда итерационные уравнения принимают вид $g_{n+1} = p(g_n + z h_n)$ и $h_{n+1} = q(h_n + z g_n)$, что позволяет выразить параметры распределения через производные производящих функций $\mu_n = y_n'(1) = 1 + (n-1)\theta$ и, при $n > 1$, $\sigma_n^2 = y_n''(1) + \mu_n - \mu_n^2 = (2n-3)\theta - (3n-5)\theta^2$, где $\theta = 2pq$ — средняя частота переключения между сериями. Промежуточные вычисления здесь довольно громоздки, поэтому мы их опустим.

Это приближение имеет существенный недостаток по сравнению с оценкой для первого теста. Здесь значимость может быть не только завышена (что допустимо), но и занижена (что крайне нежелательно). Поэтому к приближенным значениям $\tilde{\alpha}$ следует относится с осторожностью. Для рассматриваемых данных наибольшее завышение может достигать нескольких порядков (но хотя бы 1 порядка оно достигает только для $\tilde{\alpha} < 10^{-6}$), а наибольшее занижение составляет менее 6% при $\tilde{\alpha} < 10^{-1}$.

Таблицы II–IV содержат результаты точных расчетов для всех кандидатов на всех избирательных участках (см. седьмой и восьмой столбцы каждого участка).

В отличие от предыдущего теста, здесь вычисляется только одна значимость, поэтому $t = 11 \times 9 = 99$ и подозрительными становятся значимости $\tilde{\alpha} < 1/99$. Аналогичное условие $\tilde{\alpha}' < 1/9$ остается неизменным для общей значимости в таблице I.

Качественные результаты обоих тестов в целом близки для всех участков, кроме 216. Первый тест оставляет некоторые сомнения, тогда как второй успешно доказывает фальсификацию, несмотря на намеренное перемешивание бюллетеней членами избирательной комиссии (таблица I). Это иллюстрирует, что второй тест полезен. Более того, второй тест устойчив к незначительным человеческим ошибкам в стенограмме, тогда как первый тест устойчив лишь до тех пор, пока такие ошибки не влияют на наименее вероятную серию.

Однако имеется и некоторая проблема со вторым тестом. Он дает $\tilde{\alpha}_{\min} = 7{,}5 \times 10^{-4}$ на избирательном участке 215, где, скорее всего, не было фальсификаций. Конечно, такая значимость может быть и результатом неудачного стечения обстоятельств. Тем более, что она наблюдается только для 1 кандидата (для сравнения, сопоставимые подозрительные значения $\tilde{\alpha}$ на избирательном участке 217 наблюдаются для 4 кандидатов). Но другое возможное объяснение состоит в том, что последовательные бюллетени на самом деле не являются абсолютно независимыми. Возможно, они не полностью перемешиваются для одновременно голосующих избирателей. Если это так, то второй тест оказывается более чувствительным к неточности наших первоначальных предположений.

## 3. Консолидированные тесты по группам кандидатов

Чтобы ослабить гипотетический эффект, отмеченный выше, исходные данные для статистических тестов консолидируются таким образом, чтобы анализировать последовательность флагов не для каждого

Таблица V. Результаты расчетов консолидированных тестов

| Участок | $r_0$ | $s_0$ | р$\alpha_0$ | $r_1$ | $s_1$ | р$\alpha_1$ | $m$ | р$\tilde{\alpha}$ | р$\breve{\alpha}$ |
|---|---|---|---|---|---|---|---|---|---|
| 212 | 356 | 24 | 0,4 | 81 | 4 | 0,5 | 126 | 0,5 | **1,6** |
| 213 | 154 | 27 | **4,0** | 102 | 10 | **1,8** | 81 | **6,6** | **7,2** |
| 214 | 126 | 34 | **14,2** | 262 | 71 | **10,1** | 36 | **37,6** | **30,3** |
| 215 | 472 | 353 | 0,2 | 1 | 1 | 0,2 | 3 | 0,1 | 0,4 |
| 216 | 142 | 19 | **4,4** | 176 | 31 | **5,9** | 57 | **29,9** | **12,7** |
| 217 | 115 | 11 | **2,2** | 175 | 21 | **2,6** | 71 | **14,2** | **3,6** |
| 218 | 53 | 18 | **16,0** | 524 | 382 | **14,7** | 19 | **13,4** | **32,8** |
| 219 | 251 | 90 | **11,7** | 104 | 17 | **6,7** | 45 | **23,0** | **35,4** |
| 220 | 217 | 53 | **8,7** | 129 | 41 | **15,3** | 20 | **51,0** | **39,7** |

Таблица VI. Результаты расчетов адаптированных консолидированных тестов

| Участок | $r_0$ | $s_0$ | р$\alpha_0$ | $r_1$ | $s_1$ | р$\alpha_1$ | $m$ | р$\tilde{\alpha}$ | р$\breve{\alpha}$ |
|---|---|---|---|---|---|---|---|---|---|
| 212 | 197 | 9 | 0,8 | 240 | 15 | **1,6** | 176 | **4,2** | **3,9** |
| 213 | 109 | 10 | **1,6** | 147 | 15 | **1,6** | 99 | **3,2** | **4,7** |
| 214 | 79 | 10 | **4,4** | 309 | 102 | **8,3** | 62 | **8,7** | **20,3** |
| 215 | 299 | 14 | 0,6 | 174 | 6 | 0,3 | 208 | 0,8 | 0,7 |
| 216 | 120 | 16 | **4,5** | 198 | 33 | **4,8** | 67 | **18,4** | **10,9** |
| 217 | 88 | 7 | **1,3** | 202 | 26 | **2,2** | 71 | **7,7** | 0,7 |
| 218 | 35 | 7 | **5,8** | 542 | 385 | **9,4** | 33 | **3,5** | **17,4** |
| 219 | 194 | 23 | **3,9** | 161 | 17 | **3,6** | 105 | **13,7** | **22,5** |
| 220 | 152 | 15 | **3,1** | 194 | 77 | **17,3** | 98 | **14,4** | **19,3** |

кандидата в отдельности, а для всего их списка сразу. Состояние флагов теперь определяется отношением избирателя к 5 кандидатам, лидирующим на избирательном участке. Если избиратель поддержал их всех, то $f = 1$, а в противном случае $f = 0$ (как показано в предпоследнем сером столбце для каждого участка на рис. 3–5). Все вычисления выполняются так же, как и ранее, а таблица V содержит их результаты. Подозрительные значения для $t = 9 \times 2 = 18$ независимых испытаний для $\alpha_f$ и $t = 9$ независимых испытаний для $\tilde{\alpha}$ и $\breve{\alpha}$ (см. ниже) выделены жирным шрифтом.

Консолидированные тесты не только снимают сомнения относительно участка 215, но и окончательно доказывают наличие фальсификаций на участке 217. В то же время, участку 212 каким-то образом удается избежать подозрений по результату консолидированных тестов, что требует объяснения.

Проблема здесь — в неявном предположении, что фальсификаторы помогают только победителям и всем победителям. Однако, по-видимому, на участке 212 была реализована более сложная схема фальсификаций. Чтобы понять ее, требуется определенный политологический анализ, выходящий за рамки математической статистики.

Кандидат от ЛДПР Бахтинов — 6-й по результату на участке 212, в то время как на участке 213 и в ДЭГ он — 10-й (предпоследний), а на участке 214 — 11-й (последний) [31]. Необходимо пояснить, что ЛДПР не пользуется поддержкой в европейской части России, так что она выдвинула только по 1 кандидату в каждом из избирательных округов Власихи. В округе № 3 кандидат от ЛДПР Сборнов — 11-й (последний) на всех участках включая ДЭГ, так как не получил поддержки ни от избирателей, ни от фальсификаторов. В округе № 2 кандидат от ЛДПР Вавилова — 10-я (предпоследняя) на честном участке 215 и в ДЭГ, но 2-я и 5-я на участках 216 и 217, где она была в списке административных кандидатов, в пользу которых подменяли бюллетени. Примечательно, что кандидат от ЕР Булах, в итоге не включенный в этот список в округе № 2, — 11-й (последний) на всех участках, что намекает на реальную популярность самой сильной партии.

Подводя итог, можно предположить, что кандидат от ЛДПР также имел поддержку фальсификаторов на участке 212, несмотря на то, что он не был среди административных кандидатов. Следует также отметить, что избиратели, якобы поддерживавшие кандидата от ЛДПР, отдали 64,4% своих оставшихся голосов победившим административным кандидатам на участке 212. Эта величина ниже — 38,9% и 26,9% — на участках 213 и 214.

Как это случилось, мы можем только догадываться. С одной стороны, это может быть результатом тихого саботажа. Если фальсификаторы не хотели выполнять свои преступные обязанности, но боялись открыто протестовать, они могли отмечать самого слабого кандидата, делая вид, что ошиблись. С другой стороны, это могла быть попытка замаскировать фальсификацию, не рискуя помочь оппозиции. Обе описанные ситуации представляются довольно экзотическими, поэтому это явление редкое и затрагивает лишь один участок.

Ни одно из предложенных объяснений не исключает возможности того, что в фальсифицированных бюллетенях отмечены менее 5 кандидатов. Принимая это во внимание, мы адаптируем консолидированный тест следующим образом. Теперь $f = 1$, если бюллетень содержит отметки за любого из 6 самых популярных кандидатов на участке и никаких других отметок, и $f = 0$ в противном случае (как показано в последнем сером столбце для каждого участка на рис. 3–5). Это позволяет нам обнаружить фальсификации на участке 212 с $\tilde{\alpha} = 6{,}2 \times 10^{-5}$. В то же время, для других участков значимости $\alpha_f$ и $\tilde{\alpha}$ заметно возрастают, но они всё равно остаются достаточно низкими, чтобы не оставлять сомнений в фальсификациях там (за исключением участка 215, конечно) (таблица VI).

Выполненный анализ иллюстрирует некоторые важные общие соображения. Любой строгий формальный тест опирается на представление о том, как осуществляется фальсификация. Поэтому любой та-кой тест можно пройти ценой относительно небольших усилий. Однако фальсификаторы, как правило, не прилагают этих усилий. С одной стороны, они обычно уверены в своей полной безнаказанности, а с другой стороны, они глупы, как и большинство людей, совершающих преступления. Именно это позволяет нам в большинстве случаев предлагать эффективные статистические тесты без использования какой-либо политологической информации. Однако для надежного анализа необходимо использовать комбинацию всех имеющихся статистических тестов, поскольку условия применимости любого отдельного теста иногда могут нарушаться.

### 4. Тест стационарности потока голосов

Анализ также содержит еще одно неявное предположение о том, что вероятности голосования за кандидата неизменны во времени. Но, например, оппозиционные избиратели могут быть более склонны голосовать на третий день, если они знают, что их голоса, отданные в первые два дня, могут быть украдены. В этом случае поток голосов становится нестационарным. Постоянство вероятностей можно проверить следующим образом.

Разобьем последовательность флагов на $l$ отрезков одинаковой длины (если $n$ не делится на $l$ нацело, то граничные флаги распределяются между соседними отрезками в пропорциональных долях). Число единиц $k_i$ в отрезке $i = 1, 2, ..., l$ распределено биномиально с математическим ожиданием $\mu = r_1/l$ и дисперсией $\sigma^2 = r_0 r_1/nl$. Это распределение можно аппроксимировать распределением Гаусса при достаточно большой длине отрезка $n/l$. Статистика $\sum_{i=1}^{l} (k_i - \mu)^2/\sigma^2$ имеет $\chi^2$-распределение с $l-1$ степенями свободы (еще 1 степень свободы тратится на определение параметров распределения). Это позволяет вычислить значимость $\breve{\alpha}$ гипотезы о том, что все отклонения потока голосов от стационарного являются случайными. Аналогичная статистика для нулей с $\mu = r_0/l$, очевидно, та же.

Конкретный выбор $l$ не имеет большого значения, хотя наличие свободного параметра само по себе является определенным недостатком данного подхода. Мы используем значение $l = 12$. Оно одновременно дает несколько отрезков для каждого из 3 дней голосования и обеспечивает несколько десятков флагов в каждом отрезке.

Таблицы II–IV показывают результаты этих расчетов для всех кандидатов по всем избирательным участкам (см. последний столбец для каждого участка). Вероятность голосования не меняется со временем для участка 215 и, возможно, для участка 217. Она заметно меняется для участков 212 и 213 и очень сильно — для всех остальных. Однако для них уже нельзя полагаться на количественные значения $\breve{\alpha}$, поскольку тест, будучи асимптотическим, имеет тенденцию занижать значимость, причем тем сильнее, чем она меньше.

Мы не сомневаемся в итогах голосования на участке 215. Поэтому мы можем считать обсуждаемое предположение верным для рассматриваемого голосования. Это также позволяет использовать проверку стационарности потока голосов в качестве вспомогательного теста, хотя она и не дает никаких новых результатов по сравнению с двумя основными тестами. Этот тест менее ценен как доказательство, поскольку нестационарность потока голосов в принципе может иметь естественные причины. Хотя, конечно, чаще всего она является следствием фальсификаций.

Похожие результаты для консолидированного теста показаны в последнем столбце таблицы V. Разница между честным участком 215 и участком 217 с хорошим перемешиванием бюллетеней здесь становится более заметной. Для участка 212 с его необычным подходом к фальсификации вспомогательный тест едва замечает аномалии. Но если мы адаптируем консолидацию (таблица VI), то получим для этого участка значимость $\breve{\alpha} = 1{,}2 \times 10^{-4}$, достаточную, чтобы не сомневаться в нестационарности потока флагов.

### Приложение C: Методы обхода защиты сейф-пакетов от их незаметного вскрытия

Сначала поясним, как сейф-пакет используется штатным образом. Красная клейкая индикаторная лента (сейф-лента) изначально приклеена к пакету своим нижним краем, а остальная часть ее клеевой площади защищена зеркальной лентой (лайнером), поверхность которого покрыта силиконом и почти не имеет адгезии к клеевому слою [рис. 6(a)]. Чтобы запечатать пакет, лайнер снимается, а сейф-лента заклеивает разрез пакета [рис. 6(b)–(d)]. При попытке отклеивания ленты, ее сложная многослойная клеевая структура разрывается и оставляет видимые следы [рис. 6(e)]. Если ленту отдирать быстро и с силой, также растягивается и может порваться материал пакета. Обратите внимание, что уникальный номер пакета напечатан на клеевом слое защитной ленты, поэтому его нельзя бесследно удалить с ленты или заменить другим номером. Вдоль краев пакета, не защищенных лентой, напечатана линия микрошрифта (также и на другой стороне пакета, не показанной на фотографиях), чтобы сделать любую попытку несанкционированно вскрыть и повторно запечатать любой край визуально заметной [46]. Пакет вскрывают, разрезая или разрывая его в любом месте [рис. 6(f)].

Идея сейф-пакета заключается в том, что попытка доступа к его содержимому не останется незамеченной. Чем быстрее она будет обнаружена при визуальном осмотре пакета, тем лучше. Далее мы представляем 5 возможных способов скрыть несанкционированный доступ. Способы C 1, C 4 и C 5 реализуемы с помощью легкодоступных материалов и не требуют

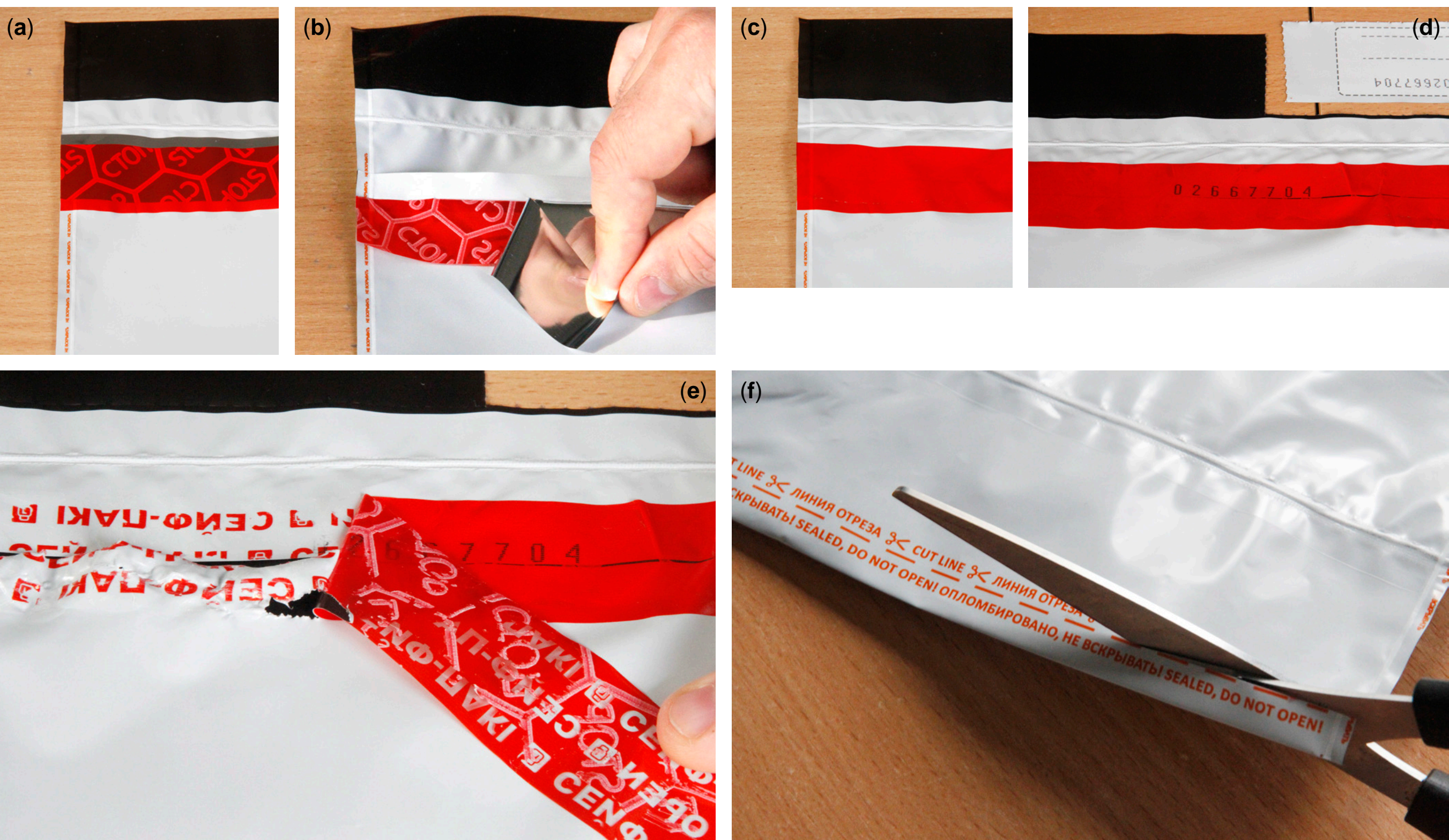

Рис. 6. Штатное использование сейф-пакета. (a) В неиспользованном пакете нижний край сейф-ленты приклеен к пакету. Металлизированный зеркальный лайнер защищает клей на части ее ширины, оставшейся не приклеенной. Часть лайнера выступает над лентой. Лента прозрачна и лайнер можно увидеть сквозь нее. (b) Чтобы запечатать пакет, неприкрепленную часть ленты с лайнером отгибают, не отсоединяя ленту от пакета, так, чтобы ее клеевой слой стал обращен наружу. Лайнер осторожно снимают, не касаясь пальцами клея. Затем ленту разгибают обратно на разрез пакета, прижимают к пакету и она приклеивается к нему намертво. (c) и (d) Сейф-лента с уникальным номером пакета запечатывает разрез. Крупный рисунок сот со словами STOP СТОП является самым внешним слоем ее многослойного клея. Этот рисунок выполнен клеем светло-серого цвета и хорошо просматривается, когда лента не приклеена к пакету или покрыта зеркальным лайнером. Как только клей входит в контакт с пакетом того же цвета, что и клей, рисунок становится почти невидим. (d) Отрывные квитанции можно отсоединить от пакета. (e) Попытка отклеить сейф-ленту разделяет многослойный клей и обнаруживает скрытые в нем надписи. (f) Обычно пакет вскрывают, разрезая его по нижнему краю сговора с производителем пакетов. Способы C 2 и C 3 требуют наличия заводских подделок с идентичными номерами, изготовленных на фабрике. Добросовестный производитель не должен выпускать дубликаты пакетов, так как их номера обязаны быть уникальными [38]. Как показывают доказательства в приложении D, в распоряжении наших избирательных комиссий были заводские подделки пакетов и они воспользовались способами C 2 и C 3 для фальсификации этих конкретных выборов.

## 1. Имитация сейф-ленты полоской красного скотча

Обычная клейкая упаковочная лента красного цвета хорошо имитирует внешний вид сейф-ленты в запечатанном состоянии (рис. 7), но не в изначальном незапечатанном состоянии. Последнее отличие легко можно не заметить, поскольку большинство людей, наблюдающих за выборами, не знакомы с сейф-пакетами.

Подготовительные операции проходят следующим образом. Оригинальная сейф-лента отклеивается заранее, ее использованная часть обрезается, а остатки ее клея смываются с пакета растворителем (рис. 8). Хотя клей устойчив ко многим обычным растворителям, мы нашли такой, который легко удаляет его, не смывая при этом микрошрифт по краям пакета. Зауженная настоящая сейф-лента сохраняется для будущего повторного использования.

Затем из подходящей упаковочной ленты красного цвета (стоимостью менее 200 ₽ за рулон) изготавливается имитация сейф-ленты. Уникальный номер пакета печатается на ее клейкой стороне с помощью ручного каплеструйного принтера (стоимостью около 10 000 ₽), который наносит маркировку быстросохну-

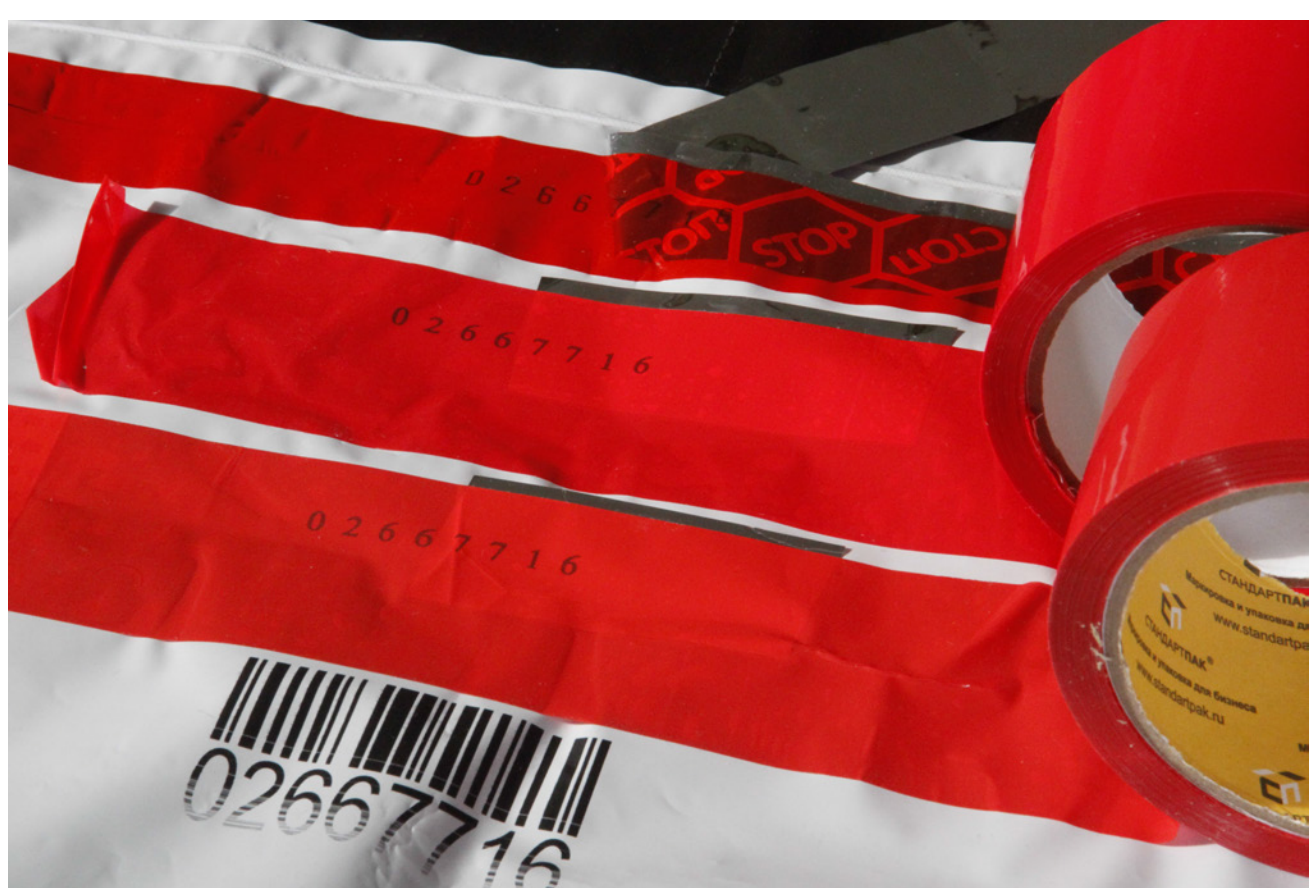

Рис. 7. Упаковочная лента красного цвета может имитировать внешний вид сейф-ленты в запечатанном состоянии. Левая половина каждой ленты полностью приклеена к пакету, в то время как правая половина все еще частично закрыта лайнером. Верхняя лента на фотографии является настоящей сейф-лентой, а две другие ленты — это упаковочный скотч разных марок. Номера на нем были напечатаны на клейкой стороне с помощью ручного каплеструйного принтера. Основа всех трех лент прозрачная, окрашен в красный цвет только их клеевой слой щими чернилами бесконтактно на произвольные поверхности (рис. 9). На нее накладывается лайнер от другого пакета. Лента обрезается до нужного размера и наклеивается на пакет.

Подготовленный таким образом пакет затем используется публично обычным способом для запечатывания в нем бюллетеней (рис. 10). Ночью поддельная лента легко удаляется, бюллетени подменяются и пакет снова запечатывается сохраненной настоящей сейф-лентой (рис. 11). Затем бюллетени публично извлекаются из выглядящего неповрежденным пакета (рис. 12). Если кто-то попытается оторвать защитную ленту в этот момент, она будет вести себя ожидаемо. Единственный видимый след несанкционированного вскрытия — уменьшенная ширина ленты, которую почти наверняка никто не заметит, если только не знает, что именно следует искать.

Мы находим этот способ вполне рабочим, но нет свидетельств, чтобы им кто-то пользовался.

## 2. Подмена сейф-пакета его заводской подделкой

Если у злоумышленников есть дубликат сейф-пакета с идентичным номером, они могут заменить им оригинальный пакет ночью. Единственной сложностью будет подделка подписей, которые могут быть оставлены на оригинальном пакете наблюдателями и членами комиссии, а также воспроизведение любых дефектов в приклеивании ленты и отрыве квитанций. Именно эти дефекты на участке 219 раскрыли всю схему преступления (приложение D 1).

## 3. Подмена сейф-ленты другой сейф-лентой

Чтобы избежать подделки подписей, злоумышленники могут заменить только ленту на ее заводской дубликат, а не подменять пакет целиком. Ночью они могут отклеить оригинальную сейф-ленту от пакета с бюллетенями, вымыть пакет растворителем, заменить бюллетени, отрезать сейф-ленту от неиспользованного номерного дубликата пакета и приклеить ее на оригинальный пакет. Если какие-либо подписи были оставлены поверх ленты, их все равно придется подделывать или оставить неполными, что может быть замечено. Также уменьшится ширина ленты. Вероятно, именно так и поступали на большинстве наших участков, см. приложение D 2 и далее.

Заметим, что, по-видимому, нет технических причин, по которым обе ленты нельзя было бы сделать одинаково узкой ширины. Для этого сейф-пакет пришлось бы подготовить заранее, отрезав и переклеив первую оригинальную ленту, прежде чем он впервые появится на публике. Однако это увеличило бы риск, что подделка будет обнаружена рано, при первом появлении пакета на публике.

При комнатной температуре полностью приклеенную ленту необходимо отлеплять осторожно и медленно, чтобы исключить риск растяжения и разрыва материала пакета, которые видны на рис. 6(e). Сильное охлаждение путем обливания ленты содержимым перевернутого вверх дном баллончика пневматического очистителя позволяет снять ее быстрее. В любом случае, затем требуется около 10 мин, чтобы смыть клей с пакета растворителем.

## 4. Понижение адгезии ленты к пакету

Покрытие основного материала пакета тонким слоем прозрачного силикона снижает адгезию ленты к нему настолько, что пакет можно открывать и закрывать многократно, не оставляя следов [45]. Хотя этот способ практиковался на многих избирательных участках в 2024 [47–49] и 2025 годах [50], во Власихе он не применялся. Отметим, что для лучшего сокрытия следов вскрытия слой силикона возможно смыть растворителем перед повторным запечатыванием пакета [45].

## 5. Малозаметный рез

Наиболее простым способом добраться до содержимого пакета является его прорезание. Затем рез заклеивается или запаивается [46]. Такой рез всегда можно обнаружить при тщательном осмотре пакета, включающем выворачивание использованного вскрытого пакета наизнанку и осмотр его изнутри. В условиях выборов, избирательные комиссии часто препятствуют полному осмотру пакета и его частей [45]. Рез может

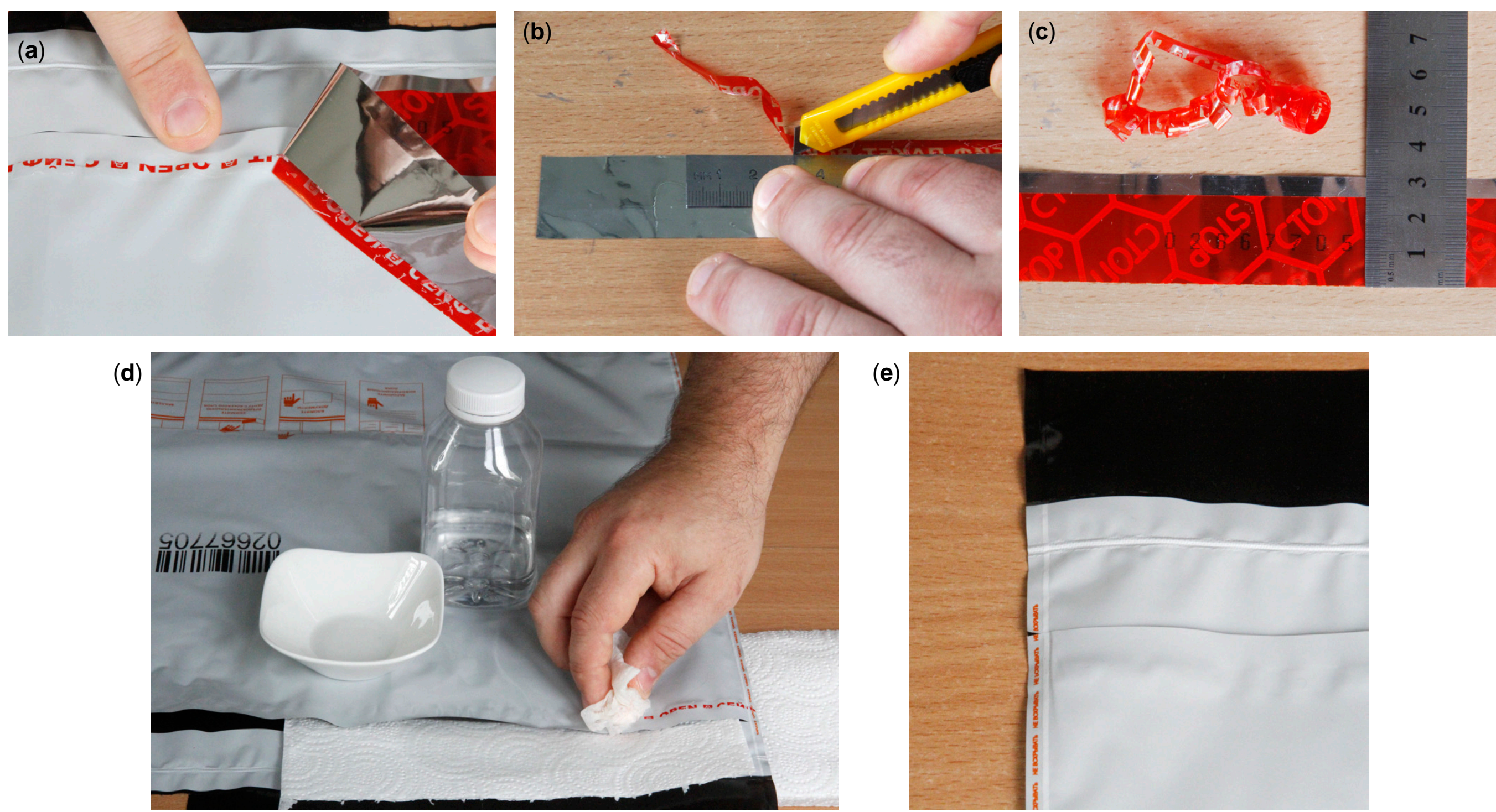

Рис. 8. Снятие оригинальной сейф-ленты с неиспользованного пакета. (a) Лента аккуратно отклеивается вместе с лайнером, оставляя на пакете узкую полоску клея. (b) Лента обрезается, чтобы удалить использованную часть ее ширины. (c) Ширина обрезанной сейф-ленты менее 25 мм. (d) Остатки клея удаляются с пакета с помощью растворителя. (e) Пакет чистый и без повреждений

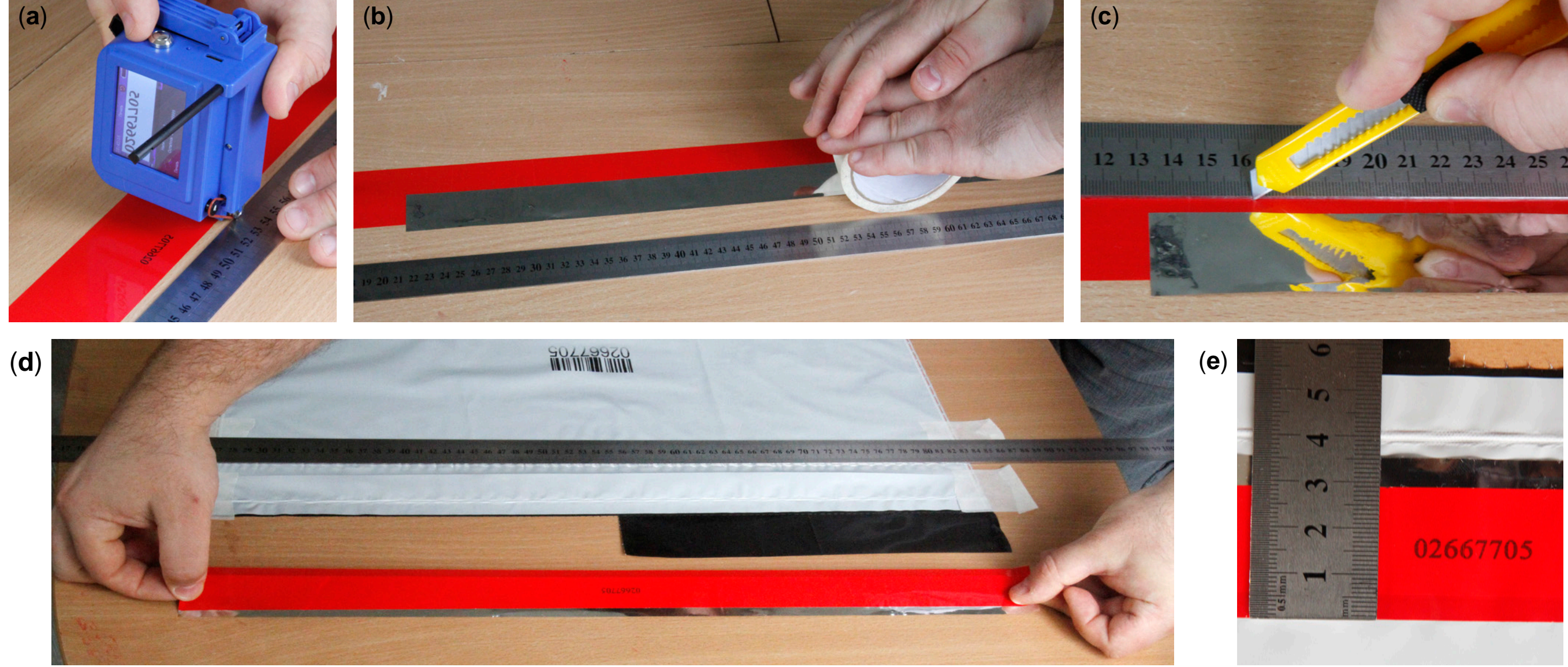

Рис. 9. Установка на пакет поддельной ленты. (a) На клейкой стороне упаковочной ленты печатается уникальный номер пакета в зеркальном отражении. (b) На клейкий слой ленты накладывается лайнер, снятый с другого пакета. По нему прокатывается круглый предмет, чтобы удалить пузырьки воздуха. (c) Лента обрезается до нужных размеров. (d) Лента с лайнером наклеивается на пакет. (e) Установленная поддельная лента имеет корректную ширину, около 30 мм

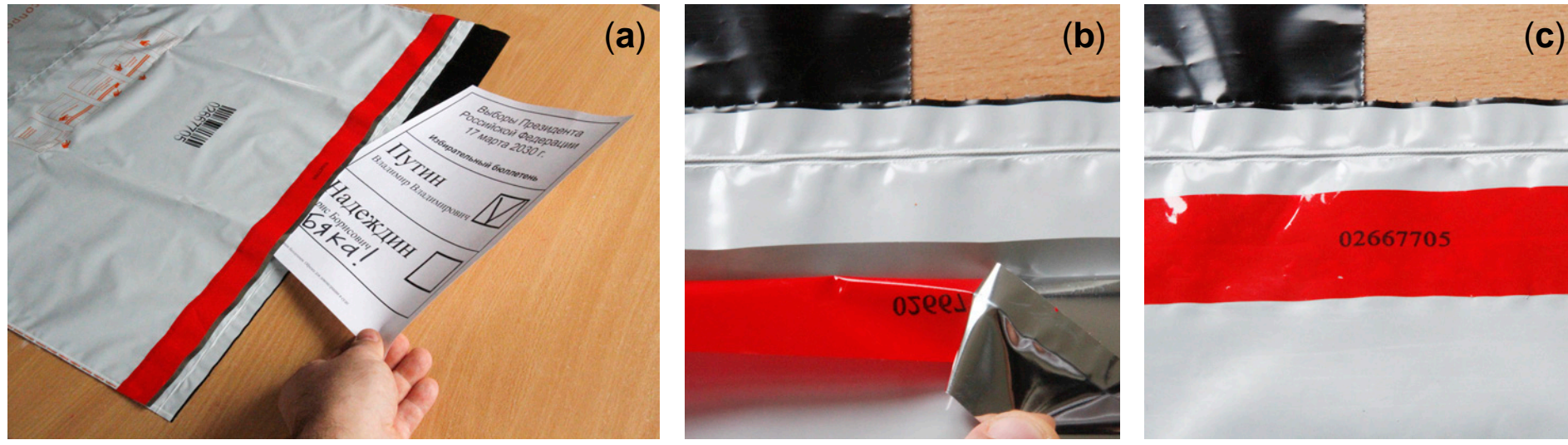

Рис. 10. Публичное запечатывание бюллетеней в сейф-пакет с поддельной лентой. (a) Подлинные бюллетени помещаются в пакет. (b) Защитный лайнер снимается обычным образом. (c) Пакет запечатывается лентой обычным образом

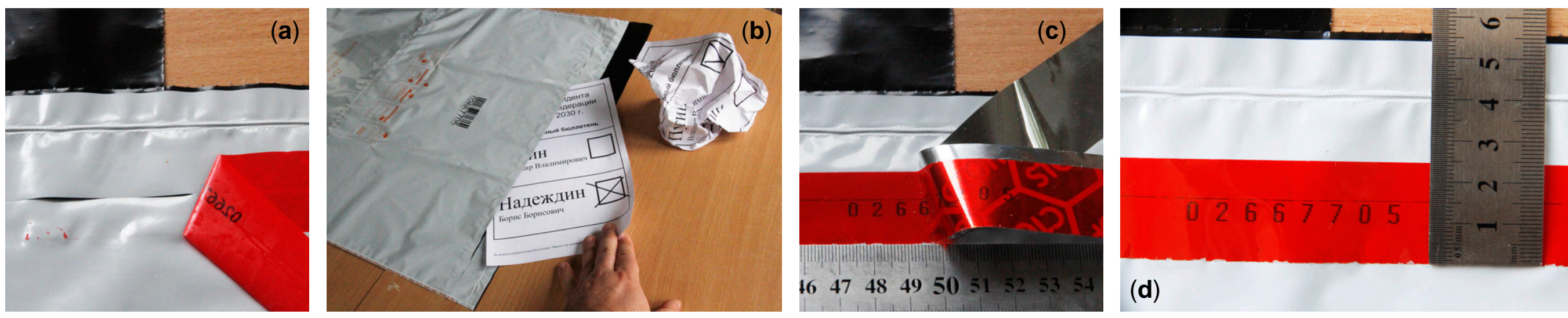

Рис. 11. Несанкционированное вскрытие пакета и подмена бюллетеней. (a) Поддельная лента легко отклеивается. Если на пакете остался ее клей, его можно смыть растворителем. (b) Бюллетени заменяются на подложные. (c) Пакет запечатывается сохраненной настоящей сейф-лентой. (d) Однако ширина приклеенной сейф-ленты составляет менее 25 мм, что является единственным видимым признаком несанкционированного вскрытия. Неровности по ее нижнему краю остались из-за нашей небольшой неаккуратности при изготовлении этого первого демонстрационного образца; обрезка на миллиметр у́же дает идеальный край

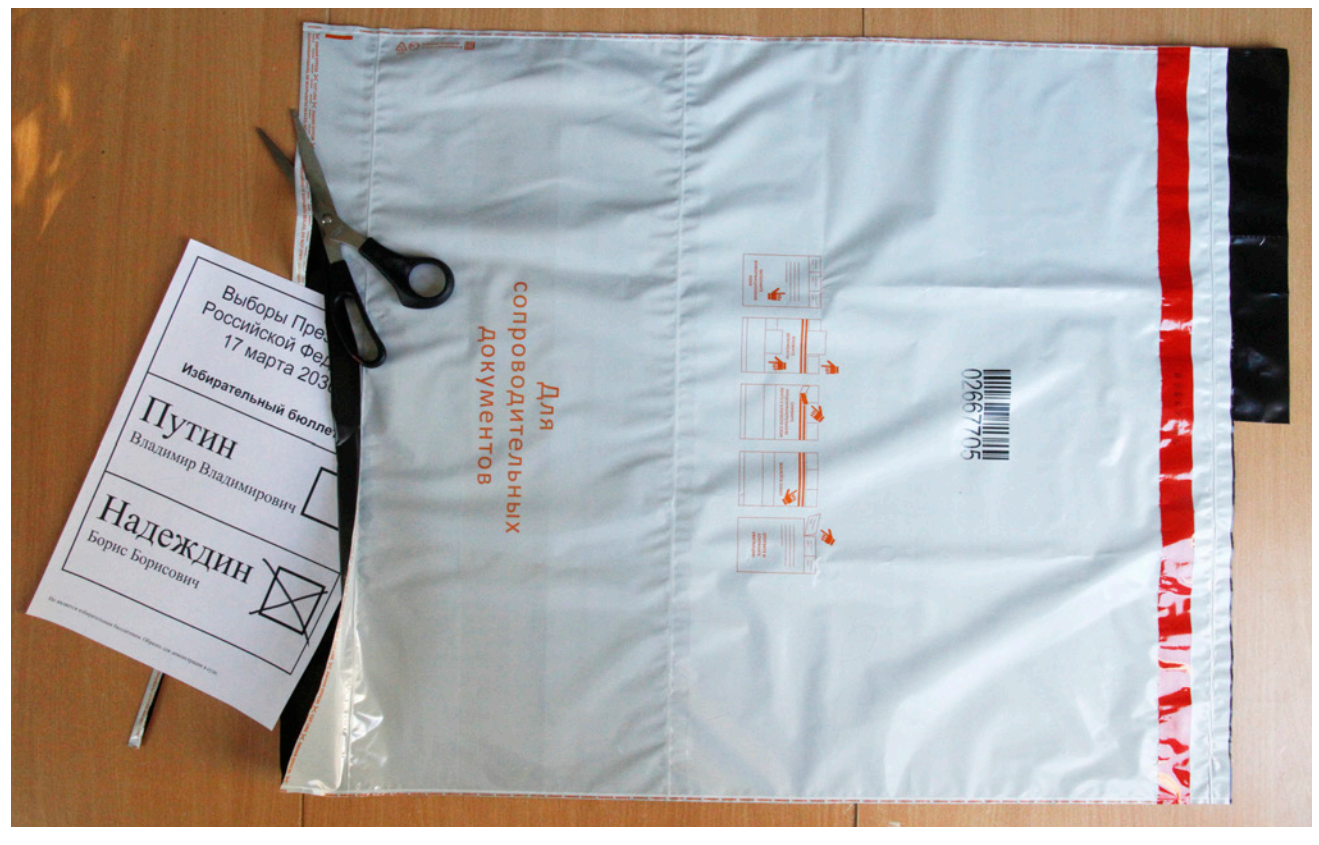

Рис. 12. Пакет вскрывается публично, а содержащиеся в нем поддельные бюллетени считаются как подлинные быть намеренно закрыт сопроводительным документом (таким, как актом о запечатывании), помещенным в наружный накладной карман пакета [51]. Рез может быть сделан практически незаметным снаружи путем расположения его вдоль линии напечатанной на пакете графики, сборочного шва или края приклеенной сейф-ленты, а заклеен клейкой лентой и хорошо заметен лишь с изнаночной стороны.

Хотя этот способ широко известен, во Власихе он не применялся.

### Приложение D: Криминалистические доказательства подмены пакетов и лент

Единственные имеющиеся в нашем распоряжении доказательства — это фотоснимки и видеозаписи, сделанные в дни голосования. Наблюдатели и кандидаты лично не изымали никаких вещественных доказательств. Вместо этого они подали заявления о совершении преступлений в местную полицию и Следственный комитет с просьбой изъять соответствующие предметы для осмотра, а также подали ходатайства на истребование вещественных доказательств судам. Все эти запросы были проигнорированы или отклонены. На настоящий момент подтверждено, что все вещественные доказательства уничтожены.

Фотографии и видеозаписи, обсуждаемые ниже, были сделаны более чем 15 людьми и 20 устройствами (в основном — смартфонами). Для ясности наших аргументов, мы используем только однородные преобразования изображений в этой статье, такие как их поворот и (только в коллажах) растяжение в одном направлении. Цветопередача на снимках варьируется в зависимости от освещения и камеры. Оригинальные файлы с камер, точное время и место каждой записи были включены в заявления о преступлениях и поданные судебные иски.

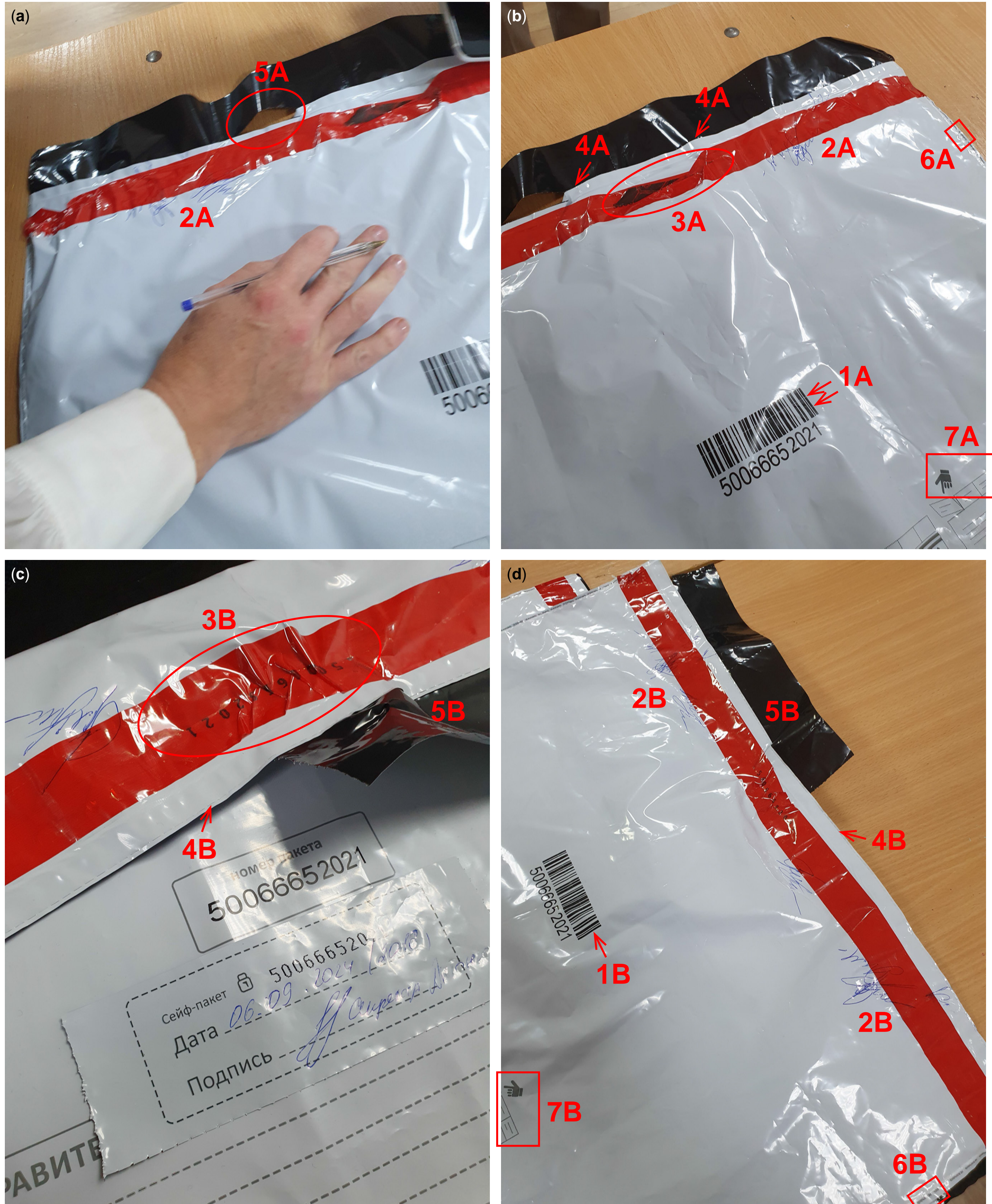

Рис. 13. Фотографии двух разных пакетов с одинаковыми номерами на избирательном участке 219. (a) и (b) Фотографии пакета A, сделанные наблюдателем 6 сентября 2024 г. после запечатывания в него бюллетеней. (c) Пакет B 7 сентября 2024 г., когда его временно извлекли из сейфа для осмотра. (d) Пакет B 8 сентября 2024 г. до того, как его вскрыли на подсчете голосов. Многочисленные различия доказывают, что не только подписи ручкой (2B) были подделаны поверх другой сейф-ленты (3B), но и весь пакет A целиком был заменен во время его хранения ночью 6 сентября на его почти идентичную заводскую подделку B с тем же серийным номером и другой сейф-лентой

### 1. Заводская пара пакетов с идентичными номерами на участке 219

На этом участке сейф-пакет с бюллетенями с первого дня голосования был подменен на его заводскую подделку в первую ночь. Фотографии высокого разрешения, сделанные до и после подмены, показывают несколько различий между пакетами, которые приводят нас к такому выводу (рис. 13). Во-первых, их серийные номера и штрих-коды, напечатанные на пакетах, совпадают (рис. 14); штрих-коды декодируются правильно. На штрих-коде хорошо заметна белая линия от битых (вышедших из строя) пикселей и еще одна более слабая линия, идущие поперек него, что однозначно идентифицирует конкретный заводской станок и примерную дату изготовления на нем пакетов. Термотрансферный принтер, наносящий этот номер и штрих-код, находится внутри поточной производственной линии, которая собирает пакет. Битые пиксели, оставляющие белые линии в наносимых принтером изображениях, — обычное явление. Их количество увеличивается от нуля до большого в случайных местах, по мере старения печатающей головки. Мы разыскали изображения штрих-кодов 5 других пакетов из той же партии [37], которые использовались по всей Московской области на выборах в сентябре 2021 г. На всех них видна либо одиночная белая линия, либо обе линии (в зависимости от разрешения изображения). Таким образом, вся производственная партия [37], а также наши дубликаты пакетов, были изготовлены на одном и том же станке.

Несмотря на одинаковый номер и графический дизайн, это разные пакеты. Линия перфорации, которая отделяет отрывные квитанции от пакета, пробивает один лист материала в пакете A, но оба листа в пакете B (рис. 16). Вторая слева квитанция частично оторвана по этой линии 6 сентября [см. особенность (5A) на рис. 13 и 16], но не оторвана на дубликате пакета в более поздние даты [см. особенность (5B)]. У нас также есть видеозапись пакета B на подсчете голосов, где его носят и переворачивают в течение нескольких минут, демонстрируя на камеру с обеих сторон, при этом не наблюдается частичного отрыва этой квитанции. Пунктирные линии рисунка, напечатанного на пакетах, имеют разную величину растекания типограф-

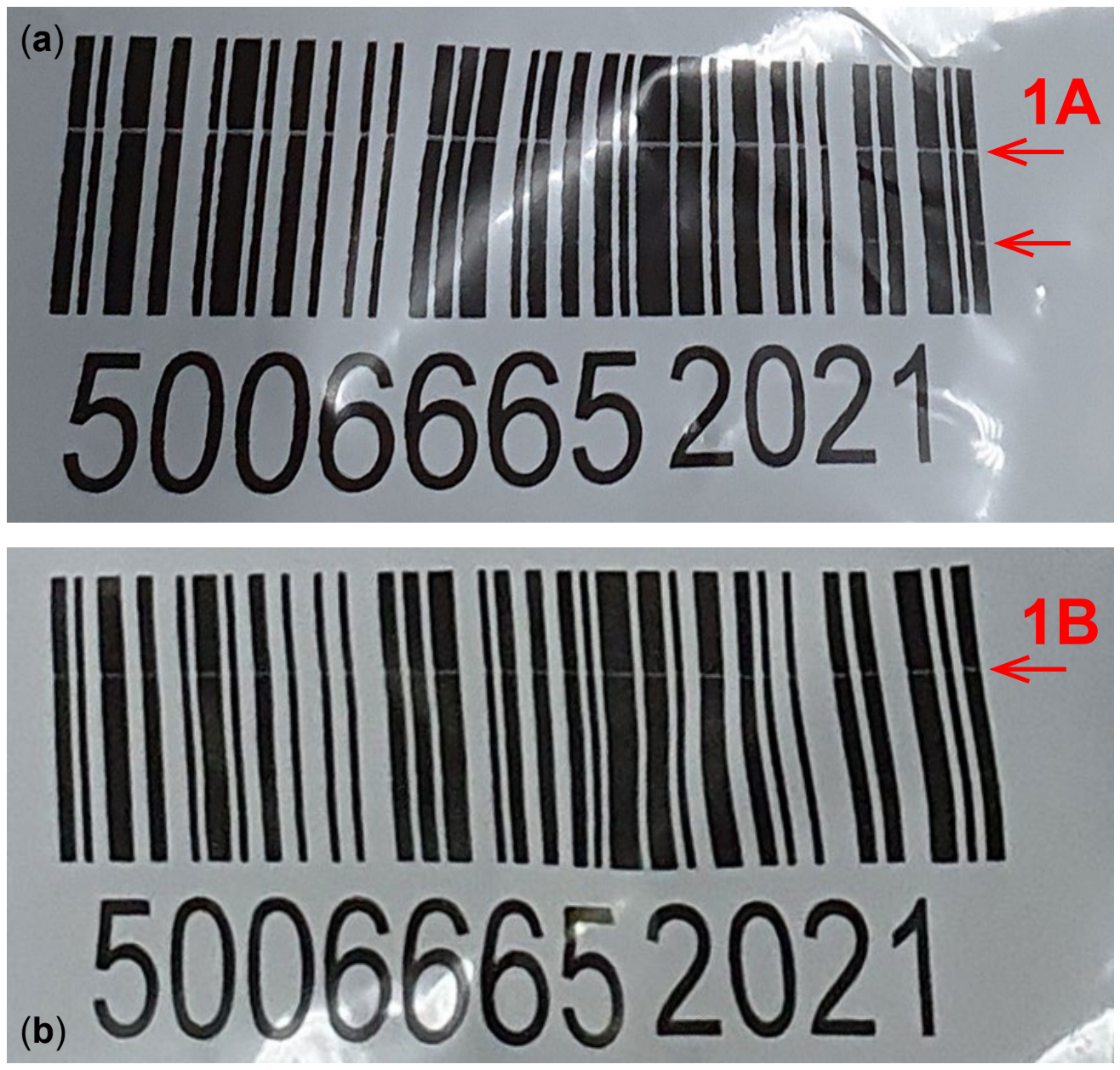

Рис. 14. Штрихкоды на обоих пакетах были напечатаны одним и тем же термотрансферным принтером, о чем свидетельствует одинаковое положение заметной белой линии битых пикселей. (a) Пакет A. (b) Пакет B. Вторая слабая линия не видна из-за более низкого разрешения последнего изображения

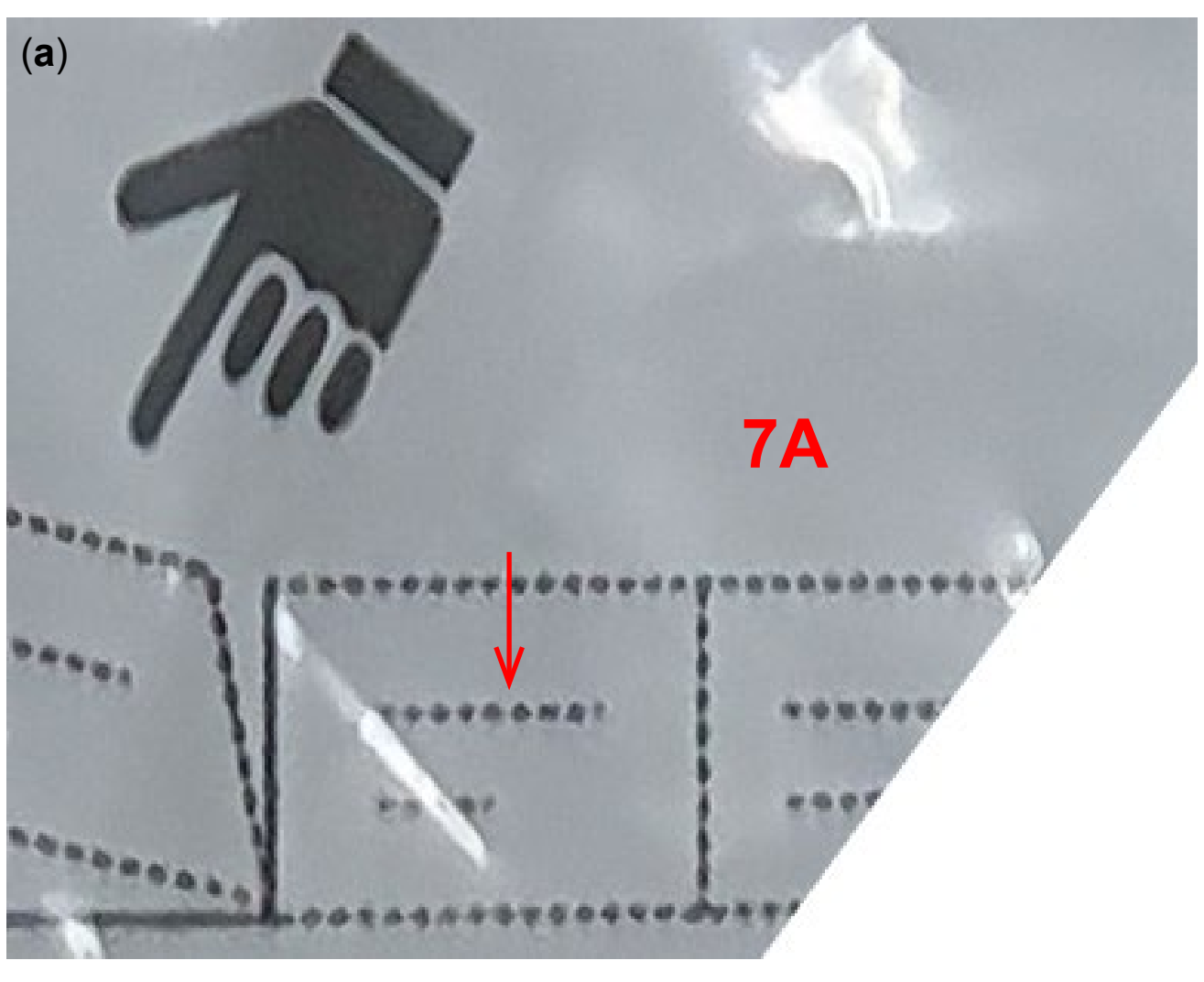

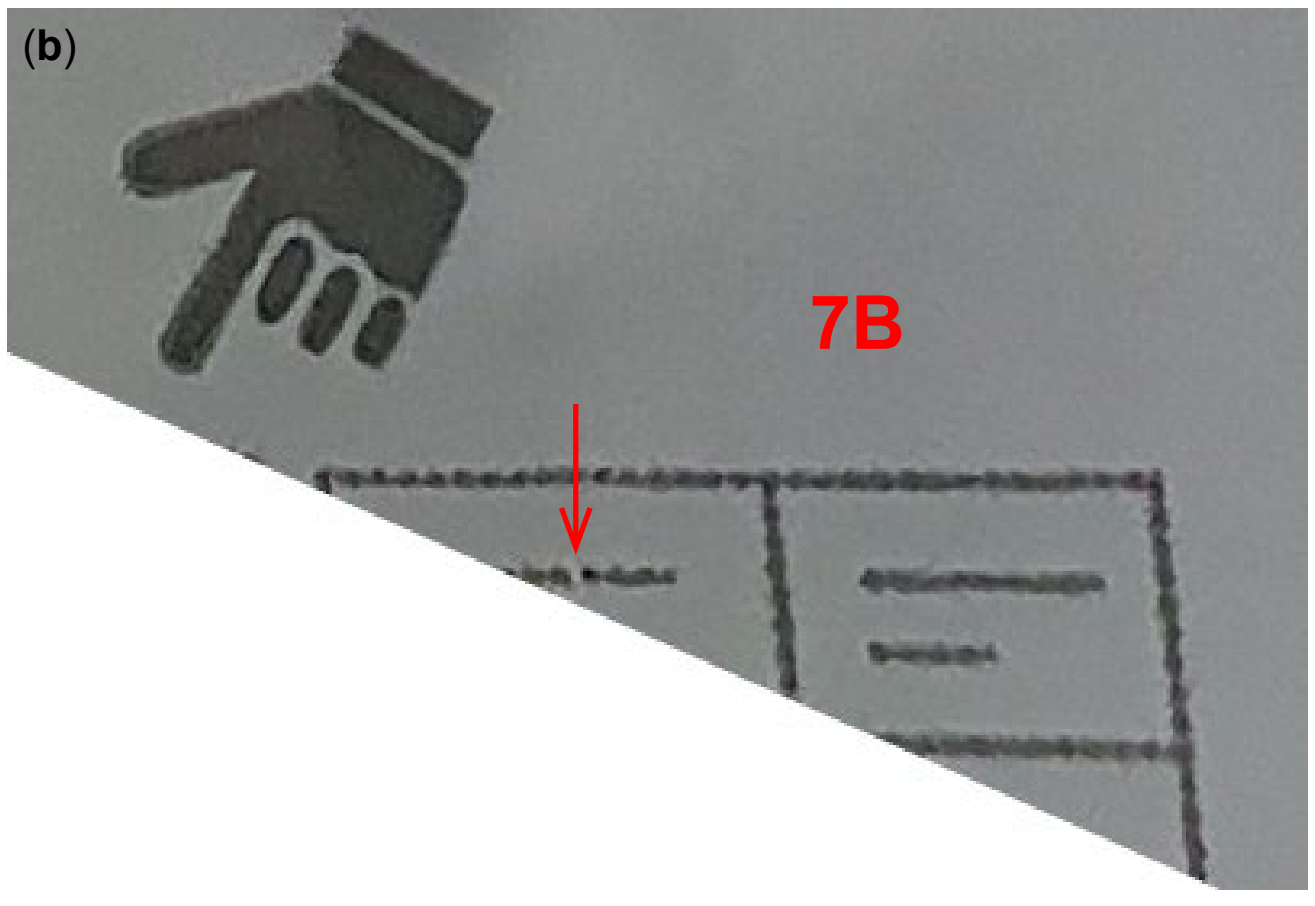

Рис. 15. Растекание типографской краски при печати линий (a) умеренное на пакете A сохраняет линии пунктирными, но (b) более сильное на пакете B превращает пунктирные линии в сплошные. Стрелка указывает на единственный зазор между регулярно расположенными точками, который не заполнен краской на пакете B, что подтверждает, что сплошные линии на изображении (b) находятся в фокусе камеры и сняты достоверно

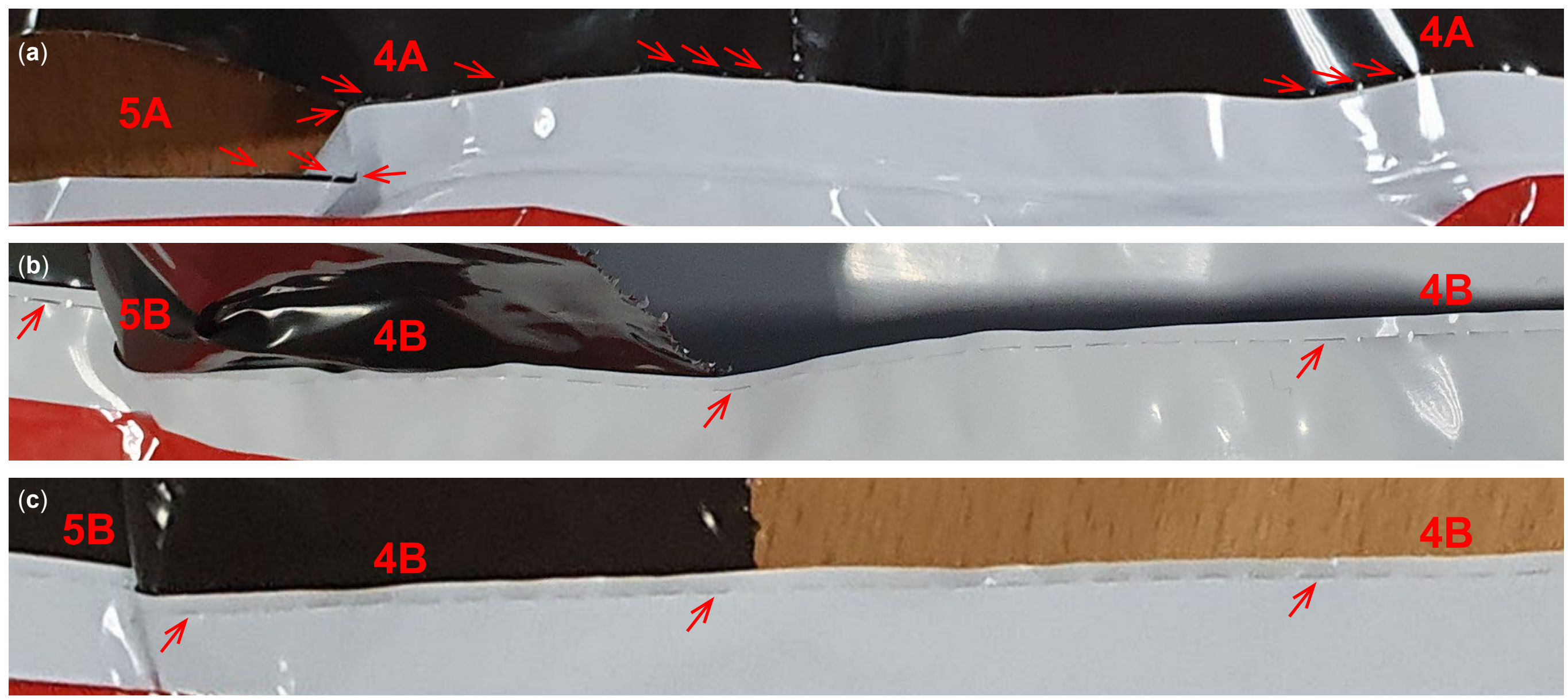

Рис. 16. Линия перфорации (отмечена стрелками) отделяет пакет от его отрывных квитанций. (a) В пакете A перфорация пробивает один лист материала — только тот, на котором напечатаны квитанции. Его черная сторона обращена к камере. (b) и (c) В пакете B перфорация пробивает оба листа материала, что является относительно часто встречающимся различием между экземплярами. На последних изображениях видна только перфорация сквозь верхний лист, но она же пробивает и нижний лист, на котором напечатаны квитанции

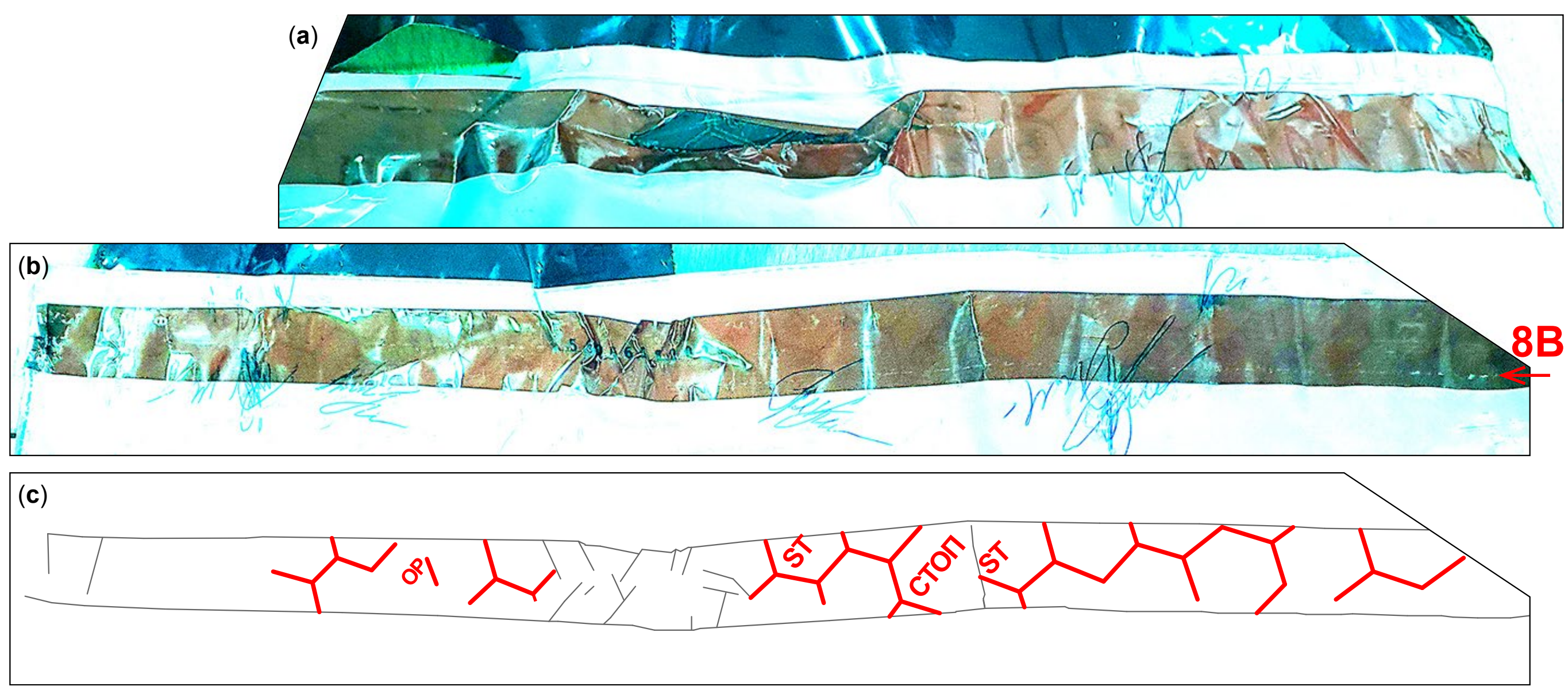

Рис. 17. Фотографии лент, запечатывающих пакеты на участке 219, с сильно увеличенным контрастом. Контраст максимизирован индивидуально для каждого их трех основных цветов (красного, зеленого и синего) с помощью инструмента «кривые» редактора фотографий, примененного ко всему изображению целиком. Затем разрешение изображения уменьшено бикубической интерполяцией с некоторым повышением локального контраста, чтобы еще лучше выделить слабо проглядывающий рисунок на ленте. (a) Пакет A. (b) Пакет B. Сравните рисунок сот, местами видный на лентах, с рис. 6(a,b) и 7. Его наличие подтверждает, что оба пакета запечатаны фабричными сейф-лентами. Если вам трудно разглядеть соты, воспользуйтесь подсказкой (c). Размер сот относительно ширины ленты подтверждает, что обе ленты имеют ширину около 30 мм. Наличие длинной прерывистой царапины (8B) примерно в 5 мм от нижнего края ленты указывает на обычную процедуру запечатывания, при которой лайнер был удален немного неправильным образом — вытянут из-под ленты, как показано на рис. 7, вместо того, чтобы отогнуть ленту клеевым слоем наружу для легкого снятия лайнера, как рекомендуется производителем и показано на рис. 6(b). Неправильное удаление лайнера местами царапает клейкий слой по этой линии. Такой след типичен при обычном использовании пакета, поскольку многие люди интуитивно удаляют лайнер таким образом, вместо того, чтобы читать заводскую инструкцию

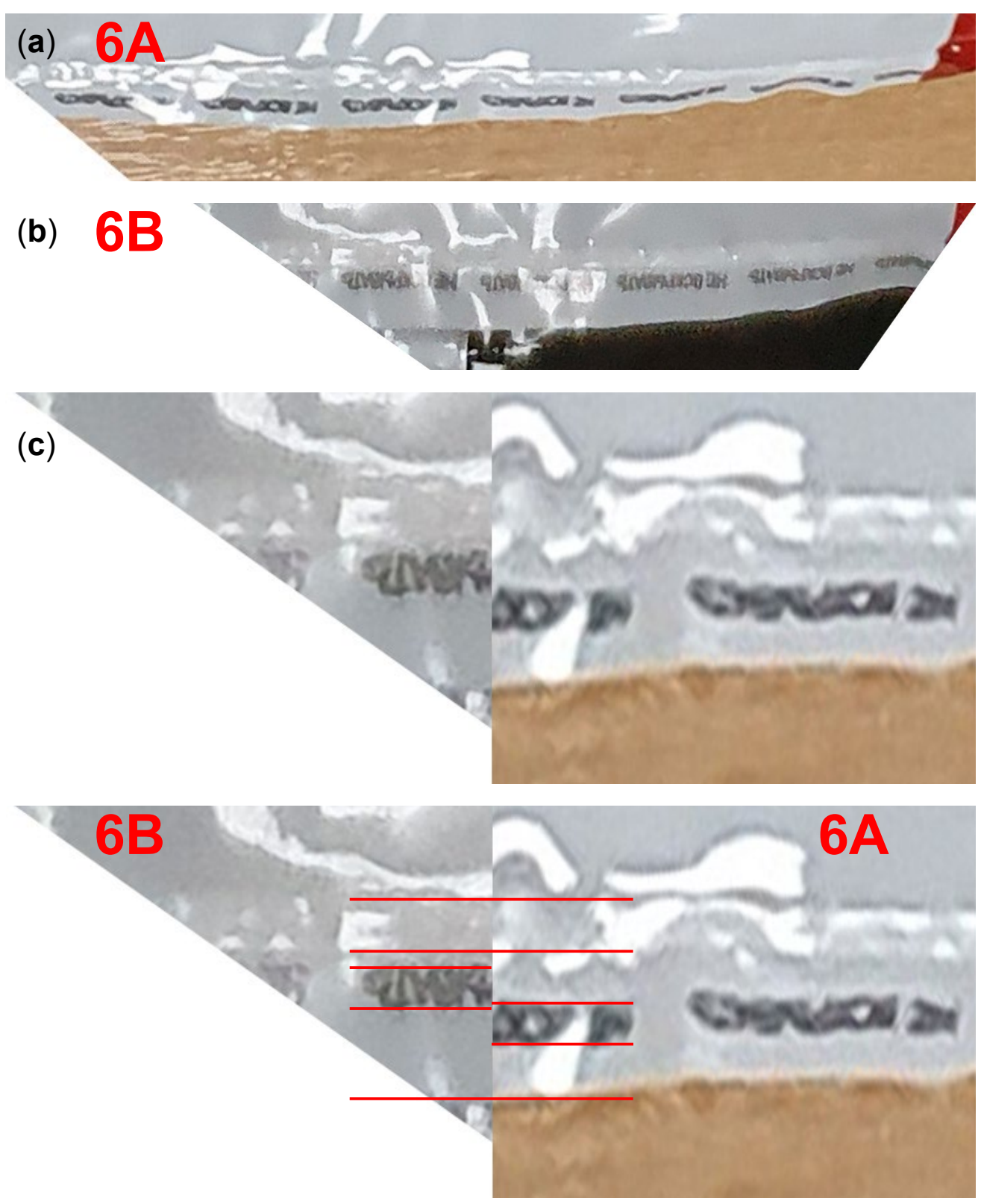

Рис. 18. Расположение строки микрошрифта относительно края пакета и паяного шва на (a) пакете A и (b) пакете B. Коллаж (c) показывает, что микрошрифт находится на разном расстоянии от края на двух пакетах. В коллаже изображение пакета B, снятое камерой под косым углом, равномерно растянуто по вертикали, чтобы точно совместить его с изображением пакета A

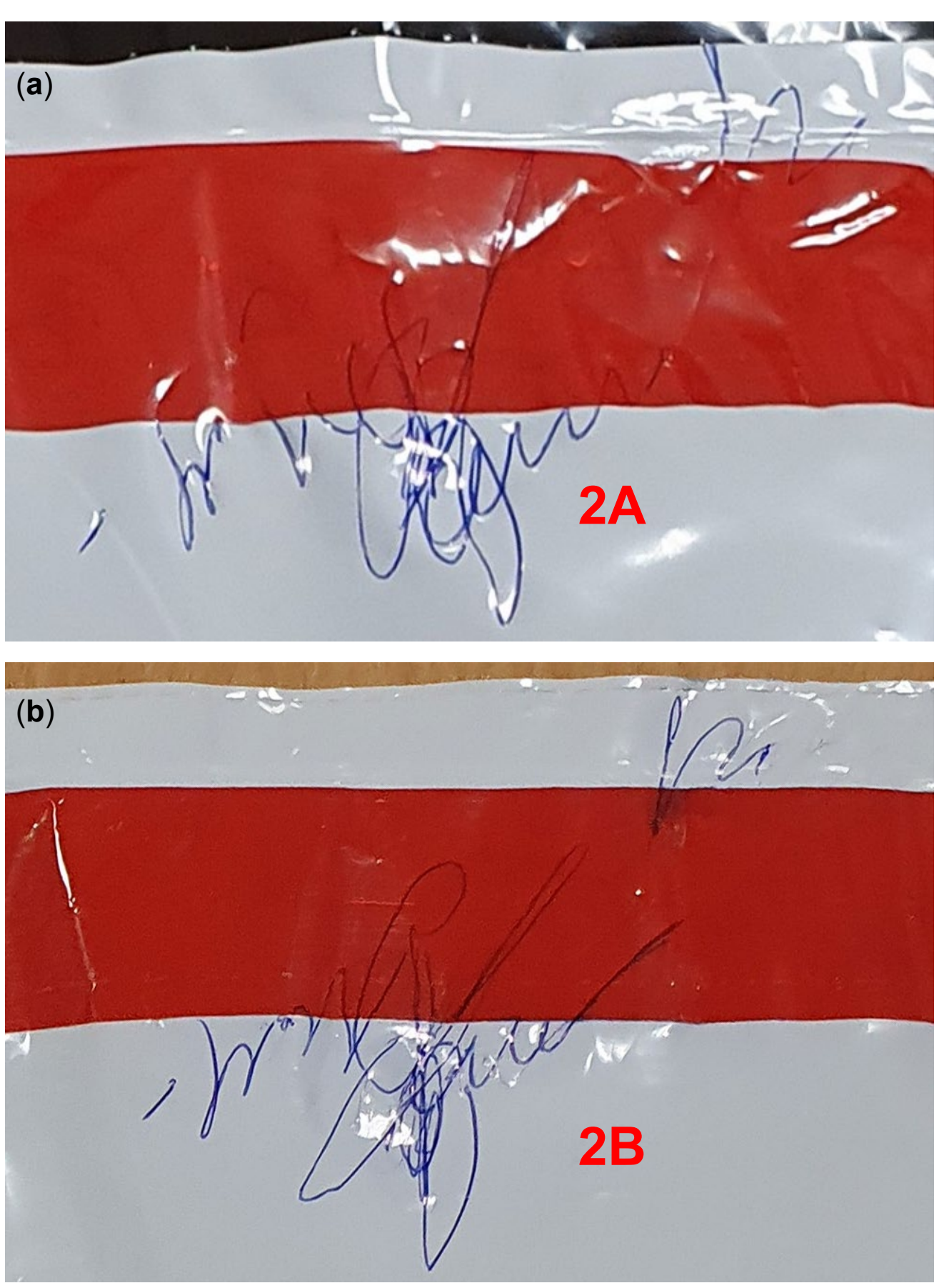

Рис. 19. Подписи наблюдателя Ушакова и кандидата Булкиной (которая наложила свою подпись на его) являются (a) подлинными на пакете A, согласно отчету Ушакова, и (b) поддельными на пакете B. Несколько других подписей на пакете B также поддельные ской краски (рис. 15). Микрошрифт по краю пакета находится дальше от края на пакете B, чем на пакете A (рис. 18). Все эти небольшие производственные различия между экземплярами позволяют нам различить два экземпляра пакета.

Как и ожидается, подписи на пакетах также отличаются (рис. 19). Подписи на пакете B были подделаны, согласно заявлениям наблюдателя Ушакова и кандидата Разиной, которые оба собственноручно расписались на пакете A 6 сентября. Сами сейф-ленты приклеены с разными заломами и складками. Тем не менее, обе являются подлинными, фабричными сейф-лентами, о чем свидетельствует наличие едва заметного рисунка сот со словами STOP СТОП и их корректная ширина 30 мм (рис. 17). Длинная прерывистая царапина (8B) на липком слое второй ленты говорит о том, что она была установлена на пакет и запечатана обычным образом.

Самая интригующая особенность этой пары пакетов — серийный номер, напечатанный на сейф-ленте. Хотя номер на ленте пакета A виден лишь частично, заметно, что его набор обычный, с небольшим интервалом между цифрами, в то время как на ленте паке-

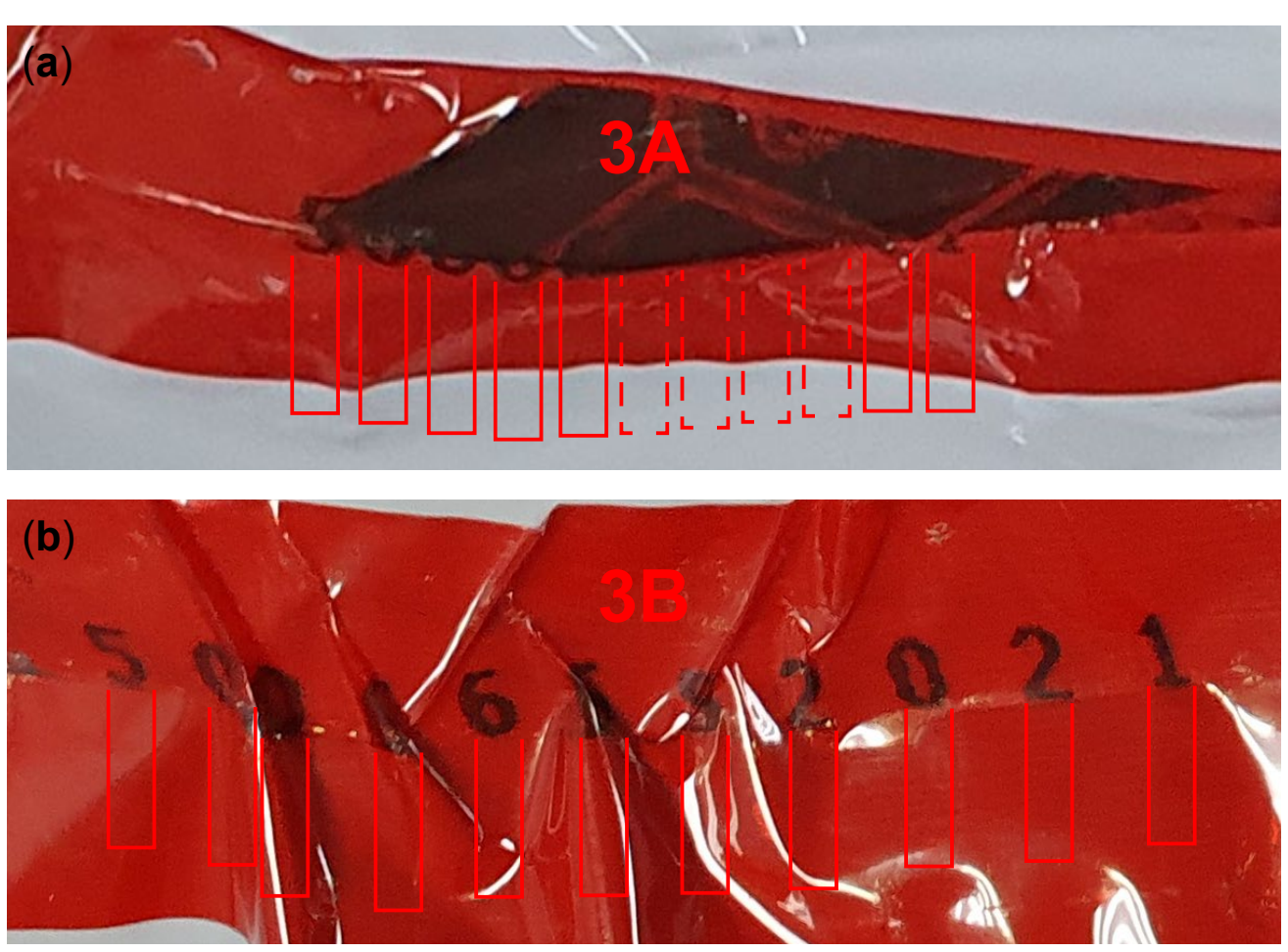

Рис. 20. 11-значный номер на сейф-ленте (a) пакета A, где семь его цифр частично видны, и (b) пакета B. Номер на пакете A набран обычным образом, с интервалом между цифрами менее половины их ширины. Номер на пакете B набран вразрядку, с межсимвольным интервалом около полутора ширины цифры

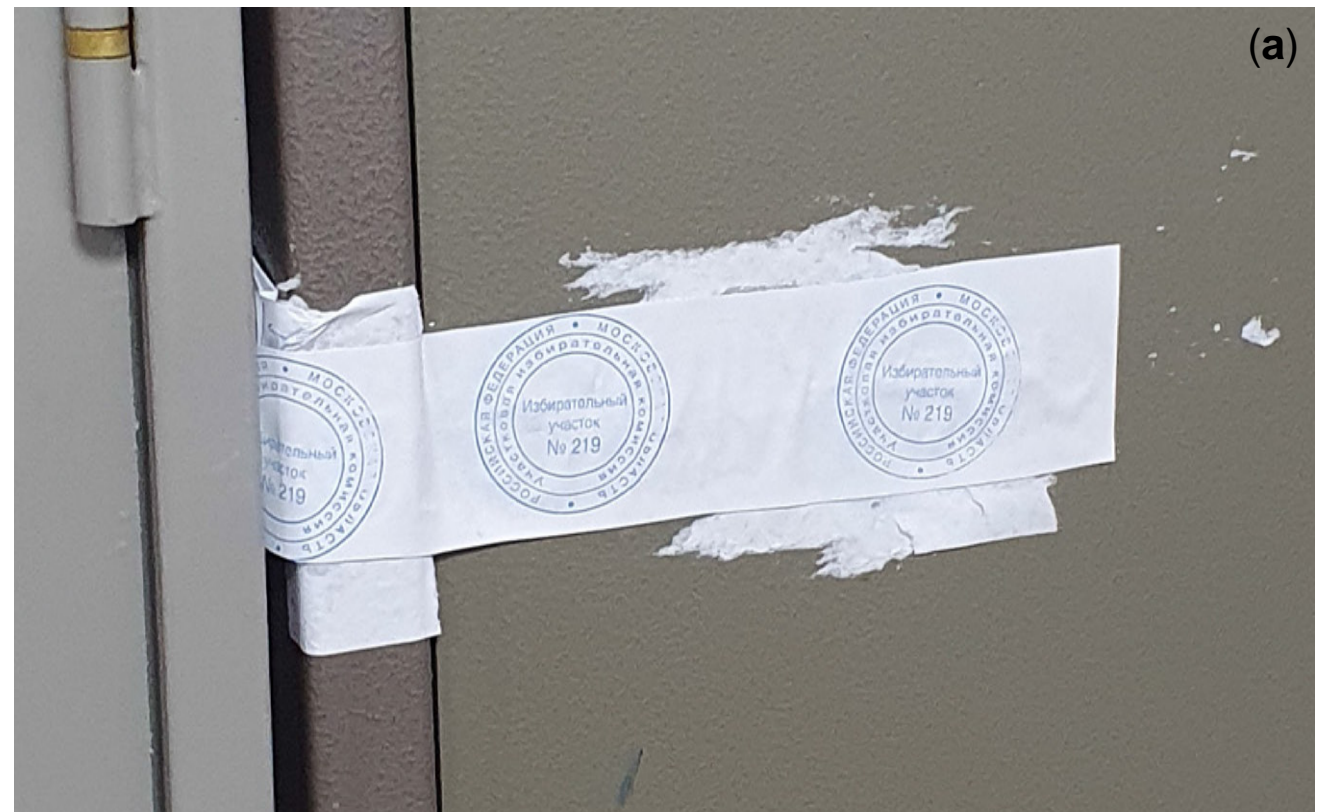

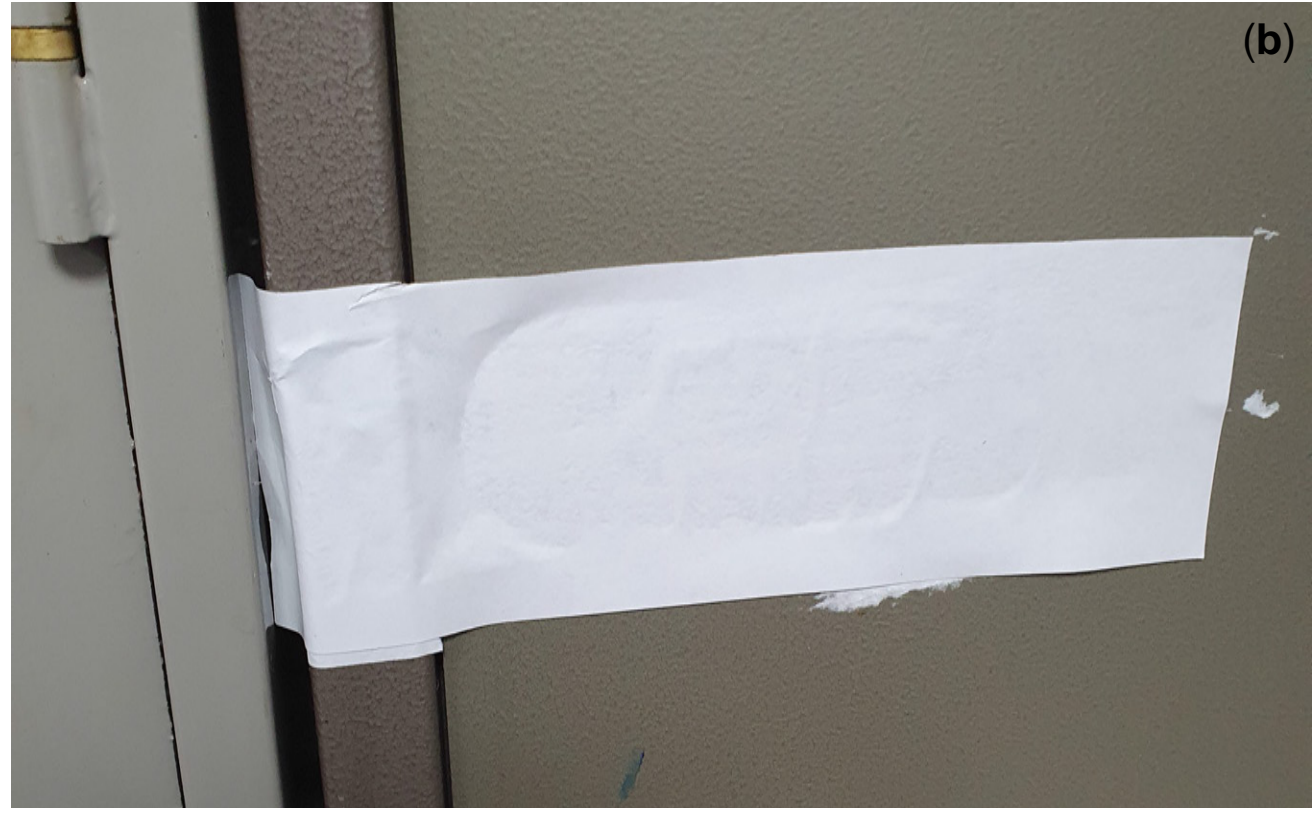

Рис. 21. Сейф в участке 219 (a) до того, как помещение для голосования закрыли на ночь 6 сентября и (b) утром следующего дня сразу после открытия помещения. Бумажная опломбировочная лента с оттисками печати участковой комиссии на сейфе была небрежно заменена ночью, пока участок был закрыт, на совершенно другую та B номер набран вразрядку (рис. 20). С какой стати производитель, изготавливая дубликаты пакетов на одной и той же производственной линии, оставил нам такую очевидную улику? Мы предполагаем, что наиболее вероятный ответ заключается в том, что на этой линии были изготовлены две производственные партии с одинаковым набором номеров для одного и того же заказчика, с небольшим временны́м интервалом. Одна из этих партий состояла из 51 000 сейф-пакетов, заказанных избирательной комиссией Московской области [37], а другая — неизвестное количество пакетов с перекрывающимися по номерному диапазону, но *по-другому набранными* номерами. К сожалению, мы видим еще несколько пар красных лент с идентичными, но по-разному набранными номерами на других наших участках.

Иные улики на этом участке включают подмененную в первую ночь бумажную опломбировочную ленту на сейфе (рис. 21). Помещение избирательного участка не было полностью опечатано на ночь: его уличная входная дверь была тщательно опечатана, но задняя дверь была просто заперта на ключ. Сейф и двери были заперты секретарем участковой комиссии при закрытии участка на ночь. На видеозаписи наружной камеры уличного видеонаблюдения видно, что она затем ждет на улице, пока наблюдатели уйдут, беседует с ночным охранником и зачем-то ненадолго возвращается к черному входу в здание.

Подмена бумажной пломбы на сейфе (но не остальные наши доказательства) была обнаружена утром второго дня и привлекла нежелательное внимание к участку. Это, возможно, привело к отказу злоумышленников от плана подменить еще один пакет в следующую ночь.

### 2. Подмена сейф-лент на участке 212

На этом участке ленты на обоих пакетах, изначально имевшие нормально набранные номера, ночью были заменены на ленты с разреженно набранными номерами (приложение C 3). Это было сделано, по-видимому, для того, чтобы избежать подделки нескольких подписей, сделанных авторучкой на обеих сторонах пакета. Ленты после подмены стали у́же (рис. 1 и 22). Микрошрифт вдоль боковых краев пакета исчез возле лент. Забавно, что злоумышленники на этом участке использовали неправильный тип растворителя, который не только смыл этот заводской шрифт, но и размазал клей по поверхности пакета и сделал ее очень липкой. Во время подсчета голосов оба пакета вытащили из сейфа сложенные пополам так, что область вокруг ленты соприкасалась с нижней частью того же пакета. Они слиплись, и члену избирательной комиссии пришлось их раздирать с видимым усилием и громким треском, записанным на нашем видео (рис. 23). На оригинальной сейф-ленте на пакете с бюллетенями первого дня голосования виден рисунок в виде сот, указывающий на ее подлинность, а также ее номер, совпадающий с номером на пакете [рис. 22(b)]. Изображения трех других лент имеют недостаточное разрешение, чтобы подтвердить наличие рисунка сот и сверить их номера.

Другой уликой на этом участке является легкое смещение положения (примерно на 1 мм) бумажной опломбировочной ленты на сейфе в течение второй ночи.

### 3. Участок 213

Изображений пакета с бюллетенями первого дня голосования у нас нет.

Пакет с бюллетенями второго дня голосования имеет ленту нормальной ширины после его запечатывания на второй день, однако на подсчете голосов лента у́же и деформирована [рис. 24(a)–(d)]. Ее расстояние от верхнего края пакета сильно варьируется [рис. 24(d)], что говорит о неудачном ее приклеивании при подмене. Один из концов ленты загибается

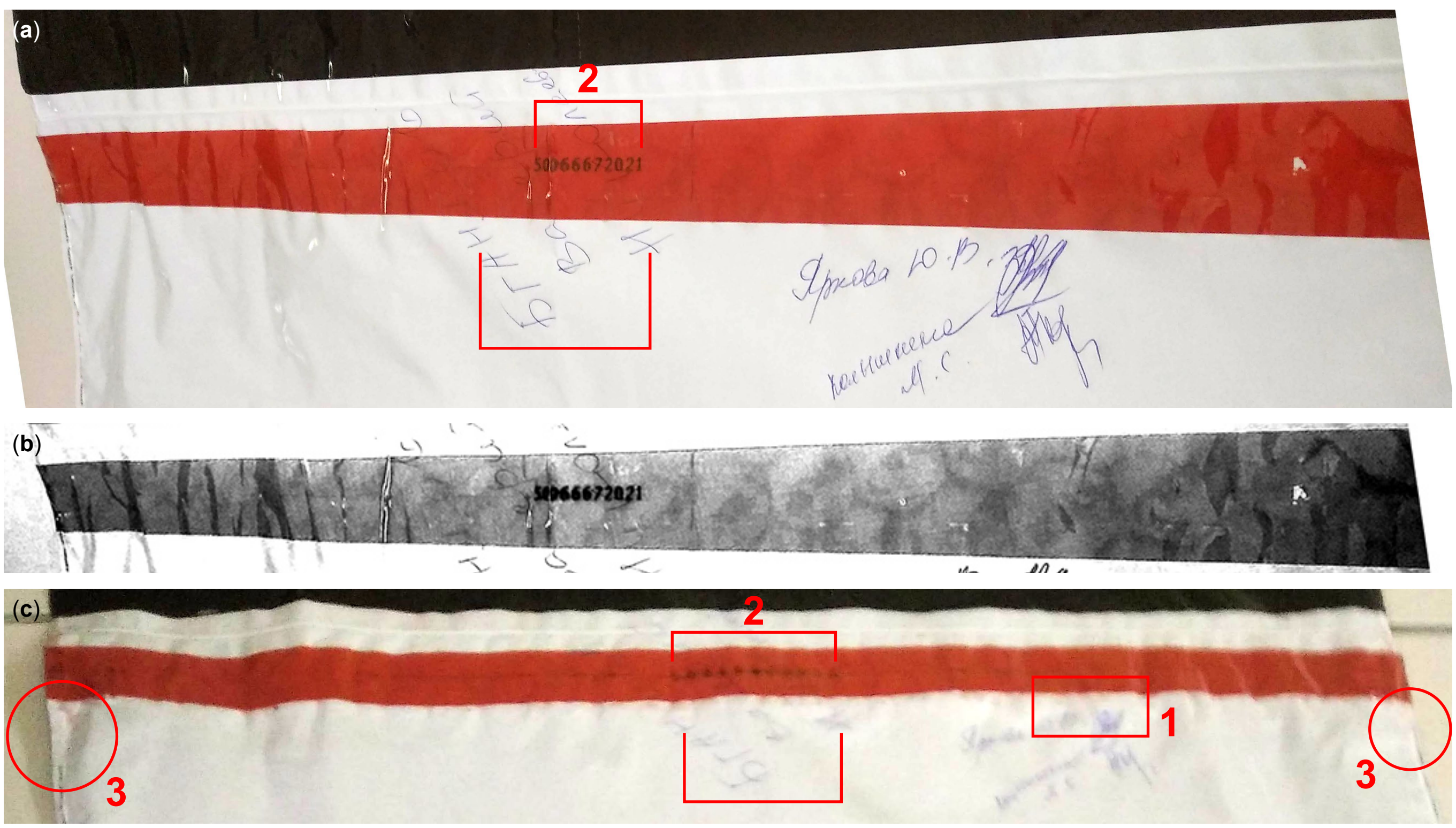

Рис. 22. Красная лента на сейф-пакете на участке 212 (a) после запечатывания пакета в первый день, (b) красный цветовой канал этого же изображения с максимизированным контрастом показывает рисунок сот на сейф-ленте, и (c) до вскрытия этого пакета во время подсчета голосов на третий день. Обратите внимание на (1) уменьшенную ширину ленты и ее смещенное расположение относительно подписей, сделанных авторучкой, (2) увеличенные межсимвольный интервал и общую ширину 11-значного номера, набор которого изменился с нормального на разреженный, что также очевидно в сравнении с шириной расположенной рядом подписи, и (3) исчезнувший фабричный микрошрифт по краям пакета, предположительно смытый растворителем

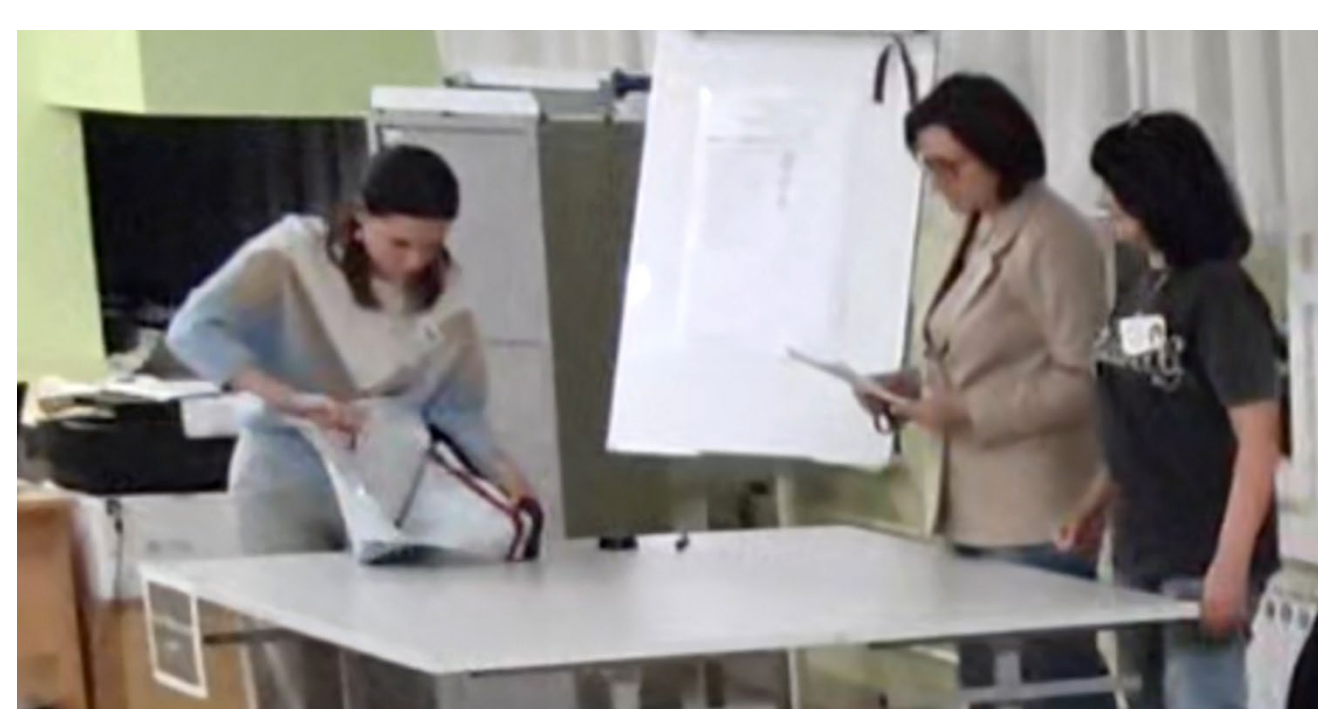

Рис. 23. Секретарь участка 212 силой отрывает липкую область вокруг ленты, прилипшую к низу пакета, в то время как председатель и заместитель председателя участковой комиссии спокойно за этим наблюдают на другую сторону пакета, образуя там прямоугольный отрезок длиной не менее 4 мм [рис. 24(e)]. Это невозможно при нормальном использовании пакета. На производственной линии непрерывная «лента» собранных пакетов с установленной сейф-лентой разрезается на отдельные пакеты вдоль этого края резаком, что является заключительным шагом их изготовления [52]. Ввиду такой технологической последовательности изготовления, конец сейф-ленты всегда обрезан вровень с боковым краем пакета и не может сдвигаться при нормальном использовании. Действительно, на кадрах видеозаписи этого же края пакета, сделанной непосредственно перед его запечатыванием, не видно выступающей на другую сторону ленты [рис. 24(f)].

## 4. Участок 214

На этом участке совокупная длина трех непрерывных и почти непрерывных последовательностей в стенограмме составляет 273 бюллетеня (рис. 3). Это говорит о том, что бюллетени в обеих сейф-пакетах были подменены, так как в них вместе находилось 285 бюллетеней (таблица I). К сожалению, имеющиеся изображения пакетов имеют недостаточное разрешение, чтобы увидеть номера на их лентах. Геометрия ленты пакета с бюллетенями первого дня голосования не изменилась к подсчету, что говорит о том, что лента не была подменена (но не исключает подмены пакета целиком). На пакете с бюллетенями второго дня голо-

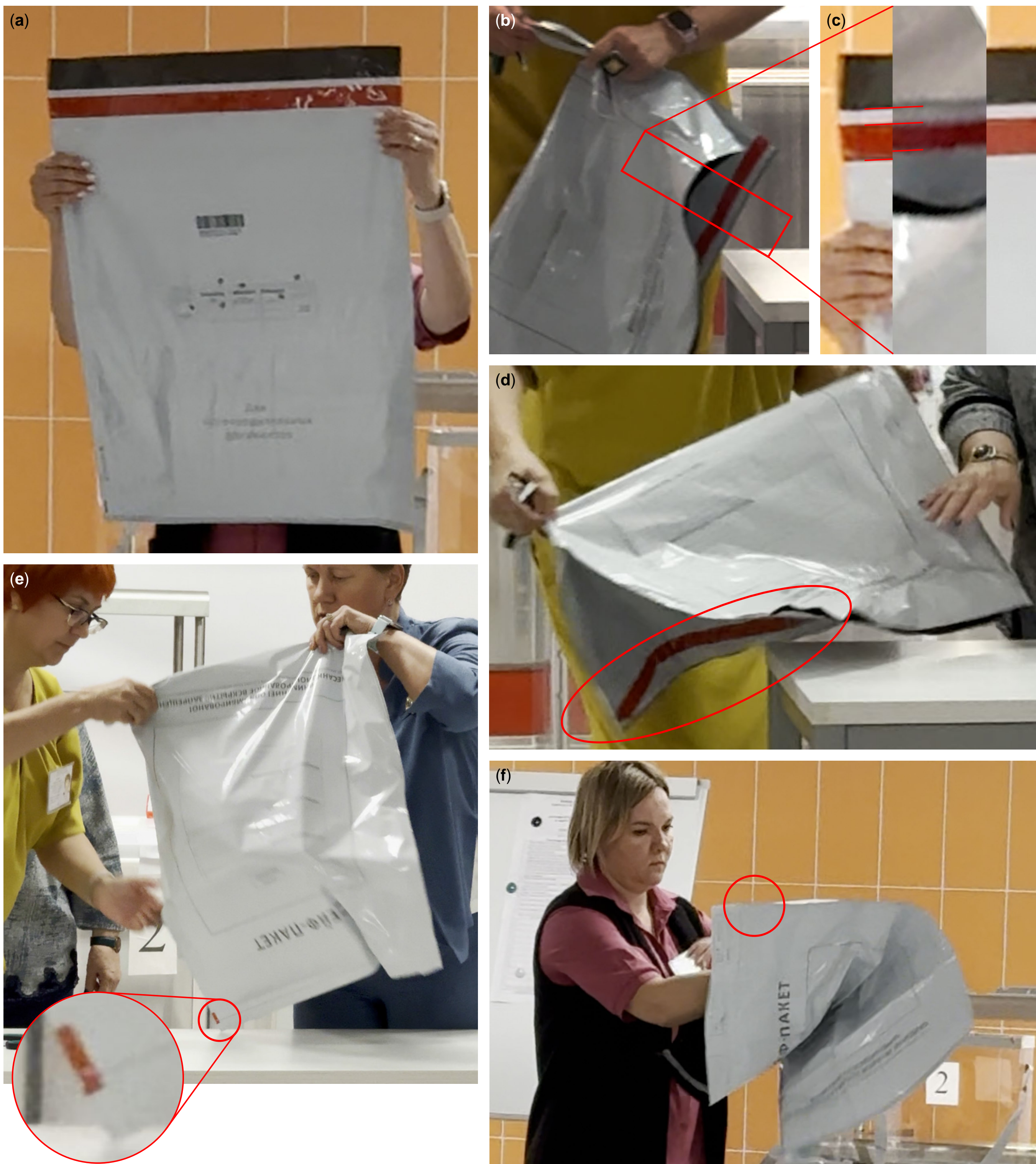

Рис. 24. Кадры видеозаписей сейф-пакета с бюллетенями второго дня голосования на участке 213 (a) после запечатывания пакета во второй день, (b) и (d) после вскрытия этого пакета во время подсчета голосов на третий день. (c) Коллаж из фрагментов кадров (a) и (b) иллюстрирует, что ширина ленты и/или ее положение относительно верхнего края пакета изменились за ночь по всей ее длине. Ширину ленты на изображении (a) можно оценить примерно в 31 мм, используя высоту штрих-кода в 20 мм в качестве размерного эталона [52]. Такая же ширина ленты получается при использовании в качестве размерного эталона высоты пакета в 695 мм (не включая отрывные квитанции). (e) На другой стороне пакета после его вскрытия виден прямоугольный конец ленты, выступающий за край и загнутый на эту сторону (см. врезку). (f) Перед запечатыванием пакета во второй день в этом же месте не видно выступающей ленты

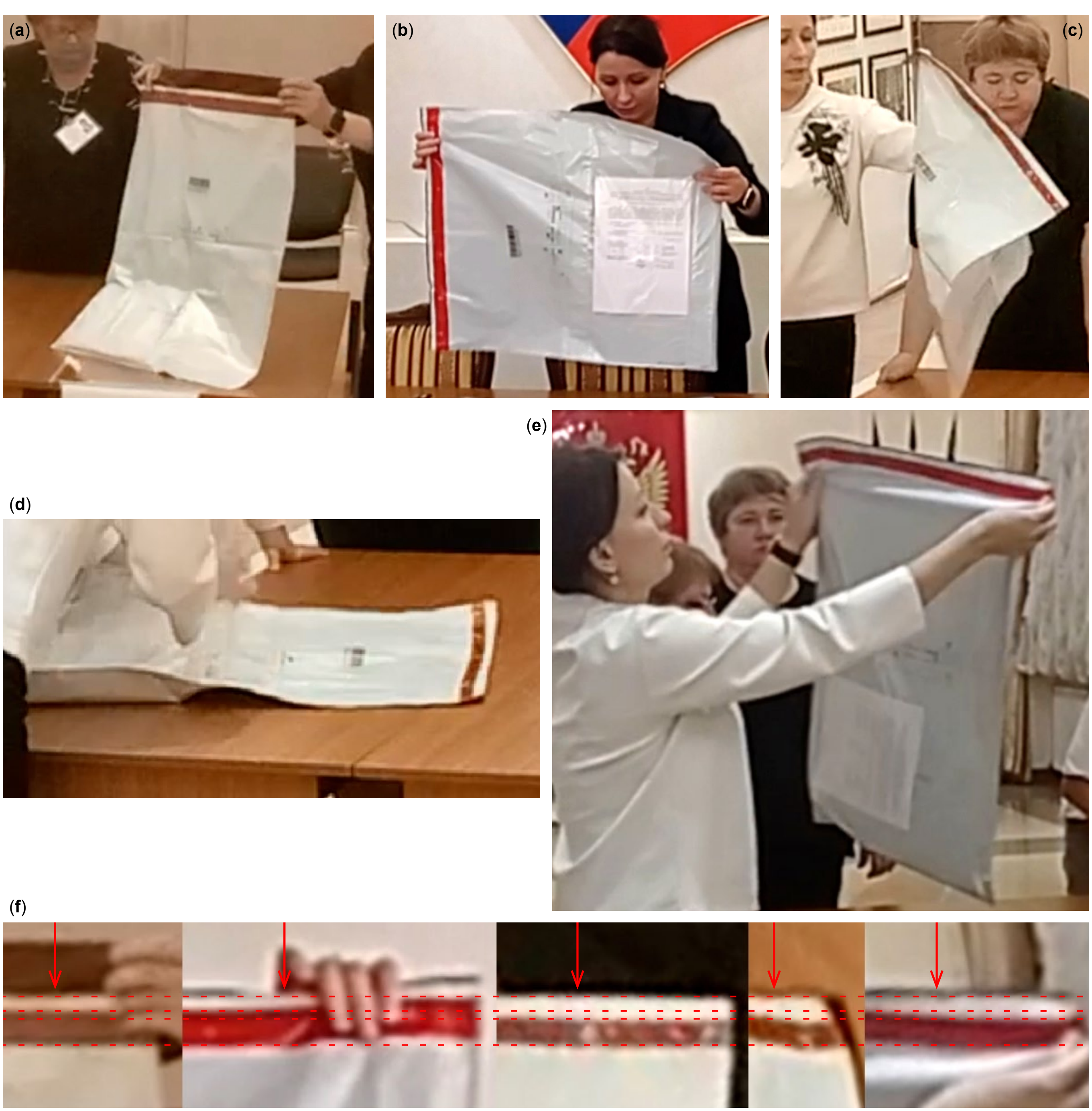

Рис. 25. Кадры видеозаписей сейф-пакетами с бюллетенями второго дня голосования на участке 214 после его запечатывания во второй день, сделанные (a) камерой 1 и (b) камерой 2, затем перед вскрытием этого пакета во время подсчета голосов на третий день, сделанные (c) камерой 1, (d) камерой 1 и (e) камерой 2. (f) Коллаж из повернутых фрагментов кадров (a)–(e) показывает изменение ширины ленты и ее расстояния от края, а также постоянство нашей оценки геометрии при использовании кадров разных камер и углов съемки. Стрелкой указана примерная точка на расстоянии 1/4 ширины пакета от его правого края в каждом сегменте коллажа. Ширину ленты также можно сравнивать с высотой штрих-кода, видимого на изображениях, которая составляет 20 мм согласно заводской спецификации [52]. Используя ее в качестве размерного эталона, можно определить, что ширина ленты на изображениях (a) и (b) составляет не менее 30 мм, тогда как на изображениях (c) и (d) она близка к 20 мм

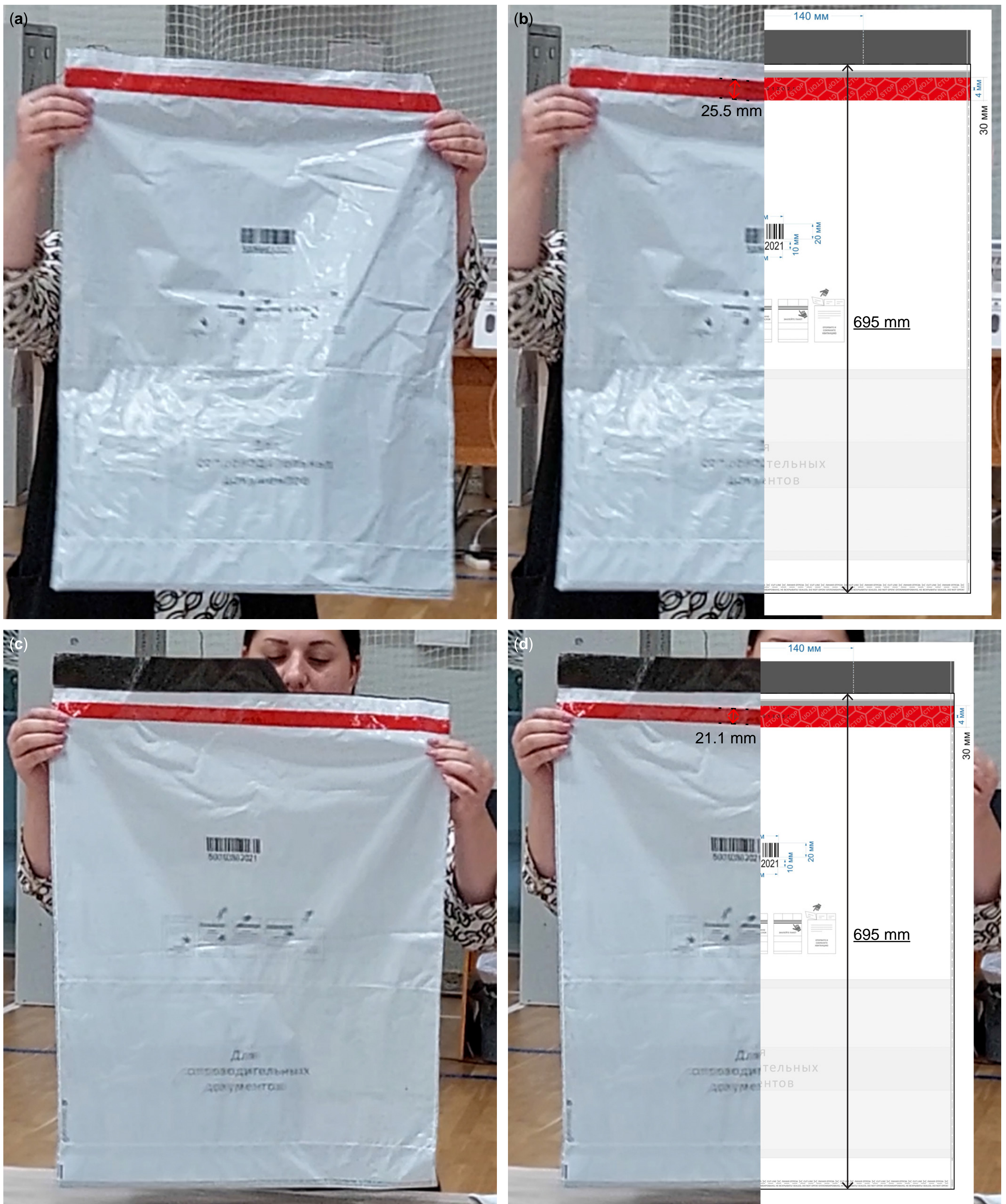

Рис. 26. Кадры видеозаписи сейф-пакетов на участке 216 перед их вскрытием на подсчете голосов. (a) Сейф-пакет от первого дня голосования, (b) с наложенной половиной его конструкторского чертежа. Ширина ленты составляет 25,5 мм при использовании высоты пакета согласно его фабричного чертежа в качестве размерного эталона. (c) Сейф-пакет от второго дня голосования, (d) с наложенной половиной его конструкторского чертежа. Ширина ленты составляет 21,1 мм при использовании того же метода ее измерения сования лента заметно изменилась, став у́же и криво приклеенной (рис. 25). Расстояние от ленты до верхнего края пакета одинаково по длине ленты во второй день, но заметно увеличено и стало неравномерным по длине ленты на подсчете голосов. Несмотря на низкое разрешение изображений, наблюдаемая разница в ширине ленты достоверна и измеряется как при сравнении бок о бок [рис. 25(f)], так и с использованием высоты штрих-кода в качестве эталона. Таким образом, лента была подменена (приложение C 3).

Другой уликой на этом участке является то, что члены избирательной комиссии многократно заходили в здание и выходили из него после закрытия участка в первый вечер, а затем вошли в помещение для голосования задолго до времени допуска наблюдателей (т. е., до 7:00) утром. Они были там без наблюдателей.

### 5. Участок 215

На этом участке, в отличие от всех остальных участков, для хранения бюллетеней использовались сейф-пакеты с 8-значными номерами. Мы не наблюдаем никаких изменений в геометрии их лент, когда сравниваем их фотографии с высоким разрешением, сделанные в первые два дня, с видеозаписью с низким разрешением, сделанной во время подсчета голосов. О существовании заводских подделок пакетов и лент с 8-значными номерами пока ничего не известно. Комиссия не перемешивала бюллетени намеренно. Стенограмма подсчета голосов и ее статистический анализ (раздел IV) не выявляют существенных аномалий. Мы делаем вывод, что на этом участке бюллетени не подменяли.

Несмотря на это, комиссия в открытую сжульничала на подсчете голосов в пользу кандидатов от ЕР. Федориву насчитали на 20, а Шик на 16 голосов больше [31], чем в нашей стенограмме, сделанной по полной видеозаписи оглашения бюллетеней. Наши комиссии неисправимы.

### 6. Участок 216

На этом участке имеющиеся видеозаписи первых двух дней имеют слишком низкое разрешение для измерения ширины ленты (рис. 27). Их просмотр дает лишь субъективное впечатление, что ленты имеют нормальную ширину. Однако на подсчете голосов оба пакета были показаны на камеру в плоскость и сняты в хорошем разрешении. Это позволяет нам сравнить их неискаженную геометрию с конструкторским чертежом пакета этой партии, предоставленным фабрикой [52] (рис. 26). Ширина ленты составляет 25,5 мм и 21,1 мм на пакетах от первого и второго дня голосования соответственно, что указывает на подмену обеих лент. Знаки номера на подменной ленте пакета первого дня слегка проглядывают на нескольких кадрах,

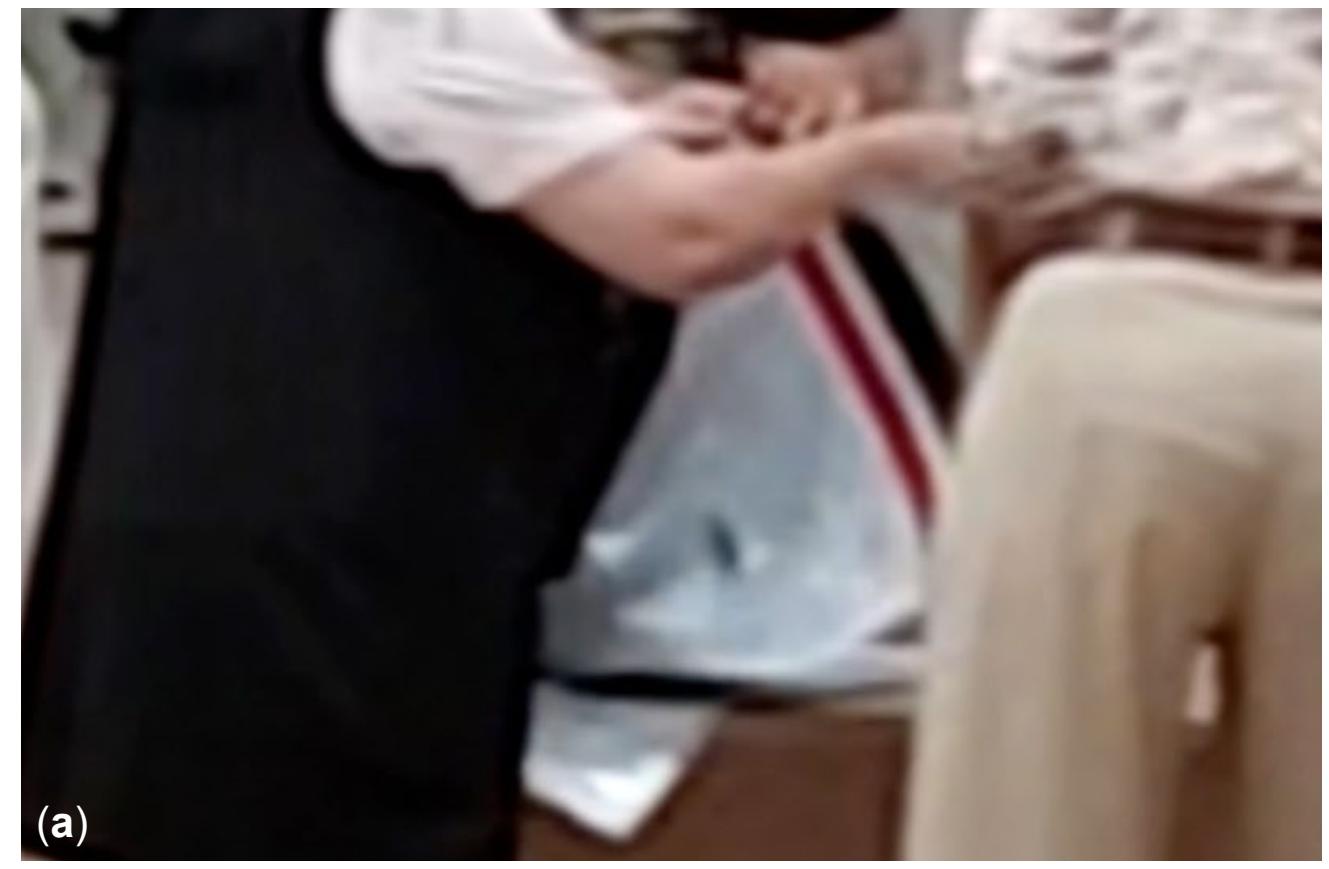

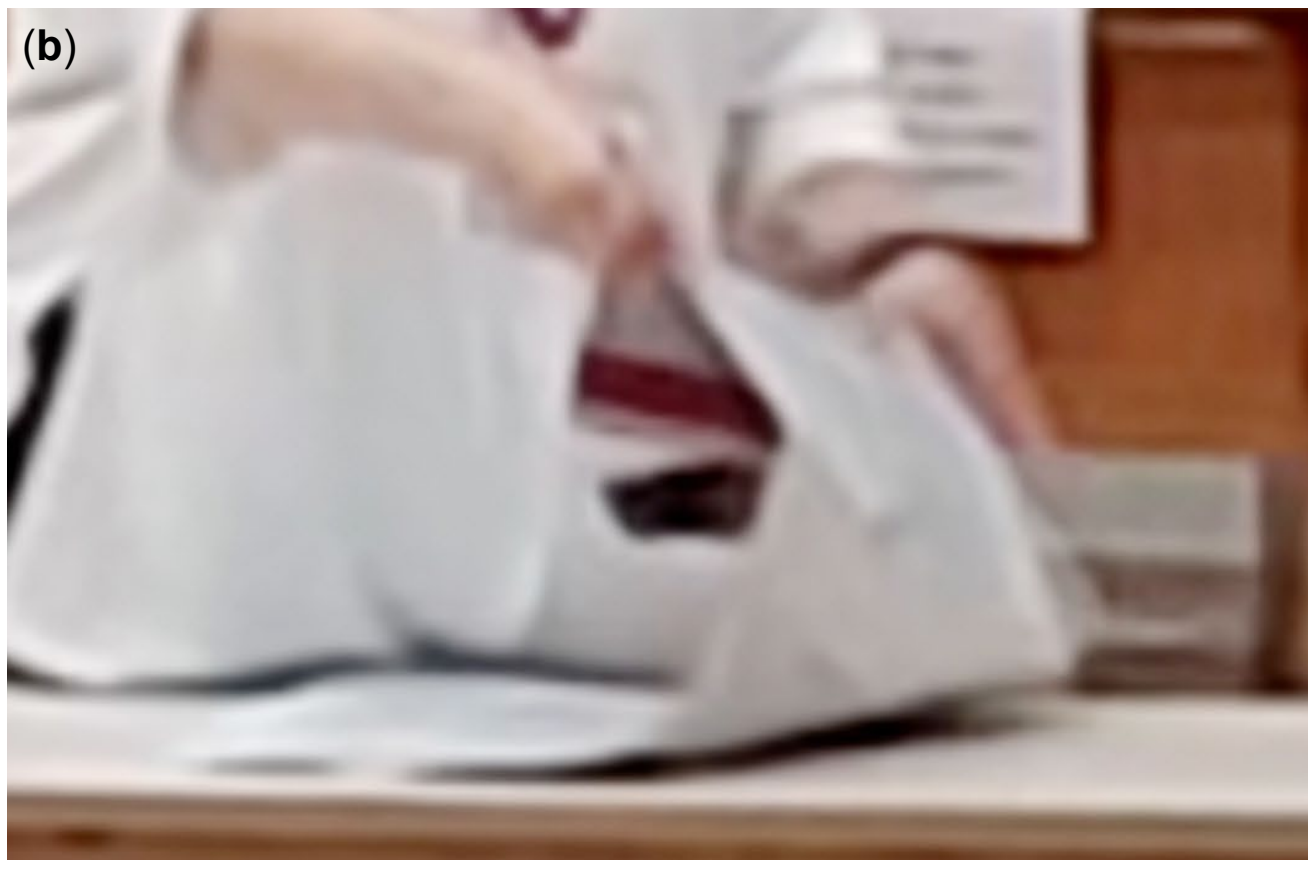

Рис. 27. Кадры видеозаписей сейф-пакетов после их запечатывания на участке 216: (a) в первый день и (b) во второй день показывая, что номер набран вразрядку (рис. 28).

Как это ни странно, этот участок в открытую «обсчитал» кандидатов от ЕР на подсчете. Федориву недосчитали 12, а Шик — 8 голосов [31] по сравнению с их количеством в нашей стенограмме.

### 7. Участок 217

Как и на участке 216, обе ленты здесь были подменены. Используя высоту пакета в качестве размерного эталона, ширина ленты на пакете от первого дня голосования оказывается менее 24,9 мм [рис. 29(a)]. Изображение этого пакета имеет заметное искажение перспективы, будучи значительной частью кадра широкоугольной камеры. Это искажение визуально сжало нижнюю часть пакета в направлении измерения, тем самым мы переоцениваем ширину ленты. Более точное измерение возможно, если применить к изображению преобразование коррекции перспективы. Однако нет необходимости прибегать к такому шагу, поскольку ширина ленты уже определена как значительно меньше 30 мм. Если использовать в качестве размерного эталона штрих-код высотой 20 мм, который стра-

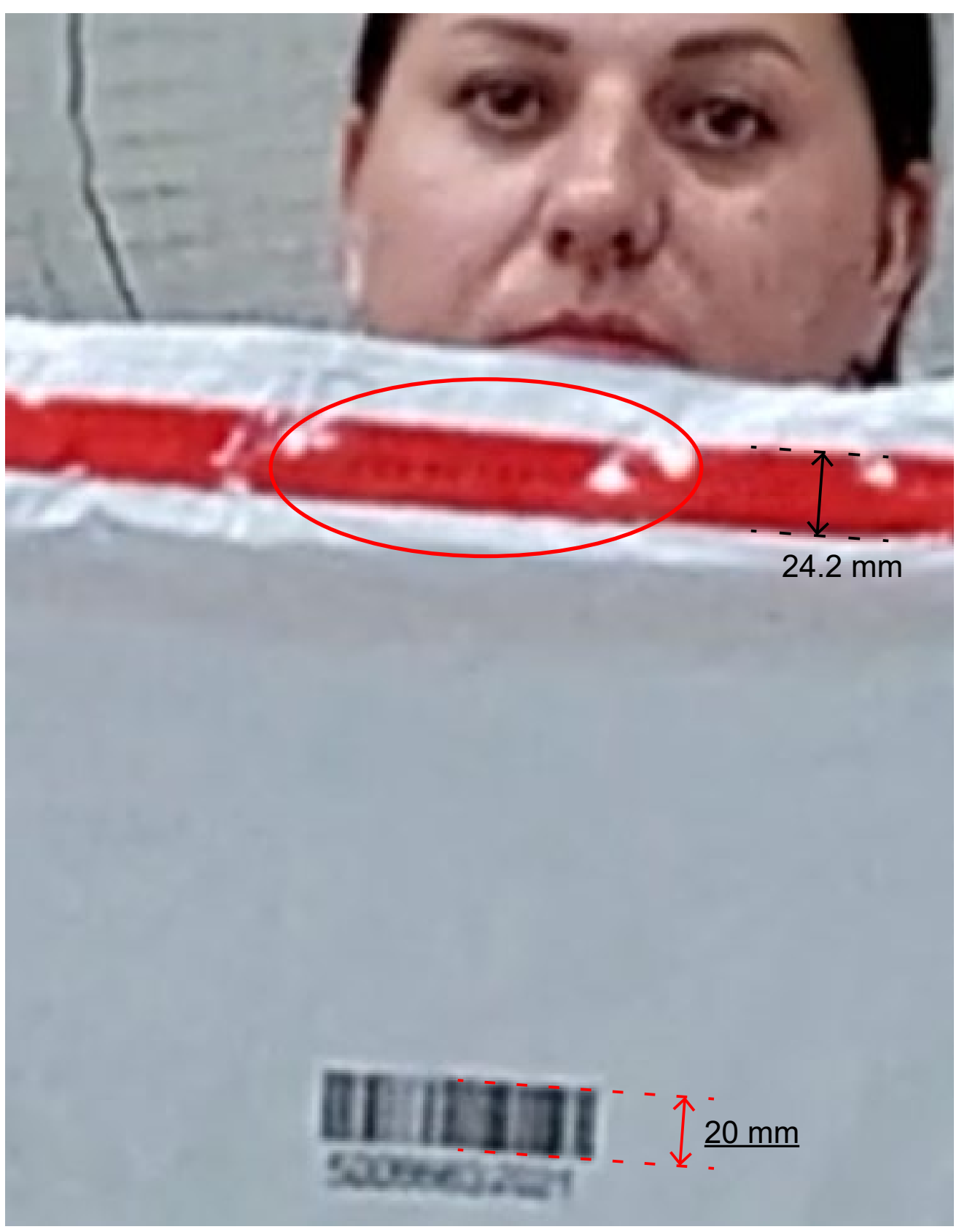

Рис. 28. Кадр видеозаписи сейф-пакета от первого дня голосования на участке 216 перед его вскрытием на подсчете голосов. Положение десяти из одиннадцати цифр номера, набранного вразрядку, различимо на ленте. Ширину ленты можно сопоставить с высотой штрих-кода, которая составляет 20 мм, согласно заводской спецификации. Используя последнюю в качестве размерного эталона, ширину ленты можно оценить в 24,2 мм дает от похожего, но меньшего искажения перспективы, ширина ленты оценивается в 24,7 мм. Коллаж [рис. 29(b)] также наглядно это иллюстрирует.

Сейф-пакет от второго дня был снят только во время подсчета голосов, когда его извлекали из сейфа (рис. 30). Используя его штрих-код в качестве размерного эталона, ширина ленты составляет 19,9 мм.

Другой уликой на этом участке является то, что члены участковой избирательной комиссии вошли в здание задолго до времени допуска наблюдателей (т. е., до 7:00) утром второго дня. Они были там без наблюдателей.

## 8. Участок 218

На этом участке, как показывают наши видеозаписи, оба пакета были помещены в сейф незапечатанными. В первый день председатель участковой комиссии публично отодрала красную ленту с пакета и выкинула ее, вместо того, чтобы правильно запечатать пакет.

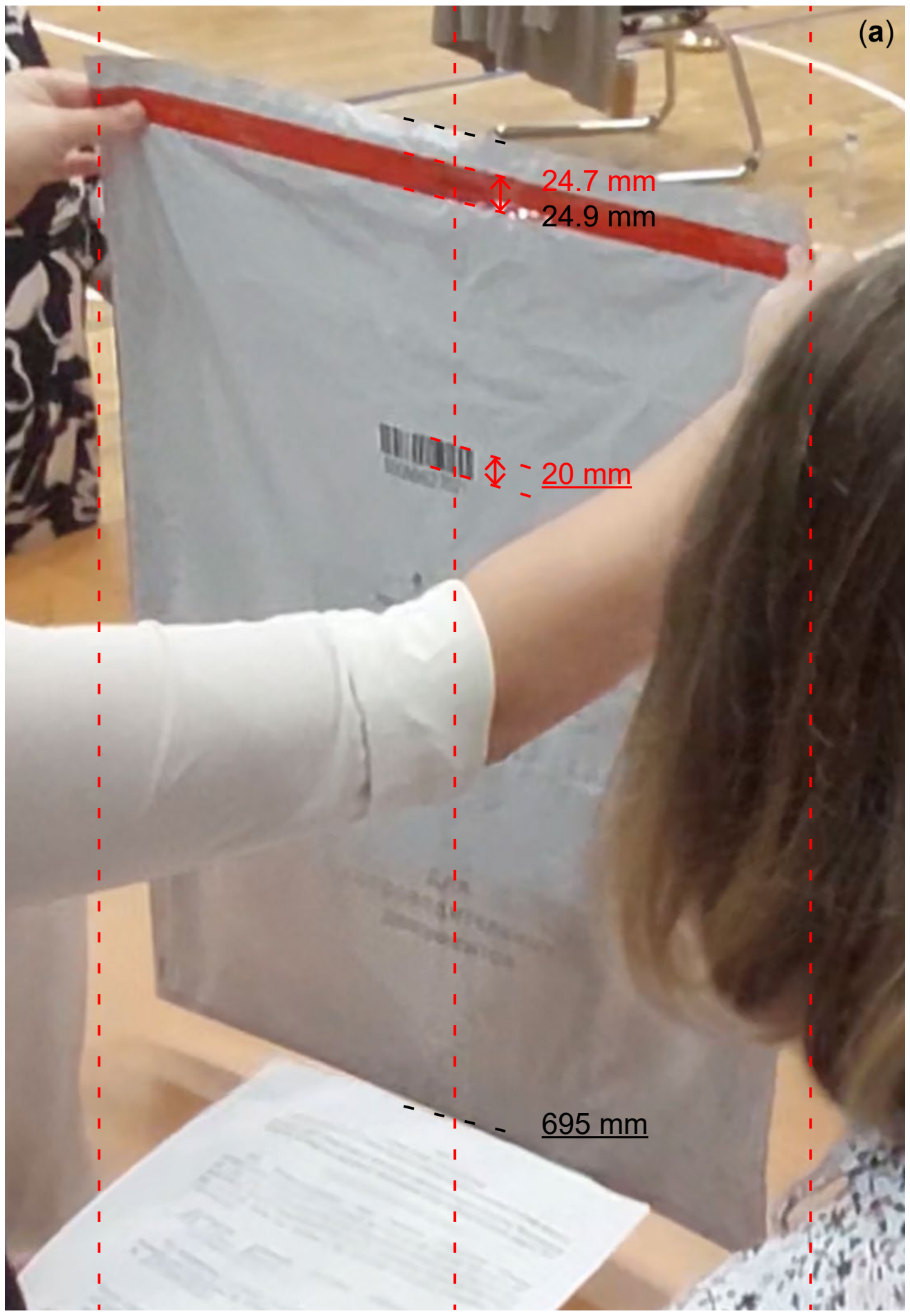

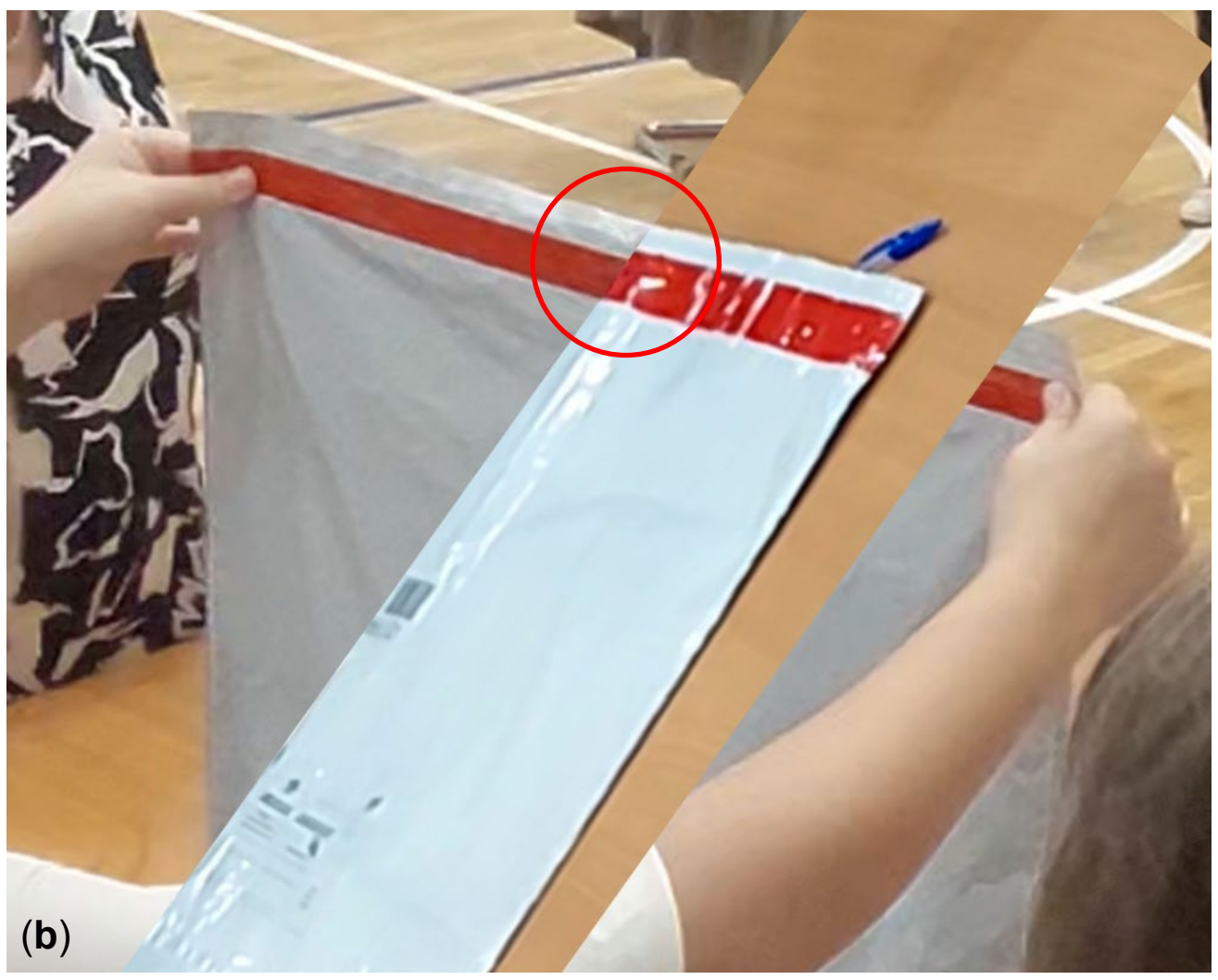

Рис. 29. Кадры видеозаписей сейф-пакета от первого дня голосования на участке 217. (a) Пакет демонстрируется присутствующим перед его вскрытием на подсчете голосов. При использовании высоты штрих-кода (пакета) в качестве размерного эталона, ширина ленты оценивается в 24,7 мм (24,9 мм). Обе оценки завышают ширину ленты из-за искажения перспективы. (b) Коллаж с повернутым кадром, сделанным в первый день после запечатывания этого пакета. Это непосредственное сравнение показывает, что лента стала у́же, а расстояние от нее до верхнего края пакета увеличилось

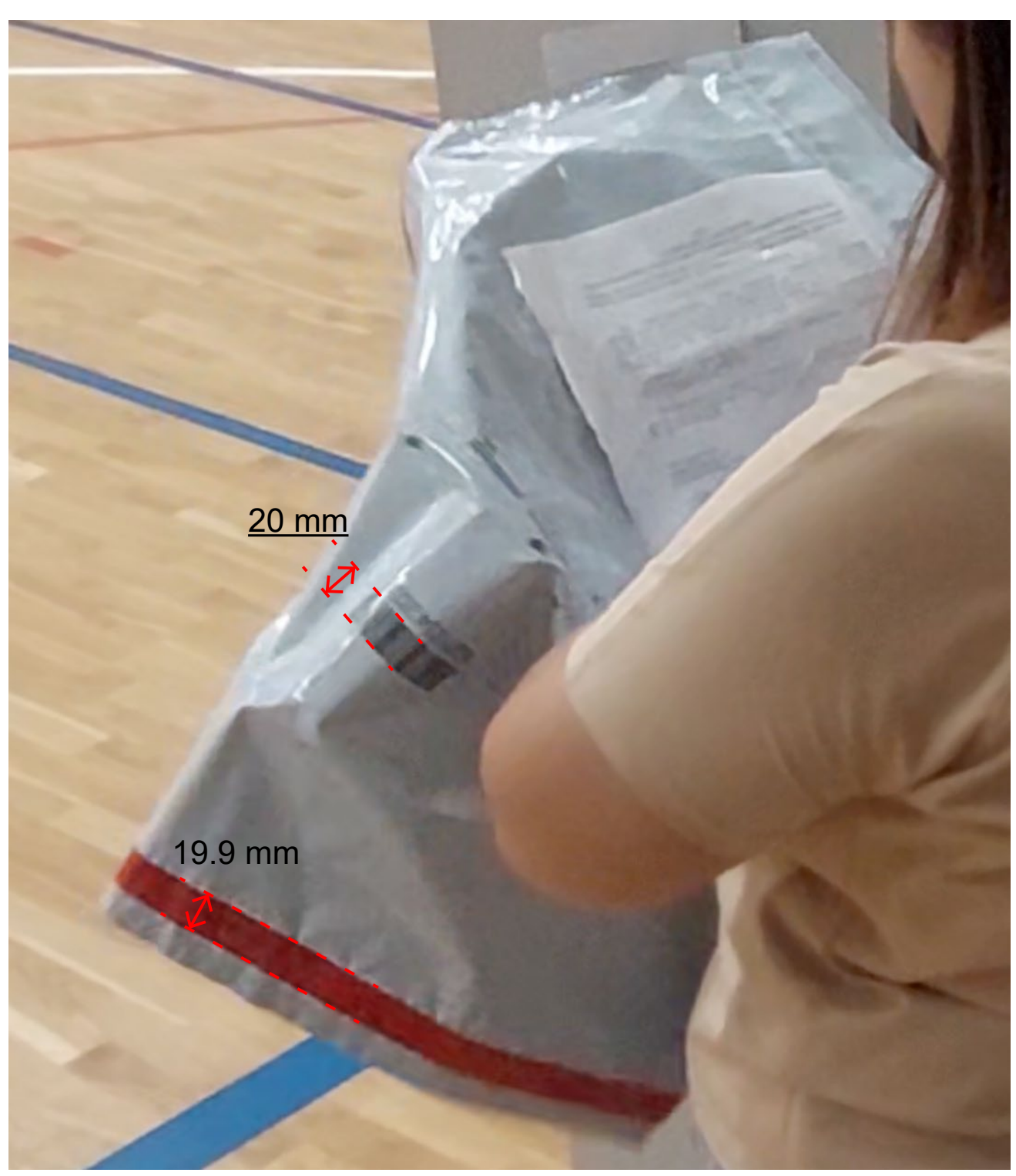

Рис. 30. Кадр видеозаписи сейф-пакета с бюллетенями второго дня голосования на участке 217 перед его вскрытием на подсчете голосов. Используя высоту штрих-кода в качестве размерного эталона, ширину ленты можно оценить в 19,9 мм

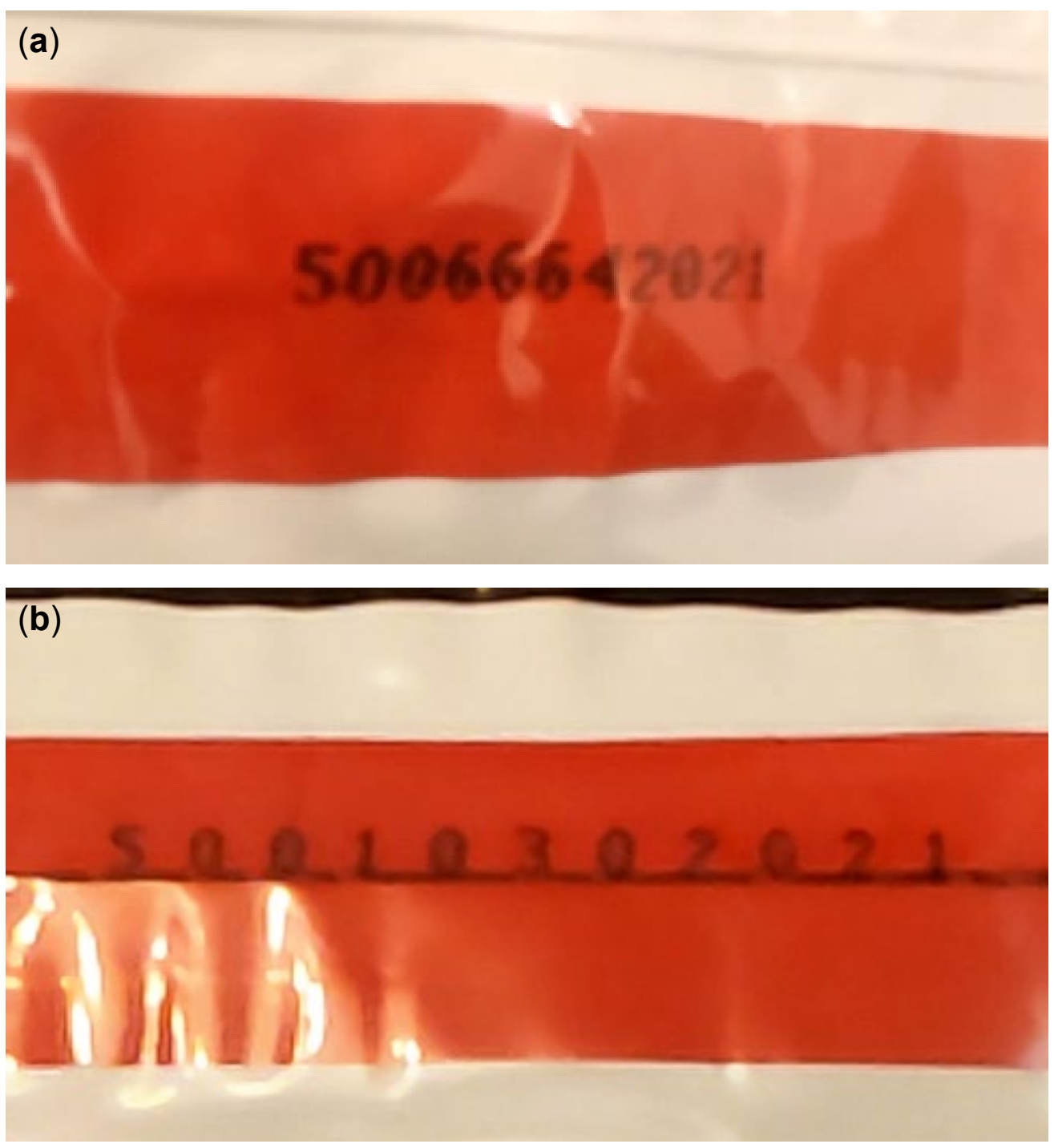

Рис. 31. Кадры видеозаписей с участка 220: (a) первый сейф-пакет после его опечатывания в первый день и (b) второй сейф-пакет после его опечатывания во второй день

Она убрала пакет с бюллетенями, ставший обычным не заклеенным полиэтиленовым пакетом, в сейф и закрыла участок на ночь. На второй день она вовсе не трогала ленту, поместив сейф-пакет с бюллетенями в сейф в незапечатанном состоянии. На подсчете голосов оба пакета были извлечены из сейфа запечатанные красной лентой надлежащим образом. Мы предполагаем, что либо первый пакет был заменен на его дубликат, либо на него ночью была приклеена другая лента. Второй пакет просто был запечатан ночью обычным образом.

Стенограмма этого участка содержит чрезвычайно длинную последовательность из 382 идентичных бюллетеней за 5 получивших мандаты кандидатов, а также две более короткие последовательности (рис. 5). Очень немногие голоса в ней выглядят случайными.

## 9. Участок 220

На этом участке бюллетени были запечатаны в пакет с нормально набранным номером на ленте в первый день, но с разреженно набранным номером во второй день (рис. 31). На подсчете голосов геометрия обеих лент выглядит неизменной, включая неизменность складок, возникших при приклеивании ленты во второй день. Разрешения съемки на подсчете недостаточно для того, чтобы увидеть на лентах их номера. Однако это не исключает подмены первого пакета целиком. Общая длина трех непрерывных последовательностей идентичных бюллетеней в стенограмме (рис. 5) равна 124, а общее количество бюллетеней за 5 победивших кандидатов равно 129. Это согласуется с возможной подменой 129 бюллетеней в первом пакете в первую ночь. Планы злоумышленников по подмене второго пакета, вероятно, были отменены из-за общественного внимания к соседнему участку 219 во второй день (приложение D 1). Последний находится на расстоянии 50 м и имеет общий с этим участком вход на огороженную забором территорию.

Вдобавок, эта избирательная комиссия не скрываясь сжульничала на подсчете голосов. Она насчитала больше голосов за 5 победивших кандидатов, чем было оглашено по бюллетеням, и недосчитала голоса за многих оппозиционных кандидатов. Например, Аникеевой (СРЗП) прибавили 24 голоса, а Федотову (КПРФ) записали на 17 голосов меньше [31], чем мы сами насчитали по стенограмме, сделанной с полной видеозаписи оглашения бюллетеней.

Таблица VII. Обнаруженные нами 11-значные номера сейф-пакетов, отсортированные по возрастанию. Переменная часть номера выделена жирным шрифтом. Номер в круглых скобках означает, что его цифры не были напрямую нами сверены на изображении красной ленты, из-за недостаточного разрешения съемки. Номер посередине между столбцами означает, что использовалась по меньшей мере одна из двух лент, но мы не знаем, какая именно. Номер в квадратных скобках означает, что у нас нет изображений этой ленты, но мы предполагаем на основе других доказательств, что она использовалась при подмене ленты или пакета

<table>
<tr><th rowspan="2">Дата выборов</th><th rowspan="2">Муниципалитет</th><th rowspan="2">Номер участка</th><th colspan="2">Номер на ленте</th></tr>
<tr><th>Набранный обычно</th><th>Набранный вразрядку</th></tr>
<tr><td>Сентябрь 2024 г.</td><td>Власиха</td><td>214</td><td>(50<b>01029</b>2021)</td><td>(50<b>01029</b>2021)</td></tr>
<tr><td></td><td></td><td>220</td><td></td><td>50<b>01030</b>2021</td></tr>
<tr><td></td><td></td><td>213</td><td>(50<b>01031</b>2021)</td><td>(50<b>01031</b>2021)</td></tr>
<tr><td></td><td></td><td>219</td><td></td><td>50<b>01032</b>2021</td></tr>
<tr><td></td><td></td><td>218</td><td colspan="2">(50<b>01033</b>2021)</td></tr>
<tr><td></td><td></td><td>217</td><td>[50<b>01034</b>2021]</td><td>(50<b>01034</b>2021)</td></tr>
<tr><td></td><td></td><td>212</td><td>(50<b>01035</b>2021)</td><td>(50<b>01035</b>2021)</td></tr>
<tr><td></td><td></td><td>216</td><td>(50<b>01036</b>2021)</td><td>(50<b>01036</b>2021)</td></tr>
<tr><td></td><td></td><td>218</td><td>(50<b>06660</b>2021)</td><td>(50<b>06660</b>2021)</td></tr>
<tr><td></td><td></td><td>213</td><td colspan="2">(50<b>06661</b>2021)</td></tr>
<tr><td></td><td></td><td>217</td><td>(50<b>06662</b>2021)</td><td>(50<b>06662</b>2021)</td></tr>
<tr><td></td><td></td><td>216</td><td>(50<b>06663</b>2021)</td><td>(50<b>06663</b>2021)</td></tr>
<tr><td></td><td></td><td>220</td><td>50<b>06664</b>2021</td><td>[50<b>06664</b>2021]</td></tr>
<tr><td></td><td></td><td>219</td><td>50<b>06665</b>2021</td><td>50<b>06665</b>2021</td></tr>
<tr><td></td><td></td><td>214</td><td>[50<b>06666</b>2021]</td><td>[50<b>06666</b>2021]</td></tr>
<tr><td></td><td></td><td>212</td><td>50<b>06667</b>2021</td><td>(50<b>06667</b>2021)</td></tr>
<tr><td>Сентябрь 2021 г.</td><td>Люберцы</td><td>1558</td><td>50<b>13317</b>2021</td><td></td></tr>
<tr><td></td><td></td><td></td><td>50<b>18815</b>2021</td><td></td></tr>
<tr><td></td><td>Балашиха</td><td>3667</td><td>50<b>23439</b>2021</td><td></td></tr>
<tr><td></td><td>Реутов</td><td>2642</td><td>50<b>43625</b>2021</td><td></td></tr>
</table>

## 10. Какое количество заводских подделок сейф-пакетов произвела фабрика?

По информации от производителя, партия, поставленная избирательной комиссии Московской области в августе 2021 г., содержала 51 000 пакетов с индивидуальными 11-значными номерами вида 50$xxxxx$2021, где $xxxxx$ — переменная часть в диапазоне от 00001 до 53749 (не все номера в этом диапазоне были поставлены) [52]. Впоследствии региональная комиссия распределила эти пакеты по своим территориальным избирательным комиссиям. Отметим, что пакеты того же размера, но с частично отличающимся графическим дизайном и 8-значными номерами, продаются в розничной торговле и также широко используются на выборах. Поэтому 11-значные номера бросаются в глаза. Мы провели поиск в дискуссионных группах наблюдателей и нашли снимки нескольких таких пакетов из других муниципалитетов Московской области. Все обнаруженные нами 11-значные номера лент перечислены в таблице VII.

Рассматривая ленты с нормально набранными номерами, снятые в достаточно высоком разрешении для прочтения номера на ленте, мы видим, что их переменная часть находится в диапазоне от 06664 до 43625. У таких же лент с набранными вразрядку номерами — в диапазоне от 01030 до 06665. Это говорит о том, что обе партии сейф-пакетов были достаточно большими. Любопытно, что в фабричной спецификации, предоставленной нам производителем, явно прописан *набранный вразрядку* номер размера $70 \times 4$ мм на ленте. Однако 4 пакета, использованные вне Власихи, имеют нормально набранные номера размера около $40{\times}4$ мм на ленте. Производитель отказался предоставлять нам дополнительную документацию, но имеющейся информации достаточно, чтобы заключить, что им были поставлены две большие партии для избирательной комиссии Московской области, в составе которых находилось порядка $10^4$ пар пакетов с попарно идентичными номерами.

Таблица раскрывает любопытный аспект организации во Власихе. У злоумышленников имелись дубликаты пакетов с двумя диапазонами последовательных номеров: 8 пар пакетов 50**01029**2021–50**01036**2021 и 8 пар пакетов 50**06660**2021–50**06667**2021. Мы предполагаем, что каждой фальсифицирующей участковой избирательной комиссии было выдано по одной паре пакетов из каждого диапазона. Не каждый выданный пакет был использован, но мы предполагаем, что каждый участок (кроме честного 215) имел в наличии две пары.

### Приложение Е: Дистанционное электронное голосование

Для участков ДЭГ стенограмм нет. Федеральная система ДЭГ (разработанная Ростелекомом), применяемая в Московской области, не дает наблюдать отметки в бюллетенях [53, 54]. Публикуются только итоговые данные. Единственная дополнительная информация, доступная по электронному голосованию, — время выдачи и получения каждого бюллетеня [54], что бесполезно для нашего анализа. Но для полноты картины мы приводим график интенсивности голосования в зависимости от времени на рис. 32. Хотя при электронном голосовании тоже существует риск фальсификаций, его итоги на наших выборах выглядят реалистично и мы не подозреваем подмен электронных бюллетеней на наших участках ДЭГ. На каждом из них избиратели отдали предпочтение 2 кандидатам от КПРФ и 3 кандидатам от ЕР.

Суждения об электронном голосовании в России неоднозначны. Эта технология практически не регулируется в правовом смысле [55]. Фактически голосование проводят различные организации, отдельные от избирательных комиссий. Известно, что результаты ДЭГ в Москве (где используется система, отличная от остальной страны) полностью сфальсифицированы [53, 56, 57]. Фальсификации на региональном ДЭГ также возможны, однако на основании имеющихся данных утверждать можно только об административном принуждении к голосованию [53, 58]. Его следы достаточно отчетливо видны и во Власихе (рис. 32), где основная масса участников ДЭГ голосовала либо в начале, либо во время обеденного перерыва первого дня голосования, который был рабочим днем. Вид наших графиков интенсивности голосования в целом не отличается от ДЭГ на других недавних выборах в Московской области [54].

Поскольку участники электронного голосования легко поддаются административному принуждению, можно сделать вывод, что они более лояльны, чем избиратели в целом. Следовательно, административные кандидаты не могут добиться лучших результатов на обычных участках, чем на участках ДЭГ. Результаты по участкам (таблица I) не согласуются с этим выводом, что само по себе может служить косвенным доказательством фальсификаций.

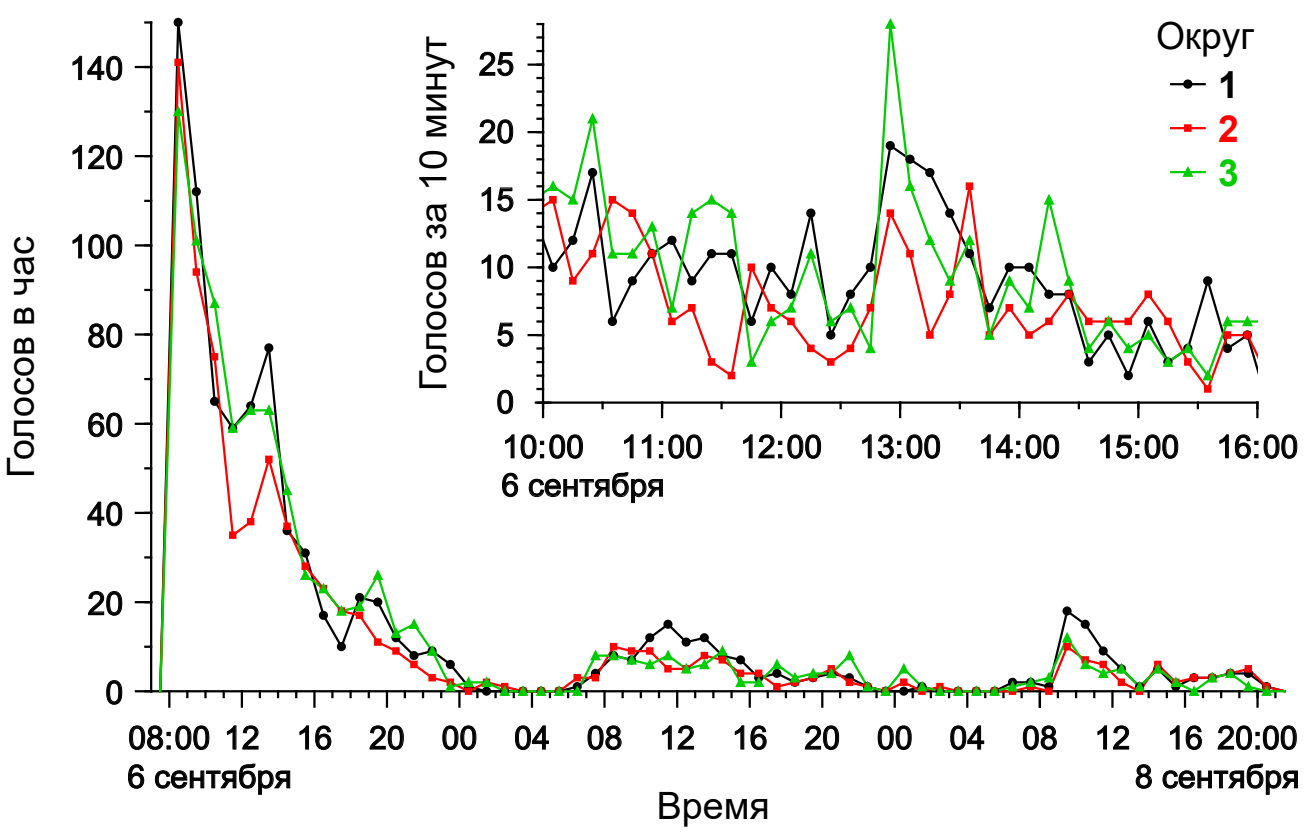

Рис. 32. Интенсивность голосования на участках ДЭГ. Онлайн-голосование началось в 08:00 6 сентября и продолжалось до 20:00 8 сентября. Два бюллетеня, поданных после этого времени, были выданы избирателям до 20:00. Часовые временны́е интервалы адекватно отображают все особенности гистограммы, за исключением узкого пика перед 13:00 в первый день в округе № 3, который отображен с 10-минутными интервалами на врезке. Большие пики могут быть объяснены принуждением избирателей к участию в голосовании

---